\documentclass[a4paper,11pt]{article} 

\usepackage[english]{babel}
\usepackage[a4paper]{geometry}
\usepackage{enumerate}  
\usepackage{amsmath,bbm}
\usepackage{graphicx}
\usepackage{color}
\usepackage{amsbsy}
\usepackage{amsthm}
\usepackage{amsfonts}
\usepackage{amssymb}
\usepackage{dsfont}
\usepackage{prettyref}   
\usepackage{float}
\usepackage{units}

\numberwithin{equation}{section}

\usepackage{hyperref}
\hypersetup{colorlinks=true, linkcolor=black, citecolor=blue}

\usepackage{tikz}
\usepackage{feynmp} 
\newcommand{\bBall}[1]{\fmfv{decor.shape=circle,decor.filled=empty, decor.size=5thick}{#1}}

\newcommand{\Square}[1]{\fmfv{decor.shape=diamond,decor.filled=empty, decor.size=4thick}{#1}}
\newcommand{\Tri}[1]{\fmfv{decor.shape=triangle,decor.filled=empty,  decor.size=4.5thick}{#1}}
\usetikzlibrary{fit}
\usetikzlibrary{shapes.geometric}
\usetikzlibrary{decorations.pathmorphing}
\usetikzlibrary{decorations.markings}
\usetikzlibrary{arrows}

\tikzset{
point/.style={circle,fill=black,inner sep=1pt},
vertex/.style={circle,fill=black,inner sep=1.5pt},   
bvertex/.style={circle,fill=black,inner sep=2.8pt},
Bvertex/.style={circle,fill=black,inner sep=4pt}, 
specialEP/.style={rectangle,fill=white,draw,inner sep=3pt},  
whitevex/.style={circle,fill=white,draw, inner sep=2pt},
linelabel/.style={sloped,above,very near start, inner sep=1pt,execute at begin node=$\scriptstyle,execute at end node=$},
baseline=(current  bounding  box.center),doubled/.style={double distance= 1pt,line width=1.5pt},
th/.style={line width=0.5 pt, gray},  
med/.style={line width=1 pt}  
}

\tikzset{arr/.style={  
        decoration={markings,
            mark= at position 0.65 with {\arrow[scale=1.3]{#1}} },
        postaction={decorate}
    }
}
\pgfdeclaredecoration{complete sines}{initial}
{
    \state{initial}[
        width=+0pt,
        next state=sine,
        persistent precomputation={\pgfmathsetmacro\matchinglength{
            \pgfdecoratedinputsegmentlength / int(\pgfdecoratedinputsegmentlength/\pgfdecorationsegmentlength)}
            \setlength{\pgfdecorationsegmentlength}{\matchinglength pt}
        }] {}
    \state{sine}[width=\pgfdecorationsegmentlength]{
        \pgfpathsine{\pgfpoint{0.25\pgfdecorationsegmentlength}{0.5\pgfdecorationsegmentamplitude}}
        \pgfpathcosine{\pgfpoint{0.25\pgfdecorationsegmentlength}{-0.5\pgfdecorationsegmentamplitude}}
        \pgfpathsine{\pgfpoint{0.25\pgfdecorationsegmentlength}{-0.5\pgfdecorationsegmentamplitude}}
        \pgfpathcosine{\pgfpoint{0.25\pgfdecorationsegmentlength}{0.5\pgfdecorationsegmentamplitude}}
}
    \state{final}{}
}
\tikzset{
gluon/.style={decorate, draw=black,
        decoration={complete sines,amplitude=4pt, segment length=6pt}}
     }

\let\a=\alpha \let\b=\beta    \let\g=\gamma     \let\d=\delta     \let\e=\varepsilon
       \let\th=\vartheta \let\k=\kappa     \let\l=\lambda
\let\m=\mu    \let\n=\nu      \let\x=\xi        \let\p=\pi        \let\r=\rho
\let\s=\sigma \let\t=\tau     \let\f=\phi    \let\ph=\varphi   
\let\ps=\psi   \let\o=\omega     
\let\G=\Gamma \let\D=\Delta       \let\L=\Lambda    
             
\let\O=\Omega

\def\PP{{\cal P}}\def\EE{{\cal E}}\def\VV{{\cal V}}
\def\FF{{\cal F}}\def\WW{{\cal W}}
\def\TT{{\cal T}}\def\NN{{\cal N}}\def\BB{{\cal B}}
\def\RR{{\cal R}}\def\LL{{\cal L}}
\def\DD{{\cal D}}\def\GG{{\cal G}}

\def\V#1{{\bf #1}}

\def\kk{{\bf k}}
\def\pp{{\bf p}}\def\qq{{\bf q}}\def\xx{{\bf x}}
\def\yy{{\bf y}}\def\nn{{\bf n}}

\def\zz{{\bf z}}

\def\bz{{\bf 0}}

\def\ff{{\mathfrak f}}  \def\fh{{\mathfrak h}}

       \def\oo{{\underline \omega}}
\def\ee{{\underline \varepsilon}}

\def\u#1{{\underline{#1}}}

 \def\vn{{\vec n}} \def\vx{{\vec x}}
\def\vy{{\vec y}} \def\vz{{\vec z}} \def\vk{{\vec k}}
\def\vq{{\vec q}} \def\vp{{\vec p}} \def\v0{{\vec 0}}

  \def\hW{{\hat W}}

  \def\hv{{\hat v}}

\def\bh{{\bar h}}

\def\tl#1{{\tilde{#1}}}

  \def\ZZZ{\mathbb{Z}} \def\RRR{\mathbb{R}}

\def\Val{{\rm Val}}
\def\indic{\hbox{\raise-2pt \hbox{\indbf 1}}}

\let\dpr=\partial

\let\==\equiv

\let\io=\infty
\let\0=\noindent

\def\pagina{{\vfill\eject}}
\def\*{{\hfill\break\null\hfill\break}}

\def\ie{\hbox{\it i.e.\ }}

\let\arr=\rightarrow

\def\lft{\left}
\def\rgt{\right}

\def\tende#1{\,\vtop{\ialign{##\crcr\rightarrowfill\crcr
             \noalign{\kern-1pt\nointerlineskip}
             \hskip3.pt${\scriptstyle #1}$\hskip3.pt\crcr}}\,}
\def\otto{\,{\kern-1.truept\leftarrow\kern-5.truept\to\kern-1.truept}\,}

\def\Tr{{\rm Tr}}

\def\sqt[#1]#2{\root #1\of {#2}}

\def\T#1{{#1_{\kern-3pt\lower7pt\hbox{$\widetilde{}$}}\kern3pt}}
\def\VVV#1{{\underline #1}_{\kern-3pt
\lower7pt\hbox{$\widetilde{}$}}\kern3pt\,}
\def\W#1{#1_{\kern-3pt\lower7.5pt\hbox{$\widetilde{}$}}\kern2pt\,}

\def\lis{\overline}

\def\indica{\leaders \hbox to 0.5cm{\hss.\hss}\hfill}
\def\guida{\leaders\hbox to 1em{\hss.\hss}\hfill}
\mathchardef\oo= "0521

\def\feyn#1#2#3{
\vskip 0.5cm
\begin{figure}[t] 
\centering{
{#1}
\caption{\small #2}  \label{#3}
}
\end{figure}
}

\def\tik#1#2#3{
\vskip 0.2cm
\begin{figure}[t] 
\centering{
{#1}
\caption{\small #2}  \label{#3}
}
\end{figure}
\vskip 0.2cm
}

\def\ins#1#2#3{\vbox to0pt{\kern-#2 \hbox{\kern#1 #3}\vss}\nointerlineskip}

\def\be{\begin{equation}}
\def\ee{\end{equation}}
\def\bea{\begin{eqnarray}}\def\eea{\end{eqnarray}}
\def\bean{\begin{eqnarray*}}\def\eean{\end{eqnarray*}}
\def\bfr{\begin{flushright}}\def\efr{\end{flushright}}
\def\bc{\begin{center}}\def\ec{\end{center}}
\def\bal{\begin{align}} \def\eal{\end{align}}

\def\[#1\]{\begin{align}#1\end{align}}

\def\ba#1{\begin{array}{#1}} \def\ea{\end{array}}
\def\bd{\begin{description}}\def\ed{\end{description}}
\def\bv{\begin{verbatim}}\def\ev{\end{verbatim}}
\def\non{\nonumber}
\def\Halmos{\hfill\vrule height10pt width4pt depth2pt \par\hbox to \hsize{}}

\usepackage{authblk}

{  

\renewcommand{\thesubsection}{\thesection.{\small\Alph{subsection}}}

\title{ Renormalization theory of a two dimensional Bose gas: \\
quantum critical point and quasi-condensed state}

\date{\small{(Dated: April 19, 2014)}}

\begin{document} 

\author[1]{S. Cenatiempo}
\author[2]{A. Giuliani}
\affil[1]{\it \small{Universit\"at Z\"urich, Winterthurerstr. 190, 8057 Z\"urich, Switzerland}}
\affil[2]{\it \small{Universit\`a degli Studi Roma Tre, L.go S. L. Murialdo 1, 00146 Roma, Italy}}

\maketitle

\begin{flushright}
\begin{minipage}{8cm}
\begin{minipage}{6cm} \it 
\small{To the memory of  Kenneth Wilson,\\ for his inspiring and influential ideas}
\end{minipage}
\begin{minipage}{2cm}
\end{minipage}
\end{minipage}
\end{flushright}

\vskip 0.5cm

\begin{center}
\begin{minipage}{12cm}
\begin{abstract}
We present a renormalization group construction of a weakly interacting Bose gas at zero temperature in the two-dimensional continuum, both in the quantum critical regime and in the presence of a condensate fraction. The construction is performed within a rigorous renormalization group scheme, borrowed from the methods of constructive field theory, which allows us to derive explicit bounds on all the orders of renormalized perturbation theory. Our scheme allows us to construct the theory of the quantum critical point completely, both in the ultraviolet and in the infrared regimes, thus extending previous heuristic approaches to this phase. For the condensate phase, we solve completely the ultraviolet problem and we investigate in detail the infrared region, up to length scales of the order $(\l^3\r_0)^{-1/2}$ (here $\l$ is the interaction strength and $\r_0$ the condensate density),  which is the largest length scale at which the problem is perturbative in nature.  We exhibit violations to the formal Ward Identities, due to the momentum cutoff used to regularize the theory, which suggest that previous proposals about the existence of a non-perturbative non-trivial fixed point for the infrared flow should be reconsidered.
\end{abstract}
\end{minipage}
\end{center}
\pagina
\tableofcontents

\pagina

\section{Introduction and main results}

The study of systems of non-relativistic bosons in three or less dimensions has been revived by the amazing experimental progresses 
of the last two decades. Starting from the mid nineties, it became possible to manipulate cold atoms in such a detailed and controlled 
way that Bose-Einstein Condensates (BEC) in optical traps and optical lattices were created, and the analogue of several quantum 
many body toy models were set up in actual experiments. It is becoming more and more possible to investigate crossover effects 
between different regimes and dimensionality. In this perspective, it is interesting to clarify and extend previous analyses about the 
low energy behavior of interacting Bose systems. 

From a theoretical point of view, the two-dimensional case is particularly challenging: two dimensions appears to be a critical case 
both for the stability of the so-called free bosons fixed point, and for the stability of Bogoliubov theory~\cite{Bogoliubov}, in the sense of the 
Renormalization Group (RG). This is {\it the} method for understanding from first principles the emergence of scaling laws in 
interacting many body systems at low or zero temperatures, as well as for computing thermodynamic and correlation functions.
Applications to the Bose gas date back to the works of Beliaev (1958,~\cite{Beliaev}) and Hugenholtz--Pines (1959,~\cite{Pines}), whose ideas were further developed in  Lee and Yang~\cite{LeeYang5}, Gavoret and Nozi\'eres~\cite{Gavoret}, 
Nepomnyashchii and Nepomnyashchii~\cite{Nepom} and  Popov and Seredniakov \cite{Popov}. 

All these works investigate the effects of corrections to Bogoliubov's theory in the condensed phase, starting from the analysis of infrared divergences 
in perturbation theory (PT). They are based on diagrammatic techniques borrowed from Quantum Field Theory and on the summations 
of {\it special classes of diagrams} selected from the divergent PT, possibly by combining them with the use of Ward identities~\cite{Gavoret}. In this sense, they do not provide a systematic study of the problem, and the resulting indications of the stability of the Bogoliubov's spectrum under perturbations are questionable. 

The first systematic  study of the whole PT in the presence of a non-zero condensate fraction in 3D, and the proof of its order by order 
convergence after proper resummations, is due to 
G. Benfatto in 1994~\cite{benfatto}. The method employed in his work is the Wilsonian RG, combined with the ideas of constructive 
RG, in the form developed by the roman school of Benfatto, Gallavotti et al since the late seventies~\cite{GalReview} and already applied to a certain number of infrared problems in condensed matter systems~\cite{BGbook,BG, BGPS-Fermi1d,BM-luttinger, BGM-Hubbard, SRGraph1,SRGraph2,EMGraph1,EMGraph2, Mastropietro}. It is worth stressing that even though the methods used in \cite{benfatto} are based on the ideas of constructive field theory, the resulting bounds derived there are not enough for 
constructing the theory in a mathematically complete form: they are enough for deriving finite bounds at all orders in renormalized 
perturbation theory, growing like $n!$ at the $n$-th order (``$n!$ bounds"), but the possible Borel summability of the series remains an outstanding open problem. A program addressing this issue has been started by T. Balaban and collaborators 
(see \cite{Balabanetal} and references therein), but the solution is still far from being reached: the idea is to apply the Wilsonian RG in the form developed by 
Balaban in a series of works dedicated to the construction of theories with a broken continuous symmetry \cite{Balaban85,Balaban87, BalabanNvec}, but the new technical 
problems arising in the case of a complex gaussian measure (as the one appearing in the Bose gas) seem to be serious obstacles to 
the solution. 

The method and the ideas of Benfatto were later extended by Pistolesi et al~\cite{CaDiC1,CaDiC2}, who implemented 
a systematic RG analysis of the Bose gas in $1<d\le 3$ dimensions, by using dimensional regularization, and by implementing local Ward identities 
at all orders of PT. The systematic use of local Ward identities allowed them to recover in a conceptually simpler and more satisfactory way some of the cancellations determined by Benfatto by inspection of PT. Moreover, they managed to extend the analysis down to $d=2$, modulo an assumption about the convergence of the flow of the particle-particle effective interaction: this was assumed to 
flow towards an $O(1)$ infrared fixed point, as  suggested by a natural truncation of PT at the one loop level. A similar conclusion for the infrared behavior of the zero temperature 
Bose gas was later recovered by Wetterich~\cite{Wetterich}, Dupuis~\cite{DupuisSengupta, DupuisBose_extended, DupuisBose_short},  and Sinner et al~\cite{Sinner2010}, via the  methods of the functional RG a'la Wetterich, which involves a different regularization and truncation scheme, combined with a numerical study of the resulting flow equations.

The main conclusion of \cite{CaDiC1,CaDiC2} (later confirmed by \cite{Wetterich, DupuisSengupta, DupuisBose_extended, DupuisBose_short,  Sinner2010}), 
which we will criticize below,  is that the Bose-Einstein condensate is stable at 
zero temperature in two dimensions, in the presence of weak repulsive interactions (and/or at low densities),
in agreement with Bogoliubov's theory.
This fact is a priori not obvious, since two dimensions is critical for the existence of BEC; e.g., there is no condensation at any finite temperature \cite{Hohenberg}. On top of that, \cite{CaDiC1,CaDiC2} find that the Bogoliubov's spectrum is stable (i.e., 
the speed of sound is finite), notwithstanding the fact that the flow of the effective constants is anomalous, in the sense that it 
is different from what suggested by naive power counting. In this respect, Ward identities play a crucial role in driving 
the flow of the effective constants towards the ``right direction". 

The zero-temperature condensed phase in two dimensions is not the only possible critical phase for non-relativistic bosons: in fact,
the condensed phase emerges in the vicinity of a quantum (i.e., zero temperature) critical point controlled by the chemical potential $\m$, see \cite{Sachdev}.
The critical point separates a Mott insulator, ``quantum disordered", phase \cite{FisherMott, Sachdev, SachdevSenthilShankar} from the BEC; the
critical theory at $\m=0$ has interesting scaling properties, with logarithmic corrections in $d=2$, which have been investigated by RG methods in \cite{FisherHohenberg, SachdevSenthilShankar}. By the same methods, Fisher and Hohenberg \cite{FisherHohenberg} 
also computed the Kosterlitz-Thouless line 
$T=T(\mu)$ separating a quasi-condensed phase (in the sense of algebraic decay of correlations) from a fully disordered one. More recently, Dupuis and Ran\c{c}on~\cite{DupuisLowT, DupuisBoseHubb_short, DupuisBoseHubb_extended} further extended the results of \cite{FisherHohenberg} by investigating 
the critical behavior at the Kosterlitz-Thouless line by functional RG methods.

In the following we shall review and extend the Wilsonian RG approach to the Bose gas in the two-dimensional continuum, in the formalism proposed by Benfatto and 
Gallavotti. We shall prove renormalizability both of the critical theory and of the condensed phase, both in the ultraviolet and 
in the infrared, and we shall 
develop a theory valid at all orders in renormalized perturbation theory, with explicit 
bounds on the generic order. 

We shall first focus on the quantum critical point, which is very simple from the perspective of power counting:
it defines a theory that is super-renormalizable in the ultraviolet and asymptotically free in the infrared; it describes the vacuum 
fluctuations of the theory, and it is in some sense the non-relativistic analogue of the dynamical vacuum of QED. Our results match with 
those of \cite{FisherHohenberg, SachdevSenthilShankar}, and extend them to all orders, in a framework that is the most promising 
for subsequent constructive approaches\footnote{It would be very interesting to see whether the ideas of Balaban et al \cite{Balabanetal} could be applied to the actual construction of this theory, which is much easier than the condensed one, and still very relevant for the physics of the Mott transition.}.

Next, we shall move to the much more interesting condensed phase. The ultraviolet is super-renormalizable, in a similar sense as the critical theory. Things start to change at the energy scale where the kinetic energy becomes comparable with $\l\rho_0$,
where $\rho_0$ is the condensate density, and $\l $ the a-dimensional interaction strength: for smaller scales the theory becomes just 
renormalizable, with 8 running coupling constants, versus one free parameter (the chemical potential, which has to be fixed so to realize the right condensate density). Compared with previous works,
we have an extra marginal coupling constant, associated with three-body interactions: 
this was previously overlooked, and its role is analyzed in our work for the first time.

Due to the presence of two relevant couplings, the flow of the effective constants is rapidly driven towards values of order 1, at which point the system leaves the perturbative regime and we stop the flow.
The critical energy scale at which 
we interrupt the flow is of the order $\l^2\r_0$ (corresponding to a length scale $(\l^3\r_0)^{-1/2}$). For higher energies
the theory is well defined at all orders and there we provide explicit bounds on the $n$-th order coefficients 
in the renormalized coupling for all the thermodynamic observables. For lower energies we cannot conclude 
anything from the  RG analysis.
The fact that the theory is well defined only in a limited range of  length scales should be interpreted as an instance of the stability of the condensate order parameter
up to length scales of the order $(\l^3\r_0)^{-1/2}$. Since we have no control on the behavior of the system on larger length scales,  we cannot exclude the possibility that the condensate coherence length is finite in the ground state: 
this is the reason of the name ``quasi-condensed state" in the title. However, we do not claim to have a proof of the absence of long-range order at zero temperature.

While we cannot investigate the region of energies smaller than $\l^2\r_0$, we have a complete control 
of the theory at higher energy scales, including the effects of cutoffs and of the irrelevant terms. 
In particular, we can verify there the validity of the Ward Identities used extensively in 
\cite{CaDiC1,CaDiC2}, as well as in \cite{Wetterich, DupuisSengupta, DupuisBose_extended, DupuisBose_short,  Sinner2010}, for extrapolating  the flow to lower energy scales
and for understanding the nature of the (putative) infrared fixed point. Quite remarkably, we 
prove that for energy scales intermediate between $\l\r_0$ and $\l^2\r_0$
some of the Ward Identities derived in \cite{CaDiC1,CaDiC2} within a dimensional regularization scheme
are explicitly violated at lowest order in the presence of a momentum regularization. We also 
identify the source of the violation in a correction term due to the cutoff function used for regularizing the 
theory. We stress that the violation appears at the one-loop level, in a region where perturbation theory is 
unambiguously valid, i.e., before we enter the non-perturbative regime. 
It may be related to the emergence of an anomaly term in the response functions. 
Therefore, the use of the Ward Identities in the deep infrared region by the authors of 
\cite{CaDiC1,CaDiC2} is questionable. We believe that our finding motivates a  reconsideration 
of the nature and the very existence of a 2D condensate at zero temperature. It would be very interesting 
to investigate in an unbiased way the deep infrared region, i.e., the one at energy scales smaller than $\l^2\r_0$,
via systematic non-perturbative simulations, possibly based on 
Borel resummations of the divergent series, which are missing so far. 
It would also be interesting to 
see whether it is possible to implement within our rigorous RG analysis the use of a different 
momentum regularization scheme that does not violate local gauge invariance at any finite scale, and to see whether in such a case the corrections to the local Ward Identities would vanish exactly or not\footnote{A possibility could be to suitably modify the infrared regulator used in \cite{DupuisBose_extended,DupuisPRE}, so to make it locally gauge invariant at any finite scale.}. We hope to come back to this issue in a future publication.

The rest of the paper is organized as follows: in Section \ref{sec1.A} we define the model and explain 
more precisely what we mean by quantum critical point or condensed state. 
In Section \ref{s2.QCP} we present the renormalization theory for the quantum critical point: first we reformulate the
model in the form of a functional integral; then we describe the multiscale integration scheme for its computation, 
including the Gallavotti-Nicol\`o tree expansion and the proof of the basic $n!$ bounds; next, we modify the expansion 
by properly renormalizing and resumming the potentially divergent contributions; finally we study the flow of the effective constants. In Section \ref{sec3} we present the renormalization theory of the condensed phase: 
as a starting point we manipulate the functional integral, so to put it into the form of a perturbed Bogoliubov's theory;
then we describe the multiscale integration of the resulting theory, first in the ultraviolet region, and then in the infrared;
finally we study the flow equations in the infrared region, we discuss their compatibility with the Ward Identities 
and compare in a more technical way our findings with those of \cite{CaDiC1,CaDiC2}. Finally, in Section \ref{sec4}, we draw the conclusions. The Appendices collect a number of (very important!) technical aspects of the construction, among which:  the one-loop beta function, the derivation of the global and local Ward Identities (including the 
correction terms due to the momentum cutoff), the verification of the Ward Identities at lowest order 
(including the proof of violation of the formal Ward Identities at lowest order).

\subsection{The model}\label{sec1.A}

We are interested in the study of the properties of a gas of non-relativistic bosons in a two-dimensional box $\O_L$ of side $L$ with periodic boundary conditions (to be eventually sent to infinity), interacting via a weak repulsive short range two-body potential. The corresponding grandcanonical Hamiltonian at chemical potential $\m$ in second quantized form is 
\be
 \label{Ha}
H_{L}= \sum_{\vk\in\mathcal D_L}  \,(\vk^{2}-\mu) \hat a^{+}_{\vk} \hat a_{\vk} \ 
+ \frac{\l}{2} \frac1{L^2}   
\sum_{\vk,\vk'\!,\, \vp\in\mathcal D_L}  \hat{v} (\vp) \ \hat a^{+}_{\vk +\vp } \hat a^{+}_{\vk'-\vp} \hat a_{\vk}
\hat a_{\vk'}\equiv H^0_L+V_L\;,
\ee
where: $\mathcal D_L=\{\vec k= 2\pi \,\vec n/L,\ \vn \in \ZZZ^2\}$; $\hat a^{+}_{\vk}$ and  $\hat a_{\vk}$ are creation and annihilation operators, satisfying the canonical commutation rules; $\hv(\vk)=\int d\vx e^{i \vk \cdot \vx} v(\vx)$ is the Fourier transform of the two-body potential $v(\vx)$, which is assumed to be positive and positive definite (for simplicity); $0<\l \ll 1$ is the strength of the interaction.
The Hamiltonian in~\eqref{Ha} acts on the Fock space $\mathcal F_L$ obtained as the direct sum of the $N$-particles Hilbert spaces of square-summable symmetric wave functions on the box $\O_L$ of side $L$.
Our goal is to construct the thermal ground state in infinite volume, defined as the limit $\b\to\infty$ of the infinite volume grand canonical Gibbs state at inverse temperature $\b$ (i.e., we want to send $L\to\infty$ first, and then $\b\to\infty$). If needed, we allow $\m$ to depend upon $L$ and $\b$, $\m=\m_{\b,L}$: the requirement is that the resulting ground state 
has total density $\r$ or total condensate density $\rho_0$; we shall be more specific in the following. 
Note that the units are chosen in such a way that $\hbar=2m=1$\footnote{With this choice, the dimensions of the physical quantities speed ($c$), momentum ($\vk$), frequency ($k_0$) and energy ($E$) are respectively: $ [c] = [|\vk|] = [L]^{-1}$ and $[E] =[k_0]= [L]^{-2}$.}.  

A basic thermodynamic quantity we are interested in is the specific ground state energy 
\be e_0(\r)=-\lim_{\b\to\infty}\lim_{L\arr\infty}\,\frac{1}{\beta L^2}\log\Tr_{\mathcal F_L}e^{-\beta(H_{L}-\m_{\b,L} N)}\;,\ee
with 
\be \m=\lim_{\b\to\infty}\lim_{L\to\infty}\m_{\b,L}=\dpr_\r e_0(\r)\;,\ee
modulo exchange of limits issues, to be discussed more carefully below. 

We are also interested in the $2n$-points correlation functions in imaginary time, also known as Schwinger functions:
given the ``imaginary time" $x_0\in[0,\b)$, and denoting by $\xx=(x_0,\vec x)\in[0,\b)\times \O_L$ the space-time coordinates, we let
\be \label{eq:schw_funct}
 S_{\s_1,\ldots,\s_{2n}}(\xx_1,\ldots,\xx_{2n})=\lim_{\b\to\infty}\lim_{L\arr\io} \frac{  \Tr_{\mathcal F_L} \left[e^{-\b (H_{L}-\m_{\b,L} N)} 
{\rm T} \lbrace a^{\s_1}_{\xx_1}\ldots a_{\xx_{2n}}^{\s_{2n}} \rbrace \right] }{ \Tr_{\mathcal F_L} e^{-\b  (H_L-\m_{\b,L} N)}}\;,
\ee
where, if 
\be \label{a}
a^\pm_\vx= \frac{1}{L} \sum_{\vk \in \DD_L} \hat a^\pm_\vk e^{\pm i \vk \cdot \vx}\;,
\ee
then the time-evolved operator $a^\pm_{\xx}$ is
\be a^\pm_{\xx}= e^{(H-\m_{\b,L} N)x_0}\, a^\pm_\vx \,e^{-(H-\m_{\b,L} N)x_0}\;.\ee
Moreover, the operator ${\rm T}$ appearing in the r.h.s. of (\ref{eq:schw_funct}) is the time-ordering operator, which 
re-arranges times in decreasing chronological order:
\be {\rm T} \lbrace a^{\s_1}_{\xx_1}\ldots a_{\xx_{2n}}^{\s_{2n}}\}=  a^{\s_{\p(1)}}_{\xx_{\p(1)}}\ldots a_{\xx_{\p(2n)}}^{\s_{\p(2n)}}\;,\ee
where $\p$ is a permutation of $\{1,\ldots,n\}$ such that $x_{\p(1),0}>\cdots>x_{\p(n),0}$\footnote{If some of the time coordinates are 
equal to each other, each group of operators with the same times is reordered in such a way that the creation operators are all on the 
left of the annihilation operators, within that group.}. In particular, the two-points Schwinger function is often called 
the propagator, or {\it interacting propagator}, to clearly distinguish it from the unperturbed one:
\be S(\xx-\yy):=S_{-+}(\xx,\yy)\;.\ee

If $\l=0$ and $\m_{\b,L}\to 0^-$ as $\b,L\to\infty$, then the ground state has propagator
\be S^0(\xx-\yy)=\r_0+\int_{\mathbb  R^3}\frac{d\kk}{(2\p)^3}\frac{e^{-i\kk(\xx-\yy)}}{-ik_0+|\vk|^2}\;,\label{1.prop}\ee
where
\be\label{eq:condensate}
\r_{0}= \lim_{\b\to\infty}\lim_{L\to\infty} \frac{1}{L^2} \frac{1}{e^{-\b  \m_{\b,L}}-1}
\ee
represents the average occupation number of the $\vec k=\vec 0$ state per unit volume (condensate density). 
Note that the propagator tends {\it polynomially} to $\r_0$, as $|\xx|\to\infty$. In this sense the theory is critical, both if $\r_0=0$ and 
$\r_0>0$. Note also that in this non-interacting case, the higher points correlations can all be reduced to the propagator, via the Wick rule, which is valid simply because the 
Hamiltonian is quadratic in the creation/annihilation operators. 

If $\l>0$, the problem is not exactly solvable anymore. For small $\l$, one can expand the thermodynamic and correlation 
functions in formal power series in $\l$, and then try to give sense to it, at least order by order. The problem is not simple, since the 
perturbation theory is affected by divergences, both in the ultraviolet and infrared side. As we shall see, the non-trivial divergences are 
the infrared ones, which need to be treated differently, depending on whether $\r_0=0$ or $\r_0>0$. 

The first case we shall treat is $\r_0=0$, which is the {\it quantum critical point}, in the sense of Fisher et al.~\cite{FisherMott} and Sachdev \cite{Sachdev}.
In this context, we expand all thermodynamic and correlation functions by decomposing $H_L$ as in the r.h.s. of (\ref{Ha}),
$H_L=H^0_L+V_L$. The main result is that the formal perturbation theory in $V_L$ with propagator (\ref{1.prop}) at $\r_0=0$ is renormalizable
at all orders. More than that: the theory is  super-renormalizable in the ultraviolet and 
asymptotically free in the infrared, with the effective two-body interaction strength flowing to zero logarithmically as the infrared cutoff 
is removed.  As a consequence, the dressed propagator has the same decay properties as the free one, modulo logarithmic 
corrections. Our method allows us to produce explicit estimates on the generic order of renormalized perturbation theory, growing like $n!$ at the $n$-th order. 

The second case is the {\it condensate state}, i.e., the analysis of the formal perturbation theory with propagator (\ref{1.prop}) at $\r_0>0$. The reference quadratic theory, which the interacting system is supposedly a perturbation of, is the Bogoliubov's Hamiltonian, which is obtained from (\ref{Ha}) by replacing $\hat a_{\v0}$ by $\sqrt{\r_0}\,L$, and by keeping all the terms that are at most quadratic in the operators $\hat a^\pm_{\vk}$ with $\vk\neq\v0$:
\be
 \label{HaB}
H_{B,L}=C_{0,L}L^2+ \sum_{\vk\in\mathcal D_L\setminus\{\v0\}} \Big[f({\vk})\hat a^{+}_{\vk} \hat a_{\vk}
+\frac12 g(\vk)(\hat a^{+}_{\vk} \hat a^+_{-\vk}+\hat a_{\vk} \hat a_{-\vk})\Big]\;,\ee
with
\bea
&& f(\vk)=\vk^2-\m_{\b,L}+\l\r_0\hat v(\v0)\big(1-\frac{1}{2\r_0 L^2}\big)+\l\r_0\hat v(\vk)\;,\quad g(\vk)=\l\r_0\hat v(\vk)\;, \label{f1.12}\\
 && C_{0,L} =-\m_{\b,L}\r_0+\frac{\l}{2}\hat v(\v0)\r_0^2\big(1-\frac{1}{\r_0 L^2}\big)\;.\eea
Correspondingly, $H_L$ can be rewritten identically as $H_L=C_{0,L}L^2+\tilde H_{B,L}+W_L$, with 
\be
 \label{HaB1}
\tilde H_{B,L}=\sum_{\vk\in\mathcal D_L\setminus\{\v0\}} \Big[F({\vk})\hat a^{+}_{\vk} \hat a_{\vk}
+\frac12 g(\vk)(\hat a^{+}_{\vk} \hat a^+_{-\vk}+\hat a_{\vk} \hat a_{-\vk})\Big]\;,
\qquad F(\vk)=\vk^2+\l\r_0\hat v(\vk)\;.\ee
Note that, up to the additive constant, $\tilde H_{B,L}$ differs from $H_{B,L}$ by a term $-\m^0_{\b,L} 
\sum_{\vec k\neq\v0}\hat a_\vk^+ \hat a_\vk$,
with $\m^0_{\b,L}=\m_{\b,L}-\l\r_0\hat v(\v0)\big(1-\frac{1}{2\r_0 L^2}\big)$, which is thought as part of the perturbation, and acts as a counterterm, 
to be fixed in such a way that the condensate density is $\r_0$. 
Note also that the perturbation $W_L$ is formally of smaller order than $\tilde H_{B,L}$. The ``naive" perturbation theory of $H_L=H^0_L+V_L$ in $V_L$
with propagator (\ref{1.prop}) at $\r_0>0$ is formally equivalent to the perturbation theory of 
$H_L=C_{0,L}L^2+\tilde H_{B,L}+W_L$ in $W_L$, with propagator of the form (in the $\b,L\to\infty$ limit):
\bea S^B(\xx-\yy)&=& \begin{pmatrix}S^B_{-+}(\xx-\yy)& S^B_{--}(\xx-\yy)\\ S^B_{++}(\xx-\yy)& S^B_{+-}(\xx-\yy)\end{pmatrix}=\\
&=&
\rho_0\mathds{1}+
\int_{\mathbb R^3} \frac{d\kk}{(2\p)^3}\frac{e^{-i\kk(\xx-\yy)}}{k_0^2+F^2(\vk)-g^2(\vk)}\begin{pmatrix}ik_0 +F(\vk)& -g(\vk)\\
-g(\vk)& -ik_0+F(\vk)\end{pmatrix}\;, \non
\eea
The expansion with respect to this propagator is expected to have better convergence properties than the naive one, at least if we 
trust Bogoliubov's theory. The very convergence of this modified perturbation theory, if valid, should be 
interpreted a posteriori as
a confirmation of Bogoliubov's picture at all orders. 

For this case, our main result is that the modified perturbation theory around Bogoliubov's Hamiltonian is renormalizable
at all orders. More precisely, it is  super-renormalizable in the ultraviolet, as for the $\rho_0=0$ case, and just renormalizable in the 
infrared, with 8 running coupling constants and one free parameter (the chemical potential). After having fixed the chemical potential, 
we are left with 7 effective parameters, whose flows need to be controlled on the basis of the study of the beta function
and of the use of Ward Identities. The presence of two marginal couplings rapidly drives the flow of the effective parameters out of the perturbative regime, from which point on we cannot conclude anything in a rigorous fashion. 
Still, at a heuristic level, one can try to guess what the system does in the deep infrared, non-perturbative, regime. 
A one-loop truncation to the beta function suggests that 
the (relevant) effective particle-particle interaction tends to an $O(1)$ non-trivial fixed point.
If this is assumed to be the case, one can then use the Ward Identities, which induce strong constraints 
on the flow of all  the other effective parameters, to understand their behavior in the deep infrared. This is 
what is done in \cite{CaDiC1,CaDiC2}, whose  conclusion is that the infrared fixed point is consistent at all orders.
The fixed point has a behavior a'la Bogoliubov with a linear spectrum of excitations, notwithstanding the presence of certain anomalous terms. While, of course, we cannot rigorously prove or disprove this picture, we can (and we shall do so in the following) investigate the validity of the Ward Identities used in 
\cite{CaDiC1,CaDiC2} in the perturbative regime, i.e., before entering the deep infrared region. In that range of scales, 
the Ward Identities have an unambiguous meaning, and we can control all the terms involved in the identities 
at all orders, with explicit bounds on the coefficients at generic order. Our finding is that some Ward Identities, 
which play a crucial role in the analysis of \cite{CaDiC1,CaDiC2} are violated already at the one-loop level, due to the presence of momentum cutoffs, which were neglected in previous analyses (the authors of \cite{CaDiC1,CaDiC2}
use dimensional regularization and extrapolation from three dimensions within a $3-\e$ expansion).
Therefore, we cannot confirm the consistency of the existence of a non-trivial fixed point in the deep infrared, and
we think that the validity of a (dressed) Bogoliubov's picture in 2D at zero temperature should be  reconsidered. We postpone further comments to Sections \ref{sec3.C.4}- \ref{sec3.C.5} below.

\section{Renormalization Group theory of the Quantum Critical Point} \label{s2.QCP}

\subsection{The functional integral representation}\label{s2Aa}

We start by analyzing the interacting theory at $\rho_0=0$ (quantum critical point). We focus on the perturbation theory for the partition 
function and the free energy. As well known, the perturbative expansion in $V_L$ with respect to the propagator (\ref{1.prop}) at 
$\rho_0=0$ can be expressed in the form of a {\it coherent state path integral representation}~\cite{NO}:
defining $\L = [0, \b) \times \O_L$, with $\O_L$ the square box of side $L$, if $Z_\L=\Tr\{e^{-\b H_L}\}$ is the interacting partition function, and $Z^0_\L=\Tr\{e^{-\b H_L^0}\}$ the non-interacting one,
\[
\frac{Z_\L}{Z^0_\L} = \int P^0_\L(d\ph)\,e^{-V_\L(\ph)}\;.\label{2.1}
\]
Here: 
\begin{enumerate}
\item 
$\ph_{\xx}^{+}=(\ph_{\xx}^{-})^{*}$ is a complex classical field, labelled by the space-time point $\xx=(x_0,\vx)\in\L$.
\item 
$P^0_\L(d\ph)$ is a complex Gaussian measure with covariance 
\[S^0_\L(\xx-\yy)=\frac1{|\L|}\sum_{\kk\in\mathcal D_\L}
\frac{e^{-i\kk(\xx-\yy)}}{-ik_0+|\vk|^2-\m^0_{\b,L}}\;,\label{2.2}
\]
where $\DD_\L=2\p\b^{-1}\ZZZ\times 2\p L^{-1}\mathbb Z^2$ and $\mu^0_{\b,L}<0$ goes to zero as 
$\b,L\to\infty$  in such a way that \eqref{eq:condensate} is zero, e.g., $\mu^0_{\b,L}=-\k_0\b^{-1}$ and 
$\k_0>0$. 
If desired, the sum over $k_0$ can be performed exactly, and leads to 
\[ \label{2.3}
S^0_\L(\xx-\yy) = \frac1{L^2}\sum_{\vk\in\DD_L} e^{-i \vk \cdot (\vx-\vy)-(x_0-y_0)\e_0(\vk)} \biggl[\frac{ \th(x_0-y_0)}{1-e^{-\b \e_0(\vk)}}+
\frac{1-\th(x_0-y_0)}{e^{\b\e_0(\vk)}-1}\biggr]\;,\]
where $\e_0(\vk)=\vk^2-\m^0_{\b,L}$ and $\th(t)$ is the step function, equal to $1$ for $t>0$ and equal to $0$ for $t\leq 0$.  
The function $S^0_\L(\xx-\yy)$ is defined a priori only for $|x_0-y_0|<\b$; if $-\b<x_0-y_0<0$, it satisfies 
$S^0_\L(x_0-y_0,\vx-\vy)=S^0_\L(x_0-y_0+\b,\vx-\vy)$. Therefore, it can be naturally extended to the whole real axis by periodicity,
and we shall indicate the resulting $\b$-periodic function by the same symbol. 
Note that the limit $\b,L\to\infty$ of $S^0_\L(\xx-\yy)$
gives \eqref{1.prop} with $\rho_0=0$. 
\item The interaction potential $V_\L(\ph)$ is 
\[ \label{2.V_QCP}
V_\L(\ph) =&\, \frac{\l}{2} \int_{\L^2} d\xx d\yy \, |\ph_{\xx}|^2 \,w(\xx-\yy)\, |\ph_{\yy}|^2   
 -\n_\L \int_\L d\xx\,|\ph_{\xx}|^2 
\]
where $w(\xx-\yy)=\d(x_0 -y_0)v(\vx -\vy)$, and $\n_\L=\m_{\b,L}-\m^0_{\b,L}$ acts as a  {\it counterterm}, to be fixed 
in such a way that the interacting propagator decays polynomially to zero at large distances. 
\end{enumerate}

We want to evaluate \eqref{2.1} via a Wilsonian RG analysis, in the form presented in~\cite{GalReview,BGbook,benfatto}. 
The idea is to first introduce an ultraviolet cutoff, in order to make the number of degrees of freedom finite, and then integrate 
them slice by slice in momentum space. In the following we will try to describe the RG construction in a way as self-consistent as possible, and we refer the reader to the review papers \cite{GalReview,BGbook,GM} for further details on the methodology. 

The ultraviolet cutoff is defined by the replacement of $S^0_\L(\xx-\yy)$ with a regularized version $S^{0}_{\L,N}(\xx-\yy)$ such that 
$\lim_{N\to\infty}S^{0}_{\L,N}(\xx-\yy)=S^0_\L(\xx-\yy)$ and defined as follows. Let $\g>1$ be a scaling parameter (fixed once and for all, 
e.g. $\g=2$). Let $\chi(t)$ be a smooth characteristic function on $\RRR$, such that\footnote{For some of the bounds in the following, it is useful to assume that $\chi$ is a function of 
$t^2$ only, i.e., $\chi(t)=\mathcal K(t^2)$ for some smooth $\mathcal K$, and we shall do so from now on.}
$\chi(t)=1$ for $|t|\leq 1$, and $\chi(t)=0$ for $|t|\geq \g$.
Then, if $\xx=(x_0,\vx)$ with $-\b<x_0<\b$ and $\vx\in\O_L$ 
\[&S^{0}_{\L,N}(\xx)=\frac1{L^2}\sum_{\vk\in\DD_L} e^{-i \vk \cdot (\vx-\vy)-x_0\e_0(\vk)} \Big[1-\chi\left(\g^{N+1}d(\xx)\right)\Big]\biggl[\frac{ \th(x_0)}{1-e^{-\b \e_0(\vk)}}+
\frac{1-\th(x_0)}{e^{\b\e_0(\vk)}-1}\biggr]\label{2.5}\]
where $d(\xx)=\sqrt{||x_0||^2_\b+||\vx||_L^4}$, and $||\cdot||_\b$ and $||\cdot||_L$  are the norms on the tori $\RRR/\b\ZZZ$ and $\RRR^2/L\ZZZ^2$, respectively.
The regularized propagator has by construction no singularity at small space-time distances. Note also that for any $\b$ finite and 
$\m^0_{\b,L}=-\k_0\b^{-1}$ the theory is automatically regular in the infrared, simply because the propagator decays exponentially 
on spatial scales larger than $O(\b^{1/2})$. 
In order to construct the thermal ground state, we will show that the regularized theory is well defined, uniformly in $N$ and $\b$,
as $N,\b\to\infty$. This is discussed in the next section.

\subsection{Multiscale decomposition, tree expansion and non-renormalized power counting}  \label{s2.multiscale}

We consider the regularized partition function:
\[
\Xi_{\L,N}:= \int P^0_{\L,N}(d\ph)\,e^{-V_\L(\ph)}\;.\label{2.6}
\]
where $P^0_{\L,N}(d\ph)$ is the gaussian integration with propagator $S^{0}_{\L,N}(\xx-\yy)$. Our purpose is to
compute $|\L|^{-1}\log \Xi_{\L,N}$ in an iterative fashion, and derive uniform bounds on the resulting expansion. 
Roughly speaking, we iteratively integrate the degrees of freedom supported on momenta of definite scale: $
|k_0|+|\vk|^2\sim \g^{h}$, starting from momenta of the same order as the ultraviolet cutoff, $h=N$, and then moving 
towards smaller and smaller scales. After having integrated the scales $N,N-1,\ldots, h+1$, the  functional integral \eqref{2.6} is 
rewritten as an integral involving only the momenta smaller than $\g^h$, and  the interaction is replaced by an ``effective'' one. This 
has the same qualitative structure as  its ``bare'' counterpart, up to a redefinition of 
the interaction strength, and modulo ``error terms", called irrelevant in the RG jargon.

The iterative 
integration is based on the following multiscale decomposition of the propagator. We first rewrite the cutoff function in \eqref{2.5} as
\[1-\chi\left(\g^{N+1}d(\xx)\right)=\sum_{h=h_\b+1}^N u_h(\xx)+[1-\chi(\g^{h_\b+1}d(\xx))]\;,\]
where  $h_\b:=\lfloor\log_\g(\k_0\b^{-1})\rfloor$ and $u_h(\xx)=\chi(\g^hd(\xx))-\chi(\g^{h+1}d(\xx))$. This induces the rewriting:
\[
S^0_{\L,N}(\xx) = \sum_{h=h_\b+1}^{N} G_{\L,h}(\xx)+R_{\L,h_\b}(\xx)\label{2.8}
\]
with
\[
 G_{\L,h}(\xx) =  \frac1{L^2}\sum_{\vk\in\DD_L} e^{-i \vk \cdot (\vx-\vy)-x_0\e_0(\vk)} u_h(\xx)
 \biggl[\frac{ \th(x_0)}{1-e^{-\b \e_0(\vk)}}+
\frac{1-\th(x_0)}{e^{\b\e_0(\vk)}-1}\biggr]\;.\label{2.9}\]
and $R_{\L,h_\b}(\xx)$ defined by a similar expression, with $u_h(\xx)$ replaced by $1-\chi(\g^{h_\b+1}d(\xx))$.
The decomposition is defined in such a way that the single-scale propagator $G_{\L,h}$ decays exponentially on
scale $\g^{-h}$ (see also \eqref{2.Gh} below), and similarly for $R_{\L,h_\b}(\xx)$. Correspondingly we rewrite the field $\ph$ as
a sum of {\it independent} fields $\ph=\sum_{h=h_\b}^N\ph^{(h)}$, with $\ph^{(h)}$, $h> h_\b$, 
a gaussian field with propagator $G_{\L,h}$, and $\ph^{(h_\b)}$
a gaussian field with propagator $R_{\L,h_\b}$,
so that 
\[
\Xi_{\L,N}:= \int \prod_{h=h_\b}^NP_{\L,h}(d\ph^{(h)})\,e^{-V_\L(\sum_{h=h_\b}^N\ph^{(h)})}\;,\label{2.10}
\]
with obvious notation. It is now clear how the iterative integration is performed: we first integrate the field $\ph^{(N)}$, and rewrite
\[
\Xi_{\L,N}:= e^{-|\L|F_{\L,N}}\int \prod_{h=h_\b}^{N-1}P_{\L,h}(d\ph^{(h)})\,e^{-V_\L^{(N-1)}(\ph^{(\le N-1)})}\;,\label{2.11}
\]
where $\ph^{(\le N-1)}=\sum_{h=h_\b}^{N-1}\ph^{(h)}$, and $V^{(N-1)}_\L$ is the effective potential on scale $N-1$, defined by 
\[|\L|F_{\L,N}+V^{(N-1)}_\L(\ph)=-\log \int P_{\L,N}(d\ph^{(N)})\,e^{-V_\L(\ph+\ph^{(N)})}\;,\quad V_\L^{(N-1)}(0)=0\;.
\]
Next we integrate $\ph^{(N-1)}$, thus defining $F_{\L,N-1}$ and $V^{(N-2)}_\L$, and so on. At each step we define
\[|\L|F_{\L,h}+V^{(h-1)}_\L(\ph)=-\log \int P_{\L,h}(d\ph^{(h)})\,e^{-V_\L^{(h)}(\ph+\ph^{(h)})}\;,\quad V_\L^{(h-1)}(0)=0\label{2.13}\]
and we proceed in this fashion until we reach the last scale $h_\b$.
As a result, we get an expansion for the specific free energy in the form $f_L(\b)=\sum_{h=h_\b}^N F_{\L,h}$.
Eventually, the specific ground state energy is obtained by taking the limit $L\to\infty$ and then $\b\to\infty$ of this 
expression. The bounds on $F_{\L,h}$ are based on simple dimensional estimates on the propagator, and on a suitable 
resummation of the ``divergent" terms, from which one can show that $F_{\L,h}$ is bounded uniformly in $N,\b,L$, for each fixed $h$;
on top of that, $F_{\L,h}$ reaches a well-defined limit as $N,\b,L\to\infty$ (to be denoted by $F_h$), and the limiting expression is absolutely summable in $h$, 
both for $h\to+\infty$ and for $h\to-\infty$. In the following we illustrate in some detail the method for deriving these bounds. In 
order to keep the exposition as simple as possible, we shall perform the 
bounds on $F_h$ directly in the 
$\b,L\to\infty$, leaving aside the issue of proving the uniform convergence of the finite $(\b,L)$-expressions to their limits\footnote{A
detailed discussion of the finite $\b$ effects for the ultraviolet integration is in \cite{B-Fermi2d}; a detailed discussion of the 
finite $(\b,L)$ effects for the infrared integration is in \cite{XYZ}.}.

Using the inductive definition \eqref{2.13}, we obtain the expansion for $F_h$ and for the kernels of the effective potentials  $V^{(h)}(\ph)=\lim_{|\L|\to\infty}V_\L^{(h)}$, where $V^{(h)}$ has the form
\be V^{(h)}(\ph)=\sum_{m\ge 1}\int_{\RRR^{6m}}d\xx_1\cdots d\yy_{m} W^{(h)}_{2m}(\xx_1,\ldots,\xx_m;\yy_1,\cdots,\yy_{m})\ph_{\xx_1}^+\cdots
\ph^+_{\xx_m}\ph^-_{\yy_{1}}\cdots\ph^-_{\yy_{m}}\label{2.14}\ee
with $W^{(h)}_{2m}$ an integral kernel that is invariant under permutations of the $\xx_i$'s among themselves and/or of the 
$\yy_i$'s among themselves, and invariant under translations and rotations. In the following we will 
also need the ``anchored" version of $V^{(h)}(\ph)$, to be denoted by $\lis V^{(h)}_{\bz}(\ph)$, which is the same as $V^{(h)}(\ph)$, 
up to  the fact that one of the space-time labels, say $\xx_1$, is fixed at an arbitrary position, say $\bz$ (due to the permutation and 
translation invariance of the kernel, the resulting expression is independent in probability of the choice of the label that is fixed and of the localization point):
\be  \lis V^{(h)}_{\bz}(\ph)=\sum_{m\ge 1}\int_{\RRR^{3(2m-1)}}\hskip-.4truecm d\xx_2\cdots d\yy_{m} W^{(h)}_{2m}(\bz,\xx_2,\ldots,\xx_m;\yy_1,\ldots,\yy_{m})\ph_{\bz}^+\ph_{\xx_2}^+\cdots\ph^+_{\xx_m}\ph^-_{\yy_{1}}\cdots \ph^-_{\yy_{m}}\;.\ee
Let us  illustrate how to derive the multiscale expansion for $F_h$ and $W^{(h)}_{2m}$. We first describe its ``naive" version,
which allows us to identify the potentially divergent terms, and suggests how to resum them in order to improve the convergence properties of the theory. The improved (resummed and renormalized) version will be described in the next sections.

Using (the limit $|\L|\to\infty$ of) \eqref{2.13}, we get
\[W^{(h-1)}_{2m}(\xx_1,\ldots,\yy_m)=\frac1{(m!)^2}\frac{\d^{2m}}{\d\ph^+_{\xx_1}\cdots\d\ph^+_{\xx_m}\d\ph^-_{\yy_1}\cdots\d\ph^-_{\yy_m}}\sum_{s\geq
1}\frac{1}{s!}\EE_{h}^{T}\big(V^{(h)} (\ph+\ph^{(h)});s\big)\Big|_{\ph=0}\label{2.weff}\]
and
\[F_{h}=\sum_{s\geq
1}\frac{1}{s!}\EE_{h}^{T}\big(\,\lis V^{(h)}_{\bz} (\ph^{(h)}), V^{(h)} (\ph^{(h)});1,s-1)\;.\label{2.fh}\]
Here $\EE_h^T$ is the truncated expectation on scale $h$, defined as
\[ &\EE_{h}^{T}(X_1(\ph^{(h)}),\ldots,X_s(\ph^{(h)});n_1,\ldots,n_s) :=\label{2.17}\\
&\hskip2.truecm=
\frac{\partial^{n_1+\cdots+n_s}}{\partial \lambda_1^{n_1}\cdots\partial\l_s^{n_s}}\log \int P_h(d\ph^{(h)})
 e^{\lambda_1
X_1(\ph^{(h)})+\cdots+\l_sX_s(\ph^{(h)})}
\Big|_{\lambda_i=0}\;\non\]
where $P_h$ is the gaussian measure with propagator $G_h$,
and $G_h$ is the 
limit $\b,L\to\infty$ of \eqref{2.9} at $h$ fixed:
\[G_h(\xx)=\lim_{|\L|\to\infty}G_{\L,h}(\xx)= u_h(\xx) \th(x_0) \int \frac{d^2\vk}{(2\pi)^2} e^{-i \vk \cdot \vx - x_0|\vk|^2}
=u_h(\xx) \th(x_0) \frac{e^{-|\vx|^2/(4x_0)}}{4\p x_0}\;,\label{2.g}\]
which is scale covariant, in the sense that 
\be G_h(x_0,\vx)=\g^h G_0(\g^hx_0,\g^{h/2}\vx)\label{2.scale}\ee
and, therefore, it satisfies the following bounds:
\[ \label{2.Gh}
||G_h||_\io \leq A \g^h\;,\qquad ||G_h||_1\leq A\g^{-h}\;,
\]
for a suitable $A>0$. Moreover, due to the presence of $u_h(\xx)\th(x_0)$ in \eqref{2.Gh}, 
\[
G_h(\xx-\yy) \neq 0 \quad \Rightarrow \quad 0 < x_0 - y_0 < \g^{-h+1}\;,\label{2.21a}
\]
which is a crucial property for the following analysis. 

It is well known that the truncated expectation in \eqref{2.17}  admits a natural graphical interpretation: 
if $X_i$ is graphically represented as a (non-local) vertex with external lines
$\ph^{(h)}$,  \eqref{2.17}
can be represented as the sum over the Feynman diagrams obtained
by contracting in all possible {\it connected} ways the lines exiting
from $n_1$ vertices of type $X_1$, $n_2$ of type $X_2$, etc, and $n_s$ of type $X_s$.
Every contraction involves a pair of fields, one of type $\ph^+$ and one of type $\ph^-$,
and  corresponds to a
propagator on scale $h$, as defined in \eqref{2.g}.

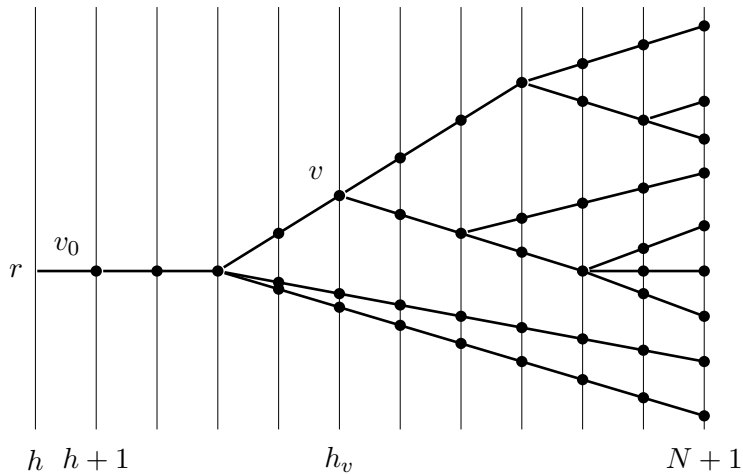
\begin{figure}[t]
\centering
\begin{tikzpicture}[x=0.8cm,y=1cm]
    \foreach \x in {0,...,11}
    {
     \draw[very thin] (\x ,1.9) -- (\x , 7.5);
    }
    	\draw (0,1.5) node {$h$};
    	\draw (1,1.5) node {$h+1$};
	\draw (11,1.5) node {$N+1$};
      \draw (5,1.5) node {$h_v$};
	\draw (-0,4) node[inner sep=0,label=180:$r$] (r) {};
	\draw[med] (r) -- ++(1,0) node [vertex,label=130:$v_0$] (v0) {};
	\draw[med]  (v0) -- ++ (1,0) node[vertex] {} -- ++ (1,0) node[vertex] (v1) {};
\draw[med]  (v1) -- ++ (1,-0.15) node[vertex] {} -- ++ (1,-0.15) node[vertex] {} -- ++ (1,-0.15) node[vertex] {} -- ++ (1,-0.15) node[vertex] {} -- ++ (1,-0.15) node[vertex] {} -- ++ (1,-0.15) node[vertex] {} -- ++ (1,-0.15) node[vertex] {} -- ++ (1,-0.15) node[vertex] {};
\draw[med]  (v1) -- ++ (1,-0.24) node[vertex] {} -- ++ (1,-0.24) node[vertex] {} -- ++ (1,-0.24) node[vertex] {} -- ++ (1,-0.24) node[vertex] {} -- ++ (1,-0.24) node[vertex] {} -- ++ (1,-0.24) node[vertex] {} -- ++ (1,-0.24) node[vertex] {} -- ++ (1,-0.24) node[vertex] {};
\draw[med]  (v1) -- ++(1,0.5) node [vertex] {} -- ++ (1,0.5) node [vertex,label=130:$v$] (v) {};
\draw[med]  (v) -- ++ (1,0.5) node [vertex] {} -- ++ (1,0.5) node [vertex] {} -- ++ (1,0.5) node [vertex] (v2) {};
\draw[med]  (v2) -- ++ (1,0.25) node [vertex] {} -- ++ (1,0.25) node [vertex] {}-- ++ (1,0.25) node [vertex] {};
\draw[med]  (v2) -- ++ (1,-0.25) node [vertex] {} -- ++ (1,-0.25) node [vertex] (v3) {} -- ++ (1,-0.25) node [vertex] {};
\draw[med]  (v3) -- ++ (1,0.25) node [vertex] {};
\draw[med]  (v) -- ++ (1,-0.25) node [vertex] {} -- ++ (1,-0.25) node [vertex] (v4) {}-- ++ (1,-0.25) node [vertex] {}-- ++ (1,-0.25) node [vertex] (v5) {}-- ++ (1,-0.3) node [vertex] {} -- ++ (1,-0.3) node [vertex] {};
\draw[med]  (v4) -- ++(1,0.2) node[vertex] {}-- ++(1,0.2) node[vertex] {}-- ++(1,0.2) node[vertex] {}-- ++(1,0.2) node[vertex] {};
\draw[med]  (v5)  -- ++(1,0.3) node [vertex] {} -- ++(1,0.3) node [vertex] {};
\draw[med]  (v5) -- ++(1,0) node [vertex] {} -- ++(1,0) node [vertex] {};
\end{tikzpicture}
\caption{A tree $\t\in {\tl{\TT}^{(h)}_{N;n}}$ with $n=9$: the root is on scale $h$ and the endpoints are on scale $N+1$.\label{naive_tree}}
\end{figure}

Of course, \eqref{2.weff} and \eqref{2.fh} can be iterated so to re-express $V^{(h)}$ in the r.h.s. in terms of $V^{(h+1)}$, and so on, until we reach scale 
$N$. On scale $N$, we let $V^{(N)}(\ph)=V(\ph)$, where $V(\ph)$ is the limit $|\L|\to\io$ of \eqref{2.V_QCP}. We shall see that 
$\n=\lim_{|\L|\to\io}\n_\L$ can be chosen equal to zero; therefore,  in order not overwhelm the notation, we shall fix it directly equal to zero:
\[V^{(N)}(\ph)=V(\ph)=\frac{\l}{2} \int_{\RRR^6} d\xx d\yy \, |\ph_{\xx}|^2 \,w(\xx-\yy)\, |\ph_{\yy}|^2\label{2.22}  \]
The final outcome is that the kernels of $V^{(h-1)}$ and the single scale contribution to the free energy, $F_h$, can be written as a
sum over connected Feynman diagrams with lines on all possible
scales between $h$ and $N$. The iteration of \eqref{2.weff} induces a
natural hierarchical organization of the scale labels of every
Feynman diagram, which can be conveniently represented in terms
of tree diagrams, first introduced by G. Gallavotti 
and F. Nicol\`o in \cite{GN}. Since then, the {\it Gallavotti-Nicol\`o} tree expansion has been 
described in detail in several papers that make use of constructive renormalization group 
methods, see e.g.\ \cite{GalReview,BGbook,GM}.
The main features of this expansion are described below. 
\begin{enumerate}[(1)]
\item Let us consider the family of all {\it unlabeled trees} which can be constructed by joining a point $r$, the {\it root}, with an ordered set of $n\ge 1$ points, the {\it endpoints} of the tree, so that $r$ is not a branching point. 

The unlabelled trees are partially ordered from the root
to the endpoints in the natural way: nodes on the left are lower than those to their right; we shall use the symbol $<$
to denote this partial ordering; $n$ is called the {\it order} of the unlabeled tree. Two unlabeled trees are identified if they can be superposed by a suitable
continuous deformation, so that the endpoints with the same index coincide.

We shall also consider the {\it labelled trees} (to be called simply trees in the following); they are defined by associating some labels with the unlabelled trees, as explained in the following items. 

\item We associate a label $h \le N$ with the root, a label $N+1$ with the endpoints and we denote by
$\tl{\TT}^{(h)}_{N;n}$ the corresponding set of labelled trees with $n$
endpoints. Moreover, we introduce a family of vertical lines,
labelled by an integer $h_v$, the {\it scale} label, taking values in $[h,N+1]$, see Fig.\ref{naive_tree}. 

We call {\it non trivial vertices} of the tree its branch points; we call {\it trivial vertices} the points where the branches intersect the family of vertical lines. The set of the {\it vertices} will be the union of the endpoints and of trivial and non trivial vertices (see the dots in Fig.\ref{naive_tree}). Note that the root is not a vertex.
Every vertex $v$ of a tree will be associated with its scale label $h_v$.
\item There is only one vertex immediately following
the root, called $v_0$ and with scale label equal to $h+1$.
\item Given a vertex $v$ of $\t\in{\tl{\TT}^{(h)}_{N;n}}$ that is not an endpoint, we can consider the subtrees of $\t$ with root $v'$
(here $v'$ is the vertex immediately preceding $v$ on $\t$), which correspond to the connected components of the restriction of $\t$ to the vertices $w\ge v$. If a subtree with root $v'$ contains, besides the root, only $v$ and one endpoint on scale $h_v+1$, it will be called a {\it trivial subtree}. Given a vertex $v$ we denote by $s_v$ the number of lines branching from $v$ (then $s_v=1$ if $v$ is a trivial vertex).
\item With each endpoint $v$ we associate $V^{(N)}(\ph)$, see \eqref{2.22}, and a set $I_v$ of four field labels $f_1,\ldots f_4$.
Given  $f\in I_v$, we let $\xx_f$ and $\s_f$ denote the space--time and the creation/annihilation label of the corresponding field:
$\ph^{\s_f}_{\xx_f}$, $\s_f=\pm$; due to the form of the interaction \eqref{2.22}, the space-time labels of the four fields are equal two by two,
and we indicate the two values by $\xx_{1,v},\xx_{2,v}$. If $v$ is not an endpoint, we
call $I_v$ the set of field labels associated with the endpoints following
the vertex $v$. Moreover, we call {\it cluster of $v$} (and indicate it by $\underline{\xx}_v$) the family of space--time points associated with all the endpoints 
following $v$, if $v$ is not an endpoint, or $v$ itself, otherwise.
In the following, we shall associate with every $\t\in\tilde\TT^{(h)}_{N;n}$ several contributions, 
distinguished by the choice of the fields that are contracted at every scale.
In order to identify these contributions, we introduce a subset $P_v$ of $I_v$, whose interpretation is that of {\it external fields} of $v$.
The $P_v$'s must satisfy various
constraints: first of all, if $v$ is not an endpoint and $v_1,\ldots,v_{s_v}$
are the $s_v\ge 1$ vertices immediately following it, then
$P_v \subseteq \cup_i
P_{v_i}$; if $v$ is an endpoint, $P_v=I_v$.
If $v$ is not an endpoint, we also denote by $Q_{v_i}$ the
intersection of $P_v$ and $P_{v_i}$, so that $P_v=\cup_i
Q_{v_i}$. The union of the subsets $P_{v_i}\setminus Q_{v_i}$
is, by definition, the set of the {\it internal fields} of $v$,
and is non empty if $s_v>1$. \end{enumerate}
In terms of these trees, the kernels of the effective potential can be written as
(defining $\underline\xx=(\xx_1,\ldots,\xx_m)$ and $\underline\yy=(\yy_1,\ldots,\yy_m)$)
\[
W^{(h)}_{2m}(\underline\xx;\underline\yy)= \sum_{n\geq 1} \sum_{\t\in\tilde\TT_{N;n}^{(h)}} \sideset{}{^{(P_{v_0})}}\sum_{{\bf P}\in\PP_\t}
\int \hskip-.2truecm\prod_{f\in I_{v_0}\setminus P_{v_0}}\hskip-.4truecm d\xx_f\
W_{}^{(h)}(\t,{\bf P};\underline{\xx}_{v_0})\label{2.23}\]
where ${\cal P}_\t$ is the family of all the choices of the sets $P_v$ compatible with the constraints illustrated above, and ${\bf P}=\{P_v\}$ are 
elements of ${\cal P}_\t$. The apex $(P_{v_0})$ on the sum indicates the constraint that we are not summing over $P_{v_0}$: rather, $P_{v_0}$ is a fixed set with $2m$ elements, $m$ of which have $\s$-label $+$ and space-time label $\xx_i$, while the remaining $m$ have $\s$-label $-$ and space-time label $\yy_i$.
In particular, $\cup_{f\in P_{v_0}}\{\xx_f\}=(\underline\xx,\underline\yy)$. Similarly, $F_{h+1}$ can be written as
\[
F_{h+1}= \sum_{n\geq 1} \sum_{\t\in\tilde\TT_{N;n}^{(h)}}\ \sideset{}{^*}\sum_{{\bf P}\in \PP_\t}\;
\int \hskip-.2truecm\prod_{f\in I_{v_0}\setminus \{f^*\}}\hskip-.2truecm d\xx_f\
W_{}^{(h)}(\t,{\bf P};\underline{\xx}_{v_0})\;,\label{2.23ab}\]
where 
$f^*$ is an arbitrary field label: the reason why there is one integral less than the total number of space-time points comes from the 
fact that $F_h$ involves one anchored effective potential $\lis V_{\bf 0}^{(h)}(\ph)$, see \eqref{2.fh}; the fact that $f^*$ is arbitrary comes from 
the translation invariance of the kernels. 
Moreover, the $*$ on the sum over ${\bf P}$ indicates the constraint that $P_{v_0}=\emptyset$ is fixed and the set of internal fields of $v_0$ is non empty.
The contribution $W_{}^{(h)}(\t,{\bf P};\u{\xx}_{v_0})$ has the form:
\[&W_{}^{(h)}(\t,{\bf P};\underline\xx_{v_0})=-(-\l/2)^{n}\Big[\prod_{v\ {\rm e.p.}} w(\xx_{1,v}-\xx_{2,v})\Big] \cdot\label{2.24}\\
&\hskip2.truecm
\cdot\Bigl[\prod_{v \text{ not e.p.}} \frac{1}{s_v!}\, \EE_{h_v}^T \lft( \ph(P_{v_1}\backslash Q_{v_1}), \ldots, \ph(P_{v_{s_v}}\backslash Q_{v_{s_v}}) \rgt)\Bigr]\;,
\non\]
where the first product runs over the endpoints of $\t$, the second over the vertices $v$ of $\t$ that are not endpoints, for each of which we indicated by $v_1,
\ldots,v_{s_v}$ the vertices immediately following $v$ on $\t$. Moreover, $\ph(P_v\setminus Q_v):=\prod_{f\in P_v\setminus Q_v}\ph^{\s_f}_{\xx_f}$. 
Note that the factors in \eqref{2.24} do not depend explicitly on $N$: therefore, the dependence of the effective potential on the ultraviolet cutoff 
is only due to the fact that the  sum over the trees is restricted to $\tl \TT^{(h)}_{N;n}$.

\feyn{
\begin{fmffile}{BoseFeyn/int_QCP}
 \unitlength = 0.8cm
\def\myl#1{3.5cm}
\fmfset{arrow_len}{3mm}
\begin{align*}
\parbox{\myl}{\centering{    
		\begin{fmfgraph*}(3.5,2)    
			\fmfleft{i1,i2}
			\fmfright{o1,o2}
			\fmf{fermion}{i1,v1,i2}
                   \fmf{wiggly, tension=1.2}{v1,v2}
                   \fmf{fermion}{o1,v2,o2}
                   \fmfdot{v1,v2}
	           \fmfv{label=$\xx$}{v1}
		\fmfv{label=$\yy$}{v2}
		\end{fmfgraph*} \\
		}} 
\end{align*}
\end{fmffile}	
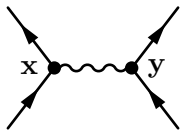
}{The vertex corresponding to the interaction $V(\ph)$.}{int_QCP}

 \vskip -0.3cm

As anticipated above, the truncated expectations in the second line of \eqref{2.24} can be expanded 
in sums over connected Feynman 
diagrams, each of which is uniquely determined by the choice of the pairing among the $\ph$ fields (each pairing is 
a grouping of the fields in pairs, each pair consisting of one field with $\s$-label equal to $+$, and one equal to $-$) and 
has a value equal to the product of the propagators associated with the paired fields. The choice of a (labelled) 
Feynman diagram is thus equivalent to the choice of a connected pairing per vertex; we indicate by 
$\tilde \G(\t,{\bf P})$ the set of connected labelled Feynman diagrams (or, equivalently, of sequences of connected pairings indexed by $v\in\t$)
compatible with the tree $\t$ and the set of field labels ${\bf P}$. Therefore, we can write:
\[&W_{}^{(h)}(\t,{\bf P};\underline\xx_{v_0})=\sum_{\GG\in\tilde\G(\t,{\bf P})}\Val(\GG)\;,\label{2.25x}\\
& \Val(\GG)
=-(-\l/2)^{n}\Big[\prod_{v\ {\rm e.p.}} w(\xx_{1,v}-\xx_{2,v})\Big] \Bigl[\prod_{v \text{ not e.p.}} \frac{1}{s_v!}\,\prod_{\ell\in\GG_v}G_{h_v}(\xx_{\ell_-}-\xx_{\ell_+})\Bigr]\;,
\non \]
where $\GG_v$ is the pairing of the internal fields of $v$ associated with $\GG$, and $\ell_\pm$ are the two 
contracted fields (with $\s$-labels $+$ and $-$, respectively) corresponding to the pair $\ell\in\GG_v$. From a graphical point of view,
the Feynman graph $\GG$ can be depicted by drawing $n$ vertices as in Fig.\ref{int_QCP} (the exiting/entering solid half-lines represent
fields of type $\ph^+$/$\ph^-$ and the ``wiggly" lines represent the interaction kernel $(\l/2)w(\xx-\yy)$), and then by joining the solid half-lines in pairs, 
in the way induced by the pairing $\{\GG_v\}$ associated with $\GG$; every contracted solid line $\ell$ has a well-defined direction (from 
$\xx_{\ell_+}$ to $\xx_{\ell_-}$), and
it carries the scale label $h_v$, where $v$ is the vertex such that $\ell\in\GG_v$: of course, it corresponds to the propagator $G_{h_v}(\xx_{\ell_-}-\xx_{\ell_+})$. 
Note that time increases along the lines in the natural direction (i.e., from $\xx_{\ell_+}$ to $\xx_{\ell_-}$), due to the time ordering condition \eqref{2.21a}. 
Let us also observe that 
every vertex $v$ corresponds to a set of propagators, those associated with the lines $\ell\in\GG_v$. If we artificially associate a label $N$ with the wiggly lines, then the union of the solid lines 
in $\cup_{w\ge v}\GG_w$, together with the wiggly lines associated with the endpoints following $v$ on $\t$, form a maximal connected set of 
lines with scale labels $\ge h_v$, called a {\it cluster}; here ``maximal" refers to the fact that any solid line exiting or entering the cluster has scale label strictly smaller than $h_v$, i.e., the cluster cannot be increased without decreasing its scale. See Fig.\ref{cluster} for an example of a labelled Feynman diagram,
together with its cluster structure and its Gallavotti-Nicol\`o tree.

\tik{
\[  \t = \hskip -0.4cm 
\parbox{5cm}{\centering{ \vskip 0.6cm
\begin{tikzpicture}[scale=0.8]
\foreach \x in {-1,...,3}   
    {
        \draw[very thin] (\x , -1.5) -- (\x , 1.5);      
    }
\draw (-1,-2) node {$h$};  
\draw (0,-2) node {};  
\draw (1,-2) node {$h_v$};
\draw (2,-2) node {$\cdots$};
\draw (3.2,-2) node {$N+1$};
\draw[med]  (-1,0) node[vertex] {}-- (0,0) node[vertex] {} -- (1,0) node[vertex] {} -- (2,0.5)node[vertex] {} -- (3,1)node[vertex, label=0:$1$] {};  
\draw[med] (2,0.5) -- (3,0)node[vertex, label=0:$2$]{};
\draw[med] (1,0) node[vertex] {} -- (2,-0.5) node[vertex] {} -- (3,-1)node[vertex,, label=0:$3$] {};
\end{tikzpicture}
}} 
\hskip  -0.2cm \Leftrightarrow \;
\parbox{3.5cm}{\centering{
\begin{tikzpicture} [scale=0.7]
\draw[med, gray]  (-0.4,0) circle(1cm);
\draw[med, gray]  (0,0) circle(2cm);
\draw (-0.4, 1.2) node[label=right:\textcolor{gray}{$N$}]{};
\draw (0, 2.2) node[label=right:\textcolor{gray}{$h_v$}]{};
\draw (-0.8, -0.1) node[vertex, label=$1$]{};
\draw (-0.1, -0.1) node[vertex, label=$2$]{};
\draw (1.2, -0.1) node[vertex, label=$3$]{};
\end{tikzpicture}
}}  
\; \Leftarrow \quad \GG= 
\parbox{3.2cm}{\centering{ 
\begin{tikzpicture} [scale=0.7]
\draw[med, gray] (-1.4,-0.45) rectangle(1.4,1.7);
\draw (1.7,1.0) node[label=left:\textcolor{gray}{$N$}]{};
\draw[med, gray] (-1.7,-1.75) rectangle(1.7,2);
\draw (1.85,-1.3) node[label=left:\textcolor{gray}{$h_v$}]{};
\draw[arr={latex},thick]  (-1.2, 2.5) --(-0.6, 1.2);
\draw[arr={latex},thick]  (1.2, 2.5) --(0.6, 1.2);
\draw (0.1,0.85) node[label=$1$]{};
\draw (-0.6,1.2) node[vertex]{};
\draw (0.6,1.2) node[vertex]{};
\draw[arr={latex},thick] (-0.6, 1.2)-- (-0.6, 0);
\draw[gluon, thick, label=$3$](-0.6, 1.2) -- (0.6, 1.2);
\draw[arr={latex},thick] (0.6,1.2) --(0.6,0);  
\draw (0,-0.2) node[label=$2$]{};
\draw (-0.6,0) node[vertex]{};
\draw (0.6,0) node[vertex]{};
\draw[arr={latex},thick] (-0.6, 0)-- (-0.6, -1.2);
\draw[gluon, thick, label=$3$](-0.6, 0) -- (0.6, 0);
\draw[arr={latex},thick] (0.6,0) --(0.6,-1.2);   
\draw (-0.6,-1.2) node[vertex]{};
\draw (0.6,-1.2) node[vertex]{};
\draw (0,-1.4) node[label=$3$]{};
\draw[arr={latex},thick]  (-0.6, -1.2) --(-1.2, -2.5);
\draw[gluon, thick, label=$3$](-0.6, -1.2) -- (0.6, -1.2);
\draw[arr={latex},thick] (0.6,-1.2) --(1.2,-2.5);   
\end{tikzpicture}
}}  \non
\]
}
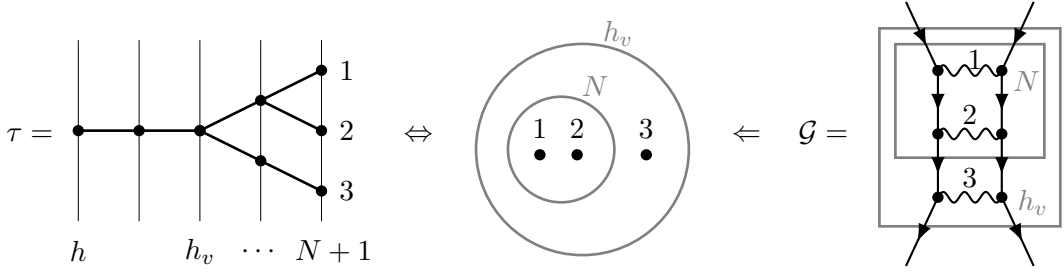
{An example of tree $\t$ of order 3 with the corresponding cluster structure, where only the non trivial vertices are depicted. The cluster structure uniquely identifies a tree $\t$ and viceversa.
Down left an element $\GG$ of the class of Feynman diagrams  compatible with $\t$. 
}{cluster}
%


By using these explicit expressions and the dimensional bounds \eqref{2.Gh} on the propagator, let us now derive a 
bound on \eqref{2.23} or, more precisely, on the $L_1$ norm of the kernel:
\[||W^{(h)}_{2m}||=\int d\xx_2\cdots d\yy_m |W^{(h)}_{2m}(\bz,\xx_2,\ldots,\xx_m;\yy_1,\ldots,\yy_m)|\;,\label{2.26}\]
as well as on $|F_{h+1}|$. Note that the choice of not integrating over $\xx_1$ in \eqref{2.26} 
and of fixing its location at $\xx_1=\bz$ is irrelevant, due to the 
translation invariance of the kernel. In order to bound \eqref{2.26} and $|F_{h+1}|$ we proceed as follows: we use the representations~\eqref{2.23}-\eqref{2.23ab}, with $W_{}^{(h)}(\t,{\bf P};\underline\xx_{v_0})$ as in \eqref{2.25x}, and we get 
\[||W^{(h)}_{2m}||\le& \sum_{n\geq 1}(|\l|/2)^{n} \sum_{\t\in\tilde\TT_{N;n}^{(h)}} \sideset{}{^{(P_{v_0})}}\sum_{{\bf P}\in\PP_\t}\sum_{\GG\in\tilde\G(\t,{\bf P})}
\int d\xx_2\cdots d\yy_m \hskip-.2truecm\prod_{f\in I_{v_0}\setminus P_{v_0}}\hskip-.4truecm d\xx_f\cdot  \label{2.29}\\
&\quad \cdot\Big[\prod_{v\ {\rm e.p.}} w(\xx_{1,v}-\xx_{2,v})\Big] \Bigl[\prod_{v \text{ not e.p.}} \frac{1}{s_v!}\,\prod_{\ell\in\GG_v}|G_{h_v}(\xx_{\ell_-}-\xx_{\ell_+})|\Bigr]\;. 
\non \]
$|F_{h+1}|$ is bounded by an analogous expression, with $P_{v_0}$ replaced by $\emptyset$ and the integration measure by 
$\prod_{f\in I_{v_0}\setminus \{f^*\}}d\xx_f$ (and, in addition, the constraint on the sum over ${\bf P}$ that the set of internal fields of $v_0$ is non empty). 
For each $\GG\in\tl{\G}(\t,{\bf P})$, we arbitrarily choose a spanning tree $T_v\subseteq \GG_v$ per vertex: $T_v$ consists 
of $s_v-1$ lines connecting in a minimal way the $s_v$ clusters $v_1,\ldots,v_{s_v}$ immediately following $v$ on $\t$. 
We rewrite $d\xx_2\cdots d\yy_m \prod_{f\in I_{v_0}\setminus P_{v_0}}\hskip-.2truecm d\xx_f$ as 
\[d\xx_2\cdots d\yy_m \prod_{f\in I_{v_0}\setminus P_{v_0}}\hskip-.4truecm d\xx_f=\Big[\prod_{v\ {\rm not}\ {\rm e.p.}}\,\,
\prod_{\ell\in T_v}d(\xx_{\ell_+}-\xx_{\ell_-})\Big]\Big[\prod_{v\ {\rm e.p.}}d(\xx_{1,v}-\xx_{2,v})\Big]\;,\]
and similarly for $\prod_{f\in I_{v_0}\setminus \{f^*\}}d\xx_f$. 
Next we bound all the propagators outside $\cup_vT_v$ by their $L_\infty$ norm, after which we are left with the product of the $L_1$ norms of the propagators
belonging to the spanning tree:
\[||W^{(h)}_{2m}||\le &\sum_{n\geq 1} \sum_{\t\in\tilde\TT_{N;n}^{(h)}} \sideset{}{^{(P_{v_0})}}\sum_{{\bf P}\in\PP_\t}
\sum_{\GG\in\tilde\G(\t,{\bf P})} |\l|^n2^{-n}\Big[\int d\xx\, w(\xx)\Big]^n\cdot\non\\
&\cdot
\Big[\prod_{v\ {\rm not}\ {\rm e.p.}}\frac1{s_v!}
\big(||G_{h_v}||_\io\big)^{\frac{\sum_{i=1}^{s_v}|P_{v_i}|-|P_v|}2-s_v+1} \big(||G_{h_v}||_1\big)^{s_v-1}\Big]\;,\label{2.31}\]
and similarly for $|F_{h+1}|$. 
Inserting \eqref{2.Gh} into this equation we find:
\[||W^{(h)}_{2m}||\le &\sum_{n\geq 1} \sum_{\t\in\tilde\TT_{N;n}^{(h)}} \sideset{}{^{(P_{v_0})}}\sum_{{\bf P}\in\PP_\t}
\sum_{\GG\in\tilde\G(\t,{\bf P})} |\l|^n2^{-n}\Big[\int d\xx\, w(\xx)\Big]^n A^{2n-m}\cdot\non\\
&\cdot\Big[\prod_{v\ {\rm not}\ {\rm e.p.}}\frac1{s_v!}
\g^{h_v(\frac{\sum_{i=1}^{s_v}|P_{v_i}|-|P_v|}2-s_v+1)} \g^{-h_v(s_v-1)}\Big]\;.\label{2.29bh}\]
Next we rewrite every $h_v$ appearing in the r.h.s. of this equation as $h+(h_v-h)$ and we use the relations:
\[
& \sum_{v\ge v_0}h\big(\sum_{i=1}^{s_v}|P_{v_i}|-|P_v|\big)=h(4n-2m)\;,\label{2.33ac}\\
& 
\sum_{v\ge v_0}(h_v-h)\big(\sum_{i=1}^{s_v}|P_{v_i}|-|P_v|\big)=\sum_{v\ge v_0}(h_v-h_{v'})(4n_v-|P_v|)\;,\\
&\sum_{v\ge v_0}h(s_v-1)=h(n-1)\;,\\
&\sum_{v\ge v_0}(h_v-h)(s_v-1)=\sum_{v\ge v_0}(h_v-h_{v'})(n_v-1)\;,\label{2.36gh}\]
where $v'$ is the vertex immediately preceding $v$ on $\t$ (so that $h_v-h_{v'}=1$), and $n_v$ is the 
number of endpoints following $v$ on $\t$. 
%
%
\feyn{
\begin{fmffile}{BoseFeyn/ladder}
 \unitlength = 1cm
\def\myl#1{8cm}
 \fmfset{arrow_len}{3mm}
\begin{align*}
\parbox{\myl}{\centering{  
\vskip 0.2cm
\begin{fmfgraph*}(8,1.2)
			\fmfleft{i1,i2}
			\fmfright{o1,o2}
			\fmf{fermion, tension=1.2}{i1,v1}
			\fmf{fermion, tension=1.2}{i2,v2}
				\fmf{fermion}{v1,v3}		
				\fmf{fermion}{v2,v4}
				\fmf{dashes, tension=0.6}{v3,v5}		
				\fmf{dashes, tension=0.6}{v4,v6}
				\fmf{fermion}{v5,v7}		
				\fmf{fermion}{v6,v8}
                  			 \fmf{wiggly, tension=0.01}{v1,v2}
					\fmf{wiggly,tension=0.01}{v3,v4}
					\fmf{wiggly, tension=0.01}{v5,v6}
					\fmf{wiggly,tension=0.01}{v7,v8}
			\fmf{fermion, tension=1.2}{v7,o1}
			\fmf{fermion, tension=1.2}{v8,o2}
                   \fmfdot{v1,v2,v3,v4,v5,v6,v7,v8}
 	        \fmfv{label=$\xx_n$,  label.angle=-90}{v1}
 		\fmfv{label=$\xx_{n-1}$,  label.angle=-90}{v3}
 		\fmfv{label=$\xx_{2}$,  label.angle=-90}{v5}
 		\fmfv{label=$\xx_1$,  label.angle=-90}{v7}
 		\fmfv{label=$\yy_n$,  label.angle=90}{v2}
 		\fmfv{label=$\yy_{n-1}$,  label.angle=90}{v4}
 		\fmfv{label=$\yy_{2}$,  label.angle=90}{v6}
 		\fmfv{label=$\yy_1$,  label.angle=90}{v8}
		\end{fmfgraph*}  }}
\end{align*}
\end{fmffile}	
}{A ladder graph of order $n$.}{ladder}
%
%
If we substitute these identities into \eqref{2.29bh}, 
we obtain
\[||W^{(h)}_{2m}||\le \g^{h(2-m)}\sum_{n\geq 1} |\l|^n C^n A^{2n-m}\sum_{\t\in\tilde\TT_{N;n}^{(h)}} \sideset{}{^{(P_{v_0})}}\sum_{{\bf P}\in\PP_\t}
\sum_{\GG\in\tilde\G(\t,{\bf P})}\prod_{v\ {\rm not}\ {\rm e.p.}}\frac1{s_v!}\g^{(h_v-h_{v'})(2-\frac{|P_v|}{2})}\label{2.34}\]
where $C=2^{-1}\int d\xx\, w(\xx)=2^{-1}\int d\vx\, v(\vx)$. Now, the number of Feynman diagrams in $\tilde\G(\t,{\bf P})$ can be bounded by 
$\prod_{v\ {\rm not}\ {\rm e.p.}}(\sum_{i=1}^{s_v}|P_{v_1}|-|P_v|)!!\le C_1^n(n!)^2$, while $\prod_{v\ {\rm not}\ {\rm e.p.}}s_v!\ge C_2^n n!$, so that 
\[||W^{(h)}_{2m}||\le \g^{h(2-m)}\sum_{n\geq 1} |\l|^n K^n n! \sum_{\t\in\tilde\TT_{N;n}^{(h)}} \sideset{}{^{(P_{v_0})}}\sum_{{\bf P}\in\PP_\t}
\prod_{v\ {\rm not}\ {\rm e.p.}}\g^{(h_v-h_{v'})(2-\frac{|P_v|}{2})}\label{2.35xs}\]
with $K=CAC_1C_2$. Of course, $|F_{h+1}|$ admits a completely analogous bound, with the only difference that $P_{v_0}$ is replaced by $\emptyset$ and, therefore, $m=0$. 
A bound like \eqref{2.35xs} is often referred to as an {\it $n!$ bound} on the perturbative expansion:
if we could show that the last product $\prod_{v\ {\rm not}\ {\rm e.p.}}\g^{(h_v-h_{v'})(2-\frac{|P_v|}{2})}$ is summable 
both over the field and the scale labels, then it would imply that the $n$-th order of perturbation theory would 
be finite and bounded explicitly by (const.)$^n|\l|^nn!$ (compatible with the Borel summability of the theory). 

Unfortunately, it is apparent that the subdiagrams with $|P_v|=2,4$ are not summable over the scale labels. This means 
that such subdiagrams need to be resummed. The iterative resummation procedure will be described in the next subsection, and it will be equivalent to 
a reorganization of the tree expansion, after which the resulting expansion will be finite at all orders, with $n!$ bounds on 
the generic order. 

\subsection{The renormalized expansion}  \label{s2.ren}

In the previous subsection we saw that the only sub-diagrams leading to possible divergences in the multiscale expansion for the free energy and for the effective potential are those with $|P_v|=2,4$. 
An important fact that will be extensively used in the following is that, thanks to the time ordering condition \eqref{2.21a}, the 
values of most of these Feynman sub-diagrams are zero. More specifically, the value of any (sub)diagram with 
$|P_v|=0,2$ is zero, and the 
value of a subdiagram with $|P_v|=4$ is different from zero only if it is a {\it ladder} graph, as the one in Fig.\ref{ladder}
(this fact is well known both in the theoretical and in the mathematical physics literature, 
see e.g. \cite{BGPS-Fermi1d, B-Fermi2d, Sachdev}). 
A few examples of diagrams 
with 2 or 4 external legs and vanishing values are shown in Fig.\ref{loops}.


\feyn{
\begin{fmffile}{BoseFeyn/loops}
 \unitlength = 0.8cm
\def\myl#1{4cm}
 \fmfset{arrow_len}{3mm}
\begin{align*}
%
\parbox{\myl}{\centering{   \fmfset{arrow_len}{3mm}
		\begin{fmfgraph*}(3.5,2)
			\fmfleft{i1}
			\fmfright{o1}
			\fmf{fermion}{i1,v1}
                   \fmf{fermion}{v2,o1}
                   \fmf{wiggly, left=0.5, tension=0.4}{v1,v2}
                   \fmf{fermion, right=0.5, tension=0.4}{v1,v2}
                   \fmfdot{v1,v2}
		\end{fmfgraph*}}}
 \hskip -0.4cm &
\parbox{\myl}{\centering{  \vskip +1.2cm
		\begin{fmfgraph*}(2.5,3.5)
			\fmfleft{i1}
			\fmfright{o1}
                    \fmftop{t}
			\fmfbottom{b}
			\fmf{fermion}{i1,v1,o1}
                   \fmf{wiggly}{v1,v2}
                   \fmf{fermion, right, tension=0.5}{v2,t}    
			\fmf{fermion, right, tension=0.5}{t,v2}  
			\fmfdot{v1,v2}           
                   \fmfv{label.angle=-160}{v2} 
			\fmf{phantom, tension=0.5}{v1,b}
		\end{fmfgraph*}}}  
\parbox{\myl}{\centering{  
\begin{fmfgraph*}(5,1.5)
			\fmfleft{i1,i2}
			\fmfright{o1,o2}
			\fmf{fermion}{i1,v1}
			\fmf{phantom}{i2,v3}
			\fmf{fermion}{v1,v2}
			\fmf{phantom}{v2,o1}
                   \fmf{wiggly, tension=0.1}{v1,v3}
			\fmf{wiggly,tension=0.1}{v2,v4}
			\fmf{fermion, tension=0.001}{v2,v3}
                   \fmf{fermion}{v3,v4,o2}
                   \fmfdot{v1,v2,v3,v4}
		\end{fmfgraph*}  }} 
\hskip -0.4cm
\parbox{\myl}{\centering{  \vskip +1.2cm
		\begin{fmfgraph*}(2.5,6)
			\fmfleft{i1}
			\fmfright{o1}
                   \fmftop{ta,tb}
			\fmfbottom{b1,b3}
			\fmf{fermion}{i1,v1}
\fmf{fermion}{v1,v3}
\fmf{fermion}{v3,o1}
                   \fmf{wiggly, tension=0.8}{v1,v2}
                   \fmf{wiggly, tension=0.8}{v3,v4}
			\fmf{fermion, right=0.3, tension=0.01}{v2,v4}
                   \fmf{fermion, right=0.3}{v4,t4}
			\fmf{phantom, tension=0.8}{ta,t2} 
\fmf{phantom, tension=0.8}{tb,t4}
                   \fmf{fermion, tension=0.5, right=0.4}{t4,t2}
                     \fmf{fermion, right=0.3}{t2,v2}
			\fmfdot{v1,v2,v3,v4}           
			\fmf{phantom, tension=0.3}{v1,b1}
		       \fmf{phantom, tension=0.3}{v3,b3}
		\end{fmfgraph*}}} 
 \\[-1.4cm]
%
&
  \parbox{\myl}{\centering{
		\begin{fmfgraph*}(4,1.2)
			\fmfleft{i1,i2}
			\fmfright{o1,o2}
 			\fmf{fermion}{i1,vl,i2}
			\fmf{wiggly, tension =1.2}{vl,v1}
                   \fmf{fermion, left, tension=0.5}{v2,v1}
                   \fmf{fermion, left, tension=0.5}{v1,v2}
                   \fmf{wiggly, tension=1.2}{v2,vr}
                   \fmf{fermion}{o1,vr,o2}
		\fmfdot{vl,v2,v1,vr}
		\end{fmfgraph*}
		}}
\parbox{\myl}{\centering{  
\begin{fmfgraph*}(4,1.2)
			\fmfleft{i1,i2}
			\fmfright{o1,o2}
			\fmf{fermion, tension=1.2}{v1,i1}
			\fmf{fermion, tension=1.2}{i2,v3}
			\fmf{fermion, tension=1.2}{o1,v2}
			\fmf{fermion}{v2,v1}		
			\fmf{fermion}{v4,o2}
                   \fmf{wiggly, tension=0.1}{v1,v3}
			\fmf{wiggly,tension=0.1}{v2,v4}
                   \fmf{fermion}{v3,v4}
                   \fmfdot{v1,v2,v3,v4}
		\end{fmfgraph*}  }}
\end{align*}
\end{fmffile}	
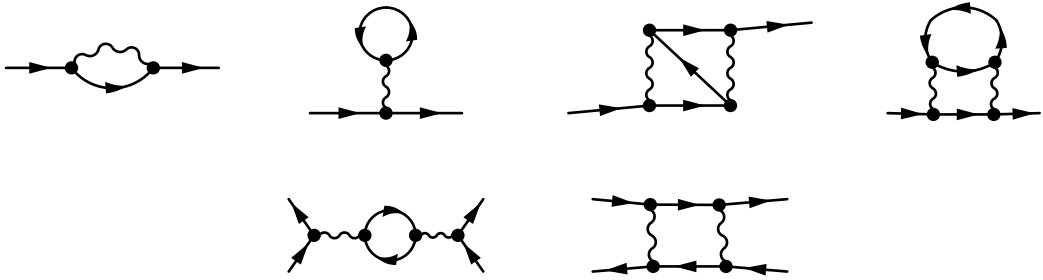
}{A few examples of diagrams with $|P_v|=2,4$ which are zero due to the time ordering condition~\eqref{2.21a}.}{loops}

The reason why the diagrams with $|P_v|=0,2$, as well as the non-ladder diagrams with $|P_v|=4$, are zero is very simple: they either contain a {\it bosonic loop},
i.e., a closed solid line, or they contain a wiggly line connecting two distinct points along the same open solid line, or both (see Fig.\ref{loops}). In any of these cases, the value of the diagram is proportional to the product 
of an order sequence of propagators $G_{h_1}(\xx_1-\xx_2)G_{h_2}(\xx_2-\xx_3)\cdots G_{h_k}(\xx_k-\xx_{k+1})$, with 
$k\ge 1$, times an interaction kernel $w(\xx_1-\xx_{k+1})=\d(x_{1,0}-x_{k+1,0})v(\vx_1-\vx_{k+1})$: such a product is zero, simply because $\d(x_{1,0}-x_{k+1,0})$ fixes the initial and final times in the sequence to be the same, while 
the time ordering condition \eqref{2.21a} requires that $x_{1,0}>x_{{k+1,0}}$ in order for the sequence to be different from zero. 

\begin{figure}[t]
\centering
\begin{tikzpicture}[x=0.8cm,y=1cm]
    \foreach \x in {0,...,11}
    {
     \draw[very thin] (\x ,1.9) -- (\x , 7.5);
    }
    	\draw (0,1.5) node {$h$};
    	\draw (1,1.5) node {$h+1$};
	\draw (11,1.5) node {$N+1$};
      	\draw (5,1.5) node {$h_v$};
	\draw (-0,4) node[inner sep=0,label=180:$r$] (r) {};
	\draw[med] (r) -- ++(1,0) node [vertex,label=130:$v_0$] (v0) {};
	\draw[med]  (v0) -- ++ (1,0) node[vertex] {} -- ++ (1,0) node[vertex] (v1) {};
\draw[med]  (v1) -- ++ (1,-0.15) node[vertex] {} -- ++ (1,-0.15) node[vertex] {} -- ++ (1,-0.15) node[vertex] {} -- ++ (1,-0.15) node[vertex] {} -- ++ (1,-0.15) node[vertex] {} -- ++ (1,-0.15) node[vertex] {} -- ++ (1,-0.15) node[vertex] {} -- ++ (1,-0.15) node[vertex] {};
\draw[med]  (v1) -- ++ (1,-0.75) node[vertex] {};  
\draw[med]  (v1) -- ++(1,0.5) node [vertex] {} -- ++ (1,0.5) node [vertex,label=130:$v$] (v) {};
\draw[med]  (v) -- ++ (1,0.5) node [vertex] {} -- ++ (1,0.5) node [vertex] {} -- ++ (1,0.5) node [vertex] (v2) {};
\draw[med]  (v2) -- ++ (1,0.5) node [vertex] {};
\draw[med]  (v2) -- ++ (1,-0.25) node [vertex] {} -- ++ (1,-0.25) node [vertex] (v3) {} -- ++ (1,-0.25) node [vertex] {};
\draw[med]  (v3) -- ++ (1,0.25) node [vertex] {};
\draw[med]  (v) -- ++ (1,-0.25) node [vertex] {} -- ++ (1,-0.25) node [vertex] (v4) {}-- ++ (1,-0.25) node [vertex] {}-- ++ (1,-0.25) node [vertex] (v5) {}-- ++ (1,-0.3) node [vertex] {};
\draw[med]  (v4) -- ++(1,0.25) node[vertex] {};
\draw[med]  (v5)  -- ++(1,0.3) node [vertex] {} -- ++(1,0.3) node [vertex] {};
\draw[med]  (v5) -- ++(1,0) node [vertex] {};
\end{tikzpicture}
\caption{An example of {\it renormalized tree} belonging to $\TT^{(h)}_{N;n}$. Even if not reported explicitly in the picture, a label $\RR$ is associated with each vertex different from $v_0$ and from the endpoints. The vertex $v_0$ is associated with an index $\LL$ or $\RR$. The endpoints at scale $h_v\in [h+2,N]$ represent $-\LL \VV^{(h_v-1)}(\ph)$,
while the endpoints at scale $N+1$ correspond as before to $V(\ph)$.\label{ren_tree}}
\end{figure}
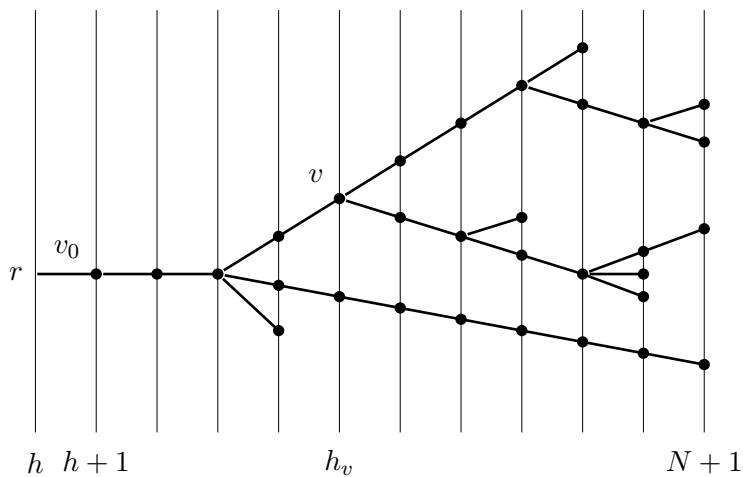

In conclusion, the only potentially dangerous subdiagrams, which make the bound \eqref{2.35xs} not uniformly summable in $\{h_v\}$
and, therefore, require a resummation, 
are the ladder diagrams as in Fig.\ref{ladder}. At the formal level, the resummation is very simple: for each $h<N$, we just split the effective potential 
as $V^{(h)}(\ph)=\LL V^{(h)}(\ph)+\RR V^{(h)}(\ph)$, where 
\[\LL V^{(h)}(\ph)=\int_{\RRR^{12}}d\xx_1\cdots d\yy_{2} \LL W^{(h)}_{4}(\xx_1,\xx_2;\yy_1,\yy_{2})\ph_{\xx_1}^+
\ph^+_{\xx_2}\ph^-_{\yy_{1}}\ph^-_{\yy_{2}}\label{2.36}\]
and $ \LL W^{(h)}_{4}$ is defined differently, depending on whether $h$ is larger or smaller than $0$: namely, 
$\LL W^{(h)}_{4}(\xx_1,\xx_2;\yy_1,\yy_{2})=
W^{(h)}_{4}(\xx_1,\xx_2;\yy_1,\yy_{2})$ if $h\ge 0$, while 
\[ \LL W^{(h)}_{4}(\xx_1,\xx_2;\yy_1,\yy_{2})& =
\int d\xx_2' d\yy_1' d\yy_2' W^{(h)}_{4}(\xx_1,\xx_2';\yy_1',\yy_{2}')\d(\xx_1-\xx_2)\d(\xx_1-\yy_1)\d(\xx_1-\yy_2)\non\\
& =:\, \l_h\,\d(\xx_1-\xx_2)\d(\xx_1-\yy_1)\d(\xx_1-\yy_2)\;,\label{2.36ac} 
\]
if $h<0$. In other words, $\LL V^{(h)}(\ph)$ 
is either the sum of all the ladder subdiagrams with propagators carrying a scale label $\ge h$, if $h\ge 0$, 
or its {\it local part} (in the sense of \eqref{2.36ac}), if $h<0$. 
Moreover, $\RR V^{(h)}(\ph)$ is the rest (the {\it irrelevant} part), i.e., 
the sum over all the field monomials of order $\ge 6$, plus possibly (if $h<0$) the non-local part of the ladder diagrams. 
$\LL V^{(h)}(\ph)$ is thought of as the effective interaction on scale $h$, 
and is treated in the same fashion as the interaction $V(\ph)$.  
The tree expansion is modified accordingly: the modified trees contributing to $V^{(h)}$ can now have endpoints on all scales between $h+2$ and $N+1$,
where an endpoint on scale $h_v$ represents an interaction vertex of type $\LL V^{(h_v-1)}(\ph)$, see Fig.\ref{ren_tree}; if such an endpoint has $h_v\le N$, then the vertex immediately preceding $v$ on $\t$ is necessarily non trivial. An action of the operator $\RR$ is associated with all the vertices 
$v>v_0$ that are not endpoints:  if $|P_v|>4$, such an action is trivial (i.e., it acts as the identity), while if $|P_v|=4$ and $h_v<0$, $\RR$ 
extracts the non-local part of the value of the subtree it acts on; if $|P_v|=4$ and $h_v\ge 0$, the action of $\RR$ kills the whole value of the subtree 
it acts on, i.e., it makes it vanish: therefore, we can freely decide that the modified trees are not allowed to have $|P_v|=4$ on vertices such that $h_v\ge 0$,
and we shall do so in the following. 

%

We denote the family of the modified trees with $n$ endpoints 
by $\TT^{(h)}_{N;n}$. In terms of these new trees, the single-scale contribution to the kernels of the effective potential 
admit a representation that is very similar to \eqref{2.23}--\eqref{2.25x}: with some abuse of notation, we write it in the form:
\[
W_{2m}^{(h)}(\underline\xx;\underline\yy)= \sum_{n\geq 1} \sum_{\t\in\TT_{N;n}^{(h)}}  \sideset{}{^{(P_{v_0})}}\sum_{{\bf P}\in \PP_\t}
\int \hskip-.2truecm\prod_{f\in I_{v_0}\setminus \{f^*\}}\hskip-.4truecm d\xx_f\
W_{}^{(h)}(\t,{\bf P};\underline{\xx}_{v_0})\label{2.37}\]
with 
\[W_{}^{(h)}(\t,{\bf P};\underline\xx_{v_0})=\sum_{\GG\in\G(\t,{\bf P})}\Val(\GG)\;,\label{2.37yt}\]
$\G(\t,{\bf P})$ the set of connected labelled Feynman diagrams compatible with the renormalized tree $\t \in \TT_{N;n}^{(h)}$ and the set ${\bf P}$, and
\[\Val(\GG)
=&(-1)^{n+1}\prod_{\substack{v \text{ not e.p.}\\ v>v_0}} \frac{\RR^{\a_v}}{s_v!}\Biggl[
\Big(\prod_{\ell\in\GG_v}G_{h_v}(\xx_{\ell_-}-\xx_{\ell_+})\Big)\cdot\non\\
&\cdot
\Big(\prod_{\substack{v^*\ {\rm e.p.}:\\ v^*>v,\ h_{v^*}=h_v+1}} \LL W_4^{(h_{v^*}-1)}(\xx_{1,v^*},\xx_{2,v^*};\yy_{1,v^*},\yy_{2,v^*})\Big)\Biggr]\;,
\label{2.38} \]
where $\a_v=0$ if $v=v_0$, and otherwise $\a_v=1$. Moreover, it is understood that the operators
$\RR$ act in the order induced by the tree ordering (i.e.,
starting from the endpoints and moving toward the root).

In the next subsection we will show that 
\be \Big| \int d\xx_2d\yy_1d\yy_2 \LL W^{(h)}_{4}(\xx_1,\xx_2;\yy_1,\yy_{2})\Big|\le 
({\rm const.})|\l|\;,\label{2.39}\ee
uniformly in $h$ and $N$. This is enough to show that the modified expansion is order by order convergent,
uniformly in the ultraviolet cutoff, with $n!$-bounds on the $n$-th order of the free energy and of the kernels of the effective potential. 
Let us explain why this is the case. 
We start from \eqref{2.37}--\eqref{2.38} and 
proceed as in the proof of \eqref{2.35xs}. As before, we need to estimate the integral over $\xx_{v_0}\setminus \xx_{f^*}$
of the values of Feynman diagrams, and then sum over
$\GG\in\G(\t,{\bf P})$, over ${\bf P}$ and over $\t$.  The contributions from the Feynman diagrams in $\G(\t,{\bf P})$, with ${\bf P}$ such that 
$|P_v|\ge 6$ for all vertices $v$ of $\t$ that are not endpoints, are the easiest to estimate. In fact, the action of the operators $\RR$ in the formula 
\eqref{2.38} for the value of any such diagram is trivial, i.e., they act as the identity on all vertices. Therefore, the estimate goes over exactly as in the
proof of \eqref{2.35xs}, with the only difference that every endpoint $v^*$ is associated with an integral like the one in the l.h.s. of \eqref{2.39}, with $h=h_{v^*}-1$, rather than with $\frac{\l}2\int d\xx w(\xx)$: of course, this makes no qualitative difference because, thanks to \eqref{2.39}, both integrals 
are bounded by (const.)$|\l|$. In conclusion, the overall contribution from these diagrams on $W_{2m}^{(h)}(\underline\xx;\underline\yy)$ (let us call it $W_{2m}^{(h;1)}(\underline\xx;\underline\yy)$)
is bounded by an expression similar to  \eqref{2.35xs}, namely 
\[ ||W^{(h;1)}_{2m}|| \le \g^{h(2-m)}\sum_{n\geq 1} |\l|^n \tilde K^n n! \sum_{\t\in\TT_{N;n}^{(h)}}\; \sideset{}{^{*(P_{v_0})}}\sum_{{\bf P}\in\PP_\t}
\prod_{v\ {\rm not}\ {\rm e.p.}}\g^{(h_v-h_{v'})(2-\frac{|P_v|}{2})}\label{2.42}\]
where the $*(P_{v_0})$ on the sum indicates the constraint that we are not summing over $P_{v_0}$ and $|P_{v_0}|=2m$, and moreover
 $|P_v|\ge 6$ for all the vertices that are not endpoints. Since 
$|P_v|\ge 6$, the exponential  factors 
$\g^{(h_v-h_{v'})(2-\frac{|P_v|}{2})}$ in the r.h.s. of \eqref{2.42} are summable both 
over $\{h_v\}$ and over $\{P_v\}$, which leads to an $n!$ bound of the form:
\[ ||W^{(h;1)}_{2m}||  \le \g^{h(2-m)}\sum_{n\geq 1} |\l|^n (K')^n n!\;,\label{2.43}\]
as desired (here and below a bound like this, involving a non-summable series like the one in the r.h.s., should be interpreted as a bound valid order by order 
on the coefficients of the perturbative expansion of the l.h.s.). We are left with the contributions from trees and field labels such that $|P_v|=4$ for some 
vertex $v$ with $h_v<0$ that is not an endpoint; such vertices are associated with a non-trivial action of $\RR$. We want to show that such an action 
corresponds, from the dimensional point of view, to a gain factor $\g^{-(h_v-h_{v'})/2}$, where $v'$ is the vertex immediately preceding $v$ on $\t$; this 
dimensional gain is enough to make the product of the exponential  factors 
$\g^{(h_v-h_{v'})(2-\frac{|P_v|}{2})}$ convergent over ${\bf P}$ and $\t$. The emergence of this gain factor has been discussed several times in the literature,
see e.g.\cite{GalReview,BGbook,GM}. Here we briefly explain the mechanism leading to it, in order to make our exposition as self-consistent as possible. 

In order to explain the effect of $\RR$, 
let us first concentrate on a vertex $v$ that is not endpoint with $|P_v|=4$ and $h_v<0$, such that there are no vertices $w>v$ with the same 
property. This means that the action of all the $\RR$'s associated with the vertices $w>v$ that are not endpoints is equal to the identity. Therefore, the contribution from the  
subtree with root $v'$, once integrated over $\tilde{\underline{\xx}}_v:=\underline{\xx}_v\setminus \cup_{f\in P_v}\{\xx_f\}$ is 
\[&F_v(\xx_1,\xx_2;\yy_1,\yy_2)=(-1)^{n_v}\int d\tilde{\underline{\xx}}_v
\prod_{\substack{w \text{ not e.p.}:\\ w>v}} \frac{1}{s_w!}\Biggl[
\Big(\prod_{\ell\in\GG_w}G_{h_w}(\xx_{\ell_-}-\xx_{\ell_+})\Big)\cdot\non\\
&\cdot
\Big(\prod_{\substack{v^*\ {\rm e.p.}:\\ v^*>w,\ h_{v^*}=h_w+1}} \LL W_4^{(h_{v^*}-1)}(\xx_{1,v^*},\xx_{2,v^*};\yy_{1,v^*},\yy_{2,v^*})\Big)\Biggr]\;,
 \]
where $\xx_1,\xx_2$ (resp. $\yy_1,\yy_2$) are the elements of $\cup_{f\in P_v}\{\xx_f\}$ such that $\s_f=+$ (resp. $\s_f=-$).
Clearly, the $L_1$ norm of this expression admits a bound analogous to $\|W_{2m}^{(h;1)}\|$, namely 
\[\int d\xx_2 d\yy_1 d\yy_2 |F_v(\xx_1,\xx_2;\yy_1,\yy_2)| \le   C^{n_v}|\l|^{n_v}n_v!\;.\label{2.43cd}\]
Similarly, we obtain 
\[&\int d\xx_2 d\yy_1 d\yy_2 \Big[\prod_{j=0}^2|x_{2,j}-x_{1,j}|^{m_j}\Big]
|F_v(\xx_1,\xx_2;\yy_1,\yy_2)| \le \non\\
&\qquad \le C_{m_0,m_1,m_2}
\g^{-h_{v}(m_0+\frac12(m_1+m_2))} C^{n_v}|\l|^{n_v}n_v!\;,\label{2.43ce}\]
which will turn out to be useful below\footnote{In order to prove \eqref{2.43ce}, it is enough to rewrite each factor $x_{2,0}-x_{1,0}$, etc., as a sum of 
differences $x_{\ell-,0}-x_{\ell+,0}$ along the spanning tree $\cup_{w>v}T_w$ (see the discussion preceding \eqref{2.31}), and to recognize that each term $x_{\ell-,0}-x_{\ell+,0}$ goes together with the corresponding propagator 
$G_{h_w}(\xx_{\ell-}-\xx_{\ell+})$; therefore, in the analogue of \eqref{2.31}, the $L_1$ norm of $G_{h_w}(\xx_{\ell-}-\xx_{\ell+})$ is replaced by 
the one of $(x_{\ell-,0}-x_{\ell+,0})G_{h_w}(\xx_{\ell-}-\xx_{\ell+})$, which is $({\rm const.})\g^{-h_{w}}||G_{h_w}||_1\le ({\rm const.})\g^{-h_{v}}||G_{h_w}||_1$.
Similar considerations are valid with $x_{2,0}-x_{1,0}$ replaced by $x_{2,1}-x_{1,1}$, etc.}.
Similar estimates are valid with $x_{2,j}-x_{1,j}$ replaced by $y_{1,j}-x_{1,j}$, etc. 
Let us now evaluate the action of  $\RR$ on $F_v(\xx_1,\xx_2;\yy_1,\yy_2)$:
\[ & \RR F_v(\xx_1,\xx_2;\yy_1,\yy_2)=F_v(\xx_1,\xx_2;\yy_1,\yy_2)-\\
&\qquad -\d(\xx_1-\xx_2)\d(\xx_1-\yy_1)\d(\xx_1-\yy_2)
\int d\xx_2'd\yy_1'd\yy_2'F_v(\xx_1,\xx_2';\yy_1',\yy_2')\;.\non
\]
On top of that, in the expression \eqref{2.38}, $\RR F_v(\xx_1,\xx_2;\yy_1,\yy_2)$ appears multiplied by the four propagators 
$G_{h_1}(\xx_{\ell_{1,-}}-\xx_1)$, $G_{h_2}(\xx_{\ell_{2,-}}-\xx_2)$, $G_{h_1'}(\yy_1-\xx_{\ell_{1,+}'})$, $G_{h_2'}(\yy_2-\xx_{\ell_{2,+}'})$
associated with the fields in $P_v$\footnote{Actually, it could also happen that some of the fields in $P_{v}$ are 
also in $P_{v_0}$, in which case the corresponding propagators should be replaced by external fields. For simplicity, 
we exclude this possibility here.}; here $h_1$ is the scale of the vertex $v_1$ such that $\ell_1\in\GG_{v_1}$ and 
$\xx_{\ell_{1,+}}\equiv \xx_1$, and similarly for the others. 
Note that $h_1,\ldots,h_4<h_v$. Moreover, when we evaluate $||W_{2m}^{(h)}||$, we also need to integrate over three out of the four coordinates $\xx_1,\ldots,\yy_2$, 
and by translation invariance we can arbitrarily choose the one 
we are not integrating over, say $\xx_1$.
In conclusion, we can isolate from the contribution under consideration to $\|W_{2m}^{(h)}\|$ the following expression:
\[ & \int d\xx_2d\yy_1 d\yy_2 G_{h_1}(\xx_{\ell_{1,-}}-\xx_1)\cdots
G_{h_2'}(\yy_2-\xx_{\ell_{2,+}'})
\RR F_v(\xx_1,\xx_2;\yy_1,\yy_2)=\label{2.49}\\
&=\int d\xx_2d\yy_1 d\yy_2 \Big[G_{h_1}(\xx_{\ell_{1,-}}-\xx_1)G_{h_2}(\xx_{\ell_{2,-}}-\xx_2)G_{h_1'}(\yy_1-\xx_{\ell_{1,+}'})
G_{h_2'}(\yy_2-\xx_{\ell_{2,+}'})\non-\\
&-
G_{h_1}(\xx_{\ell_{1,-}}-\xx_1)G_{h_2}(\xx_{\ell_{2,-}}-\xx_1)G_{h_1'}(\xx_1-\xx_{\ell_{1,+}'})
G_{h_2'}(\xx_1-\xx_{\ell_{2,+}'})\Big]
F_v(\xx_1,\xx_2;\yy_1,\yy_2)\;.\non\]
Note that the expression in square brackets is the difference between the original product of propagators and a similar product where 
all the space-time points $\xx_1,\xx_2,\yy_1,\yy_2$ have been ``localized" to $\xx_1$. Such expression can be equivalently rewritten as
\[\Big[G_{h_1}(\xx_{\ell_{1,-}}-\xx_1)\Big(G_{h_2}(\xx_{\ell_{2,-}}-\xx_2)-G_{h_2}(\xx_{\ell_{2,-}}-\xx_1)\Big)
G_{h_1'}(\yy_1-\xx_{\ell_{1,+}'})
G_{h_2'}(\yy_2-\xx_{\ell_{2,+}'})+\non\\
+G_{h_1}(\xx_{\ell_{1,-}}-\xx_1)G_{h_2}(\xx_{\ell_{2,-}}-\xx_1)\Big(
G_{h_1'}(\yy_1-\xx_{\ell_{1,+}'})-G_{h_1'}(\xx_1-\xx_{\ell_{1,+}'})\Big)
G_{h_2'}(\yy_2-\xx_{\ell_{2,+}'})
+\non\\
+G_{h_1}(\xx_{\ell_{1,-}}-\xx_1)G_{h_2}(\xx_{\ell_{2,-}}-\xx_1)
G_{h_1'}(\xx_1-\xx_{\ell_{1,+}'})\Big(G_{h_2'}(\yy_2-\xx_{\ell_{2,+}'})-G_{h_2'}(\xx_1-\xx_{\ell_{2,+}'})\Big)\Big]\non
\]
and the differences in parentheses can be further rewritten in interpolated form as
\[\Big(G_{h_2}(\xx_{\ell_{2,-}}-\xx_2)-G_{h_2}(\xx_{\ell_{2,-}}-\xx_1)\Big)=-\int_0^1 ds\,(\xx_2-\xx_1)\cdot \boldsymbol{\partial}G_{h_1}(\xx_{\ell_{2,-}}-\xx_{12}(s))
\;,\]
where $\xx_{12}(s):=\xx_1+s(\xx_2-\xx_1)$ and similar expressions are valid for the other two parentheses. Plugging this back into \eqref{2.49},
we put the factor $x_{2,j}-x_{1,j}$ together with $F_v(\xx_1,\xx_2;\yy_1,\yy_2)$, from which we see that dimensionally it corresponds to a factor 
$\g^{-h_v}$ or $\g^{-h_v/2}$, depending on whether $j=0$ or $j>0$, see \eqref{2.43ce}. Moreover, the derivative $\dpr_j$ acting on $G_{h_1}$ dimensionally corresponds to a factor $\g^{h_1}$ or $\g^{h_1/2}$, depending on whether $j=0$ or $j>0$, simply because 
$||\partial_0 G_h||_1\le ({\rm const.})\g^{h}||G_h||_1$,
$||\partial_1 G_h||_1\le ({\rm const.})\g^{h/2}||G_h||_1$, etc. 
Putting things together, we see that the action of $\RR$ is dimensionally bounded by a factor that is at least $\g^{-(h_v-\max_i\{h_i\})/2}$, as desired. The argument 
explained here for an operator $\RR$ acting on a vertex $v$ that is not followed by other vertices such that $\RR\neq1$, can then be iterated until the root is reached. 
In this way, each vertex that is renormalized non-trivially gains a factor $\g^{(h_v-h')/2}$, where $h'<h_v$. Note that 
in the iterative procedure described above 
 it may happen that some of the derivatives coming from the interpolations act on the external fields. This means that 
the expansion \eqref{2.14} should be replaced by 
\[ V^{(h)}(\ph)=&\sum_{m\ge 1}\sum_{{\boldsymbol{\a}}_{1},\ldots,{\boldsymbol{\alpha}}_m'}\int_{\RRR^{6m}}d\xx_1\cdots d\yy_{m} 
W^{(h)}_{2m;{\boldsymbol{\alpha}}_1,\ldots,{\boldsymbol{\a}}_m'}(\xx_1,\ldots,\xx_m;\yy_1,\cdots,\yy_{m})\cdot\non\\
&\cdot\dpr_{\xx_{1}}^{{\boldsymbol{\a}}_{1}}\ph_{\xx_1}^+\cdots
\dpr_{\xx_{m}}^{{\boldsymbol{\a}}_{m}}\ph^+_{\xx_m}\dpr_{\yy_{1}}^{{\boldsymbol{\a}}_{1}'}\ph^-_{\yy_{1}}\cdots
\dpr_{\yy_{m}}^{{\boldsymbol{\a}}_{m}'}
\ph^-_{\yy_{m}}\label{2.54y}\]
where $\boldsymbol{\a}_i=(\a_i^0,\a_i^1,\a_i^2)$, and $\dpr_{\xx_{i}}^{{\boldsymbol{\a}}_{i}}=\dpr_{x_{i,0}}^{\a_i^0}\dpr_{x_{i,1}}^{\a_i^1}\dpr_{x_{i,2}}^{\a_i^2}$, 
and similary for $\dpr_{\yy_{i}}^{{\boldsymbol{\a}}_{i}'}$. As a side remark, it can be shown that the summation over ${\boldsymbol{\a}}_i,{\boldsymbol{\a}}_i'$
can be restricted to the indices such that $\a_i^j,\a'^{j}_i\in\{0,1\}$, see \cite[Section 3.3]{XYZ}. The result is 
\[ ||W^{(h)}_{2m;{\boldsymbol{\alpha}}_1,\ldots,{\boldsymbol{\a}}_m'}|| \le  & \; 
 \g^{\lft(2-m- ||{\boldsymbol{\a}}^0||
-\frac12 ||\vec{\boldsymbol{\a}}||) \rgt)h} \label{2.55} \\
 & \cdot \sum_{n\geq 1} |\l|^n \tilde K^n n! \sum_{\t\in\TT_{N;n}^{(h)}} \sideset{}{^{(P_{v_0})}}\sum_{{\bf P}\in\PP_\t}\hskip -0.3cm
\prod_{v\ {\rm not}\ {\rm e.p.}}\g^{(h_v-h_{v'})(2-\frac{|P_v|}{2}-z(P_v))}\;, \non \]
where $||{\boldsymbol{\a}}^0||:= \sum_{i} (\a^0_i +\a'^0_i)$,
$||\vec {\boldsymbol{\a}}||:= \sum_{i}\sum_{j=1,2} (\a^j_i +\a'^j_i)$, and
\[ z(P_v)=\begin{cases} 1/2\ {\rm if}\ |P_v|=4,\\
0\ {\rm otherwise}\end{cases}\label{2.54}\]
and we recall that $v'$ is the vertex immediately preceding $v$ on $\t$\footnote{
A complete proof of \eqref{2.55} requires the discussion of a few other technical points, including the change of variables from the original to the interpolated variables, as well as the possible accumulation of the derivatives coming from the interpolation procedure on a given propagator (which may a priori
worsen the combinatorial factors in \eqref{2.55}). For these and other closely related issues we refer the reader to previous literature, see in particular \cite[Section 3]{XYZ}.}. Note that now the renormalized scaling dimension $2-\frac{|P_v|}{2}-z(P_v)$ appearing at exponent in \eqref{2.55} is strictly negative, for all admissible $P_v$'s; therefore, the r.h.s. of \eqref{2.55}  is summable both over $\{h_v\}$ and over $\{P_v\}$, and we get
\[  ||W^{(h)}_{2m;{\boldsymbol{\alpha}}_1,\ldots,{\boldsymbol{\a}}_m'}|| \le 
 \g^{\lft(2-m- ||{\boldsymbol{\a}}^0||
-\frac12 ||\vec{\boldsymbol{\a}}||) \rgt)h}  \sum_{n\geq 1} |\l|^n \tilde K^n n! ;,\label{2.54x}\]
as desired. 

\subsection{The flow of the effective interaction}\label{s2.fl}

In this section we prove \eqref{2.39}, by distinguishing the ultraviolet ($h\ge 0$) and infrared ($h<0$) regimes. 

\subsubsection*{The ultraviolet regime}

We recall that if $h\ge 0$ then $\LL W^{(h)}_{4}(\xx_1,\xx_2;\yy_1,\yy_{2}):= W^{(h)}_{4}(\xx_1,\xx_2;\yy_1,\yy_{2})$ and that $W^{(h)}_4$ 
is the sum of all the ladder diagrams as in Fig.\ref{ladder} with scale labels larger than $h$ and lower than $N+1$. In this particular case, $W^{(h)}_4$
admits an explicit expression in the following form:
\[ &W^{(h)}_{4}(\xx_1,\yy_1;\xx,\yy)=\frac{\l}2w(\xx-\yy)\d(\xx-\xx_1)\d(\yy-\yy_1)-\frac12\sum_{n\ge 2}(-\l)^{n}\int d\xx_2\cdots d\xx_n\cdot\non\\
&\cdot\int d\yy_2 \cdots d\yy_n
\d(\xx-\xx_n)\d(\yy-\yy_n)\,G_{[h+1,N]}(\xx_1-\xx_2)\cdots G_{[h+1,N]}(\xx_{n-1}-\xx_n)\cdot\non\\
&\cdot G_{[h+1,N]}(\yy_1-\yy_2)\cdots G_{[h+1,N]}(\yy_{n-1}-\yy_n)
w(\xx_1-\yy_1)\cdots w(\xx_n-\yy_n) \;,\label{2.58}\]
where $G_{[h+1,N]}(\xx):=\sum_{h<k\le N}G_k(\xx)$. The replacement of the single scale propagator by $G_{[h+1,N]}(\xx)$ results from the sum over the 
scale labels of the special class of labelled diagrams we are looking at. 
Using the fact that $w(\xx)=\d(x_0)v(\vx)$, we can immediately integrate out the $y_{i,0}$'s, so that 
\[||W^{(h)}_4||\le \frac12\sum_{n\ge 1}|\l|^n ||v||_1\cdot ||v||^{n-1}_\infty\cdot ||G_{[h+1,N]}||_1^{n-1} \big[\sup_{x_0}\int d\vx\, G_{[h+1,N]}(x_0,\vx)\big]^{n-1}\;,
\label{2.59}\]
where $||v||_1=\int d\vx\, v(\vx)$ and $||v||_\infty=\sup_{\vx}|v(\vx)|$. From \eqref{2.Gh} we see that 
\[||G_{[h+1,N]}||_1\le A'\g^{-h}\,, \label{2.GhN1}\] 
for a suitable $A'>0$. 
Moreover, using \eqref{2.g}, we also find 
\[ \label{2.GhN2}
||G_{[h+1,N]}(x_0,\cdot)||_1:=\int d\vx\, G_{[h+1,N]}(x_0,\vx)\le C\,,
\] 
uniformly in $x_0$, so that 
\[||W^{(h)}_4||\le \frac{|\l|}2||v||_1\sum_{n\ge 1} (K|\l|)^{n-1}\g^{-h(n-1)}\;,\label{2.60}\]
for $K=A'C$, which is the desired estimate for $h\ge 0$. 


\feyn{
\begin{fmffile}{BoseFeyn/ladder_gain}
 \unitlength = 1cm
\def\myl#1{7cm}
 \fmfset{arrow_len}{3mm}
\begin{align*}
\parbox{\myl}{\centering{  \hskip -1cm
\begin{fmfgraph*}(8,1.2)
			\fmfleft{i1,i2}
			\fmfright{o1,o2}
			\fmf{fermion, tension=1.2}{i1,v1}
			\fmf{fermion, tension=1.2}{i2,v2}
				\fmf{fermion}{v1,v3}		
				\fmf{fermion, foreground=(0.6,,0.6,,0.6)}{v2,v4}
				\fmf{dashes, tension=0.6}{v3,v5}		
				\fmf{dashes, tension=0.6, foreground=(0.6,,0.6,,0.6)}{v4,v6}
				\fmf{fermion}{v5,v7}		
				\fmf{fermion, foreground=(0.6,,0.6,,0.6)}{v6,v8}
                  			 \fmf{wiggly, tension=0.01}{v1,v2}
					\fmf{wiggly,tension=0.01}{v3,v4}
					\fmf{wiggly, tension=0.01}{v5,v6}
					\fmf{wiggly,tension=0.01}{v7,v8}
			\fmf{fermion, tension=1.2}{v7,o1}
			\fmf{fermion, tension=1.2}{v8,o2}
                   \fmfdot{v1,v2,v3,v4,v5,v6,v7,v8}
 	        \fmfv{label=$\xx_n$,  label.angle=-90}{v1}
 		\fmfv{label=$\xx_{n-1}$,  label.angle=-90}{v3}
 		\fmfv{label=$\xx_{2}$,  label.angle=-90}{v5}
 		\fmfv{label=$\xx_1$,  label.angle=-90}{v7}
 		\fmfv{label=$\yy_n$,  label.angle=90}{v2}
 		\fmfv{label=$\yy_{n-1}$,  label.angle=90}{v4}
 		\fmfv{label=$\yy_{2}$,  label.angle=90}{v6}
 		\fmfv{label=$\yy_1$,  label.angle=90}{v8}
		\end{fmfgraph*}  }}
\parbox{\myl}{\centering{  
\begin{fmfgraph*}(8,1.2)
			\fmfleft{i1,i2}
			\fmfright{o1,o2}
			\fmf{fermion, tension=1.2}{i1,v1}
			\fmf{fermion, tension=1.2}{i2,v2}
				\fmf{fermion}{v1,v3}		
				\fmf{fermion}{v2,v4}
				\fmf{dashes, tension=0.6}{v3,v5}		
				\fmf{dashes, tension=0.6}{v4,v6}
				\fmf{fermion}{v5,v7}		
				\fmf{fermion}{v6,v8}
                  			 \fmf{wiggly, tension=0.01}{v1,v2}
					\fmf{wiggly,tension=0.01, foreground=(0.6,,0.6,,0.6)}{v3,v4}
					\fmf{wiggly, tension=0.01, foreground=(0.6,,0.6,,0.6)}{v5,v6}
					\fmf{wiggly,tension=0.01, foreground=(0.6,,0.6,,0.6)}{v7,v8}
			\fmf{fermion, tension=1.2}{v7,o1}
			\fmf{fermion, tension=1.2}{v8,o2}
                   \fmfdot{v1,v2,v3,v4,v5,v6,v7,v8}
 	        \fmfv{label=$\xx_n$,  label.angle=-90}{v1}
 		\fmfv{label=$\xx_{n-1}$,  label.angle=-90}{v3}
 		\fmfv{label=$\xx_{2}$,  label.angle=-90}{v5}
 		\fmfv{label=$\xx_2$,  label.angle=-90}{v7}
 		\fmfv{label=$\yy_n$,  label.angle=90}{v2}
 		\fmfv{label=$\yy_{n-1}$,  label.angle=90}{v4}
 		\fmfv{label=$\yy_{2}$,  label.angle=90}{v6}
 		\fmfv{label=$\yy_1$,  label.angle=90}{v8}
		\end{fmfgraph*}  }}
\end{align*}
\end{fmffile}	
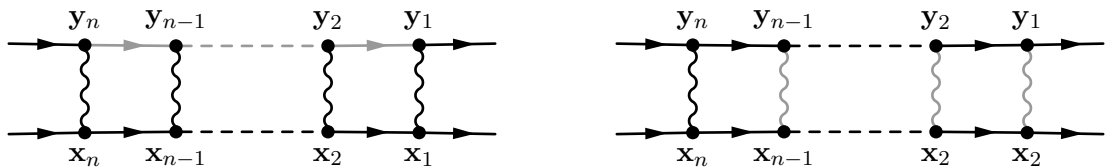
}{Two different choices of the spanning tree for a ladder graph of order $n$. The lines not belonging to the spanning tree are depicted in gray. In the left side we depict the usual choice of the spanning tree, as described after~\eqref{2.29}. In the right side we depict an alternative choice, which is convenient in the ultraviolet regime $h\ge 0$.
}{ladder_gain}


The idea behind the bound \eqref{2.60}  is that, due to the structure of the ladder graphs, 
we can choose a ``non usual'' spanning tree which allows us to improve the naive dimensional estimate \eqref{2.35xs}, whenever $h\geq0$. According to the ``usual'' procedure for the derivation of the dimensional bounds, \ie the one presented after \eqref{2.29}, the wiggly lines all belongs to the spanning tree; this means that we get a factor  $||v||_1$ for each interaction $\o(\xx - \yy)$, while the propagators not belonging to the spanning tree are bounded by 
their $L_\io$ norm, see \eqref{2.31}. 
From \eqref{2.59} it is apparent that in bounding the left side we followed a different procedure: 
only one of the wiggly lines was chosen to belong to the spanning tree, while the remaining $(n-1)$ interaction terms $v(\vx_i - \vy_i)$ were bounded by their $L_\io$ norms. The connection among the remaining space-time points of the ladder is guaranteed by the propagators, see the l.h.s. of Fig.\ref{ladder_gain}. We conclude by noticing that the dependence of $W_4^{(h)}$ on the ultraviolet cutoff $N$  is exponentially small, thanks to the dimensional gain $\g^{-h}$ 
appearing in the previous bounds, see \eqref{2.60}.

\subsubsection*{The infrared regime}

%
\feyn{
\begin{align*}
\parbox{3.2cm}{\centering{ 
\begin{tikzpicture} [scale=0.7]
\draw[med]  (0,0) circle(1cm);
\draw (-0.1,0.1) node[label=$\xx_1$]{};
\draw (0.5,-0.4) node[label=$\xx_2$]{};
\draw (-0.45,0.35) node[vertex](x1){};  
\draw (0.4,0.55) node[vertex](x2){};  
\draw (-0.4,-0.75) node[label=$\yy_1$]{};
\draw (0.2,-1.2) node[label=$\yy_2$]{};
\draw (-0.4,-0.5) node[vertex](y1){};   
\draw (0.4,-0.25) node[vertex](y2){};   
\draw[arr={latex},thick]  (x1) --(-1.6, 1.4);
\draw[arr={latex},thick]  (x2) --(1.6, 1.4);
\draw[arr={latex},thick]  (y1) --(-1.6, -1.4);
\draw[arr={latex},thick]  (y2) --(1.6, -1.4);
\end{tikzpicture} \\[0.2cm]
$h\geq0$
}}
\begin{fmffile}{BoseFeyn/loc_QCP}
 \unitlength = 0.8cm
\fmfset{arrow_len}{3mm}
\parbox{5cm}{\centering{   \skip 0.6cm 
		\begin{fmfgraph*}(3,1.8)   
			\fmfleft{i1,i2}
			\fmfright{o1,o2}
			\fmf{fermion}{i1,v1}
                    \fmf{fermion}{i2,v1}
                   \fmf{fermion}{v1,o1}
                   \fmf{fermion}{v1,o2}
	           \fmfdot{v1}
		\end{fmfgraph*} \\[0.4cm]
             $h < 0$
		}} 
\end{fmffile}	
\end{align*}
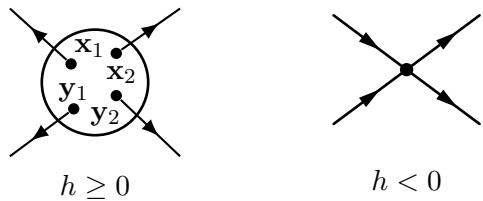
}{Graphical representation of the effective interactions at scale $h$, in the ultraviolet (l.h.s) and infrared (r.h.s.) regimes, see \eqref{2.effIR} and \eqref{2.effUV} respectively.}{loc_QCP}
%


\vskip -0.5cm

If $h<0$ the action of $\LL$ on $W^{(h)}_4$ is non trivial, see \eqref{2.36ac}. In order to prove \eqref{2.39}, we focus on the  {\it effective interaction}
\[\l_h=\int d\xx_2d\yy_1 d\yy_2 W^{(h)}_4(\xx_1,\xx_2;\yy_1,\yy_2)\;,\qquad h\le 0\;,\]
defined in \eqref{2.36ac}, 
and we note that the very definition of $W^{(h)}_4$ induces a recursive equation (``flow equation") on $\l_h$:
\[\l_{h-1}=\l_h+\b^\l_h(\l_h,\ldots,\l_{-1},\l)\;,\qquad h\le 0\;,\label{2.62}\]
where $\b^\l_h$ is the local part (i.e., the integral over $\xx_2,\yy_1,\yy_2$) of the sum of the values of the diagrams associated with renormalized trees and 
field labels such that: 
the root is on scale $h-1$, $|P_{v_0}|=4$ and $\cup_{i=1}^{s_{v_0}}P_{v_i}\neq P_{v_0}$, where $v_1,\ldots,v_{s_{v_0}}$ are the vertices immediately following $v_0$ on $\t$ (note that these conditions guarantee that the number of endpoints is $\ge 2$). Every endpoint $v^*_i$ on a tree 
contributing to $\b^\l_h$ is associated with an interaction \[
\l_{h_{v^*_i}-1}\int d\xx |\ph_\xx|^4 \qquad {\rm if }\; h_{v^*_i} \le 0\;, \label{2.effUV}\]
 or with 
\[\int d\xx_1\cdots d\yy_{2} W^{(h_{v^*_i}-1)}_{4}(\xx_1,\xx_2;\yy_1,\yy_{2})\ph_{\xx_1}^+
\ph^+_{\xx_2}\ph^-_{\yy_{1}}\ph^-_{\yy_{2}} \qquad {\rm if}\; h_{v^*_i}> 0\;, \label{2.effIR} 
\]
see Fig.\ref{loc_QCP} for a graphical representation of the two vertices. In the latter equation $W^{(h)}_{4}$ is the explicit function of $\l$ in \eqref{2.58}. Therefore, every tree contributing to $\b^\l_h$ carries a natural 
dependence on $\l_{h_{v^*_i}-1}$, coming from the endpoints on scale $\le 0$, and on $\l$, coming from the endpoints on scale $>0$: therefore, 
as indicated in 
\eqref{2.62}, we think of $\b^\l_h$ as a function of $\l_{h},\ldots,\l_{-1},\l$. 
In the RG jargon, $\b^\l_h$ is called the {\it beta function}
for $\l_h$.


\feyn{
\begin{align*}
\b_h^{\l;(2)} & = 
\parbox{4.5cm}{\centering{ \vskip 0.6cm
\begin{tikzpicture}[scale=0.8]
\foreach \x in {-1,...,2}   
    {
        \draw[very thin] (\x , -1.2) -- (\x , 1.2);      
    }
\draw (-0.3,-0.3) node {$v_0$};
\draw (-1,-1.8) node {$$};  
\draw (0,-1.8) node {$h$};  
\draw (1,-1.8) node {$$};
\draw (2,-1.8) node {$h+2$};
\draw (-0.25,0.3) node {$\LL$};
\draw[med]  (-1,0) node[vertex] {}-- (0,0) node[vertex] {} -- (1,0.5) node[vertex, label=0:$\l_{h}$] {};  
\draw[med] (0,0) node[vertex] {} -- (1,-0.5) node[vertex, label=0:$\l_{h}$]{};
\end{tikzpicture}
}}  \hskip -0.5cm +
\parbox{5cm}{\centering{ \vskip 0.6cm
\begin{tikzpicture}[scale=0.8]
\foreach \x in {-1,...,2}   
    {
        \draw[very thin] (\x , -1.2) -- (\x , 1.2);      
    }
\draw (-0.3,-0.3) node {$v_0$};
\draw (-1,-1.8) node {$$};  
\draw (0,-1.8) node {$h$};  
\draw (1,-1.8) node {$$};
\draw (2,-1.8) node {$h+2$};
\draw (-0.25,0.3) node {$\LL$};
\draw (0.75,0.3) node {$\RR$};
\draw[med]  (-1,0) node[vertex] {}-- (0,0) node[vertex] {} -- (1,0) node[vertex] {} -- (2,0.5)node[vertex, label=0:$\l_{h+1}$] {};  
\draw[med] (1,0) node[vertex] {} -- (2,-0.5) node[vertex, label=0:$\l_{h+1}$]{};
\end{tikzpicture}
}} \\[0.5cm]
%
%
& = 
\begin{fmffile}{BoseFeyn/beta_QCP}
 \unitlength = 0.8cm
\def\myl#1{3.5cm}
\fmfset{arrow_len}{3mm}
\parbox{4cm}{\centering{ 
		\begin{fmfgraph*}(3.5,1.8) 
			\fmfleft{i1,i2}
			\fmfright{o1,o2}
                   \fmf{fermion}{i1,v1}
			\fmf{fermion}{i2,v1}
			\fmf{fermion, right=0.7, tension=0.5,label=$h$}{v1,v2}
                    \fmf{fermion, left=0.7, tension=0.5,label=$h$}{v1,v2}
			\fmf{fermion}{v2,o1}	
			\fmf{fermion}{v2,o2}				
                    \fmfdot{v1,v2}
                   \fmfv{label=$\;\l_{h}\;$, label.angle=158}{v1}
			\fmfv{label=$\;\l_{h}\;$, label.angle=-22}{v2}
		\end{fmfgraph*}
	}} + \hskip 0.2cm
\parbox{4cm}{\centering{    
		\begin{fmfgraph*}(3.5,1.8)    
			\fmfleft{i1,i2}
			\fmfright{o1,o2}
                   \fmf{fermion}{i1,v1}
			\fmf{fermion}{i2,v1}
			\fmf{fermion, right=0.7, tension=0.5,label=$h+1$}{v1,v2}
                    \fmf{fermion, left=0.7, tension=0.5,label=$h$}{v1,v2}
			\fmf{fermion}{v2,o1}	
			\fmf{fermion}{v2,o2}				
                   \fmfdot{v1,v2}
			\fmfv{label=$\;\l_{h+1}\;$, label.angle=158}{v1}
			\fmfv{label=$\;\l_{h+1}\;$, label.angle=-22}{v2}
		\end{fmfgraph*} 
		}} 
\end{fmffile}	
\end{align*}
}{Labeled trees and Feynman diagrams representation of the second order contribution in $(\l_h,\ldots,\l_{-1},\l)$ to the beta function of $\l_{h-1}$, for $h\leq -2$. The index $\LL$ associated with $v_0$ recalls the  fact that $\l_{h-1}$ is the integral of $\LL W^{h-1}_4$.}  {beta_QCP}


In order to control the flow of $\l_h$ with $h\le -1$, we explicitly compute the second order contribution in $(\l_h,\ldots,\l_{-1},\l)$ to the beta function, see Fig.\ref{beta_QCP}. If $h\le -2$: 
\[
\b^{\l;(2)}_h(\l_h,\ldots,\l_{-1},\l)=-2\int d\xx\, G_h(\xx)\big(\l_h^2G_h(\xx)+2\l_{h+1}^2G_{h+1}(\xx)\big)\;,\label{2.64}
\]
where we used the fact that $u_h(\xx)u_k(\xx)\equiv 0$, $\forall h,k: |h-k|>1$. A similar formula is valid for $h=-1,0$:
\[&
\b^{\l;(2)}_{-1}(\l_{-1},\l)=-2\int d\xx\, G_{-1}(\xx)\big(\l_{-1}^2G_{-1}(\xx)+\frac{\l^2}2\tilde G_{0}(\xx)\big)\;,\label{2.65}\\
& \b^{\l;(2)}_{0}(\l)=-2\int d\xx\, G_{0}(\xx)\big(\frac{\l^2}4\tilde G_{0}(\xx)+\frac{\l^2}2\tilde G_{1}(\xx)\big)\;,\label{2.66}
\]
where $\tilde G_h(\xx):=\int d\vy d\vz\, v(\vy) G_h(x_0,\vy-\vz) v(\vz-\vx)$, and we used the fact that \[W^{(h)}_{4}(\xx_1,\yy_1;\xx,\yy)=\frac{\l}2w(\xx-\yy)\d(\xx-\xx_1)\d(\yy-\yy_1)+O(\l^2)\;,\] as it follows from
\eqref{2.58}--\eqref{2.60}. From the definition of $G_h$, \eqref{2.g}, it is apparent that the second order beta 
function is strictly negative, $\forall h\le 0$. Moreover, using the fact that $\l_{h}=\l_{h+1}+\b^\l_{h+1}=\l_{h+1}+O(\e_{h+1}^2)$, where 
$\e_h:=\max_{k\ge h}\max\{|\l_k|,|\l|\}$, we see that in \eqref{2.64} we can replace $\l_{h+1}$ by $\l_h$, up to higher order corrections; moreover, using the 
scaling property \eqref{2.scale}, we find:
\[
\b^{\l;(2)}_h(\l_h,\ldots,\l_{-1},\l)=-\b_2\l_h^2+O(\e_{h+1}^3)\;,\]
with
\[
\qquad \b_2:=2\int d\xx\, G_0(\xx)\big(G_0(\xx)+2G_{1}(\xx)\big)>0\;.
\]
In conclusion, 
\[\l_{h-1}-\l_{h}=-\b_2\l_h^2+O(\e_h^3)\;,\label{2.70}\]
with initial datum $\l_0=\frac{\l}2(1+O(\l))$. Eq.\eqref{2.70} admits a solution that is bounded by (const.)$|\l|$ uniformly in $h\le 0$, and going to zero as 
$\sim |h|^{-1}$ as $h\to-\infty$. To see this, note that \eqref{2.70} is the finite-difference version of the equation $\dot\l=\b_2\l^2$, up to errors of the order
$O(\l^3)$; 
since the latter ODE has solution $\l(h)=\l_0/(1-\b_2 h\l_0)$ for $h\le 0$, it is easy to conclude that the solution to \eqref{2.70}
has a qualitatively similar behavior. In the RG jargon, such behavior is referred to as {\it asymptotic freedom}.


\pagina

\section{Renormalization group theory of the Condensed State}\label{sec3}

\subsection{The functional integral representation for the perturbation theory around Bogoliubov Hamiltonian}

From now on, we focus on the construction of the interacting condensed state, \ie the interacting theory with propagator \eqref{1.prop} at $\r_0>0$, introduced at 
the end of Section \ref{sec1.A}. As discussed there, it is convenient to perform and analyze the perturbation theory of interest around the Bogoliubov Hamiltonian,
which is supposedly the infrared fixed point for our theory (in the following we shall see that such an expectation must be reconsidered and corrected).
With this purpose in mind, here we re-describe the perturbation theory around Bogoliubov's Hamiltonian in the language of the  functional integral.

We start from the representation \eqref{2.1} for the interacting partition function, with $P^0_\L(d\ph)$ the gaussian measure with propagator \eqref{2.2}
and $\m^0_{\b,L}$ chosen in such a way that $\lim_{\b\to\infty}\lim_{L\to\infty}\b L^2(-\m^0_{\b,L})=1/\r_0$. We write the bosonic field $\ph^\pm_\xx$ as the sum of two independent fields: the first corresponding to its mean value, $\xi^\pm:=|\L|^{-1}\int_\L \ph^\pm_\xx d\xx$, whose interpretation 
is that of  the {\it amplitude} of the condensate; the second  representing the fluctuations around the condensate:
\[ \label{3.1}
\ph^\pm_\xx:= \x^\pm + \ps_\xx^\pm\;,\qquad \ps_\xx^\pm=\frac1{|\L|} \sum_{\kk\in\DD_\L: \kk\neq {\bf 0}}e^{\pm i\kk\xx}\hat \ph^{\pm}_\kk\;,\]
where $\DD_\L$ was defined after \eqref{2.2}. This decomposition induces a corresponding splitting of  
$P^0_\L(d\ph)$ in the form of a product of a gaussian measure on $\xi$ and a gaussian measure on $\psi$.   
After this substitution the interaction potential takes the form 
\[
V_\L(\x+\ps)=  |\L|\Big(\frac{\l}2 \hat{v}(\v0) |\x|^4-\bar\m^B_\L |\x|^2\Big) + \VV_\L(\ps, \x)
\]
with
\[ \label{3.V}
\VV_\L(\ps, \x) & = \frac{\l}{2} \int_{\L^2}\hskip-.2truecm  d\xx d\yy\,w(\xx-\yy) \,\big[ (\x^+)^2 \ps^-_\xx \ps^-_\yy + (\x^-)^2 \ps^+_\xx \ps^+_\yy +2|\x|^2 \ps^+_\xx \ps^-_\yy \big]\\
&-(\bar\m^B_\L-\l|\x|^2\hat v(\v0))\int_\L d\xx|\ps_{\xx}|^2\non\\
&+\frac{\l}{2} \int_{\L^2} d\xx d\yy \, |\ps_{\xx}|^2 \,w(\xx-\yy)\, |\ps_{\yy}|^2   + \l \int_{\L^2} d\xx d\yy\, \big( \ps^+_\xx \x^- +  \x^+\ps^-_\xx \big)\, w(\xx - \yy) |\ps_{\yy}|^2\non\\
&+\bar \n_\L\int_\L d\xx(|\ps_{\xx}|^2+|\x|^2)\;.\non
\]
Here $\hat{v}(\v0)=\int v(\vx) d\vx$ and $\bar \n_\L=-\m_{\b,L}+\m^{0}_{\b,L}+\bar\m^B_\L$ must be chosen in order to fix the condensate density to $\r_0$.
In this language, the Bogoliubov approximation consists in neglecting the last two lines in \eqref{3.V}. Rather than neglecting them, 
we combine the first two lines in the r.h.s. of \eqref{3.V} (which are quadratic in 
$\ps$) with the $\psi$-dependent part of $P^0_\L$: this defines a new reference gaussian measure $P^B_{\x,\L}(d\ps)$, with 
modified propagator. The constant $\bar\m^B_\L$ is another free parameter, which we can play with.
The remaining terms in \eqref{3.V} are treated perturbatively around the new reference measure. More explicitly, 
denoting by $\VV^B_\L(\ps, \x)$ the sum of terms on the last two lines of \eqref{3.V} 
we can rewrite \eqref{2.1} as:
\[
\frac{Z_\L}{Z^0_\L} = \int P_\L(d\x) e^{-|\L|f^B_\L(\x)}
\int P^B_{\x,\L}(d\ps)\,e^{-\VV^B_{\x,\L}(\ps)}\;,\label{3.4}
\]
where (if we define $\m_{\L}^B:=\bar\m^B_\L+\m^0_{\b,L}$)
\[P_\L(d\x)=\frac{d^2\x}{\NN_0}e^{|\L|\big( \m^B_{\L} |\x|^2-\frac{\l}{2}\hv(0) |\x|^4 \big)}\;.\]
Here $d^2\x=d({\rm Re}\,\xi)d({\rm Im}\,\xi)$ and
\[\NN_0=\frac{\p^{3/2}}2\sqrt{\frac{2}{\l\hat v(\v0)|\L|}}e^{|\L|(\m_{\L}^B)^2/(2\l\hat v(\v0))}\;.\]
Moreover, the function $f^B_\L(\x)$ in the r.h.s of  \eqref{3.4} arises from the ratio of the normalizations associated with the gaussian measures $P^0_\L(d\psi)$ and 
$P^B_{\x,\L}(d\psi)$, and is given by (defining $F(\vk):=|\vk|^2-\m^B_\L+\l\hat v(\v0)|\x|^2+\l\hat v(\vk)|\x|^2$ and
$\e'(\vk):=\sqrt{F(\vk)^2-(\l\hat v(\vk)|\x|^2)^2}$)
\[ e^{-|\L|f^B_\L(\x)}=\mathcal N_0\frac{Z^B_{\x,\L}}{Z^0_\L}\frac{\e'(\v0)}{(-\m^0_{\L})}\;, \label{3.7}\]
where
\[ 
Z^B_{\x,\L}=\prod_{\vk\in\DD_L} \frac{e^{\b(F(\vk)-\e'(\vk))/2}}{1-e^{-\b\e'(\vk)}}\]
and $\DD_L$ was defined after \eqref{Ha}. In the rewritings above, $P_\L(d\x)$ is thought of as 
the ``bare" measure associated with $\xi$, while $P_\L(d\x) e^{-|\L|f^B_\L(\x)}$ is its dressed measure in the Bogoliubov's approximation, 
and 
\[P_\L(d\x)e^{-|\L|f_\L(\x)}:=  P_\L(d\x)e^{-|\L|f^B_\L(\x)}
\int P^B_{\x,\L}(d\ps)\,e^{-\VV^B_{\x,\L}(\ps)}\label{3.9a}\]
is its fully dressed measure. It is apparent from its definition that $P_\L(d\x)$ tends to concentrate on the circle
$|\x|=\m_\L^B/(\l\hat v(\v0))$ as $|\L|\to\infty$. A similar property is valid for $P_\L(d\x) e^{-|\L|f^B_\L(\x)}$. It is then natural to assume that 
also the fully dressed measure $P_\L(d\x)e^{-|\L|f_\L(\x)}$ concentrates on a circle $|\x|=\r_0$ as $|\L|\to\infty$. The counterterm $\bar\n_\L$ must be fixed in such a way that $\r_0$ corresponds to the actual condensate density. The consistency of this natural assumptions should be checked a posteriori of the construction of 
$f_\L(\x)$. By (gauge) symmetry, $f_\L(\x)$ is a radial function, i.e., it only depends upon $|\x|$. Therefore, with no loss of generality, we can 
pick $\x$ to be real. 

From now on, motivated by the considerations above, 
we shall limit ourselves to describe the construction of $f_\L(\x)$ in the case that $\x^+=\x^-=\sqrt{\r_0}$, with $\r_0$ fixed in such a way that $f(\x)=\lim_{|\L|\to\infty}
f_\L(\x)$ has a critical point at $\r_0$. We will also make the convenient choice that $\lim_{|\L|\to\infty}\m^B_\L=
\l\hat v(\v0)\r_0$. A global control on $f(\x)$ is beyond the purpose of this paper. Note that the replacement of the fluctuating field $\x$ by the
constant $\sqrt{\r_0}$ is the analogue of the well known {\it c--number substitution} for the creation and annihilation operators 
associated with the zero mode, which is rigorously known to be correct as far as the computation of the pressure is concerned 
\cite{Book-Lieb, Ginibre}); in this sense, our assumption of restricting our analysis to $\x^-=\sqrt{\r_0}$ is not expected to be a serious limitation, and we expect 
that it leads to the correct result {\it in the thermodynamic limit} $L\to\infty$. There are of course issues to be discussed about exchanging the limit $L\to\infty$ with 
the replacement $\xi\to\sqrt{\r_0}$ (particularly as far as the computation of correlations are concerned), but also these issues are beyond the purpose of this paper. 


\feyn{
\begin{fmffile}{BoseFeyn/int_BEC}
 \unitlength = 0.8cm
\def\myl#1{3.5cm}
\fmfset{arrow_len}{3mm}
\begin{align*}
& 
\parbox{\myl}{\centering{    
		\begin{fmfgraph*}(3,2)    
			\fmftop{i1,i2}
			\fmfbottom{o1,o2}
			\fmf{fermion}{i1,v1}
                   \fmf{fermion}{v1,i2}
                   \fmf{wiggly, tension=0.1}{v1,v2}
                   \fmf{fermion}{o1,v2,o2}
                   \fmfdot{v1,v2}
	           \fmfv{label=$\xx$}{v1}
		\fmfv{label=$\yy$, label.angle=45}{v2}
		\end{fmfgraph*} \\
$4$
		}} 
\parbox{\myl}{\centering{    
		\begin{fmfgraph*}(3,2)    
			\fmftop{i1,i2}
			\fmfbottom{o1,o2}
			\fmf{dots}{i1,v1}
                   \fmf{fermion}{v1,i2}
                   \fmf{wiggly, tension=0.1}{v1,v2}
                   \fmf{fermion}{o1,v2,o2}
                   \fmfdot{v1,v2}
	           \fmfv{label=$\xx$}{v1}
		\fmfv{label=$\yy$, label.angle=45}{v2}
		\end{fmfgraph*} \\
$3^+$
		}}  
\parbox{\myl}{\centering{    
		\begin{fmfgraph*}(3,2)    
			\fmftop{i1,i2}
			\fmfbottom{o1,o2}
			\fmf{fermion}{i1,v1}
                   \fmf{dots}{v1,i2}
                   \fmf{wiggly, tension=0.1}{v1,v2}
                   \fmf{fermion}{o1,v2,o2}
                   \fmfdot{v1,v2}
	           \fmfv{label=$\xx$}{v1}
		\fmfv{label=$\yy$, label.angle=45}{v2}
		\end{fmfgraph*} \\
$3^-$
		}} 
\parbox{\myl}{\centering{ \vskip 0.2cm
		\begin{fmfgraph*}(3,1.5)
			\fmfleft{i1}
			\fmfright{o2}
			\fmf{fermion}{i1,v1,o2}
                   \fmfdot{v1}
			\fmfv{label=$\xx$, label.angle=-90}{v1}
		\end{fmfgraph*} \\ \vskip 0.2cm
$2$
		}}  
\end{align*}
\end{fmffile}	\vskip -0.2cm
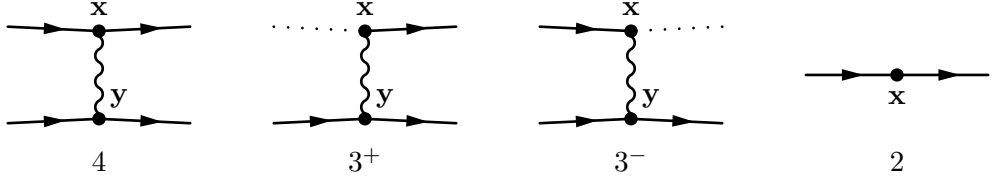
}{The graphical interpretation of the interaction terms in \eqref{3.V2}. The dotted lines correspond to the (original) zero momentum fields $\x^\pm$, then substituted by $\sqrt{\r_0}$. }{int_BEC}


After having fixed $\x^+=\x^-=\sqrt{\r_0}$, the problem becomes that of constructing the theory associated with the functional integral 
\be \Xi_\L=e^{-|\L|(f^B_\L(\sqrt{\r_0})+\bar\n_\L\r_0)}\int P^B_{\L}(d\ps)\,e^{-\bar  V_{\L}(\ps)}\;,\label{3.10}\ee
where
\[ \label{3.V2}
\bar  V_\L(\ps) & = \frac{\l}{2} \int_{\L^2} d\xx d\yy \, |\ps_{\xx}|^2 \,w(\xx-\yy)\, |\ps_{\yy}|^2   + \l\sqrt{\r_0} \int_{\L^2} d\xx d\yy\, \big( \ps^+_\xx +\ps^-_\xx \big)\, w(\xx - \yy) |\ps_{\yy}|^2\non\\
&+\bar\n_\L\int_\L d\xx |\ps_{\xx}|^2\;,
\]
see Fig.\ref{int_BEC}. Moreover,  $P^B_{\L}(d\ps)$ is the same as $P^B_{\x,\L}(d\ps)$ above, with $\x$ replaced by $\sqrt{\r_0}$. 
We recall that $\m^B_\L$ is chosen in such a way that $\m^B=\lim_{|\L|\to\infty}\m^B_\L=\l\r_0\hat v(\v0)$.  In the limit 
$\b,L\to\infty$ the propagator associated with this modified measure is 
\[& g^B(\xx-\yy)= \begin{pmatrix}g^B_{-+}(\xx-\yy)& g^B_{--}(\xx-\yy)\\ g^B_{++}(\xx-\yy)& g^B_{+-}(\xx-\yy)\end{pmatrix}=\\
&=
\int_{\mathbb R^3} \frac{d\kk}{(2\p)^3}\frac{e^{-i\kk(\xx-\yy)}}{k_0^2+(\e'(\vk))^2}\begin{pmatrix}ik_0 +F(\vk)& -\l\r_0\hat v(\vk)\\
-\l\r_0\hat v(\vk)& -ik_0+F(\vk)\end{pmatrix}=\label{3.13}\\
&= \int\frac{d\vk}{(2\p)^2} \frac{e^{-i\vk(\vx-\vy)-|x_0-y_0|\e'(\vk)}}{2\e'(\vk)}
\begin{pmatrix} F(\vk)+\s(x_0-y_0)\e'(\vk)& -\l\r_0\hat v(\vk)\non\\
-\l\r_0\hat v(\vk)& F(\vk)+\s(y_0-x_0)\e'(\vk)\end{pmatrix}\;,
\non
 \]
where 
$F(\vk)=|\vk|^2+\l\hat v(\vk)\r_0$, 
$\e'(\vk)=\sqrt{F(\vk)^2-(\l\hat v(\vk)\r_0)^2}$, and $\s(x_0-y_0)$ is a sign function, equal to $1$ if $x_0-y_0>0$, and equal to $-1$ if $x_0-y_0\le 0$. 
In the following, and more precisely in the ultraviolet integration described in the following section, 
it will be convenient to think of $g^B$ as $g^B=\bar g+r$, where 
\[\bar g(\xx-\yy)=  \frac{e^{-|\vx|^2/(4|x_0-y_0|)}}{4\p |x_0-y_0|}
\begin{pmatrix} \th(x_0-y_0)& 0\\
0& \th(y_0-x_0)\end{pmatrix}\equiv\begin{pmatrix} G(\xx-\yy)& 0\\
0& G(\yy-\xx)\end{pmatrix}
\label{3.14}\]
corresponds to the free propagator, which is (as we already saw)
time-preserving and particle-number-preserving,  while
\[&r(\xx-\yy)=- \int_{\mathbb R^3} \frac{d\kk}{(2\p)^3}e^{-i\kk(\xx-\yy)}\frac{\l\r_0\hat v(\vk)}{k_0^2+(\e'(\vk))^2}\begin{pmatrix}\frac{ik_0 +|\vk|^2}{-ik_0+|\vk|^2}& 1\\
1& \frac{-ik_0 +|\vk|^2}{ik_0+|\vk|^2}\end{pmatrix}\;.\label{3.15}
\]
In order to compute the free energy associated with the functional integral \eqref{3.10}, we proceed as in Section \ref{s2.QCP}, 
namely we regularize the theory by 
introducing an ultraviolet cutoff, to be eventually removed. 
The ultraviolet regularization is implemented by replacing the propagator $g^B_\L$ 
by its cutoffed version $\sum_{h\le N}g^B_{\L,h}$, with $N$ a cutoff parameter to be eventually sent to $+\infty$. This replacement is the analogue of the replacement of $S^0_\L$ in \eqref{2.3} by its regularized version \eqref{2.5} or, equivalently, \eqref{2.8}. The limit $\b,L\to\infty$ of the single scale propagator $g^B_{\L,h}$ is analogous to 
\eqref{2.g}: we write it as $g^B_h=\bar g_h+r_h$, where
\[\bar g_h(\xx-\yy)=\begin{pmatrix} G_h(\xx-\yy)& 0\\
0& G_h(\yy-\xx)\end{pmatrix}\;,\]
with $G_h$ given by \eqref{2.g}, and 
\[r_h(\xx-\yy)&= -
 \int\frac{d\kk}{(2\p)^3} \ff_h(\kk)e^{-i\kk(\xx-\yy)}\frac{\l\r_0\hat v(\vk)}{k_0^2+(\e'(\vk))^2}\begin{pmatrix}\frac{ik_0 +|\vk|^2}{-ik_0+|\vk|^2}& -1\\
-1& \frac{-ik_0 +|\vk|^2}{ik_0+|\vk|^2}\end{pmatrix}
  \label{3.16}\]
where $\ff_h(\kk)=\chi(\g^{-h}(k_0^2+|\vk|^4)^{\nicefrac{1}{2}})-\chi(\g^{-h+1}(k_0^2+|\vk|^4)^{\nicefrac{1}{2}})$.
It is implicit that all these regularizations have their natural counterpart at finite space-time volume $\L$, 
which we will denote by $g^B_{\L,h}$, $\bar g_{\L,h}$, $r_{\L,h}$.

The replacement of $g^B$ by its regularized version corresponds to the replacement of the partition function \eqref{3.10} by 
\be \Xi_{\L,N}=e^{-|\L|(f^B_\L(\sqrt{\r_0})+\bar\n_\L\r_0)}\int \prod_{h\le N}P^B_{\L,h}(d\ps^{(h)})\,e^{-\bar  V_{\L}(\ps^{(\le N)})}\;,\label{3.10bis}\ee
and the idea now is to compute the r.h.s. by iteratively integrating the degrees of freedom on scale $N$, $N-1$, etc, thus rewriting the 
finite volume free energy as a series $f^B_\L(\sqrt{\r_0})+\bar\n_\L\r_0+\sum_{h\le N}F_{\L,h}$, with $F_{\L,h}$ representing the 
contribution from the single-scale integration $P^B_{\L,h}(d\ps^{(h)})$. Sending $N\to\infty$ corresponds to the ultraviolet limit, 
while the control of the series as $h\to-\infty$ corresponds to the infrared limit. As we will explain in detail below, the iterative 
computation proceeds smoothly (in a way very similar to the one described in Section \ref{s2.QCP}) for all scales $h\ge \bar h$, where $\g^{\bar h}=C_0\l$,
with $C_0$ a (finite but sufficiently large) positive constant. Note that, for $h\ge \bar h$, the term $|\vk|^4$ entering the definition of $\e'(\vk)=
\sqrt{|\vk|^4+2\l\r_0\hat v(\vk)|\vk|^2}$ dominates the second term under the square root sign: $|\vk|^4\ge ({\rm const.})\l\r_0\hat v(\vk)|\vk|^2$; the opposite inequality is 
valid for lower scales. This implies that the Bogoliubov's propagator satisfies different dimensional estimates in the two regimes. For $h\ge \bar h$, the 
theory looks like a small perturbation of the theory of the quantum critical point, see next section for a discussion of this regime. 
For $h<\bar h$ the multiscale integration procedure must be modified and it becomes much more involved (and interesting!), see Section \ref{sec3.c} below.

\subsection{The ultraviolet integration}  \label{s.BEC_UV}

In this subsection we describe how to integrate the degrees of freedom corresponding to the scales $\bar h\le h\le N$ 
(where $\bar h$ is fixed so that $\g^{\bar h}=C_0\l$)
in the r.h.s. of \eqref{3.10bis}. Proceeding in a way analogous to that described in Section \ref{s2.multiscale}, we 
construct the effective potential $V^{(h)}(\psi^{(\le h)})$ via the analogue of \eqref{2.14} and \eqref{2.weff}, and the 
single-scale 
contribution to the free energy, $F_h$, via the analogue \eqref{2.fh}, with two main differences: 
(1) the symbol
$\EE^T_h$ in \eqref{2.weff} and \eqref{2.fh} should now be re-interpreted as the truncated expectation with respect to 
$P^B_{h}(d\ps^{(h)})$; (2) at the first step, $h=N$, the effective potential $V^{(N)}(\psi)$ should now be re-interpreted as 
being equal to $\bar V(\psi)$ in \eqref{3.V2}, rather than to $V(\psi)$ in \eqref{2.22}. Note that 
neither the potential $\bar V(\psi)$ nor the propagator $g^B_h$ are now particle-number conserving: therefore, 
the effective potential on scale $h$ is not going to be particle-conserving either, and the analogue of \eqref{2.14}
should now allow for a number $m_+$ of $\psi^+$ fields different in general from the number $m_-$ of $\psi^-$ fields. 
We shall denote by $W_{(m_+,m_-)}^{(h)}$ the kernel of the effective potential on scale $h$ with $m_+$ (resp. $m_-$)  
external legs of type $\psi^+$ (resp. $\psi^-$). 

The outcome of the iterative construction can be expressed again in terms of a tree expansion, completely analogous to 
that described in Section \ref{s2.multiscale}, which implies the analogues of \eqref{2.23}, \eqref{2.23ab}
and \eqref{2.24}. Once again, the only differences between the current formulas and those of Section \ref{s2.multiscale} 
are that both  the potential on scale $N$ 
and the propagator are different. In particular, the trees contributing to $V^{(h)}$ can have endpoints of different types,
either of type $4$, or $3^+$, or $3^-$, or $2$, see Fig.\ref{int_BEC}: we shall indicate by $n_4$ the number of endpoints of type $4$, by $n_3^+$ the number of endpoints of type $3^+$, etc. Moreover, we let $n_3=n_3^++n_3^-$. 
If ${\bf n}=(n_2,n_3,n_4)$, we  indicate by $\tl{\TT}^{(h)}_{N;{\bf n}}$ (resp. ${\TT}^{(h)}_{N;{\bf n}}$) the set of non-renormalized (resp. renormalized) trees contributing to 
$V^{(h)}$ with a specified number of endpoints of different types. 

Regarding the new propagator $g^B_h=\bar g_h+r_h$, we note that while $\bar g_h$ is related in a trivial way with $G_h$ and, therefore, it is scale invariant (see \eqref{2.scale}), this is not the case for $r_h$. Still, $r_h$ can be written as $r_h(x_0,\vx)=\int d\vy\, v(\vx-\vy) \tilde r_h(x_0,\vy)$, where $\tilde r_h$ is defined by an expression similar 
to the r.h.s. \eqref{3.16}, with $\hat v(\vk)$ replaced by $1$ under the integral sign. Moreover, $\tilde r_h$ is 
essentially scale invariant, i.e., $\tilde r_h\simeq \g^\bh \tilde r_0(\g^h x_0,\g^{h/2}\vx)$, 
up to 
corrections of relative size $\g^{\bar h-h}$, which are negligible for $h\gg \bar h$. In particular, for all $h\ge \bh$, 
$||r_h||_\io\le C\g^h\g^{\bh-h}\min\{1,\g^{-h}\}$ and $||r_h||_1\le C\g^{-h}\g^{\bh-h}$. Therefore, the same dimensional 
bounds 
\eqref{2.Gh} remain valid both for $g^B_h$ and for $\bar g_h$. On top of that, the rest $r_h$ 
is
better behaved from a dimensional point of view, i.e., 
the $L_1$ (resp. $L_\io$) norm of $r_h$ has an extra $\g^{\bar h-h}$ (resp. $\g^{\bh-h}\min\{1,\g^{-h}\}$)
as compared to the $L_1$ (resp. $L_\io$) norm of  $\bar g_h$. In the following, it will be useful to distinguish the contributions to the effective potential  coming from the time-ordered, particle-conserving propagators $\bar g_k$
from those coming from the rest $r_k$. To this purpose, we write the analogue of \eqref{2.23} as
\[
W^{(h)}_{(m_+,m_-)}(\underline\xx;\underline\yy)= \sum_{{\bf n}>\V0} \sum_{\t\in\tilde\TT_{N;{\bf n}}^{(h)}} \sideset{}{^{(P_{v_0})}}\sum_{{\bf P}\in\PP_\t}\hskip-.4truecm\sum_{\tilde{\bf n}\in\NN({\bf P})}
\int \hskip-.2truecm\prod_{f\in I_{v_0}\setminus P_{v_0}}\hskip-.4truecm d\xx_f\
W_{}^{(h)}(\t,{\bf P},\tilde{\bf n};\underline{\xx}_{v_0})\label{2.23new}\]
where: (1) ${\bf n}=(n_2,n_3,n_4)>\V0$ means that $n_2,n_3,n_4$ are all non-negative, but not simultaneously zero;
(2) $\tilde {\bf n}=\{\tilde{n}_{r,v}\}_{v\in\t}$ with $\tilde n_{r,v}$ the number of propagators of type $r_{h_v}$ 
contained in $v$ [we say that a propagator is contained in $v$ if it associated with a pair 
of fields in $P_w\setminus Q_w$ for some $w> v$]; (3) $\NN({\bf P})$ denotes the set of values of $\tilde{\bf n}$ compatible with 
${\bf P}$. Of course, $F_{h+1}$ can be expanded in a similar way. 

By proceeding exactly as in Section 
\ref{s2.multiscale} (in particular, by following the procedure described after \eqref{2.26}), we obtain the analogue of 
\eqref{2.35xs} (details are left to the reader):
\[
||W^{(h)}_{(m_+,m_-)}||&\le \g^{h (2-m/2)}\sum_{{\bf n}>\V0} |\l|^{n_3+n_4}|\bar \n|^{n_2} K^{n} (n_4+\tfrac{n_3}2-\tfrac{m}{2})! \,\g^{-h(n_2 + n_3/2)}
\cdot\nonumber\\\
&\cdot \sum_{\t\in\tilde\TT_{N;{\bf n}}^{(h)}} \sideset{}{^{(P_{v_0})}}\sum_{{\bf P}\in\PP_\t}\hskip-.3truecm\sum_{\tilde{\bf n}\in\NN({\bf P})}\g^{(\bh-h)\tilde n_r}
\prod_{v\,{\rm not}\,{\rm e.p.}}\g^{(h_v-h_{v'})\,d_v}
\;,\label{3.9}\]
where $n=n({\bf n})=n_2+n_3+n_4$, $m=m_++m_-$, and the {\it vertex dimension} $d_v=d_v({\bf n}_{v},\tilde{\bf n}_{v}, |P_v|)$ is given by
\[ \label{3.dim}
d_v= 2 - \frac{|P_v|}{2}   -n_{2,v}  -\frac{1}{2} n_{3,v} -\tilde n_{r,v}\;.
\]
Here and in the following $n_{q,v}$ denotes the number of endpoints with $q$ external legs following $v$ on $\t$ and 
$n_q:=n_{q,v_0}$; similarly, $\tilde n_{r}:=\tilde n_{r,v_0}$. 
Moreover, ${\bf n}_v=(n_{2,v},n_{3,v},n_{4,v})$. The vacuum contribution, $F_{h+1}$, admits a similar bound. 

The potentially divergent contributions to \eqref{3.9} come from the subdiagrams such that 
$d_v\ge 0$, namely by the subdiagrams such that:
\[ \label{3.divergences}
{\rm (a)}& \quad |P_v| =4  \quad {\rm with}\quad (n_{2,v},n_{3,v},\tilde n_{r,v})
=(0,0,0) \;;  \non \\
{\rm (b)}& \quad |P_v| =3  \quad {\rm with}\quad (n_{2,v},n_{3,v},\tilde n_{r,v})
=(0,1,0) \;; \\
{\rm (c)}& \quad |P_v| = 2 \quad {\rm with}\quad (n_{2,v},n_{3,v},\tilde n_{r,v})
=(0,0,0),(1,0,0),(0,2,0), (0,0,1)\;. \non
\]
Note that the cases $|P_v|=3$ with $(n_{2,v},n_{3,v},\tilde n_{r,v})=(0,0,0)$, and 
$|P_v|=2$ with $(n_{2,v},n_{3,v},\tilde n_{r,v})=(0,1,0)$, which in principle would lead to $d_v>0$,
are impossible. In addition to this, since $\bar g_h(\xx)$ preserves the time-ordering, some of the marginal or relevant 
sub-diagrams are identically zero, as it was the case in the theory of the quantum critical point. More specifically, it is
easy to check that there exist no non-vanishing diagrams with $|P_v| = 2$ and $(n_{2,v},n_{3,v},\tilde n_{r,v})=(0,0,0),(1,0,0)$\footnote{The vanishing of the non-trivial 
diagrams with $|P_v| = 2$ and $(n_{2,v},n_{3,v},\tilde n_{r,v})=(1,0,0)$ is valid if the 2-legged vertices are particle-conserving, i.e., if they are proportional to 
$\int d\xx|\psi_\xx|^2x$, as in our case, see \eqref{3.V2}. However, we will see in a moment that non-particle-conserving 2-legged vertices, proportional to $\int d\xx(\psi^+_\xx
\psi^+_\xx+h.c.)$, are generated by the iterative 
integration, as soon as we reach scales $\bh\le h\le 0$. Therefore, a posteriori we will also have non-vanishing contributions with $|P_v| = 2$ and $(n_{2,v},n_{3,v},\tilde n_{r,v})=(1,0,0)$, with the 2-legged vertex 
that is necessarily of type $\int d\xx(\psi^+_\xx\psi^+_\xx+h.c.)$.}, except the trivial one with $n_2=1$ and $n_3=n_4=0$.
Moreover, the only non-vanishing diagrams among the potentially divergent ones have a ladder structure, similar 
to the one we found in the theory of the quantum critical point. The list of the non-vanishing diagrams is shown in Fig.\ref{ladder3}, \ref{ladder2} and \ref{ladder2bis}.

\vskip -1.5cm


\feyn{
\begin{fmffile}{BoseFeyn/ladder3}
 \unitlength = 1cm
\def\myl#1{7.5cm}
 \fmfset{arrow_len}{3mm}
\begin{align*}
& \parbox{\myl}{\centering{  \hskip -0.5cm
\begin{fmfgraph*}(8,1.2)
			\fmfleft{i1,i2}
			\fmfright{o1,o2}
			\fmf{fermion, tension=1.2}{i1,v1}
			\fmf{dots, tension=1.2}{i2,v2}
				\fmf{fermion}{v1,v3}		
				\fmf{fermion}{v2,v4}
				\fmf{dashes, tension=0.6}{v3,v5}		
				\fmf{dashes, tension=0.6}{v4,v6}
				\fmf{fermion}{v5,v7}		
				\fmf{fermion}{v6,v8}
                  			 \fmf{wiggly, tension=0.01}{v1,v2}
					\fmf{wiggly,tension=0.01}{v3,v4}
					\fmf{wiggly, tension=0.01}{v5,v6}
					\fmf{wiggly,tension=0.01}{v7,v8}
			\fmf{fermion, tension=1.2}{v7,o1}
			\fmf{fermion, tension=1.2}{v8,o2}
                   \fmfdot{v1,v2,v3,v4,v5,v6,v7,v8}
 	        \fmfv{label=$\xx_n$,  label.angle=-90}{v1}
 		\fmfv{label=$\xx_{n-1}$,  label.angle=-90}{v3}
 		\fmfv{label=$\xx_{2}$,  label.angle=-90}{v5}
 		\fmfv{label=$\xx_1$,  label.angle=-90}{v7}
 		\fmfv{label=$\yy_n$,  label.angle=90}{v2}
 		\fmfv{label=$\yy_{n-1}$,  label.angle=90}{v4}
 		\fmfv{label=$\yy_{2}$,  label.angle=90}{v6}
 		\fmfv{label=$\yy_1$,  label.angle=90}{v8}
		\end{fmfgraph*}  }}
\parbox{\myl}{\centering{  \hskip -0.5cm
\begin{fmfgraph*}(8,1.2)
			\fmfleft{i1,i2}
			\fmfright{o1,o2}
			\fmf{fermion, tension=1.2}{i1,v1}
			\fmf{fermion, tension=1.2}{i2,v2}
				\fmf{fermion}{v1,v3}		
				\fmf{fermion}{v2,v4}
				\fmf{dashes, tension=0.6}{v3,v5}		
				\fmf{dashes, tension=0.6}{v4,v6}
				\fmf{fermion}{v5,v7}		
				\fmf{fermion}{v6,v8}
                  			 \fmf{wiggly, tension=0.01}{v1,v2}
					\fmf{wiggly,tension=0.01}{v3,v4}
					\fmf{wiggly, tension=0.01}{v5,v6}
					\fmf{wiggly,tension=0.01}{v7,v8}
			\fmf{fermion, tension=1.2}{v7,o1}
			\fmf{dots, tension=1.2}{v8,o2}
                   \fmfdot{v1,v2,v3,v4,v5,v6,v7,v8}
 	        \fmfv{label=$\xx_n$,  label.angle=-90}{v1}
 		\fmfv{label=$\xx_{n-1}$,  label.angle=-90}{v3}
 		\fmfv{label=$\xx_{2}$,  label.angle=-90}{v5}
 		\fmfv{label=$\xx_1$,  label.angle=-90}{v7}
 		\fmfv{label=$\yy_n$,  label.angle=90}{v2}
 		\fmfv{label=$\yy_{n-1}$,  label.angle=90}{v4}
 		\fmfv{label=$\yy_{2}$,  label.angle=90}{v6}
 		\fmfv{label=$\yy_1$,  label.angle=90}{v8}
		\end{fmfgraph*}  }}  
\end{align*}
\end{fmffile}	
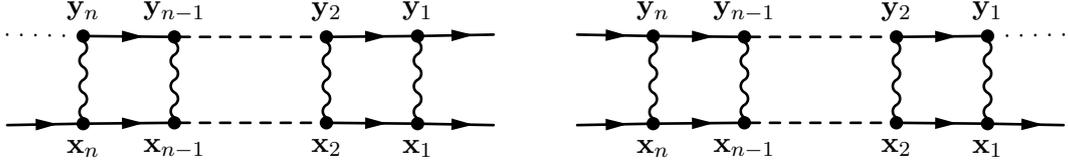
}{The three legged ladder graphs of order $n$.}{ladder3}


\feyn{
\begin{fmffile}{BoseFeyn/ladder2}
 \unitlength = 1cm
\def\myl#1{7.5cm}
 \fmfset{arrow_len}{3mm} 
\begin{align*} 
%
&
\parbox{\myl}{\centering{ \vskip 0.2cm \hskip -0.5cm  
\begin{fmfgraph*}(8,1.2)
			\fmfleft{i1,i2}
			\fmfright{o1,o2}
			\fmf{fermion, tension=1.2}{i1,v1}
			\fmf{dots, tension=1.2}{i2,v2}
				\fmf{fermion}{v1,v3}		
				\fmf{fermion}{v2,v4}
				\fmf{dashes, tension=0.6}{v3,v5}		
				\fmf{dashes, tension=0.6}{v4,v6}
				\fmf{fermion}{v5,v7}		
				\fmf{fermion}{v6,v8}
                  			 \fmf{wiggly, tension=0.01}{v1,v2}
					\fmf{wiggly,tension=0.01}{v3,v4}
					\fmf{wiggly, tension=0.01}{v5,v6}
					\fmf{wiggly,tension=0.01}{v7,v8}
			\fmf{dots, tension=1.2}{v7,o1}
			\fmf{fermion, tension=1.2}{v8,o2}
                   \fmfdot{v1,v2,v3,v4,v5,v6,v7,v8}
 	        \fmfv{label=$\xx_n$,  label.angle=-90}{v1}
 		\fmfv{label=$\xx_{n-1}$,  label.angle=-90}{v3}
 		\fmfv{label=$\xx_{2}$,  label.angle=-90}{v5}
 		\fmfv{label=$\xx_1$,  label.angle=-90}{v7}
 		\fmfv{label=$\yy_n$,  label.angle=90}{v2}
 		\fmfv{label=$\yy_{n-1}$,  label.angle=90}{v4}
 		\fmfv{label=$\yy_{2}$,  label.angle=90}{v6}
 		\fmfv{label=$\yy_1$,  label.angle=90}{v8}
		\end{fmfgraph*}  }}
\parbox{\myl}{\centering{   \vskip 0.2cm \hskip -0.5cm  
\begin{fmfgraph*}(8,1.2)
			\fmfleft{i1,i2}
			\fmfright{o1,o2}
			\fmf{fermion, tension=1.2}{i1,v1}
			\fmf{dots, tension=1.2}{i2,v2}
				\fmf{fermion}{v1,v3}		
				\fmf{fermion}{v2,v4}
				\fmf{dashes, tension=0.6}{v3,v5}		
				\fmf{dashes, tension=0.6}{v4,v6}
				\fmf{fermion}{v5,v7}		
				\fmf{fermion}{v6,v8}
                  			 \fmf{wiggly, tension=0.01}{v1,v2}
					\fmf{wiggly,tension=0.01}{v3,v4}
					\fmf{wiggly, tension=0.01}{v5,v6}
					\fmf{wiggly,tension=0.01}{v7,v8}
			\fmf{fermion, tension=1.2}{v7,o1}
			\fmf{dots, tension=1.2}{v8,o2}
                   \fmfdot{v1,v2,v3,v4,v5,v6,v7,v8}
 	        \fmfv{label=$\xx_n$,  label.angle=-90}{v1}
 		\fmfv{label=$\xx_{n-1}$,  label.angle=-90}{v3}
 		\fmfv{label=$\xx_{2}$,  label.angle=-90}{v5}
 		\fmfv{label=$\xx_1$,  label.angle=-90}{v7}
 		\fmfv{label=$\yy_n$,  label.angle=90}{v2}
 		\fmfv{label=$\yy_{n-1}$,  label.angle=90}{v4}
 		\fmfv{label=$\yy_{2}$,  label.angle=90}{v6}
 		\fmfv{label=$\yy_1$,  label.angle=90}{v8}
		\end{fmfgraph*}  }} \\[36pt]
&
\parbox{\myl}{\centering{  \hskip -0.5cm
\begin{fmfgraph*}(8,1.2)
			\fmfleft{i1,i2}
			\fmfright{o1,o2}
			\fmf{fermion, tension=1.2}{i1,v1}
			\fmf{fermion, tension=1.2}{i2,v2}
				\fmf{fermion}{v1,v3}		
				\fmf{fermion}{v2,v4}
				\fmf{dashes, tension=0.6}{v3,v5}		
				\fmf{dashes, tension=0.6}{v4,v6}
				\fmf{fermion}{v5,v7}		
				\fmf{fermion}{v6,v8}
                  			 \fmf{wiggly, tension=0.01}{v1,v2}
					\fmf{wiggly,tension=0.01}{v3,v4}
					\fmf{wiggly, tension=0.01}{v5,v6}
					\fmf{wiggly,tension=0.01}{v7,v8}
			\fmf{phantom, tension=1.2}{v7,o1}
			\fmf{phantom, tension=1.2}{v8,o2}
 			\fmf{phantom, tension=1.2}{o1,m1,o2}
			\fmf{phantom, tension=0.15}{m1,m2}
			\fmf{fermion, tension=0.01, right=0.4}{v7,m2}
			\fmf{fermion, tension=0.01, left=0.4}{v8,m2}
                   \fmfdot{v1,v2,v3,v4,v5,v6,v7,v8}
 	        \fmfv{label=$\xx_n$,  label.angle=-90}{v1}
 		\fmfv{label=$\xx_{n-1}$,  label.angle=-90}{v3}
 		\fmfv{label=$\xx_{2}$,  label.angle=-90}{v5}
 		\fmfv{label=$\xx_1$,  label.angle=-90}{v7}
 		\fmfv{label=$\yy_n$,  label.angle=90}{v2}
 		\fmfv{label=$\yy_{n-1}$,  label.angle=90}{v4}
 		\fmfv{label=$\yy_{2}$,  label.angle=90}{v6}
 		\fmfv{label=$\yy_1$,  label.angle=90}{v8}
		\end{fmfgraph*}  }}
\parbox{\myl}{\centering{  \hskip -0.5cm
\begin{fmfgraph*}(8,1.2)
			\fmfleft{i1,i2}
			\fmfright{o1,o2}
                   \fmf{phantom, tension=1.2}{i1,v1}
			\fmf{phantom, tension=1.2}{i2,v2}
 			\fmf{phantom, tension=1.2}{i1,m1,i2}
			\fmf{phantom, tension=0.15}{m1,m2}
			\fmf{fermion, tension=0.01, right=0.4}{m2,v1}
			\fmf{fermion, tension=0.01, left=0.4}{m2,v2}
				\fmf{fermion}{v1,v3}		
				\fmf{fermion}{v2,v4}
				\fmf{dashes, tension=0.6}{v3,v5}		
				\fmf{dashes, tension=0.6}{v4,v6}
				\fmf{fermion}{v5,v7}		
				\fmf{fermion}{v6,v8}
                  			 \fmf{wiggly, tension=0.01}{v1,v2}
					\fmf{wiggly,tension=0.01}{v3,v4}
					\fmf{wiggly, tension=0.01}{v5,v6}
					\fmf{wiggly,tension=0.01}{v7,v8}
			\fmf{fermion, tension=1.2}{v7,o1}
			\fmf{fermion, tension=1.2}{v8,o2}
                   \fmfdot{v1,v2,v3,v4,v5,v6,v7,v8}
 	        \fmfv{label=$\xx_n$,  label.angle=-90}{v1}
 		\fmfv{label=$\xx_{n-1}$,  label.angle=-90}{v3}
 		\fmfv{label=$\xx_{2}$,  label.angle=-90}{v5}
 		\fmfv{label=$\xx_1$,  label.angle=-90}{v7}
 		\fmfv{label=$\yy_n$,  label.angle=90}{v2}
 		\fmfv{label=$\yy_{n-1}$,  label.angle=90}{v4}
 		\fmfv{label=$\yy_{2}$,  label.angle=90}{v6}
 		\fmfv{label=$\yy_1$,  label.angle=90}{v8}
		\end{fmfgraph*}  }}
\end{align*}
\end{fmffile}	
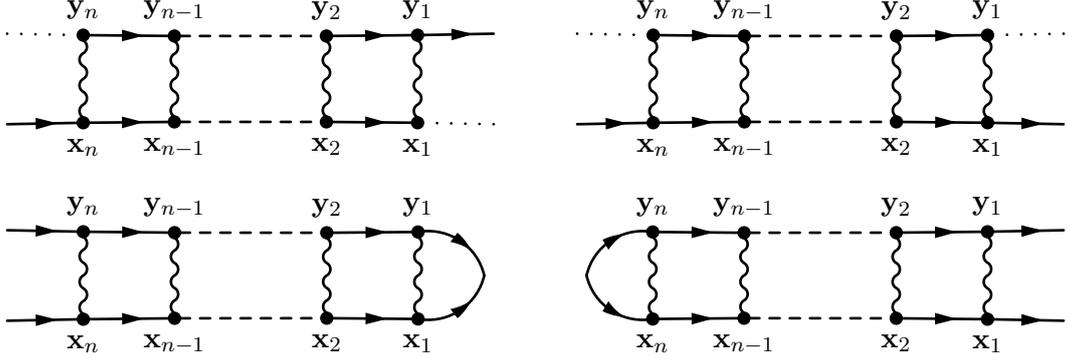
}{Two legged ladder graphs of order $n$. In the first line the particle--conserving ladders with $(n_{2,v},n_{3,v},\tilde n_{r,v})=(0,2,0)$. Two types of two legged graphs not--particle--conserving, missing in the initial potential \eqref{3.V2}, are generated by the iterative integration when  $(n_{2,v},n_{3,v},\tilde n_{r,v})=(0,0,1)$, as shown in the second line.}{ladder2}


\feyn{
\begin{fmffile}{BoseFeyn/ladder2bis}
 \unitlength = 1cm
\def\myl#1{7.5cm}
 \fmfset{arrow_len}{3mm}
\begin{align*}
\parbox{\myl}{\centering{  \hskip -0.5cm
\begin{fmfgraph*}(8,1.2)
			\fmfleft{i1,i2}
			\fmfright{o1,o2}
			\fmf{fermion, tension=1.2}{i1,v1}
			\fmf{fermion, tension=1.2}{i2,v2}
				\fmf{fermion}{v1,v3}		
				\fmf{fermion}{v2,v4}
				\fmf{dashes, tension=0.6}{v3,v5}		
				\fmf{dashes, tension=0.6}{v4,v6}
				\fmf{fermion}{v5,v7}		
				\fmf{fermion}{v6,v8}
                  			 \fmf{wiggly, tension=0.01}{v1,v2}
					\fmf{wiggly,tension=0.01}{v3,v4}
					\fmf{wiggly, tension=0.01}{v5,v6}
					\fmf{wiggly,tension=0.01}{v7,v8}
			\fmf{phantom, tension=1.2}{v7,o1}
			\fmf{phantom, tension=1.2}{v8,o2}
 			\fmf{phantom, tension=1.2}{o1,m1,o2}
			\fmf{phantom, tension=0.15}{m1,m2}
			\fmf{fermion, tension=0.01, right=0.4}{v7,m2}
			\fmf{fermion, tension=0.01, left=0.4}{v8,m2}
                   \fmfdot{v1,v2,v3,v4,v5,v6,v7,v8,m2}
 	        \fmfv{label=$\xx_n$,  label.angle=-90}{v1}
 		\fmfv{label=$\xx_{n-1}$,  label.angle=-90}{v3}
 		\fmfv{label=$\xx_{2}$,  label.angle=-90}{v5}
 		\fmfv{label=$\xx_1$,  label.angle=-90}{v7}
 		\fmfv{label=$\yy_n$,  label.angle=90}{v2}
 		\fmfv{label=$\yy_{n-1}$,  label.angle=90}{v4}
 		\fmfv{label=$\yy_{2}$,  label.angle=90}{v6}
 		\fmfv{label=$\yy_1$,  label.angle=90}{v8}
             \fmfv{label=$\zz$, label.angle=-60}{m2}
		\end{fmfgraph*}  }}
\parbox{\myl}{\centering{  \hskip -0.5cm
\begin{fmfgraph*}(8,1.2)
			\fmfleft{i1,i2}
			\fmfright{o1,o2}
                   \fmf{phantom, tension=1.2}{i1,v1}
			\fmf{phantom, tension=1.2}{i2,v2}
 			\fmf{phantom, tension=1.2}{i1,m1,i2}
			\fmf{phantom, tension=0.15}{m1,m2}
			\fmf{fermion, tension=0.01, right=0.4}{m2,v1}
			\fmf{fermion, tension=0.01, left=0.4}{m2,v2}
				\fmf{fermion}{v1,v3}		
				\fmf{fermion}{v2,v4}
				\fmf{dashes, tension=0.6}{v3,v5}		
				\fmf{dashes, tension=0.6}{v4,v6}
				\fmf{fermion}{v5,v7}		
				\fmf{fermion}{v6,v8}
                  			 \fmf{wiggly, tension=0.01}{v1,v2}
					\fmf{wiggly,tension=0.01}{v3,v4}
					\fmf{wiggly, tension=0.01}{v5,v6}
					\fmf{wiggly,tension=0.01}{v7,v8}
			\fmf{fermion, tension=1.2}{v7,o1}
			\fmf{fermion, tension=1.2}{v8,o2}
                   \fmfdot{v1,v2,v3,v4,v5,v6,v7,v8,m2}
 	        \fmfv{label=$\xx_n$,  label.angle=-90}{v1}
 		\fmfv{label=$\xx_{n-1}$,  label.angle=-90}{v3}
 		\fmfv{label=$\xx_{2}$,  label.angle=-90}{v5}
 		\fmfv{label=$\xx_1$,  label.angle=-90}{v7}
 		\fmfv{label=$\yy_n$,  label.angle=90}{v2}
 		\fmfv{label=$\yy_{n-1}$,  label.angle=90}{v4}
 		\fmfv{label=$\yy_{2}$,  label.angle=90}{v6}
 		\fmfv{label=$\yy_1$,  label.angle=90}{v8}
             \fmfv{label=$\zz$, label.angle=150}{m2}
		\end{fmfgraph*}  }}
\end{align*} 
\end{fmffile}	
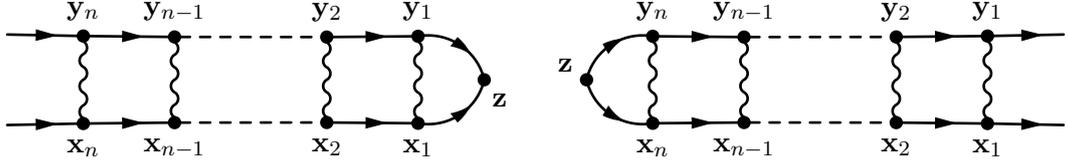
}{Two legged ladder graphs with $(n_{2,v},n_{3,v},\tilde n_{r,v})=(1,0,0)$.}{ladder2bis}


Due to the presence of these relevant and marginal contributions, we need to define a proper localization procedure, 
and correspondingly to reorganize the expansion in a way similar to the one discussed in Sections \ref{s2.ren} and
\ref{s2.fl}. In particular, at each iteration step, we define split 
$V^{(h)}(\ps)=\LL V^{(h)}(\ps) +\RR V^{(h)}(\ps)$ with
\[
& \LL V^{(h)}(\ps) =\int_{\RRR^{12}}d\xx_1\cdots d\yy_{2} \LL W^{(h)}_{(2,2)}(\xx_1,\xx_2;\yy_1,\yy_{2})\ps_{\xx_1}^+
\ps^+_{\xx_2}\ps^-_{\yy_{1}}\ps^-_{\yy_{2}} \label{3.11}\\
& + \int_{\RRR^{9}}\hskip -0.2cm d\xx_1 d\xx_2 d\xx_3 \Big[\LL W^{(h)}_{(2,1)}(\xx_1,\xx_2;\xx_3)\ps_{\xx_1}^+ \ps^+_{\xx_2}\ps^-_{\xx_3} 
+  \LL W^{(h)}_{(1,2)}(\xx_1;\xx_2, \xx_3)\ps_{\xx_1}^+
\ps^-_{\xx_2}\ps^-_{\xx_3}\Big] \non \\
& + \int_{\RRR^6} d\xx_1 d\xx_2  \lft[ \LL W^{(h)}_{(2,0)}(\xx_1,\xx_2)\ps_{\xx_1}^+\ps_{\xx_2}^+ 
+\LL W^{(h)}_{(1,1)}(\xx_1;\xx_2)\ps_{\xx_1}^+\ps_{\xx_2}^- 
+\LL W^{(h)}_{(0,2)}(\xx_1,\xx_2)\ps_{\xx_1}^-\ps_{\xx_2}^-  \rgt]\;.
\non
\]
The localized kernels $\LL W^{(h)}_{(2,2)}$, $\LL W^{(h)}_{(2,1)}$, etc, are defined differently, 
depending on whether $h$ is larger or smaller than $0$. In particular, if $h\ge 0$,
\[\LL W^{(h)}_{(2,2)}(\xx_1,\xx_2;\yy_1,\yy_{2})=
W^{(h)}_{(2,2)}(\xx_1,\xx_2;\yy_1,\yy_{2})\Big|_{(0,0,0)}\;,\] 
where $[\cdot]\big|_{(0,0,0)}$ means that we are taking the contribution corresponding to $(n_{2,v},n_{3,v}$, $\tilde n_{r,v})=(0,0,0)$.
Moreover, if $h<0$,
\[ \LL W^{(h)}_{(2,2)}(\xx_1,\xx_2;\yy_1,\yy_{2}) =\bar \l_h\,\d(\xx_1-\xx_2)\d(\xx_1-\yy_1)\d(\xx_1-\yy_2)\label{3.25x}\]
with 
\[\bar \l_h:=\int d\xx_2' d\yy_1' d\yy_2' W^{(h)}_{(2,2)}(\xx_1,\xx_2';\yy_1',\yy_{2}')\Big|_{(0,0,0)}\;.\label{3.26x}\]
In other words, $\LL W^{(h)}_{(2,2)}$ 
is equal either to the sum of all the ladder sub-diagrams built out of 4-legged vertices and propagators of type $\bar g$ carrying 
a scale label $\ge h$, if $h\ge 0$, 
or to its {\it local part} (in the sense of \eqref{3.25x}-\eqref{3.26x}), if $h<0$. This is the same definition that we had 
in the theory of the quantum critical point. 
The local parts of the 3- and 2-legged 
kernels are defined similarly. In particular, for $h<0$, we introduce the {\it running coupling constants} 
\[& \bar\m_h:=\int d\xx_2' d\xx_3' W^{(h)}_{(2,1)}(\xx_1,\xx_2';\xx_3')\Big|_{(0,1,0)}=
\int d\xx_2' d\xx_3' W^{(h)}_{(1,2)}(\xx_1;\xx_2',\xx_3')\Big|_{(0,1,0)}\non\\
& \bar z_h:=\int d\yy \Big[W^{(h)}_{(2,0)}(\xx,\yy)\Big|_{(0,0,1)}+
W^{(h)}_{(2,0)}(\xx,\yy)\Big|_{(1,0,0)}\Big]\label{3.barz}\\
&\bar \n_h:=\bar \n+\int d\yy W^{(h)}_{(1,1)}(\xx,\yy)\Big|_{(0,2,0)}\equiv \bar\n+\d\bar\n_h\non
\]
where in the second line the contribution corresponding to $(n_{2,v},n_{3,v}$, $\tilde n_{r,v})=(1,0,0)$
comes from graphs as in Fig.\ref{ladder2bis} (i.e., the 2-legged vertex is necessarily of type $\bar z_h$). Moreover, by symmetry,
the second line would be the same even if we changed the label $(2,0)$ to $(0,2)$.
By proceeding in a way completely analogous to Section \ref{s2.fl}, we find that, for 
$(m_+,m_-)=(2,2),(2,1),(1,2)$, 
%
\[ 
\Big| \int d\xx_2 \cdots d\xx_m\, \LL W^{(h)}_{(m_+,m_-)}(\xx_1,\cdots, \xx_m)\Big|\le 
({\rm const.})\l\;. \label{3.12x}
\]
Moreover, if $(m_+,m_-)=(1,1),(2,0),(0,2)$ similar bounds are valid (see Appendix \ref{UVflow}): 
\[ 
&\Big|\bar \n- \int d\xx_2 \, \LL W^{(h)}_{(1,1)}(\xx_1, \xx_2)\Big|\le ({\rm const.})\l^2\min\{\g^{-h},-h\}\;,\label{3.12y}\\
&\Big| \int d\xx_2 \, \LL W^{(h)}_{(2,0)}(\xx_1, \xx_2)\Big|\le ({\rm const.})\l^2 \min\{\g^{-h},-h\}\;. \label{3.12z}
\]
After this resummation,  $F_{h+1}$ and $W_{(m_+,m_-)}^{(h)}$ are expressed as sums over trees where all the vertices 
which are not endpoints have negative dimension. Every endpoint $v^*$ is associated with one of the 
terms in \eqref{3.11} with $h=h_v^*-1$, whose kernels are all bounded as in  \eqref{3.12x}. In conclusion, after the resummation, we find, similarly to \eqref{2.55}:
\[ |F_{h+1}|&\le \g^{2h}\sum_{{\bf n}>\V0} \l^{n_3+n_4}\tilde z^{n_z}\tilde \n^{n_\n} \bar K^{n} (n_4+\tfrac{n_3}2-\tfrac{m}{2})! \,\g^{-h(n_\n + n_3/2)}\g^{(\bh-h)n_z}
\cdot\nonumber\\\
&\cdot \sum_{\t\in\TT_{N;{\bf n}}^{(h)}} \sideset{}{^{*}}\sum_{{\bf P}\in\PP_\t}
\sum_{\tilde{\bf n}\in\NN({\bf P})}\g^{(\bar h-h)\tilde n_r}
\prod_{v\,{\rm not}\,{\rm e.p.}}\g^{(h_v-h_{v'})\,(d_v-z_v)} \min\{1,\g^{-h^*_\t}\}\label{3.25first}\]
where: (i) $\tilde\n:=\sup_{h\ge \bar h}|\bar\n_h|$, which is smaller than $|\bar\n|+({\rm const.})\l^2|\log\l|$,
and $\tilde z$ is chosen in such a way that 
\[\Big| \int d\xx_2 \, \LL W^{(h)}_{(2,0)}(\xx_1, \xx_2)\Big|\le  \tilde z\g^\bh\min\{1,\g^{-h}\},\label{3.rt}\] i.e., using \eqref{3.12z}, we see that $\tilde z$ is smaller than (const.)$\l |\log\l|$;
(ii)
the $*$ on the sum indicates the constraints that $P_{v_0}=\emptyset$ is fixed and the set of internal fields of $v_0$ is non empty;
(iii) the symbol $n_\n$ (resp. $n_z$) denotes the number of endpoints of type $(m_+,m_-)=(1,1)$ (resp. $(m_+,m_-)=(2,0),(0,2)$), 
$n_2=n_\n+n_z$ and $n=n_2+n_3+n_4$; (iv) $h^*_\t$ is the highest among the scales of the endpoints of type 
$(2,0)$, if any, or of the  propagators of type $r_h$, otherwise (note that the vacuum diagrams have at least two ``non-particle-conserving" 
endpoints or propagators, i.e., endpoints of type $\bar z_h$ or propagators of type $r_h$; otherwise, their value vanishes).
The factor $\min\{1,\g^{-h^*_\t}\}$ either comes from \eqref{3.rt}, or from the remark that  $||r_{h}||_\io\le C\g^h\g^{\bh-h}\min\{1,\g^{-h}\}$ (if  the tree has at least one propagator of type $r_h$, in bounding the tree value via \eqref{2.31} we can decide to take the $L_\io$ norm of the propagator of
type $r_h$, which produces the desired gain factor).
We also recall that $\TT_{N;{\bf n}}^{(h)}$ indicates the set of renormalized trees. 
Moreover, $z_v$ is the dimensional gain induced by the renormalization 
procedure, which is a function of $|P_v|$, ${\bf n}_v$ and $\tilde{\bf n}_v$. More precisely, 
$z_v$ is equal to $1/2$ in the cases listed in \eqref{3.divergences}, and zero otherwise.  
Since the renormalized vertex dimension is always negative, the exponential factors $\g^{(h_v-h_{v'})(d_v-z_v)}$ 
in the r.h.s. of \eqref{3.25first} are summable both  over $\{h_v\}$ and over $\{P_v\}$. 
After these summations, 
we are led to an $n!$ bound of the form ($[\cdot]_+$ indicates the positive part):
\[ & |F_{h+1}|\le  \label{3.26}\\
& \g^{2h}\sum_{{\bf n}>\V0} \l^{n_3+n_4} \tilde z^{n_z}\tilde \n^{n_\n} & \bar K^{n} (n_4+\tfrac{n_3}2-\tfrac{m}{2})! 
 \g^{-h(n_{\n} + n_3/2)}\g^{(\bh -h)(n_z+[2-n_z]_+)}\min\{1,\g^{-\theta h}\}\;, \non \]
for some $\theta>0$. Eq.\eqref{3.26} is acceptable as long as $h\ge \bar h$ and $\tilde\n\g^{-\bar h}=:\d$ is small enough, as we shall assume from now on. 
In a similar way, we derive an $n!$ bound for the kernels of the (renormalized) effective potential, 
whose external legs can now be associated with the action of derivative operators, as in \eqref{2.54y}. Using a notation 
analogous to \eqref{2.55}, we can write the result as:
\[& \|W^{(h)}_{(m_+,m_-);{\boldsymbol\a}_1,\ldots,{\boldsymbol\a}_m}\|\le \label{3.26bis}\\
&\quad\le \g^{h(2-\frac{m}2-\|{\boldsymbol\a}^0\|-\frac12\|\vec {\boldsymbol\a}\|)}\sum_{{\bf n}>\V0}^*
\l^{n_3+n_4}\tilde z^{n_z}\tilde \n^{n_\n} \bar K^{n} (n_4+\tfrac{n_3}2-\tfrac{m}{2})! \,\g^{-h(n_\n + n_3/2)}\g^{(\bh-h)n_z}
\;,\non\]
where the $*$ on the sum recalls that the number of endpoints must be compatible with the number of external legs, namely 
$n_4+n_3/2\ge m/2-1$. 

In conclusion, we can use the iterative integration procedure above for all scales 
$h\ge \bar h$. For smaller scales we need to modify the multiscale integration procedure, as described in the following sections. 
Note that at scale $\bar h$ the bound \eqref{3.26bis} leads to the following estimate on the kernels with $m$ external legs:
\[ \|W^{(\bar h)}_{(m_+,m_-);{\boldsymbol\a}_1,\ldots,{\boldsymbol\a}_m}\|\le 
 \g^{\bar h(2-\frac{m}2-\|{\boldsymbol\a}^0\|-\frac12\|\vec {\boldsymbol\a}\|)}\sum_{{\bf n}>\V0}^*
\l^{\frac{n_3}2+n_4}\tilde z^{n_z}\d^{n_\n} \tilde K^{n} (n_4+\tfrac{n_3}2-\tfrac{m}{2})! \;.\label{3.26tris}\]
%
Of course, since $\g^{\bar h}$ is of the order $\l$, the factor $ \g^{\bar h(2-\frac{m}2-\|{\boldsymbol\a}^0\|-\frac12\|\vec {\boldsymbol\a}\|)}$ could be partially 
simplified with $\l^{\frac{n_3}2+n_4}$: however, for the subsequent bounds, it is conceptually more transparent to think of them 
as two separate factors.

\pagina


\subsection{The infrared integration}\label{sec3.c}

In the previous section we discussed the integration of the ultraviolet scales, up to a scale $\bh$ such that $\g^\bh$ is of the order $\l$. We are now  left with
\[ \Xi=e^{-|\L|(f^B(\sqrt{\r_0})+\bar\n\r_0 +\sum_{h > \bh} F_{h})}\int P^B_{\le \bar h}(d\ps^{(\le \bar h)})\,e^{-V^{(\bh)}(\ps^{(\le \bh)})}\;,\label{3c.1}\]
where we dropped the labels $\L$ for simplicity, and $P^B_{\le \bar h}(d\ps^{(\le\bar h)}):=\prod_{h\le  \bh}P^B_{h}(d\ps^{(h)})$. In order to perform the infrared integration we rewrite the propagator $g^B_{\le\bar h}$ of $P^B_{\le \bar h}(d\ps^{(\le \bar h)})$ as $g^B_{\le\bar h}=\tilde g_{\bar h}  +g_{\le \bar h}$, where $\tilde g_\bh=\tilde g_\bh^{(1)}+\tilde g_\bh^{(2)}$ and
\[ \label{eq3.36}
\tilde g_\bh^{(1)}(\xx-\yy) = \int \frac{d\kk}{(2\pi)^3} e^{-i\kk(\xx-\yy)}& \begin{pmatrix}\tfrac1{-ik_0+|\vk|^2}&0\\0& 
\frac1{ik_0+|\vk|^2}\end{pmatrix}\cdot\\
&\cdot
\big[1-\chi \big(\g^\bh(|x_0|^2+|\vx|^4)^{\nicefrac{1}{2}} \big)-\chi \big(\g^{-\bh}(|k_0|^2+|\vk|^4)^{\nicefrac{1}{2}}\big) \big]\non
\]
and 
\[ \tilde g^{(2)}_{\bar h}(\xx-\yy)=\int_{\mathbb R^3} \frac{d\kk}{(2\p)^3}\frac{e^{-i\kk(\xx-\yy)}}{k_0^2+(\e'(\vk))^2}& \begin{pmatrix}ik_0 +F(\vk)& -\l\r_0\hat v(\vk)\\
-\l\r_0\hat v(\vk)& -ik_0+F(\vk)\end{pmatrix}\cdot \non \\
& 
\cdot\big[\chi(\g^{-\bar h}(|k_0|^2+|\vk|^4)^{\nicefrac{1}{2}})-\chi(\g^{-\bar h}\|\kk\|)\big]\label{eq3.36bis}
\]
with $\|\kk\|^2:= k_0^2 + 2\l \r_0 \hat{v}(\vec{0})|\vk|^2 $. The propagator  $g_{\le \bar h}$ is defined by an expression similar to \eqref{eq3.36bis}, with 
the cutoff function $\big[\chi(\g^{-\bar h}(|k_0|^2+|\vk|^4)^{\nicefrac{1}{2}})-\chi(\g^{-\bar h}\|\kk\|)\big]$  replaced by $\chi(\g^{-\bar h}\|\kk\|)$ under the integral sign.
In Appendix \ref{bounds} we show that $\tilde g_{\bar h}$ admits qualitatively the same dimensional bound as $g^B_{\bar h}$, namely
\[|\dpr_{x_0}^{n_0}\dpr_{\vx}^{\vn}\tilde g_{\bar h}(\xx)|\le \frac{C_{N,n_0,\vn}\g^{\bar h(1+n_0+|\vn|/2)}}{1+\big[\g^{\bar h}(|x_0|+|\vx|^2)\big]^N}
\label{s3.37}\]
for all $N,n_0,n_1,n_2\ge 0$ (here $\vn=(n_1,n_2)$) and suitable constants $C_{N,n_0,\vn}>0$. Therefore, we can rewrite the functional integral in the r.h.s. of \eqref{3c.1} as
\[\int P_{\le \bar h}(d\ps^{(\le \bar h)})\int \tilde P_{\bar h}(d\tilde \psi^{(\bar h)})\,e^{-V^{(\bh)}(\tilde\ps^{(\bh)}+\ps^{(\le \bar h)})}\label{3.34}\]
where $\tilde P_{\bar h}$ has propagator $\tilde g_{\bar h}$ and $P_{\le\bar h}$ has propagator $g_{\le\bar h}$. Next we integrate out the 
field $\psi^{(\bar h)}$ and we end up with a new effective potential $\bar V^{(\bar h)}$ admitting the same dimensional estimates as $V^{(\bar h)}$. 
The basic idea for integrating the lower scales is to rewrite $\chi(\g^{-\bar h}\|\kk\|)$ as $\sum_{h\le \bar h}f_h(\kk)$, with $f_h(\kk)=\chi(\g^{-h}\|\kk\|)
-\chi(\g^{-h+1}\|\kk\|)$, into the definition of $g_{\le \bar h}$: such a rewriting induces a multiscale resolution of $g_{\le \bar h}$ in the form 
$g_{\le\bar h}=\sum_{h\le \bar h}g^{(h)}$, which 
is used to integrate step by step the functional integral, in a way analogous to what we discussed so far. The outcome of this 
new multiscale integration can again be expressed in terms of new trees, whose endpoints represent the effective interaction $\bar V^{(\bar h)}$ on scale $\bar h$.
The point is that 
$g^{(h)}$ satisfies new dimensional estimates, which are qualitatively different from \eqref{s3.37}. Therefore, the kernels of the effective potentials on scale $h<\bar h$
satisfy 
new dimensional estimates, which force us to change the definition of localization and renormalization. 
In order to define the infrared integration and localization procedure in the most transparent way, it is convenient to re-express the 
infrared field $\ps^{(\le \bar h)}$ in terms of its real and imaginary parts, as first suggested by G. Benfatto
\cite{benfatto} and later used in \cite{CaDiC1,CaDiC2}:
\[
\ps^{\pm(\le \bar h)}_\xx=\frac{1}{\sqrt2} \lft(\ps_\xx^{l(\le \bar h)} \pm i\ps^{t(\le \bar h)}_\xx \rgt)\;.  \label{newvar}
\]
These new fields have natural scaling properties, as we shall see in a moment, for the good reason that they 
represent the longitudinal and transverse components of $\ps^{(\le \bar h)}$, with respect to the set of stationary points of 
$f(\xi)$, see \eqref{3.9a} and the following comments. 
In terms of the new ``basis" $\psi^{l},\psi^{t}$, the propagator $g_{\le \bar h}$ takes the form 
\[ g_{\le \bar h}(\xx-\yy)=
\int_{\mathbb R^3} \frac{d\kk}{(2\p)^3}\frac{e^{-i\kk(\xx-\yy)}}{k_0^2+(\e'(\vk))^2}\begin{pmatrix}|\vk|^2 & k_0\\
-k_0 & 2\l\r_0\hat v(\vk)+|\vk|^2\end{pmatrix}
\chi(\g^{-\bar h}\|\kk\|)\;,\label{3.36}
\]
where the first row and column now correspond to the index $l$, while the second row and column to the index $t$.

\subsubsection{Non-renormalized bounds}\label{s3.C.1}

Let us briefly describe here the naive (i.e., non-renormalized) infrared multi-scale procedure that one would get by decomposing the propagator as suggested after 
\eqref{3.34}. This digression will be helpful in order to compute the scaling dimensions and to define a proper localization procedure. 
We write $g_{\le\bar h}(\xx)=\sum_{h\le\bar h}g^{(h)}(\xx)$, with $g^{(h)}(\xx)$ that, in the basis $\psi^l,\psi^t$, is given by an expression similar to  
\eqref{3.36}, with $\chi(\g^{-\bar h}\|\kk\|)$ replaced by $f_h(\kk)$. The single-scale propagator has the following scaling property, which can be derived in a way analogous to the proof of \eqref{s3.37}
(see Appendix \ref{bounds}): if $\a,\a'\in\{l,t\}$ 
\[|\dpr_{x_0}^{n_0}\dpr_{\vx}^{\vn}g^{(h)}_{\a,\a'}(\xx)|\le C_{N,n_0,\vn} \g^{h(1+n_0+|\vn|)}\g^{-\bar h|\vn|/2}
\frac{\g^{(h-\bar h)(\d_{\a,l}+\d_{\a',l})}}{1+\big[\g^{h}(|x_0|+\g^{-\bar h/2}|\vx'|)\big]^N}\label{3.37}\]
for all $N,n_0,n_1,n_2\ge 0$ and suitable constants $C_{N,n_0,\vn}>0$. Moreover $\vx':=\vx/\sqrt{\r_0}$, which has the same physical dimensions as $x_0$.
Eq.\eqref{3.37} makes apparent that the matrix elements 
of $g^{(h)}$ in the basis $l,t$ have well-defined scaling properties (while, of course, in the basis $\pm$ they have not). 
Eq.\eqref{3.37} induces a dimensional estimate on the kernels of the effective potential, 
via the same procedure used in the previous sections. More precisely, we first 
rewrite the effective potential $\bar V^{(\bar h)}$ in the basis $l,t$, then we 
proceed as in the derivation of \eqref{2.29}, thus finding (details are left to the reader):
\[ &
\|W^{(h)}_{m_l,m_t;{\boldsymbol\a}_1,\ldots,{\boldsymbol\a}_m}\|  \le\non \\
&\le \sum_{{\bf n}>0} \sum_{\t\in\tilde\TT_{\bar h;\nn}^{(h)}} 
K^n
\sideset{}{^{(m_l, m_t)}}  \sum_{{\bf P}\in\PP_\t} \hskip -0.2cm \sum_{\GG\in\tilde\G(\t,{\bf P})}\Big[\prod_{v\ {\rm not}\ {\rm e.p.}}\frac1{s_v!}  
 \g^{\bar h[-(\sum_{i=1}^{s_v}|P_{v_i}^l|-|P_v^l|)+s_v-1-\frac12(\sum_{i=1}^{s_v}\vq(P_{v_i})-\vq(P_v))]} 
 \non \\
&    \label{3.38}
  \g^{h_v\big[\frac12(\sum_{i=1}^{s_v}|P_{v_i}^t|-|P_v^t|)+\frac32(\sum_{i=1}^{s_v}|P_{v_i}^l|-|P_v^l|)
-3(s_v-1)+(\sum_{i=1}^{s_v}q(P_{v_i})-q(P_v))\big]}  \Big] \\
&
 \Big[\prod_{v\ {\rm e.p.}}\g^{\bh(2 -\frac12|P_v|-q_0(P_v)-\frac12\vq(P_v))} 
\l^{\frac12n_{3,v}+n_{4,v}}\tilde z^{n_{z,v}}\d^{n_{\n,v}}  (n_{4,v}+\tfrac{n_{3,v}}2-\tfrac{|P_v|}{2})! \Big]  \non
\]
where:\begin{itemize}
\item $\tilde\TT_{\bar h;\nn}^{(h)}$ is a family of trees with endpoints on scale $\bar h$. Each endpoint represents one of the contributions to $\bar V^{(\bar h)}$ generated by the 
ultraviolet integration described in the previous section; note that now there are infinitely many different types of endpoints, labelled by the number and types of external legs, 
and by their {\it order}: an endpoint $v$ with $n_{ext}^l$ (resp. $n_{ext}^t$) external legs of type $l$ (resp. $t$) 
of order ${\bf n}_v=(n_{z,v},n_{\n,v},n_{3,v},n_{4,v})$ is by definition the sum of the (ultraviolet) trees with the proper number of external legs and
$n_\n$ endpoints of type $(1,1)$, etc., in the sense of Eqs.\eqref{3.25first}
and \eqref{3.26tris}. The label ${\bf n}$ attached to $\tilde\TT_{\bar h;\nn}^{(h)}$ refers to the total order of the tree, which is the sum of the orders of its endpoints. 
\item $P_v$ is a set of indices labelling the fields ``exiting" from the vertex $v$, in the sense described in Section \ref{s2.multiscale}. We assume that $P_v$ also carries the information 
about the type (either $l$ or $t$) of the fields
and the number of derivatives acting on them: given a field label $f\in P_v$, we denote by $\a(f)\in\{l,t\}$ its type, and by $q_i(f)$ the number of derivatives $\dpr_i$ acting on it.
Moreover, $P^l_v=\{f\in P_v: \a(f)=l\}$, $P^t_v=\{f\in P_v: \a(f)=t\}$, and $q_i(P_v)=\sum_{f\in P_v}q_i(f)$. Finally, $\vq(f)=\sum_{i=1}^2q_i(f)$, $q(f)=\sum_{i=1}^3q_i(f)$,
$\vq(P_v)=\sum_{i=1}^2q_i(P_v)$ and $q(P_v)=\sum_{i=1}^3q_i(P_v)$. 
\item The two dimensional factors in the second and third lines collect all the dimensional factors coming from the estimates of the propagators (see \eqref{3.37}) and 
the effect of the integrals over $\xx$ along the lines of the spanning tree (see comments after \eqref{2.29}): roughly speaking every propagator on scale $h$ obtained by 
contracting two fields $f_1$ and $f_2$ carries a dimensional factor
\[\prod_{f\in\{f_1,f_2\}}\g^{h(\frac12+q(f))}\g^{-\frac12{\bar h}\vq(f)}\g^{(h-\bar h)\d_{\a(f),l}}\] and every integral carries a factor $\g^{-3h+\bar h}$.
Finally, the factor in the last line comes from the dimensional estimates of the endpoints, see \eqref{3.26tris}.
\end{itemize}
Using the analogues of \eqref{2.33ac}--\eqref{2.36gh} we find
\[ &
\|W^{(h)}_{m_l,m_t;{\boldsymbol\a}_1,\ldots,{\boldsymbol\a}_m}\|  \le\g^{\bar h(-1 +m_l + \frac12 \| \vec {\boldsymbol  \a}\| )}
\g^{h (3 -\frac12 m_t -\frac32 m_l  - \|{\boldsymbol \a}^0\| - \| \vec {\boldsymbol \a}\|)}  
\non \\
& \sum_{{\bf n}>0} \sum_{\t\in\tilde\TT_{\bar h;\nn}^{(h)}} 
K^n
\sideset{}{^{(P_{v_0})}}\sum_{{\bf P}\in\PP_\t} \sum_{\GG\in\tilde\G(\t,{\bf P})}\Big[\prod_{v\ {\rm not}\ {\rm e.p.}}\frac1{s_v!}  
 \g^{(h_v - h_{v'})d_v} \Big] \label{3.44}\\
& \Big[\prod_{v\ {\rm e.p.}}\g^{(\bh-h_{v'})d_v}
\l^{\frac12n_{3,v}+n_{4,v}}\tilde z^{n_{z,v}}\d^{n_{\n,v}}  (n_{4,v}+\tfrac{n_{3,v}}2-\tfrac{|P_v|}{2})! \Big]  \non
\]
where the scaling dimension $d_v$ is
\[  \label{nonrend}
d_v = 3  - \frac32 |P_v^l|-\frac12 |P^t_v| - q(P_v)
\]
On the basis of \eqref{3.44}-\eqref{nonrend}, we see that there is a finite number of 
relevant and marginal sub-diagrams (once again, the relevant sub-diagrams are those with $d_v>0$,
while the marginal ones are those with $d_v=0$). A (almost complete) list of the relevant and marginal terms 
is shown in Fig.\ref{locBEC}.

%
\feyn{
\begin{fmffile}{BoseFeyn/locBEC}
\unitlength = 1cm  
\def\myl#1{2.5cm} 
\[
& \parbox{\myl}{\centering{	 	   
		\begin{fmfgraph*}(2,1.25)
			\fmfleft{i1,i2,i3}
			\fmfright{o1,o2,o3}
			\fmf{plain}{i1,v,o2}
			\fmf{plain}{i2,v,o1}
			\fmf{plain}{i3,v,o3}
			\bBall{v}
		\end{fmfgraph*}  \\ $0$
	}}
	\quad \parbox{\myl}{\centering{	 	   
		\begin{fmfgraph*}(2,1.25)
			\fmfleft{i1,i2}
			\fmfright{o1,o2}
			\fmf{plain}{i1,v,o2}
			\fmf{plain}{i2,v,o1}
			\bBall{v}
		\end{fmfgraph*}    \\ $1$
	}}
	\quad \parbox{\myl}{\centering{
		\begin{fmfgraph*}(2,1.25)
			\fmfright{i1,i2}
			\fmfleft{o1}
			\fmf{plain}{i1,v,i2}
			\fmf{dashes, tension=1.5}{o1,v}
			\bBall{v}
		\end{fmfgraph*}   \\ $\nicefrac{1}{2}$
	}}  
	 \quad \parbox{\myl}{\centering{	
		\begin{fmfgraph*}(2 ,1.25)
			\fmfleft{i1}
			\fmfright{o1}
			\fmf{plain}{i1,v1,o1}
			\bBall{v1}
		\end{fmfgraph*}    \\ $2$
	}} 
	 \quad \parbox{\myl}{\centering{	
		\begin{fmfgraph*}(2 ,1.25)
			\fmfleft{i1}
			\fmfright{o1}
			\fmf{dashes}{i1,v1,o1}
			\bBall{v1}
		\end{fmfgraph*}     \\ $0$
	}}   \non \\[12pt] &
 \hskip 1.5cm \parbox{\myl}{\centering{	 	   
		\begin{fmfgraph*}(2,1.25)
			\fmfleft{i1,i2}
			\fmfright{o1,o2,o3}
			\fmf{plain, tension=1.5}{i1,v,i2}
			\fmf{plain}{o2,v,o1}
			\fmf{plain}{v,o3}
			\bBall{v}
		\end{fmfgraph*}     \\ $\nicefrac{1}{2}$
	}} \quad
 \parbox{\myl}{\centering{	 	   
		\begin{fmfgraph*}(2,1.25)
			\fmfleft{i1}
			\fmfright{o1,o2,o3}
			\fmf{dashes, tension=2}{i1,v}
			\fmf{plain}{o2,v,o1}
			\fmf{plain}{v,o3}
			\bBall{v}
		\end{fmfgraph*}       \\ $0$
	}} \quad
\parbox{\myl}{\centering{	 	   
		\begin{fmfgraph*}(2,1.25)
			\fmfleft{i1}
			\fmfright{o1,o2}
			\fmf{plain, tension=1.5}{i1,v}
			\fmf{plain}{o2,v,o1}
			\bBall{v}
		\end{fmfgraph*}   \\ $\nicefrac{3}{2}$
	}} \quad
 \parbox{\myl}{\centering{	
		\begin{fmfgraph*}(2 ,1.25)    
			\fmfleft{i1}
			\fmfright{o1}
			\fmf{plain}{i1,v1}
			\fmf{dashes, tension=0.8}{v1,o1}	
                    \bBall{v1}
		\end{fmfgraph*}     \\ $1$
           }} \non
\]
\end{fmffile}
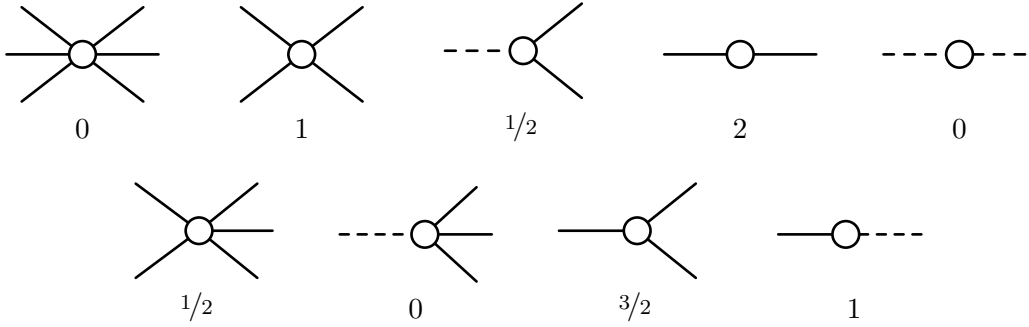
}{A list of sub-diagrams with non-negative scaling dimensions (the
scaling dimension is indicated under each diagram). 
Solid lines correspond to fields of type $t$, while dashed lines to fields of type $l$.
The figure lists all possible relevant and marginal diagrams {\it without} derivatives acting on the external lines. 
The reader can easily reconstruct from \eqref{nonrend} the other relevant and marginal couplings with $q(P_v)>0$.}{locBEC}

It should now be clear that the natural localization procedure needed for renormalizing the infrared theory requires the introduction 
of an $\LL$ operator acting non-trivially on all the kernels with $d_v\ge 0$, i.e., those in Fig.\ref{locBEC} (plus the few others
with derivatives acting on the external fields). At each step we iteratively ``dress" the propagator by combining the marginal quadratic 
terms with the gaussian reference measure. This means that the renormalized single-scale propagator is not going to be the same $g^{(h)}$ introduced above, but rather 
a dressed version of it (i.e., a similar propagator, but defined in terms of a few renormalized parameters). This and other details are 
described in the next section.

\subsubsection{Renormalized bounds}\label{s3.C.2}

Motivated by the discussion in Section \ref{s3.C.1}, we define a modified multiscale integration procedure of the degrees of freedom at scales 
$h\le \bar h$, along the lines sketched at the end of the previous subsection. After the integration of the fields on 
scales $\bar h,\bar h-1,\ldots, h+1$, we rewrite \eqref{3c.1} as
\[ \Xi=e^{-|\L|(f^B(\sqrt{\r_0})+\bar\n\r_0 +\sum_{k \ge h} F_{k})}\int P_{\le h}(d\ps^{(\le h)})\,e^{-\VV^{(h)}(\ps^{(\le h)})}\;,\label{IRpot}\]
where $P_{\le h}(d\ps^{(\le h)})$ has propagator (in the basis $l,t$):
\be \label{3.25}
g^{(\le h)}(\xx)=  \int  \frac{d\kk}{(2\pi)^3}\, \chi_{h}(\kk)\,\frac{\, e^{-i\kk \cdot \xx}}{\DD_h(\kk)}\left(\begin{array}{cc}
\tl{A}_h(\kk)\, |\vk|^{2} + \tl B_h(\kk)\, k_0^2   & \tl E_h(\kk)\, k_{0}\\
-\tilde E_h(\kk)\, k_{0} &   \tl Z_h(\kk)
\end{array}\right)
\ee
with 
\[\DD_h(\kk) = \tl Z_h(\kk)\big(\tl C_h(\kk) k_0^2 +\tl A_h(\kk)  |\vk|^2\big)\;,\qquad 
\tl C_h(\kk)\tl Z_h(\kk):=\tl E_h^2(\kk) +\tl B_h(\kk)\tl Z_h(\kk)\;,\label{3.48b}\]
and, defining $A_h:=\tilde A_h(\V0)$, $B_h:=\tilde B_h(\V0)$, etc, the cutoff function $\chi_h(\kk)$ is:
\[\chi_h(\kk):=\chi(\g^{-h}\|\kk\|_h)\;,\quad{\rm with}\quad \|\kk\|_h^2:=k_0^2+(A_h/C_h)|\vk|^2\;.\label{3.49c}\] 
Moreover, 
\[ \VV^{(h)}(\ps)=&\sum_{m_l,m_t\ge 0}^*
\sum_{{\boldsymbol\a},{\boldsymbol \alpha'}}\int d\xx_1\cdots d\yy_{m_t} 
W^{(h)}_{m_l,m_t;{\boldsymbol\alpha},{\boldsymbol\a'}}(\xx_1,\ldots,\xx_{m_l};\yy_1,\cdots,\yy_{m_t})\cdot  \non \\
&\quad \cdot
\Big[\prod_{i=1}^{m_l}\dpr_{\xx_{i}}^{{\boldsymbol{\a}}_{i}}\ps_{\xx_i}^l\Big]\,\Big[\prod_{i=1}^{m_t}
\dpr_{\yy_{i}}^{{\boldsymbol{\a}}_{i}'}
\ps^t_{\yy_{i}}\Big]\label{2.54yh}\]
with the $*$ on the sum indicating the constraint that $m:= m_l+m_t>0$, and ${\boldsymbol\a}$  (resp. ${\boldsymbol\a'}$)
being a shorthand for $({\boldsymbol\a}_1,\ldots,{\boldsymbol\a}_{m_l})$ (resp. $({\boldsymbol\a}_1',\ldots,{\boldsymbol\a}_{m_t}')$). If $h=\bar h$, the effective potential $\VV^{(\bar h)}$ coincides with the 
function $\bar V^{(\bar h)}$ introduced after \eqref{3.34}.
For what follows, it is also convenient to re-express \eqref{2.54yh}  in momentum space:
\[
\VV^{(h)}(\ps) = &\sum^*_{m_{l}, m_{t} \geq 0} 
\int \frac{d \kk_1}{(2\pi)^{3}} \ldots \frac{d\kk_{m}}{(2\pi)^{3}}\,
\hW^{(h)}_{m_l, m_t}\bigl(\kk_2, \cdots, \kk_m\bigr)\,(2\p)^3
\d\Bigl(\sum_{i=1}^{m} \kk_i \Bigr)\cdot\non\\
&\cdot \Big[ \prod_{i=1}^{m_l} \hat \ps^l_{\kk_i} \Big] \Big[ \prod_{i=m_l+1}^{m} \hat \ps^t_{\kk_i} \Big]
\]
where $m=m_l+m_t$.
Finally, $F_h$, $\tl{A}_h$, $\tl{B}_h$, $\tl{E}_h$, $\tl{Z}_h$ and $W^{(h)}_{m_l,m_t;{\boldsymbol\alpha},{\boldsymbol\a'}}$ are defined recursively via the inductive construction described below. 

In order to inductively prove (\ref{IRpot}), we split $\VV^{(h)}$ as
$\LL \VV^{(h)} + \RR \VV^{(h)}$, where $\RR =
1-\LL$ and $\LL$, the {\em localization operator}, is a linear
operator on functions of the form (\ref{2.54yh}), defined by its action on the
kernels $\hW^{(h)}_{m_l,m_t}$ in the following way:
\[ \label{loc1}
\LL \hW^{(h)}_{0,6}(\kk_2, \ldots, \kk_6) & := \hW^{(h)}_{0,6}(\bz, \ldots, \bz)  \non \\
\LL \hW^{(h)}_{0,5}(\kk_2, \ldots, \kk_5) &  := \hW^{(h)}_{0,5}(\bz, \ldots, \bz) \non \\
\LL \hW^{(h)}_{0,4}(\kk_2, \kk_3,\kk_4) & := \hW^{(h)}_{0,4}(\bz, \bz,\bz)  + \sum_{i=2}^4 \kk_i \dpr_{\kk_i}\hW^{(h)}_{0,4}(\bz, \bz,\bz)\non  \\
\LL \hW^{(h)}_{0,3}(\kk_2, \kk_3) & := \hW^{(h)}_{0,3}(\bz, \bz) + \sum_{i=2}^3 \kk_i \dpr_{\kk_i} \hW^{(h)}_{0,3}(\bz,\bz)
 \\
\LL \hW^{(h)}_{0,2}(\kk) & := \hW^{(h)}_{0,2}(\bz) + 
 \kk \cdot \dpr_{\kk} \hW^{(h)}_{0,2}(\bz) +
\frac{1}{2} \sum_{i,j=0}^2 k_i k_j \dpr_{k_i} \dpr_{k_j} \hW^{(h)}_{0,2}(\bz) \non \\
\LL \hW^{(h)}_{1,3}(\kk_2,\kk_3,\kk_4) & := \hW^{(h)}_{1,3}(\bz,\bz,\bz) \non  \\
\LL \hW^{(h)}_{1,2}(\kk, \pp) & := \hW^{(h)}_{1,2}(\bz,\bz) \non  \\
\LL \hW^{(h)}_{1,1}(\kk) & := \hW^{(h)}_{1,1}(\bz)+ \kk\cdot \dpr_{\kk} \hW^{(h)}_{1,1}(\bz) \non\\
\LL \hW^{(h)}_{2,0}(\kk) & := \hW^{(h)}_{2,0}(\bz) \non \]
and $\LL \hW^{(h)}_{m_l, m_t}:=0$ otherwise. Of course, if desired, one could translate these definitions 
in real rather than momentum space, in which case they would take a form analogous to those given above, in the
theory of the quantum critical point, or in the ultraviolet integration of the condensed phase. In momentum space, 
the definition of localization should be understood as follows. When we Taylor expand $\hat W^{(h)}_{m_l,m_t}$ 
with respect to the momenta, the term of order $n$ in the Taylor expansion has an improved scaling dimension,
as compared to $\hat W^{(h)}_{m_l,m_t}$ itself: more precisely, if $\hat W^{(h)}_{m_l,m_t}$ has scaling dimension 
$d(m_l,m_t)=3-\frac32m_l-\frac12m_t$, then the $n$-th order term in the Taylor expansion in $\kk_i$ has 
dimension $d(m_l,m_t)-n$. The reason is that each operator $\kk_i\dpr_{\kk_i}$ dimensionally corresponds to a scaling 
factor $\g^{h-h'}$, where $h'>h$: in fact, when $\dpr_{\kk_i}$ acts on $\hat W^{(h)}_{m_l,m_t}(\kk_2,\ldots,\kk_m)$,
which is a sum of Feynman diagrams, the derivative can act onto one of the propagators appearing in 
such diagrams, each of which has a scale label strictly larger than $h$; from a dimensional point of view, a derivative $\dpr_{\kk_i}$ 
acting on a propagator $\hat g^{(h')}(\kk_i+\qq)$ on scale $h'> h$ behaves dimensionally (up to $\l$-dependent factors)
as a multiplication by $\g^{-h'}$, see \eqref{3.37}. Similarly, the factor $\kk_i$ in $\kk_i\dpr_{\kk_i}$ can be thought of as being attached to one 
of the external legs, whose scale label is $\le h$, which means that it will be contracted in the multiscale integration 
process in the form of a propagator $\hat g^{(\le h)}(\kk_i)$; from a dimensional point of view, 
$\kk_i\cdot\hat g^{(\le h)}(\kk_i)$ behaves like $\hat g^{(\le h)}(\kk_i)$ times $\g^h$. Therefore, the rationale behind the definition of $\LL \hat W^{(h)}_{m_l,m_t}$ 
is that we set it equal to its Taylor series in $\kk_i$ truncated at order $n$, where $n$ is such that $d(m_l,m_t)-n\ge 0$ and $d(m_l,m_t)-n-1<0$. In this way, the 
rest of the Taylor expansion, which is part of $\RR \VV^{(h)}$, is irrelevant in the sense that its scaling dimension 
is negative. 

A priori, the definitions \eqref{loc1} produce as many {\it running coupling constants} as the number of terms 
appearing in the r.h.s. However, luckily enough, not all those terms are really there: many of them are vanishing by symmetry. More precisely, using the parity properties of the propagator and of the interaction, one 
easily sees that
\[
&\hW^{(h)}_{0,5}(\bz,\cdots,\bz) = 0 \;,
\quad   \dpr_{\kk}\hW^{(h)}_{0,4}(\bz, \bz,\bz)=0\;,\quad \hW^{(h)}_{0,3}(\bz,\bz) =0 \;,\\
&\dpr_{\kk}\hW_{0,2}^{(h)}(\bz)=0\;,\quad  \hW^{(h)}_{1,3}(\bz,\bz,\bz) =0\;,  \quad 
\hW^{(h)}_{1,1}(\bz)=0\;.
\]
Moreover
\[ & 
\dpr_{k_j} \hW_{1,1}^{(h)}(\bz) =0\;,\quad {\rm if}\quad    j=1,2   \non \\
& \dpr_{k_i} \dpr_{k_j}  \hW_{0,2}^{(h)}(\bz)=0\;,\quad {\rm if}\quad  i\neq j   \;,
\]
and $\dpr_{k_1}^2 \hW_{0,2}^{(h)}(\bz)=\dpr_{k_2}^2 \hW_{0,2}^{(h)}(\bz)$. In addition to these parity cancellations,
we can also use the conservation of momentum and the 
permutation symmetry 
between the external fields to infer that 
\[\int \frac{d \kk_1\,d\kk_2\,d\kk_3}{(2\pi)^{6}} (\kk_2\dpr_{\kk_2}+\kk_3\dpr_{\kk_3}) \hW_{0,3}^{(h)}(\bz,\bz)
\hat \psi^t_{\kk_1}\hat \psi^t_{\kk_2}\hat \psi^t_{\kk_3}\cdot \d(\kk_1+\kk_2+\kk_3)=0\;. 
\]
We now let 
\[ \label{cc}
 \g^{-\bh}\l_{6,h}   :=\hW^{(h)}_{0,6}(\bz,\cdots,\bz) \;,  &&   \g^{h/2}\m_h & :=\hW^{(h)}_{1,2}(\bz,\bz)  \non \\
\g^{h-\bh}\l_h  :=\hW^{(h)}_{0,4}(\bz,\cdots,\bz)\;, && 
\g^{2h-\bh}\n_h  &:= \hW^{(h)}_{0,2}(\bz)\;.
\]
The constants $\l_{6,h}, \l_{h}, \m_h,\n_h$ are called the {\it running coupling constants} (see Fig.\ref{CC}), and they are 
all {\it real}, as it follows from the reality properties of the propagator and of the interaction. 
The dimensional factors in the l.h.s. of the definitions \eqref{cc} are all of the form $\g^{(h-\bh)(3-\frac32m_l-\frac12m_t)}\g^{\bh(2-\frac12(m_l+m_t))}$, and are 
introduced in order to compensate a product of bad factors of the form $\g^{(\bh-h)d_v}$ associated with the endpoints with $d_v=3-\frac32m_l-\frac12m_t\ge 0$,
as those in the third line of \eqref{3.44}; the choice of these scaling factors is justified a posteriori by the fact that in the 
dimensional estimate of the renormalized kernels (see \eqref{3.73} below) every marginal or relevant endpoint contributes with a factor $\l_{6,h}$, $\l_h$, $\m_h$, or $\n_h$,
depending on its type, without any other extra bad dimensional factor. Using \eqref{3.26tris}, we see that $\l_{6,\bar h}$ is of order $\l^3$, 
$\l_{\bar h}$ is of order $\l$, $\m_{\bar h}$ is of order $\l^{1/2}$, and $\n_{\bar h}$ is of order $\d$.

We also define the {\it wave function renormalization constants}  $a_{h}$,  $b_h$, $e_h$ and $z_h$ as follows:
\[ & a_h:=\dpr^2_{k_1} \hW^{(h)}_{0,2}(\bz)=\dpr^2_{k_2} \hW^{(h)}_{0,2}(\bz)\;,\\
& b_h:=\dpr^2_{k_0} \hW^{(h)}_{0,2}(\bz)\;,\label{3.59b}\\
& e_h:=-\dpr_{k_0}\hW^{(h)}_{1,1}(\bz)\;,\label{3.60e}\\
& z_h:= 2\hW^{(h)}_{2,0}(\bz)\;,\label{3.60z}
\]
and remarkably also these constants are all {\it real}. The quadratic part of $\LL \VV^{(h)}$ associated with these 
constants is:
\[ &
\LL_Q \VV^{(h)}(\ps) = \frac12\int \frac{d\kk}{(2\p)^3}(\hat \psi^{l}_{-\kk},\ 
\hat \psi^{t}_{-\kk})\hat M^{(h)}_Q(\kk)\begin{pmatrix}\hat \psi^l_\kk \\ \hat \psi^t_\kk\end{pmatrix}\equiv \frac12\big(\psi,M_Q^{(h)}\psi\big)\;,\label{e3.62}\\
& \hat M^{(h)}_Q(\kk):=\begin{pmatrix} z_h & -e_h k_0\\
e_h k_0 & a_h|\vk|^2+b_hk_0^2\end{pmatrix}\;,\]
which has exactly the same symmetry and reality structure as the exponent of the gaussian weight in the reference
Bogoliubov's measure. Therefore, we can combine $\LL_Q \VV^{(h)}$ with the gaussian measure at scale $h$,
by proceeding as follows. We let
\[  \label{m1}
e^{-|\L|t_h} 
 \tl P_{\le h}(d\ps^{(\le h)}):=
 P_{\le h}(d\ps^{(\le h)})\,e^{-\LL_Q \VV^{(h)}(\ps^{(\le h)})} 
\]
where $t_h$ accounts for the change in the normalization of the two gaussian measures. The ``dressed" measure 
$\tl P_{\le h}(d\ps^{(\le h)})$ has propagator
\be \label{3.25ren}
\tl g^{(\le h)}(\xx)=  \int  \frac{d\kk}{(2\pi)^3}\, \chi_{h}(\kk)\,\frac{\, e^{-i\kk \cdot \xx}}{\DD_{h-1}(\kk)}\left(\begin{array}{cc}
\tl{A}_{h-1}(\kk)\, |\vk|^{2} + \tl B_{h-1}(\kk)\, k_0^2   & \tl E_{h-1}(\kk)\, k_{0}\\
-\tilde E_{h-1}(\kk)\, k_{0} &   \tl Z_{h-1}(\kk)
\end{array}\right)
\ee
where
\[
\tl A_{h-1}(\kk) =  \tl A_h(\kk) + a_h \chi_h(\kk)\;, &&  \tl B_{h-1}(\kk) =  \tl B_h (\kk)+ b_h \chi_h(\kk) \;, \non \\
 \tl E_{h-1}(\kk) =  \tl E_h (\kk) + e_h \chi_h(\kk) \;, && \tl Z_{h-1}(\kk) =  \tl Z_h(\kk) + z_h \chi_h(\kk)\;,\label{e3.65}
\]
and ${\mathcal D}_{h-1}(\kk)$ is defined as in \eqref{3.48b} with $h$ replaced by $h-1$. The initial data at scale 
$\bar h$ for these renormalization constants are:
\[ \label{3.61bis}
\tl A_\bh(\kk) \equiv A_\bh= 1  && \tl B_\bh(\kk) \equiv B_\bh= 0 && \tl E_\bh(\kk)\equiv E_\bh = 1 && \tl Z_\bh(\kk) = 
2\l\r_0\hat v(\vk)+|\vk|^2\;,
\]
so that $Z_{\bh}=2\l\r_0\hat v(\v0)$. Next we split the cutoff function in \eqref{3.25ren} as $\chi_h(\kk)= \chi_{h-1}(\kk) + 
\tilde f_h(\kk)$, where $\chi_{h-1}(\kk)$ is defined as in \eqref{3.49c}, with $h$ replaced by $h-1$,
and we define
\[\LL_C \VV^{(h)}(\psi)&=\g^{-\bh}\l_{6,h}\int d\xx (\psi^t_\xx)^6+\g^{h/2}\m_h\int d\xx \psi^l_\xx(\psi^t_\xx)^2\\
&+
\g^{h-\bh}\l_h\int d\xx (\psi^t_\xx)^4+\g^{2h-\bh}\int d\xx (\psi^t_\xx)^2\;,\label{3.LC}\]
(the label $C$ stands for ``couplings") so that $\LL \VV^{(h)}(\psi)=\LL_Q \VV^{(h)}(\psi)+\LL_C \VV^{(h)}(\psi)$. 
Then we rewrite \eqref{IRpot} as
\[&\Xi=e^{-|\L|(f^B(\sqrt{\r_0})+\bar\n\r_0 +\sum_{k \ge h} F_{h}+t_h)}
\int P_{\le h-1}(d\ps^{(\le h-1)})\cdot\label{3.68}\\
&\qquad \cdot\int \tilde P_{h}(d\ps^{(h)}) e^{-\LL_C \VV^{(h)}(\ps^{(\le h-1)}+ \ps^{(h)})-
\RR  \VV^{(h)}(\ps^{(\le h-1)}+ \ps^{(h)})}\non
\]
where $P_{\le h-1}(d\ps^{(\le h-1)})$ has the same propagator as \eqref{3.25}, with $h$ replaced by $h-1$, 
while $\tilde P_{h}(d\ps^{(h)}) $ has propagator $\tilde g^{(h)}(\xx)$ given by an expression analogous to the r.h.s. of 
\eqref{3.25ren}, with $\chi_h(\kk)$ replaced by $\tilde f_h(\kk)=\chi_h(\kk)-\chi_{h-1}(\kk)$. 

At this point, we integrate the field on scale $h$ and define:
\[ \label{3.64}
|\L| \tl F_h + \VV^{(h-1)}(\ps) := -\log\int \tilde P_{h}(d\ps^{(h)}) e^{-\LL_C \VV^{(h)}(\ps^{(\le h-1)}+ \ps^{(h)})-
\RR  \VV^{(h)}(\ps^{(\le h-1)}+ \ps^{(h)})}\;.
\]
By plugging \eqref{3.64} into \eqref{3.68}, we reproduce our inductive assumption \eqref{IRpot} at scale $h-1$, with
$F_{h-1}:= t_h + \tl F_h$.  

The inductive integration procedure described above gives rise to a new family of renormalized trees, 
analogous to those described in Section \ref{s3.C.1}. The renormalized trees contributing to $\VV^{(h)}$ are 
now denoted by $\TT^{(h)}_{\bh; \nn}$. The action of an operator $\RR$ is associated with all the vertices $v>v_0$ that 
are not endpoints.
The trees in $\TT^{(h)}_{\bh; \nn}$ can have endpoints on all scales between $h+2$ and $\bh+1$, and the endpoints 
can be either of type $\LL$ (marginal or relevant) or of type $\RR$ (irrelevant). The marginal or relevant endpoints 
can live on all scales between $h+2$ and $\bh+1$, and they represent an interaction term of  type 
$\LL_C\VV^{(h_v-1)}(\ps)$. Depending on the specific monomial in $\LL_C \VV^{(h_v-1)}$ that is associated with the endpoint, we shall say that the endpoint is of type $\l_6$, or $\m$, or $\l$, or $\n$, with obvious convention. 
An endpoint $v$ of type $\LL$ is necessarily contracted on scale $h_{v}-1$, i.e., $P_v\neq P_{v'}$, where $v'$ is the
unique node of the tree preceding  $v$ on $\t$, with $h_{v'}=h_{v-1}$. Finally, the irrelevant endpoints are necessarily on scale 
$\bh+1$ and they can be contracted on any of the lower scales. They are associated with one of the contributions 
to $\RR\VV^{(\bh)}$, where $\VV^{(\bh)}=\bar V^{(\bh)}$ was defined after 
\eqref{3.34}.

%
\feyn{
\begin{fmffile}{BoseFeyn/CC}
\unitlength = 1cm  
\def\myl#1{2.5cm}
\[
& \parbox{\myl}{\centering{	 	   
		\begin{fmfgraph*}(2,1.25)
			\fmfleft{i1,i2,i3}
			\fmfright{o1,o2,o3}
			\fmf{plain}{i1,v,o2}
			\fmf{plain}{i2,v,o1}
			\fmf{plain}{i3,v,o3}
			\fmfdot{v}
		\end{fmfgraph*}  \\ $\g^{-\bh}\l_{6,h}$
	}}
	\quad \parbox{\myl}{\centering{
		\begin{fmfgraph*}(2,1.25)
			\fmfright{i1,i2}
			\fmfleft{o1}
			\fmf{plain}{i1,v,i2}
			\fmf{dashes, tension=1.5}{o1,v}
			\fmfdot{v}
		\end{fmfgraph*}   \\ $\g^{h/2}\m_h$
	}}  
	\quad \parbox{\myl}{\centering{	 	   
		\begin{fmfgraph*}(2,1.25)
			\fmfleft{i1,i2}
			\fmfright{o1,o2}
			\fmf{plain}{i1,v,o2}
			\fmf{plain}{i2,v,o1}
			\fmfdot{v}
		\end{fmfgraph*}    \\ $\g^{h-\bh}\l_h$
	}}
	 \quad \parbox{\myl}{\centering{	
		\begin{fmfgraph*}(2 ,1.25)
			\fmfleft{i1}
			\fmfright{o1}
			\fmf{plain}{i1,v1,o1}
		\fmfdot{v1}
		\end{fmfgraph*}    \\ $\g^{2h-\bh}\n_h$
	}}  \nonumber
\]
\end{fmffile}
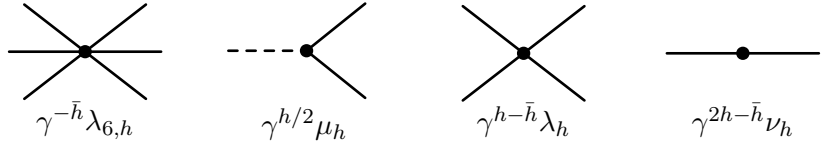
}{Renormalization group analysis for the interacting condensed state. Running coupling constants in the infrared region, $h\leq \bh$.
}{CC}

The renormalized single-scale propagator defined in \eqref{3.25} satisfies a modified dimensional bound, as compared to \eqref{3.37}, depending on the renormalization constants: if $\a,\a'\in\{l,t\}$:
\[  \label{3.37bis}
|\dpr_{x_0}^{n_0}\dpr_{\vx}^{\vn}g^{(h)}_{\a,\a'}(\xx)| & \le 
C_{N,n_0,\vn} \g^{h(1+n_0+|\vn|)}\g^{-\bar h|\vn|/2}
\frac{\g^{(h-\bar h)(\d_{\a,l}+\d_{\a',l})}}{1+\big[\g^{h}\big(|x_0|+\sqrt{\frac{C_h}{A_h}}|\vx'|\big)\big]^N}\cdot\\
&\cdot\frac1{A_h} \Big(\g^{\bh}\frac{C_h}{A_h}\Big)^{|\vn|/2}\, \Big(\frac{E_h}{\sqrt{C_h Z_h}}\Big)^{(\d_{\a,l}\d_{\a',t}+\d_{\a,t}\d_{\a',l})}
\Big(\g^\bh\sqrt{\frac{C_h}{Z_h}}\,\Big)^{(\d_{\a,l}+\d_{\a',l})}\;,\non
\]
for all $N,n_0,n_1,n_2\ge 0$ and suitable constants $C_{N,n_0,\vn}>0$. In deriving \eqref{3.37bis} we assumed that 
$B_h\ge 0$ (a fact that will be proved below), so that $C_h\ge B_h$. 

Using this dimensional estimate, and the fact that an $\RR$ operator acts on all the vertices $v>v_0$ of the tree that are not endpoints (so that all the scaling dimensions of such vertices are automatically negative), we find the (renormalized) analogue of \eqref{3.38} (details are left to the reader):
\[ &
\|W^{(h)}_{m_l,m_t;{\boldsymbol\a}_1,\ldots,{\boldsymbol\a}_m}\|  \le\g^{\bar h(-1 +m_l + \frac12 \| \vec {\boldsymbol  \a}\| )}
\g^{h (3 -\frac12 m_t -\frac32 m_l  - \|{\boldsymbol \a}^0\| - \| \vec {\boldsymbol \a}\|)}  
\non \\
& \sum_{{\bf n}>0} \sum_{\t\in \TT_{\bar h;\nn}^{(h)}} 
K^n
\sideset{}{^{(P_{v_0})}}\sum_{{\bf P}\in\PP_\t} \sum_{\GG\in \G(\t,{\bf P})}\Big[\prod_{v\ {\rm not}\ {\rm e.p.}}\frac1{s_v!}  
 \g^{(h_v - h_{v'})(d_v-z_v)}U_v \Big] \Big[\prod_{v\ {\rm e.p.}}W_v\Big]\;, \label{3.73}\]
 where:
\begin{itemize}
\item the contribution $W_v$ associated with the endpoints is equal to: \\
(1) $\l_{6,h_{v-1}}$ (resp. $\l_{h_v-1}$, or 
$\m_{h_v-1}$, or $\n_{h_v-1}$) if the endpoint is marginal of type $\l_6$ (resp. relevant of type $\l$, or $\m$, or $\n$); \\
(2) $\g^{(\bh-h_{v'})(d_v-z_v)}
\l^{\frac12n_{3,v}+n_{4,v}}\tilde z^{n_{z,v}}\d^{n_{\n,v}}  (n_{4,v}+\tfrac{n_{3,v}}2-\tfrac{|P_v|}{2})!$ if the endpoint is irrelevant; 
here $v'$ is the scale at which the endpoint is contracted, i.e., the scale of the first node preceding $v$ on $\t$ such that 
$P_v\neq P_{v'}$.  
\item  The function $U_v$ appearing in the product over the vertices that are not endpoints collects 
the constants depending on the renormalization constants $A_h, B_h$, etc, arising from the term in the second line 
of \eqref{3.37bis}, and from the integrations along the spanning tree. It is defined as:
\[U_v&=A_{h_v}^{-\frac12(\sum_{i=1}^{s_v}|P_{v_i}|-|P_v|)}
\Big(\g^{\bh}\frac{C_{h_v}}{A_{h_v}}\Big)^{\frac12(\sum_{i=1}^{s_v}\vec q(P_{v_i})-\vec q(P_v))-(s_v-1)}\cdot\non\\
&\quad \cdot\Big(\frac{E_{h_v}}{\sqrt{C_{h_v} Z_{h_v}}}\Big)^{\tilde n^v_{lt}}
\Big(\g^\bh\sqrt{\frac{C_{h_v}}{Z_{h_v}}}\,\Big)^{\sum_{i=1}^{s_v}|P^l_{v_i}|-|P_v^l|}\;,\label{3.74}
\]
where $n_{lt}^v $ is the number of propagators of type $(l,t)$ or $(t,l)$ 
internal to $v$, but not in any other cluster $w$ following $v$ on $\t$ (i.e., it is the number of propagator of type $lt$ obtained by contracting two fields
internal to $v$, in the sense of item (5) before \eqref{2.23}).
\item The dimensional gain $z_v$ appearing at exponent in the factors $\g^{(h_v-h_{v'})(d_v-z_v)}$ 
is equal to $z_v=\lceil d_v\rceil$ (here $\lceil\cdot\rceil$ indicates the integer part plus 1), if $d_v\ge 0$, 
and $z_v=0$ otherwise. By construction, $d_v-z_v\le -1/2$ for all the vertices in the tree. 
\end{itemize}

Since the renormalized scaling dimension $d_v - z_v$ in the r.h.s. of \eqref{3.73} is negative for all the vertices 
of the tree, we can sum over the scale labels and, by proceeding as in the previous sections, obtain an $n!$ bound 
analogous to (say) \eqref{3.26} for the renormalized kernels of the effective potential. An immediate consequence of the proof leading to \eqref{3.73} is that contributions from trees $\t\in \TT_{\bar h;\nn}^{(h)}$ with a vertex $v$ on scale $h_v=k>h$ admit an improved bound with respect to \eqref{3.73}, with an extra factor $\g^{\theta(h-k)}$, for any $0<\theta<1/2$; this factor can be thought as a dimensional gain with respect to the ``basic'' dimensional bound in \eqref{3.73}. This improved bound is usually referred to as the {\it short memory} property (\ie long trees are exponentially suppressed); it is due to the fact that the renormalized scaling dimensions $(d_v-z_v)$ in \eqref{3.73} are all negative and smaller equal than $-1/2$, and can be obtained by taking a fraction of the factors $\g^{(h_v-h_{v'})(d_v-z_v)}$ associated with the branches of the tree $\t$ on the path connecting the root with the vertex on scale $k$.

Of course, the bound makes 
sense as long as the factors $U_v$ and $W_v$ in \eqref{3.73} remain bounded, at least order by order 
in renormalized perturbation theory. Boundedness of the renormalization and running coupling constants 
is a non trivial fact, which can be analyzed in terms of the flow of such constants under the iterations 
of the multiscale expansion. As mentioned above, the function controlling the flow of the effective constants is called the
beta function, and will be discussed in the next sections. It must be stressed that, due to the large number of 
renormalization and running coupling constants, a brute force study of the beta function is very hard,
particularly if one wants to push the study to the whole infrared limit $h\to-\infty$.
The hope would be to take advantage of a number of remarkable exact relations between the effective constants, which
reduce the number of independent effective constants to be controlled. These exact relations, which can be thought 
of as cancellations in the beta function, follow from Ward Identities, see next section. Let us also anticipate 
the fact that in the presence of a momentum regularization, as the one we are using here, 
the Ward Identities are affected by finite correction terms, due to the cutoffs, which change them as compared with 
the formal expression one would get by neglecting the regularization effects. The implications 
of these correction (anomaly) terms  are dramatic, since they do not allow to prove the exact vanishing of 
the ``bad" terms in the beta function, i.e., of those terms that drive  the effective constants to $+\infty$.
These correction terms are already visible at the one-loop level, as discussed in the next sections. 

\subsubsection{The flow equations}

The $n!$ bounds in \eqref{3.73} allow us to give a meaning at all orders to the renormalized theory, as long as the 
running coupling constants remain small, and the renormalization constants entering the dressed propagator 
are such that the factors $U_v$ in \eqref{3.74} remain of order 1 (as they are on scale $\bh$). The 
iterative construction described above induces a flow equation for the running coupling and renormalization constants, of the form:
\[ & \l_{6,h-1}=\l_{6,h}+\b^{\l_6}_h\;,\qquad \l_{h-1}=\g(\l_{6,h}+\b^{\l}_h)\label{3.75}\\
&\m_{h-1}=\g^{1/2}(\m_h+\b^{\m}_h)\;,\qquad \n_{h-1}=\g^2(\n_h+\b^\n_h)\label{3.77}\\
&A_{h-1}=A_h+\b^A_{h+1}\;,\qquad B_{h-1}=B_h+\b^B_{h+1}\label{3.77b}\\
&E_{h-1}=E_h+\b^E_{h+1}\;,\qquad Z_{h-1}=Z_h+\b^Z_{h+1}\;.\label{3.78}
\]
where the {\it beta functions} $\b^{\#}_h$ are
related to the kernels $W^{(h)}_{m_l,m_t;{\boldsymbol\a}_1,\ldots,{\boldsymbol\a}_m}$ (see definitions from \eqref{cc} to \eqref{3.60z}) 
and, therefore, they are expressed by series
in the running coupling and renormalization constants admitting the same bound \eqref{3.73}. If under the evolution
induced by the flow equations \eqref{3.75}--\eqref{3.78} we reach a scale at which one or more of the 
running coupling constants become of order one, then we stop the flow at that scale, which we denote by $h^*$. 
Otherwise, i.e., if the running coupling constants remain small for all scales, 
we say that the theory is well defined in the infrared, in which case we set $h^*=-\io$. 

Note that, as long as the theory makes sense (i.e., as long as the running coupling constants remain small), the flow equation is dominated by the 
first non-trivial truncation of the series defining the beta function. If the approximate flow obtained by such a truncation drives the constants towards smaller values, then we are in good shape, because the higher order contributions to the beta function will be truly negligible; in such a situation, it is easy to show by a standard stability analysis that the complete flow stays close to its lowest order truncation.
If, on the contrary, the first non-trivial truncation to the beta function drives the running coupling constants towards larger values, then $h^*$ is finite, and we cannot conclude anything 
about the infrared behavior of the system at lower scales\footnote{A special but important case realizes when the first non-trivial truncation of the beta function 
displays a cancellation, which makes the beta function zero at that order: in such a case we 
need to go to higher orders in order to see whether the beta function is truly zero or not. Remarkable examples where this happens are models of
spinless fermions in one dimension \cite{BGPS-Fermi1d,GM,BM-luttinger}, for which  
it is possible to prove that the beta function is zero at all orders, by making use of remarkable, very subtle, cancellations at all orders in perturbation theory, following from 
the Schwinger-Dyson equation combined with local Ward Identities, see \cite{BM-luttinger}.}.
In fact, in such a case, if $h< h^*$ the higher order contributions to the beta function tend to dominate, and we cannot conclude anything sensible from finite truncations 
to the beta function. Unfortunately, the present case belongs to the latter category. In fact, if we truncate the flow equations \eqref{3.75}--\eqref{3.78} 
at the lowest non-trivial order
we find (see Appendix \ref{appB}):
\[&\l_{h-1} = \g\l_h - 2 \g \frac{1}{A_h C_h \g^\bh}\Big( 18 \,\l^2_h 
-12\,\l_h \m_h^2\frac{\g^\bh }{Z_h} +2\m_h^4 \, \frac{\g^{2\bh}}{Z_h^2} \Big) \b_0^{(2)} \non \\
& \hskip 1cm + 4 \g \frac{1}{A_h C_h \g^\bh}\,\lft[\Big(-6\,\l_h \m_h^2\frac{\g^\bh }{Z_h}  +  \m_h^4\, \frac{\g^{2\bh}}{Z_h^2} \Big) \b_0^{(2,\chi)} +  \m_h^4\,\frac{\g^{2\bh}}{Z_h^2} 
{\b}_0^{(3,\chi)}  \rgt] \label{3.80}\\[3pt]
& \m_{h-1} = \g^{1/2}\m_{h}  -2\,\frac{\g^{1/2}}{A_h C_h \g^\bh} \, \m_h \lft[\Big(6 \l_h -   2   \m_h^2\,\frac{\g^\bh}{Z_h}\Big) \,\b_0^{(2)} +2\, \m_h^2\frac{\g^\bh}{Z_h}\, \b_0^{(2,\chi)}\rgt] \label{3.81}\\[3pt]
& \n_{h-1} =\g^2\n_h + \g^2\frac{1}{A_h}\lft[ \Big( 6\l_h - 2\m^2_h\frac{\g^{\bh}}{Z_h}  \Big)\, \b_0^{(1)}  - 2 \m^2_h\,\frac{ \g^{\bh}}{Z_h}\,\b_0^{(1,\chi)}\rgt]   \label{3.82}
\]
where, denoting $\ff_h(\rho)=\chi(\g^{-h}\r)-\chi(\g^{-h+1}\r)$,
\[
&\b_0^{(1)} = \frac{1}{2\pi^2} \int  d\r\, \ff_0(\r)\,,  \label{3.83} \\
&  \b_0^{(2)} = \frac{1}{2\pi^2} \int \frac{d\r}{\r^2}\, \big(\ff^2_0(\r) +2\ff_0(\r) \ff_1(\r) \big)\,,  \label{3.83b}\\
& \b_0^{(1,\chi)} =   \frac{1}{2\pi^2}(1-\g^{-1})\int_{1}^\g d\r\, \chi(\r) \big(1-\chi(\r) \big)\,, \label{3.83c} \\
& \b_0^{(2,\chi)} = \frac{1}{2\pi^2} (\g-1) \int_{1}^\g \frac{d\r}{\r^2}\, \chi(\r) (1 -\chi(\r))^2\,, \label{3.83d} \\
& 
{\b}_0^{(3,\chi)} = \frac{1}{2\pi^2}(\g-1) \int_{1}^\g \frac{d\r}{\r^2}\, \chi(\r) (1 -\chi(\r))^3 \label{3.83e}
\] 
Moreover, the flow of $\l_{6,h}$ has the form $\l_{6,h-1}=\l_{6,h}+O(\l_h^3)$, while the flow of the renormalization constants reads:
\[&Z_{h-2} - Z_{h-1} = -2 \frac{1}{A_h C_h}\, \m_h^2\,\b_0^{(2)} \label{3.84}\\
&E_{h-2} - E_{h-1} = -2\,\frac{1}{A_h C_h }\, \m_h^2\, \frac{E_h}{Z_h}  \b^{(2)}_0 \label{3.85}\\
&B_{h-2}-B_{h-1} = 2 \frac{\m^2_h}{A_h Z_h}  \Big(\frac{E^2_h}{C_h Z_h} \b_0^{(2)}+\frac13 \b_0^{(\chi')}\Big) \label{3.87}\\
& A_{h-2}-A_{h-1}  =\frac23 \frac{\m^2_h}{C_h Z_h}  \b_0^{(\chi')}\;,\label{3.87a}\]
where 
\[\b_0^{(\chi')}=(\g-1) \frac{1}{2\pi^2}\int_{1}^\g \frac{d\r}\r\, \Big[\r\big(\chi'(\r)\big)^2 + 2 \chi'(\r)\, \big(1-\chi(\r)\big) \Big]\;.\label{33.9944}
\]
The initial data are all fixed but $\n_{\bar h}$, which can be freely adjusted in order to control the flow of $\n_h$.
In order to keep $\n_h$ small  for all scales $h\le \bh$, we invert the flow equation for $\n_h$ (see \eqref{3.77})
in the form
\[\n_h=\g^{2(k-h)}\n_k-\sum_{k<k'\le h}\g^{2(k'-h)}\b^\n_{k'}\]
and then impose  $\lim_{k\to\-\io}\g^{2k}\n_k=0$, so that 
\[\n_h=-\sum_{k\le h}\g^{2(k-h)}\b^\n_{k}
\label{3.89}\]
which should be interpreted as a fixed point equation for the sequence $\{\n_k\}_{k\le \bh}$ (and, therefore, for $\n_{\bh}$ itself, because the sequence $\{\n_k\}_{k\le \bh}$ is uniquely determined by the choice of $\n_\bh$). At lowest order
(see \eqref{3.82}), the solution of \eqref{3.89} has the form
\[\n_h=-\sum_{k\le h}\g^{2(k-h)}\frac{1}{A_k}\lft[ \Big( 6\l_k - 2\m^2_k\frac{\g^{\bh}}{Z_k}  \Big)\, \b_0^{(1)}  - 2 \m^2_k\,\frac{ \g^{\bh}}{Z_k}\,\b_0^{(1,\chi)}\rgt]\;.\]
which tells us that $\n_h$ is of the order $O(\l_h)$. 
While the flow of $\n_h$ can be controlled  by properly fixing $\n_\bh$ (that is, by properly fixing $\bar \n$),
there is nothing we can do for controlling the flows
of $\l_h$ and $\m_h$: they are driven by the linear term, which induces an exponentially fast divergence in 
$\bh-h$. The result is that (recall that $\l_{\bh}$ is of order $\l$ and $\m_{\bh}$ is of order $\sqrt\l$):
$$\l_h=\g^{\bh-h}\l_{\bh}(1+O(\g^{\bh-h}\l_{\bh}))\;,\qquad \m_h=\g^{(\bh-h)/2}\m_{\bh}(1+O(\g^{\bh-h}\l_{\bh})),$$ 
which is valid as long as $\g^{\bh-h}\l_{\bh}$ is small enough, 
say smaller than a suitable constant $\varepsilon_0$ (independent of $\l$),  i.e., 
up to a scale $h^*$ of the order $\log_\g\l^2$. 
In the range of scales $h^*\le h\le \bh$, the marginal constants $A_h, E_h$ and $Z_h$ remain close to their values at scale $\bar h$, that is they are changed at most by a factor $1+O(\varepsilon_0)$. Similarly, $B_h$ grows from zero to a value of the order $O(\lambda^{-1}\varepsilon_0)$, so that $C_h$ remains close to its value at scale $\bar h$, up to a relative error of the order $O(\varepsilon_0)$. Finally $\l_{6,h}$ grows from its initial datum at scale $\bh$, which is of order $\l^3$, to a value of the order $O(\e_0^3)$. 


For smaller scales we cannot conclude anything sensible from the renormalized perturbative expansion discussed here. In the literature \cite{CaDiC1,CaDiC2,Wetterich, DupuisSengupta, DupuisBose_extended, DupuisBose_short,  Sinner2010} there are 
a few heuristic claims about the nature of the infrared theory, which are based on an extrapolation of the flow
to the non-perturbative region $h\le h^*$ and on the 
implementation of some remarkable identities and cancellations, known as Ward Identities \cite{CaDiC1,CaDiC2}, which are supposedly ``non-perturbative" in nature.
Therefore, it has some interest to discuss here in a non-ambiguous way the predictions of these Ward Identities in the perturbative regime $h^*\le h\le \bh$,
where the Bogoliubov's scaling is visible but the theory is still perturbative (and, therefore, our analysis at all orders not only makes sense, but it 
gives a precise meaning to the approximate schemes described in \cite{CaDiC1,CaDiC2} and in \cite{Wetterich, DupuisSengupta, DupuisBose_extended, DupuisBose_short,  Sinner2010}).

The comparison of the predictions of the formal Ward Identities first derived in \cite{CaDiC1,CaDiC2} with our exact findings in the regime $h^*\le h\le \bh$ is discussed in the next section.
Quite interestingly, we find a violation of the predictions of one of the (local) Ward Identities at the level of the one-loop beta function. 
This violation, or anomaly, was completely overlooked in the literature so far. We also identify the 
source of the violation, in the form of correction due to the momentum cutoff, whose effect can be consistently treated in our 
multiscale integration procedure. The anomaly that we find forces us to reconsider the analysis of the infrared fixed point of the theory proposed in \cite{CaDiC1,CaDiC2}, see 
Section \ref{sec3.C.5} below for a discussion of this point.

\subsubsection{Ward Identities and anomaly}\label{sec3.C.4}

The Ward Identities (WIs) are identities between correlation functions, which can be derived by a ``change of coordinates" in the functional integral of the form 
$\psi^\pm_\xx\to e^{\pm i\a(\xx)}\psi^\pm_\xx$ (phase change, or gauge transformation). If $\a(\xx)$ is independent of $\xx$, then the corresponding WI is usually referred to as a {\it global} WI, while it is referred to 
as {\it local}, otherwise. In order to derive the WIs in a correct and non-ambiguous way, one first needs to define the quantities of interest (say, the partition function and the generating function for correlations) in terms of a well-defined functional integral, that is a functional integral regularized by the presence of suitable ultraviolet and infrared 
cut-offs. Next, one performs the aforementioned change of variables in the regularized functional integral, thus deriving exact identities among regularized 
correlation functions. Finally, whenever possible, one removes the regularization and derives the limiting expression of the regularized WIs. In some cases, the extra correction terms 
due to the presence of the cutoff in the WIs {\it do not vanish} in the limit where we remove the cut-offs, in which case such terms are called {\it anomalies}. 
They are often interpreted as ``quantum violations to the conservation laws". 

The WIs can be used to derive convenient identities among the beta functions of different running
coupling constants, or to identify subtle cancellations in the beta function. In the present context, the use of WIs for controlling the flow of the running coupling or renormalization 
constants is due to \cite{CaDiC1,CaDiC2}. Using a dimensional regularization scheme and an $\e$ expansion around dimension $d=3$ (in the form $d=3-\e$) they derived a number of 
remarkable WIs among the running coupling and the renormalization constants, thus reducing the number of independent constants to just one\footnote{Actually, by taking into account the presence of the marginal coupling
$\l_{6,h}$, neglected in \cite{CaDiC2}, the number of independent couplings should be two, see comment 1 in the itemized list at the end of Section \ref{sec3.C.5}.}. 
Moreover, they argued that the 
remaining constant (which can be chosen to be $\l_h$) reaches a non-trivial fixed point in the infrared. Once this is assumed, the flow of all the 
other constants is driven by the WIs and suggests that the infrared theory still displays a linear excitation spectrum a'la Bogoliubov, without anomalous dimensions in the physical response functions. 

In this section we want to reconsider and criticize this picture, by explicitly showing the existence of a non-vanishing correction term 
in the identities suggested by the formal WIs (i.e., those in which the effects of cutoffs are a priori neglected). We will compare explicitly our findings only with those of  \cite{CaDiC1,CaDiC2}, but similar comments 
apply for \cite{Wetterich, DupuisSengupta, DupuisBose_extended, DupuisBose_short,  Sinner2010}.

Let us first focus on the global WIs relating among each other the running coupling and renormalization constants
at scale $h$.
As mentioned above, they can be obtained by performing a global phase transformation in an auxiliary functional integral, which we choose as follows: 
\[
e^{-|\L|\mathcal F_{h-1}-\WW^{(h-1)}(\phi)} = \int P_{\geq h}^B(d\ps) e^{-\bar V(\ps +\f)}\;.\label{C3.1}
\]
Here $\mathcal F_{h-1}$ is a normalization constant and $P_{\geq h}^B(d\ps)$ is a modified version of our reference Bogoliubov measure,  with an extra infrared cutoff at scale $h$, see Appendix \ref{appDd} for details. The local parts of the kernels of $\WW^{(h-1)}(\phi)$ are related in a simple way to the local parts of the kernels of $\VV^{(h-1)}(\phi)$, as proved in Appendix \ref{appDd.3}, see \eqref{C.54}. Therefore, by deriving  $\WW^{(h-1)}(\phi)$ with respect to $\phi^l,\phi^t$ and then setting $\phi=0$,
we can get several useful identities among these local kernels, including the following (see \eqref{C.20} and \eqref{C.21} 
for a derivation):
\[& Z_{h-1}-2 \g^{2h-\bh} \n_h -2\sqrt{2\r_0}\g^{h/2}\m_h =\hat{W}^{(h)}_{1,1;J^\D}(\bz, \bz)\label{wi1}\\
&  \g^{h/2}\m_h-4\sqrt{2\r_0}\g^{h-\bh}\l_h=\hat{W}^{(h)}_{0,3;J^\D}(\bz, \bz)\label{wi2}\]
The terms in the right sides, defined via \eqref{C.17} (see also \eqref{C.20}-\eqref{C.21} and \eqref{ref.assurda}) are the correction terms due to the momentum cutoffs. They are due to the finite infrared cutoff on scale $h$ appearing in the reference gaussian measure in \eqref{C3.1}. In principle, there could also be effects from the ultraviolet momentum cutoff on scale $N$ (to be eventually removed) that we need to introduce in order 
to give a meaning to the right side of \eqref{C3.1}. However, the super-renormalizability of the ultraviolet theory 
proved in Section \ref{s.BEC_UV}\footnote{Here by super-renormalizability of the ultraviolet theory we mean that 
all the interactions are effectively irrelevant: even those that are superficially marginal have dimensional gains in the ultraviolet that make them exponentially insensitive to the ultraviolet cutoff as $N\to\infty$.} implies that the corrections 
due to a finite ultraviolet cutoff on scale $N$ vanish exponentially as $N\to\infty$. This can be proved 
by a simple modification of the analysis in Section \ref{s.BEC_UV}, along the lines of e.g. \cite[Appendix A.2]{GMPAnnPhys}, which we do not belabor here. If we neglect the right sides of \eqref{wi1}-\eqref{wi2}, we obtain 
two {\it formal} WIs that coincide with those derived by Pistolesi et al. in a dimensional 
regularization scheme, see \cite[Eq.(3.18)-(3.19)]{CaDiC2}.
Note that by using dimensional regularization one neglects essentially by construction any anomaly term induced by the momentum cutoffs, and the resulting flow equations may 
have in general a different qualitative behavior.
If our correction terms were dimensionally sub-dominant with respect to the left sides for $h\ll\bh$, then we could say 
that the formal WIs are asymptotically correct in the infrared regime. However, this is {\it not} the case: the one-loop 
computation shows that at lowest non trivial order (defining $\b_0^{(2,\chi)}$ and $\b_0^{(3,\chi)}$ as in \eqref{3.83d}-\eqref{3.83e})
\[&\hat{W}^{(h)}_{1,1;J^\D}(\bz, \bz)=4\frac{\m_{h}^2}{A_{h} C_{h}} \frac{\g}{\g -1}\, \b_0^{(2,\chi)}\;,\\
&\hat{W}^{(h)}_{0,3;J^\D}(\bz, \bz)=\frac1{\sqrt{2\r_0}}\frac{\m_{h}^2}{A_{h} C_{h}} \frac{\g}{\g -1}\, (6\b_0^{(2,\chi)}-4\b_0^{(3,\chi)})\]
which are valid as long as $h^*\le h\le \bh$, 
up to higher order corrections in $\l$ and in $\g^{h-\bh}$, see Appendix \ref{appEe.1} for a proof. Therefore, strictly speaking,
the formal global WIs are violated already at lowest order in perturbation theory. 
Nevertheless, it is 
apparent from the definitions of $\b_0^{(2,\chi)}$ and $\b_0^{(3,\chi)}$ that these terms vanish in the sharp cutoff limit, i.e., 
if we let the smooth cutoff function $\chi(t)$ that enters all the definitions of our cutoffs tend to a step function that is equal 
to $1$ for $t< \g$ and equal to $0$ for $t>\g$. In this sense, the correction terms to the global WIs are ``trivial" and we can conclude that the formal global WIs are asymptotically correct in the infrared regime, in the sharp cutoff limit. 

The problem is more serious for the local WIs, which can be used to relate among each other the renormalization constants $Z_h,E_h,A_h,B_h$. 
Two key local WIs are the following: 
\[\frac{Z_{h-1}}{Z_\bh}&\simeq E_{h-1}+\frac1{\sqrt{2\r_0}}\big[\dpr_{p_0}\hat W^{(h)}_{1,0;J^\D}(\V0)-\dpr_{p_0}\hat W^{(h)}_{1,0;J^{\d T}}(\V0)\big]\;,\label{3.103}\\
 \frac{E_{h-1} - 1}{Z_\bh}&\simeq -B_{h-1}-\frac1{2\sqrt{2\r_0}}\big[\dpr^2_{p_0}\hat W^{(h)}_{0,1;J^\D}(\V0)-\dpr^2_{p_0}\hat W^{(h)}_{0,1;J^{\d T}}(\V0)\big]\;,\label{3.104}\]
where ``$\simeq$" means ``up to dimensionally negligible corrections", i.e., up to errors of relative size $\g^{\theta(h-\bh)}$
for some $\theta>0$. The two identities are an equivalent restatement of the equations \eqref{C.37} and \eqref{C.40}
proved in Appendix \ref{appC.2}, and the terms in square brackets are the corrections due to the infrared cutoff in \eqref{C3.1}. They are the analogues of \cite[Eq.(4.14)-(4.17)]{CaDiC2}, which they reduce to by neglecting the correction terms. 
As for the global WIs, the identities \eqref{3.103}-\eqref{3.104} can be checked at lowest non-trivial order in the renormalized expansion, and the correction terms computed explicitly.
This is done in Appendices \ref{appE.2} and \ref{appE.3}. 
While the correction in \eqref{3.103} vanishes at lowest order
(even though we see no reason why it should vanish exactly at all orders),
the correction in \eqref{3.104} is non-trivial. At the one-loop level it is (see Appendix \ref{appE.3})
\[-\frac1{2\sqrt{2\r_0}}\big[\dpr^2_{p_0}\hat W^{(h)}_{0,1;J^\D}(\V0)-\dpr^2_{p_0}\hat W^{(h)}_{0,1;J^{\d T}}(\V0)\big]
=\frac23 \frac{\m^2_h}{A_h Z_h} \b_0^{(\chi')}\frac{\g}{\g-1}\;, \]
where $\b_0^{(\chi')}$ is defined in \eqref{33.9944}. Quite surprisingly, not only the correction term does not vanish in the 
sharp cutoff limit, but in 
such a limit it gives a divergent contribution to the flow. In this sense, the correction  strongly depends on the specific 
shape of the cutoff function; it can be checked that even its sign depends on the choice of $\chi(t)$ and/or of the scaling 
parameter $\g$ (e.g., take a smoothened version of the function that is $=1$ for $t\le 1$, $=0$ for $t\ge \g$, and linear in 
between; by varying $\g$, the correction term passes from negative to positive values). This indicates that 
the cutoff affects substantially the computation of the infrared-regularized 
thermodynamic observables and that, 
therefore, the multi-scale scheme at hand may not be  reliable at sufficiently low energies. 
Of course, the 
fact that the flow of $B_h$ depends strongly on the (arbitrary) shape of the cutoff function does not mean that the 
thermodynamic observables are ill defined: it just means that the contribution from the energy scales below the cutoff 
is also dependent on the shape $\chi$ and it is of comparable size with that from larger scales. It is hard to interpret
the physical meaning of this phenomenon, but it may indicate that Bogoliubov theory (even if properly renormalized)
is unstable at zero temperature in two dimensions, or the emergence of (possibly non-perturbative)
anomaly terms in the response functions. 
The reader should compare this result with the three dimensional 
case \cite{benfatto}, where the theory is asymptotically free in the infrared and, correspondingly,  
the correction terms in the local WIs are asymptotically vanishing in the infrared limit (see \cite[Section 4.2]{TesiSerena} for a proof). 

It should be stressed that in a Renormalization Group 
treatment of the theory based on dimensional regularization, no correction terms appear in the WIs and, therefore, 
they have no effect on the infrared flow of the coupling constants, 
irrespective of the dimensionality of the system. The mismatch between the predictions of dimensional regularization 
and those based on a constructive scheme with momentum regularization, like ours, instills the doubt that 
dimensional regularization may not be a reliable method in the current context. In particular, it suggests that 
any extrapolation of the flow to the deep infrared (i.e., to scales lower than $h^*$) based on the use 
of local WIs is of doubtful validity. We comment more on this issue in the following section.

\subsubsection{Some heuristic considerations on the nature of the infrared theory}\label{sec3.C.5}

As already mentioned, a Renormalization Group treatment of the two-dimensional condensate based on 
a dimensional regularization scheme and on extrapolation from $d=3$ to lower dimensions was discussed in 
\cite{CaDiC2}. There the authors argue that at scales smaller than $h^*$ the beta function flow drives the 
running coupling and renormalization constants towards a non-trivial fixed point, which can be understood as follows. 
Neglecting the correction terms, and using systematically the WIs discussed above at lowest non-trivial order 
(in particular, using the replacements \eqref{eD.3} which can be inductively justified at lowest order), the beta function for $\l_h$ can be rewritten as:
\be \l_{h-1} = \g\l_h - 4 \g \frac{\l_h^2}{A_h C_h \g^\bh}\,\b_0^{(2)}  \ee
that, if taken (too) seriously, implies that the flow drives $\l_h$ towards the fixed point $\l^*=A^*C^*\g^\bh/(4\b_0^{(2)})$,
provided that also $A_h$ and $C_h$ reach two fixed points, called $A^*$ and $C^*$.
The formal WIs suggest that the fixed points $A^*$ and $C^*$ exist and are of order $1$ and $\l^{-1}$, respectively.
Therefore, the fixed point $\l^*$ is expected to be of order 1 with respect to $\l$ and, 
therefore, any analytical ``proof" of its existence  is inevitably heuristic (if $\l_h$ becomes of order $1$, the use of the truncated equations is not justified). Of course one could hope that the $O(1)$ fixed point exists and its numerical
value is sufficiently small, so that the truncation is a posteriori justified, but a possible proof of this fact would
certainly require extensive numerical simulations in addition to analytical arguments. 
Still, it makes sense to assume that such a fixed point exists for $\l_h$ and, under this hypothesis, ask 
about the behavior of all the other running coupling and renormalization constants.  
The authors of \cite{CaDiC2} proposed a very nice and non-trivial argument for controlling the flow of all the 
other constants in terms of that of $\l_h$, via a smart combination of the WIs at disposal, discussed in 
\cite[Section IV.B]{CaDiC2} and reproduced in the language of the
current paper in \cite[Chapters 3 and 4]{TesiSerena}. The use of WIs  implies remarkable cancellations at all orders in the beta function: in particular they show that the contributions that potentially 
drive $A_h$ and $C_h$ to infinity are zero at all orders. The physical consequence of this fact is that 
the spectrum of excitation of the interacting theory remains linear and the physical response functions (e.g. density-density correlations) have the same qualitative behavior as predicted by Bogoliubov's theory, notwithstanding a non trivial renormalization of the $l-l$ component of the propagator.
The cancellation required for this argument to work can of course be checked explicitly at lowest order, 
as done in \cite[Appendix B and C]{CaDiC2}, see in particular \cite[(C.6)-(C.8)]{CaDiC2}, which should be compared with our Eqs.\eqref{3.84}-\eqref{3.87}.

While very interesting and inspiring, this scheme seems to rely too heavily on the dimensional regularization scheme: 
as mentioned above, in our momentum cutoff scheme, which allows us to study two dimensions directly (rather than 
dimension $3-\e$ with $\e$ a small parameter), the WIs are violated already at the one-loop level.
In particular, the cancellations proved by the authors of \cite{CaDiC2} at all orders, are just false in 
a momentum cutoff scheme. Therefore, the key ingredient for the stability of the (putative) infrared fixed point 
is missing here, and this may indicate the emergence of non-perturbative anomaly terms, neglected in the 
analysis of \cite{CaDiC2}, which may qualitatively change the nature of the ground state of the system. 
Of course, our analytical methods do not allow to investigate the existence of non-perturbative anomaly terms arising 
in the regime $h\le h^*$. However, our finding calls for a reconsideration of the assumptions in \cite{CaDiC2}
and for a non-perturbative numerical analysis of the model, which may help in understanding better the 
qualitative features of the ground state. 

Let us conclude this section by mentioning that even if the cancellations proposed by \cite{CaDiC2} took place,
there would still be a few extra issues to discuss in order to fully control the infrared theory, neglected in the analysis of
\cite{CaDiC2}. These other issues, even if overlooked in \cite{CaDiC2}, as well as in later works \cite{Wetterich, DupuisSengupta, DupuisBose_extended, DupuisBose_short,  Sinner2010}, can all be solved via closer inspection, as studied in detail in \cite{TesiSerena}:\begin{enumerate}
\item In two dimensions there is the extra marginal coupling 
constant $\l_{6,h}$, which is absent, since irrelevant, in $3-\e$ dimensions. Therefore, the hypothesis that $\l_h$ 
reaches a fixed point  should be supplemented by the assumption that also $\l_{6,h}$ reaches one.
The existence of a fixed point for the pair $(\l_h,\l_{6,h})$ is in fact compatible with a
low-order truncation of the (coupled) flows of $\l_h,\l_{6,h}$, see \cite[Section 3.5.2]{TesiSerena}.
\item As proved in \cite{CaDiC2}, the assumption  that $\l_h,\l_{6,h}$ reach a fixed point implies that $Z_h$ and 
$\m_h$ go to zero very fast in the infrared, like $\g^h$ and $\g^{h/2}$, respectively. The fast vanishing of $Z_h$ 
implies that the factors $U_v$ in \eqref{3.74} are dimensionally unbounded in general as $h\to-\infty$: 
this effectively changes the scaling dimensions of the operators involved, and requires the introduction 
of three more running coupling constants, whose flows can however be reduced again to that of $(\l_h,\l_{6,h})$ via the use of three novel global WIs. See \cite[Section 3.1]{TesiSerena}.
\end{enumerate}

Of course, it is not worth entering the details of this discussion here, since the basic assumption for the 
study of the flow in the deep infrared region $h\le h^*$, i.e., the validity of the local WIs for controlling the flow of the 
renormalization constant, is explicitly violated already at the one-loop level.

\section{Conclusions}\label{sec4}

We presented a renormalization group construction of a weakly interacting two dimensional Bose gas,
both in the quantum critical regime (zero condensate and zero temperature) and in the presence of a 
condensate fraction. The construction is performed within a rigorous renormalization group scheme, borrowed
from the methods of constructive field theory, which allows us to derive explicit bounds on all the orders 
of renormalized perturbation theory. Contrary to other heuristic renormalization group approaches, our scheme allows us to evaluate and bound explicitly the effects of the irrelevant terms, without the need of neglecting them. 

This scheme allows us to construct completely, at all orders, the theory of the quantum critical point, both in the ultraviolet and in the 
infrared. The theory turns out to be super-renormalizable in the ultraviolet (thanks to the finite range of the interaction 
potential between particles), and asymptotically free (marginally irrelevant) in the infrared, as expected on the basis of approximate renormalization schemes \cite{FisherHohenberg,Wetterich, DupuisLowT}.

The application of the same scheme to the condensate phase is more subtle: while the ultraviolet regime
corresponding to length scales smaller than the range of the potential $R_0$, and the crossover regime 
corresponding to length scales intermediate between $R_0$ and $(\l\r_0)^{-1/2}$, are fully controllable, in the same 
fashion as the quantum critical point, the study of the infrared regime of larger length scales is much harder. In that regime there appear three relevant and five marginal effective constants; their flow is driven to larger values by the presence of two of the three relevant couplings and, therefore, we are forced to stop the flow at length scales 
of the order $(\l^3\r_0)^{-1/2}$. For larger scales, non-perturbative approaches are required; they go beyond what can 
we do by our analytical methods and we cannot reach a definite conclusion about the nature of the infrared theory. 

Still, interestingly enough, we explicitly exhibit violations to the formal Ward Identities in the third regime, the one 
corresponding to length scales between $(\l\r_0)^{-1/2}$ and $(\l^3\r_0)^{-1/2}$. In this range the renormalized theory is 
perturbative and predictions based on low order truncations reliable. Therefore, the fact that some of the formal Ward 
Identities are violated at the one-loop level, as we prove, suggests the possibility that (non-perturbative) anomaly terms 
appear in the deep infrared regime. Certainly, it puts on shaky grounds their application to the deep infrared 
regime, which played a key role in previous proposals that the renormalization group flow reaches a non-perturbative 
infrared fixed point, characterized by a linear spectrum of excitations analogous to Bogoliubov's one. 
We hope that future research, both from the analytical and numerical side, will help to clarify the nature of the 
infrared theory and to confirm or dismiss the Bogoliubov's picture in two dimensional systems.  \\

{\bf Acknowledgements.} {The research leading to these results has received funding from the European Research Council under the European Union's Seventh Framework Programme ERC Starting Grant CoMBoS (grant agreement
n$^o$ 239694). We thank Giuseppe Benfatto and Vieri Mastropietro for several discussions and illuminating comments. S.C. acknowledges the Hausdorff Center for Mathematics in Bonn for financial support.}

\pagina
\appendix 

{\renewcommand{\thesubsection}{\thesection.{\small\arabic{subsection}}}


\feyn{
\begin{align*}
\b_h^{\bar z} & = 
\begin{fmffile}{BoseFeyn/zhUV}
 \unitlength = 0.8cm
\def\myl#1{3.5cm}
\fmfset{arrow_len}{3mm}
\parbox{4cm}{\centering{ 
		\begin{fmfgraph*}(3.5,2.8) 
			\fmfleft{i1}
			\fmfright{o1}
                    \fmftop{t} \fmfbottom{b}
                    \fmf{phantom, tension=3.5}{t,vt}
                    \fmf{phantom, tension=0.5}{b,v1}
                   \fmf{fermion}{i1,v1}
			\fmf{fermion,label=$h$, right=0.7, tension=0.3}{v1,vt}
                    \fmf{fermion, left=0.7, tension=0.3}{v1,vt}
			\fmf{fermion}{o1,v1}					
                    \fmfdot{v1}
                   \fmfv{label=$\;\l_{h}\;$, label.angle=-90}{v1}
		\end{fmfgraph*}  
	}}  
\end{fmffile}	
\end{align*}   \vskip -0.8cm
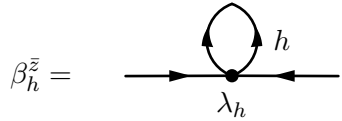
}{Feynman diagrams representation of the first non trivial contribution  to the beta function of $\bar z_{h-1}$, for $h\leq -1$.  }{zhUV}


\section{Ultraviolet flow in the condensate phase} \label{UVflow}

In this section we prove Eqs.\eqref{3.12y}-\eqref{3.12z}. We denote by $\bar z_h$ and $\d\n_h$ 
the quantities defined via the second and third lines of \eqref{3.barz}, both for $h\ge 0$ and $\bh\le h<0$. 
The graphs 
contributing to $\d\n_h$ are all and only ladder diagrams as in the first line of Fig.\ref{ladder2}, for all $h\ge \bh$;
those contributing to $\bar z_h$ are all and only  ladder diagrams as in the second line of Fig.\ref{ladder2},
if $h\ge 0$, while they can be either as in the second line of Fig.\ref{ladder2} or as in Fig.\ref{ladder2bis}, if
$\bh\le h<0$. For both couplings, we distinguish the regime $h\ge 0$ from $\bh\le h<0$.

Let us first consider $\d\bar\n_h$. If $h\ge 0$ we proceed as
in the ``ultraviolet regime" subsection of Section \ref{s2.fl}, thus obtaining a bound on the $n$-th order diagrams of the 
same form as the $n$-th order term in the right side of \eqref{2.60}. Since $n\ge 2$, we find that $|\d\bar \n_h|\le C
\l^2\g^{-h}$ for all $h\ge 0$. For $h<0$ we write the beta function equation for $\d\bar\n_h$, analogous  to \eqref{2.62},
with initial datum $\d\bar\n_0$ of the order $\l^2$: $\d\bar\n_{h-1}=\d\bar\n_h+\b_h^{\d\bar\n}$, with $h\le 0$.
Using the (renormalized analogoue of) \eqref{3.9}, we see that $|\b_h^{\d\bar\n}|\le C\l^2$, $\forall \bh\le h\le 0$,
which implies that $|\d\bar\n_h|\le C\l^2|h|$, $\forall \bh\le h\le 0$, as desired.

We now apply the same strategy to the study of the flow of $\bar z_h$. If $h\ge 0$, we find that the diagrams of order $n$ are bounded by $\l(K\l)^{n-1}\g^{\bh-h}$, with $n\ge 1$; here the factor $\g^{\bh-h}$ is a dimensional estimate on the $L_\io$ norm
of the off diagonal propagator $r_h$. Therefore,
$|\bar z_h|\le ({\rm const.})\l\g^{\bh-h}$, for all $h\ge 0$. In particular, $\bar z_0$ is of order $\l^2$. From smaller scales,
we study again the beta function equation $\bar z_{h-1}=\bar z_h+\b_h^{\bar z}$, where $|\b_h^{\bar z}|\le C\l^2$,
for all $\bh\le h\le 0$ (see Fig.\ref{zhUV} for a graphical representation of the first non trivial contribution to $\b_h^{\bar z}$). Therefore, as for $\d\bar \n_h$, we find $|\bar z_h|\le C\l^2|h|$, $\forall \bh\le h\le 0$,
which concludes the proof of \eqref{3.12y}-\eqref{3.12z}.
\section{Bounds on the propagators}  \label{bounds}
 
In this section we prove the decay bound \eqref{s3.37}. We first focus on $\tl g^{(1)}_\bh$. Of 
course, it is enough that we restrict to its $-+$ component, which we rewrite as the sum of two terms, 
whose definition is induced by the following rewriting of the cutoff function appearing under the integral sign in  \eqref{eq3.36}:
\[  \label{p2}
& 1-\chi \big(\g^\bh(x_0^2+|\vx|^4)^{1/2} \big)-\chi \big(\g^{-\bh}(k_0^2+|\vk|^4)^{\nicefrac{1}{2}}\big) \equiv (1-\fh_{\bh}(\xx))\big(1 -\tl\chi_\bh(\kk)\big)-\fh_{\bh}(\xx) \tl \chi_\bh(\kk)
\]
where 
\[
\fh_{\bh}(\xx)  =  \chi \big(\g^\bh(x_0^2+|\vx|^4)^{1/2} \big)\;, && 
\tl{\chi}_\bh(\kk)  = \chi \big(\g^{-\bh}(k_0^2+|\vk|^4)^{1/2} \big)\;.
\]
The corresponding decomposition for the $-+$ component of $\tilde g^{(1)}_\bh$ 
 is $[\tl g^{(1)}_{\bh}(\xx)]_{-+}= g_\bh^{(1a)}(\xx) - g_\bh^{(1b)}(\xx)$. We now show that the two terms separately satisfy the same bound as \eqref{s3.37}.  The easiest term to treat 
is $g_\bh^{(1b)}(\xx)$, which we  write as
\[ g_\bh^{(1b)}(\xx)=\fh_\bh(\xx)\int \frac{d\kk}{(2\p)^3} e^{-i\kk\xx}\frac{\tl \chi_\bh(\kk)}{-ik_0+|\vk|^2}.\label{p33}\]
Using the compact support properties of $\tl \chi_\bh(\kk)$, it is immediate to check that 
\[  \label{A.8tl}
\big| \tl g_{\bh}^{(1b)}(\xx) \big| & =\fh_\bh(\xx) \bigg| \int \frac{d^{3}\kk}{(2\pi)^{3}} e^{-i \kk \cdot \xx} 
\frac{\tl \chi_\bh(\kk)}{-ik_0+|\vk|^2}\bigg| \leq ({\rm const}.)  \g^{\bh}\fh_\bh(\xx)\,,
\]
which implies \eqref{s3.37} for $n_0=|\vn|=0$. In order to estimate the derivatives of 
$\tl g_{\bh}^{(1b)}(\xx)$, note that each derivative $\dpr_{x_0}$ (resp. $\dpr_{x_i}$ with $i=1,2$) 
acting on $\fh_\bh(\xx)$ produces a factor proportional to $\g^{\bh}$ (resp. $\g^{\bh/2}$), as desired. Moreover, each 
 derivative $\dpr_{x_0}$ (resp. $\dpr_{x_i}$ with $i=1,2$) 
acting on the integral in the right side of \eqref{p33} produces a factor $-ik_0$ (resp. $-ik_i$)
under the integral sign, which is bounded proportionaly to $\g^{\bh}$ (resp. $\g^{\bh/2}$), thanks to the 
compact support properties of $\tilde \chi_\bh$. This concludes the proof that $\tilde g^{(1b)}_\bh$ satisfies a bound like
\eqref{s3.37}.

Let us now focus on $\tilde g_\bh^{(1a)}(\xx)$. In order to bound it, we rewrite the cutoff function 
$(1-\tl \chi_\bh(\kk))$ appearing in its definition as  $\sum_{h> \bh} \ff_h(\kk)$, with $\ff_h(\kk)$ defined after \eqref{3.16} and note that, thanks to compact support properties of $\ff_h$,
\[
\bigg| \dpr_{x_0}^{n_0}\dpr_{\vx}^{\vn}\int \frac{d^{3}\kk}{(2\pi)^{3}} e^{-i \kk \cdot \xx} \frac{\ff_h(\kk)}{-ik_0+|\vk|^2}\bigg| \leq C_{n_0,\vn} \g^{h(1+n_0+|\vn|/2)}
\,.
\]
Moreover, integrating by parts and using again the compact support properties of $\ff_h$, we find
\[
|x_0|^N
\bigg| \int \frac{d^{3}\kk}{(2\pi)^{3}} e^{-i \kk \cdot \xx} \frac{\ff_h(\kk)}{-ik_0+|\vk|^2}\bigg|=
\bigg| \int \frac{d^{3}\kk}{(2\pi)^{3}} e^{-i \kk \cdot \xx}\dpr_{k_0}^N \frac{\ff_h(\kk)}{-ik_0+|\vk|^2}\bigg|
 \leq C_{N} \g^{h(1-N)}\]
 and,
 similarly,
\[
|\vx|^N
\bigg| \int \frac{d^{3}\kk}{(2\pi)^{3}} e^{-i \kk \cdot \xx} \frac{\ff_h(\kk)}{-ik_0+|\vk|^2}\bigg|
 \leq C_{N} \g^{h(1-N/2)}\;.\]
Combining the previous three equations we find
\[
\bigg| \dpr_{x_0}^{n_0}\dpr_{\vx}^{\vn}\int \frac{d^{3}\kk}{(2\pi)^{3}} e^{-i \kk \cdot \xx} \frac{\ff_h(\kk)}{-ik_0+|\vk|^2}\bigg| \leq C_{N,n_0,\vn} \frac{\g^{h(1+n_0+|\vn|/2)}}{1+[\g^h(|x_0|+|\vx|^2)]^N}
\,,\label{p88}
\]
 which leads to 
\[ \big| \dpr_{x_0}^{n_0}\dpr_{\vx}^{\vn}\tilde g^{(1a)}_\bh(\xx)\big|\le C_{N,n_0,\vn}(1-\fh_\bh(\xx))\sum_{h>\bh}
 \frac{\g^{h(1+n_0+|\vn|/2)}}{1+[\g^h(|x_0|+|\vx|^2)]^N}\;.\]
 Now, on the support of $(1-\fh_\bh(\xx))$, the combination $(|x_0|+|\vx|^2)$ is larger than (const.)$\g^{\bh}$ and, therefore, for all $N_1,N_2\ge 0$ such that $N_1+N_2=N$,
\[ \big| \dpr_{x_0}^{n_0}\dpr_{\vx}^{\vn}\tilde g^{(1a)}_\bh(\xx)\big|\le C_{N,n_0,\vn}'(1-\fh_\bh(\xx))\sum_{h>\bh}
 \frac{\g^{(\bh-h)N_1}\g^{h(1+n_0+|\vn|/2)}}{1+[\g^h(|x_0|+|\vx|^2)]^{N_2}}\;,\]
 which is summable over $h$ as soon as $N_1>n_0+|\vn|/2$. By picking such an $N_1$ we obtain the desired bound on 
 $\tilde g^{(1a)}_\bh$. 
 
 The proof of the desired decay bound for $\tilde g^{(2)}_\bh$ is completely analogous to 
 the proof of \eqref{p88} and, therefore, we do not give its details here. The same is true for the proof of \eqref{3.37},
 which we also  leave to the reader. 
 

\section{Lowest order computations}\label{appB}

In this section we prove \eqref{3.80}--\eqref{3.82} and \eqref{3.84}--\eqref{3.87a}, i.e., we explicitly compute the beta functions for 
$\l_h$, $\m_h$, $\n_h$, $Z_h$, $E_h$, $B_h$ and $A_h$ at the lowest non-trivial order (which turns out to coincide with the one-loop computation). 
Rather than presenting the computations in the same order as
we presented the formulas after \eqref{3.80}, we proceed in order of increasing difficulty, starting from the easiest computation, which is the beta function of $Z_h$, then moving 
to the beta function of $\n_h$, etc.  

In the following, in the computation of the one-loop contributions to the beta function, we systematically perform a number of approximations, which either 
induce corrections of higher order than the one we are computing, or are dimensionally irrelevant for $h\le \bh$. The approximations we do are the following: (1)
we systematically replace $\g^{h+1}\l_{h+1}$ by $\g^{h}\l_h$, 
$\g^{(h+1)/2}\m_{h+1}$
by $\g^{h/2}\m_h$,  $\g^{2(h+1)}\n_{h+1}$ by $\g^{2h}\n_h$, $A_{h+1}$ by $A_h$, $B_{h+1}$ by $B_h$, $E_{h+1}$ by $E_h$, and $Z_{h+1}$ by $Z_h$, since these replacements induce an error
that is of higher order in $\l$; (2) similarly, we replace the renormalization functions $\tilde A_{h-1}(\kk)$, etc, appearing in the definition of the propagator on scale $h$ by $A_h$, etc;
moreover, recalling the definition of $\chi_h(\kk)$ in \eqref{3.49c}, 
we replace the support function 
$\tilde f_h(\kk)=\chi_h(\kk)-\chi_{h-1}(\kk)$ by $\ff_h(|\kk'|):=\chi(\g^{-h}|\kk'|)-\chi(\g^{-h+1}|\kk'|)$, where with some abuse of notation we let 
$\kk':=(k_0,\sqrt{A_h/C_h}\,\vk)\equiv(k_0,\vk')$; note that also 
these replacements induce errors of higher order in $\l$;
 (3) finally, 
we systematically neglect the terms  coming from trees with 
at least one endpoint on scale $\bh$, since these
terms have relative size $\g^{\theta(h-\bh)}$ with $0<\theta<1$ as compared to the main contributions to the beta function (see \eqref{3.73}).

\feyn{
\begin{fmffile}{BoseFeyn/zeta2d}
 \unitlength = 0.8cm
\def\myl#1{2.7cm}
\[
  \b_h^Z  = 2\;
	\parbox{\myl}{\centering{ 
			\begin{fmfgraph*}(2.8,2) 
			\fmfleft{i1}
			\fmfright{o1}
			\fmftop{t}
			\fmfbottom{b}
			\fmf{phantom, tension=1.8}{t,v3}    
			\fmf{phantom, tension=1.8}{b,v4}    
            \fmf{dashes, tension=1.5}{i1,v1}			
			\fmf{plain,left=0.3}{v1,v3}
			\fmf{plain,left=0.3}{v3,v2}			
			\fmf{plain,right=0.3}{v1,v4}
			\fmf{plain,right=0.3}{v4,v2}
			\fmf{dashes, tension=1.5}{o1,v2}
		\fmfdot{v1,v2}
		\end{fmfgraph*}   
			}}     \nonumber
\]  
\end{fmffile}  
}{Leading order beta function for $Z_h$. The graph represents the lowest order contribution to $\hW^{(h)}_{2,0}(\bz)$, and the two solid lines (associated with two propagators of type 
$\tilde g_{tt}^{(h_i)}$) come with two labels $h_1,h_2$, which we need to sum over, with the constraints that $\min\{h_1,h_2\}=h$ and $|h_1-h_2|\le 1$.
}{zeta2d}

\subsection{Lowest order beta function for $Z_h$} \label{s.zeta}

The lowest order contribution to $\b_{h}^{Z}$ (see \eqref{3.78} and \eqref{3.60z}) is of order $\l_h$ (or, equivalently, of order $\m_h^2$) and 
can be represented graphically by diagrams of the form in Fig.\ref{zeta2d}, computed at zero external momentum, 
with the two propagators labelled by two scale labels $h_1,h_2$ such that $\min\{h_1,h_2\}=h$. Note that by the compact support properties of the propagator $|h_1-h_2|\le 1$. The sum of the values of these diagrams is 
\[
\b_{h}^{Z}  = -2 \int \frac{d\kk}{(2\p)^3}\Big[\g^h\m_h^2\big(\hat{\tilde g}_{tt}^{(h)}(\kk)\big)^2+2\g^{h+1}\m_{h+1}^2\hat{\tilde g}^{(h)}_{tt}(\kk)\hat{\tilde g}^{(h+1)}_{tt}(\kk)\Big]\;,
\label{B.3}\]
where 
\[
\hat{\tl g}^{(h)}_{tt}(\kk)=  \frac{  \tl f_{h}(\kk)}{\tilde C_{h-1}(\kk)
\big[k_0^2+(\tl{A}_{h-1}(\kk)/\tl{C}_{h-1}(\kk))|\vk|^2\big]}\;,
\]
where $\tl Z_{h-1}(\kk)\tilde C_{h-1}(\kk)=\tl Z_{h-1}(\kk)\tilde B_{h-1}(\kk)+\tilde E_{h-1}^2(\kk)$. 
Under the approximations explained above, this propagator can be replaced by
\[
\bar g^{(h)}_{tt}(\kk')=  \ff_{h}(|\kk'|)\,\frac{ 1}{C_h|\kk'|^2}\;.
 \]
For future reference, let us write down the analogues of this equation, as far as the propagators with labels $lt$ and $ll$ are concerned: 
\[
\bar g^{(h)}_{ll}(\kk')=  \ff_{h}(|\kk'|)\,\frac{B_h k_0^2+C_h|\vk'|^2}{Z_hC_h|\kk'|^2}\;,\qquad \bar g^{(h)}_{lt}(\kk)=\ff_{h}(|\kk'|)\,\frac{E_h k_0}{Z_hC_h|\kk'|^2}\;.
\]
Using these replacements, as well as the ones spelled above, we find that \eqref{B.3} is equal (up to higher order corrections) to 
\[\b_{h}^{Z} = -2\g^h\m_h^2\frac1{A_h C_h}\int \frac{d\kk}{(2\p)^3}\frac{\ff^2_h(|\kk|) + 2\ff_h(|\kk|)\ff_{h+1}(|\kk|)}{|\kk|^4}\;,
\]
so that after rescaling and after the change of variables $k_0=\r \cos \th$, $|\vk|=\r \sin \th$, with $\r \geq 0$ and $\th \in [0, \pi]$, we get 
\[
\b_{h}^{Z} = -2 \frac{1}{A_h C_h}\, \m_h^2\, \frac{1}{2\pi^2} \int \frac{d\r}{\r^2}\, \big(\ff^2_0(\r) +2\ff_0(\r)\ff_1(\r) \big)\;,\]
which is the same as \eqref{3.84}.

\subsection{Lowest order beta function for $\n_h$}  \label{s.nu}


\feyn{
\begin{fmffile}{BoseFeyn/nu2d}
 \unitlength = 0.8cm
\def\myl#1{2.5cm}
\begin{align*}
\g^{2h-\bh}\b^{\n}_h =
	\parbox{2.5cm}{\centering{ 
		\begin{fmfgraph*}(2,2)
			\fmfleft{i1}
			\fmfright{o1}
			\fmf{plain, tension=0.8}{i1,v}
			\fmf{plain}{v,v}
			\fmf{plain, tension=0.8}{v,o1} 
			\fmfdot{v}
		\end{fmfgraph*} 
	}}			\;+\;
	\parbox{\myl}{\centering{ 
			\begin{fmfgraph*}(2.8,1.4) 
			\fmfleft{i1}
			\fmfright{o1}
			\fmf{plain, tension=1.5}{i1,v1}
			\fmf{plain,left=0.7, tension=0.4}{v1,v2}
			\fmf{dashes,right=0.7, tension=0.4}{v1,v2}
			\fmf{plain, tension=1.5}{o1,v2}
			\fmfdot{v1,v2}
			\end{fmfgraph*}  
			}}  	\;+\;
	\parbox{\myl}{\centering{ 
			\begin{fmfgraph*}(2.8,1.4) 
            \fmfleft{i}
			\fmfright{o}
			\fmftop{t}
			\fmfbottom{b}
			\fmf{phantom, tension=4}{t,v3}    
			\fmf{phantom, tension=4}{b,v4}   
            \fmf{plain, tension=1.5}{i,v1}			
			\fmf{plain,left=0.4, tension=1.2}{v1,v3}
			\fmf{dashes,left=0.4}{v3,v2}			
			\fmf{dashes,right=0.4}{v1,v4}
			\fmf{plain,right=0.4}{v4,v2}
			\fmf{plain, tension=1.5}{o,v2}
			\fmfdot{v1,v2}			
			\end{fmfgraph*}  
			}} 
\end{align*}
\end{fmffile}	
}{Leading order beta function for $\n_h$. }{nu2d}

The lowest order contributions to $\b_h^\n$ (see \eqref{3.77} and \eqref{cc})
are of the order $\l_h$ and can be represented graphically as in Fig.\ref{nu2d}. Once again, the propagators are 
associated with scale labels such that the minimum scale is equal to $h$. We denote by $\b^{\n,(1)}_h$ the contribution from the first diagram in Fig.\ref{nu2d}, 
and by $\b^{\n,(2)}_h$ those from the last two diagrams. Proceeding as in the previous section we find that, up to higher order corrections,
\[ \label{nu1}
\g^{2h -\bh}\b^{\n,(1)}_h = 6\, \frac{1}{A_h}\, \g^{h -\bh } \l_h \int  \frac{d^3 \kk}{(2\pi)^3} \frac{\ff_h(|\kk|)}{|\kk|^2} =   6\, \frac{1}{A_h}\, \g^{2h -\bh} \l_h  \,\frac{1}{2\pi^2} \int d\r\, 
\ff_0(\r)\;.
\]
In order to calculate $\b^{\n,(2)}_h$ we notice that the sum of the remaining two diagrams in Fig.\ref{nu2d} involves the following combinations of propagators
(after the usual replacements):
\[
\bar g^{(h)}_{tt}(\kk)\bar g^{(h)}_{ll}(\kk) + \bigl(\bar g^{(h)}_{tl}(\kk)\bigr)^2  &=  \frac{\ff^2_h(|\kk|) }{Z_h C_h |\kk|^2} \label{B.7} \\
\bar g^{(h)}_{tt}(\kk)\bar  g^{(h+1)}_{ll}(\kk)  +\bar g^{(h+1)}_{tt}(\kk)\bar  g^{(h)}_{ll}(\kk) + 2\bar g^{(h)}_{tl}(\kk)\bar  g^{(h+1)}_{tl}(\kk) 
& = \frac{ 2 \ff_h(|\kk|)  \ff_{h+1}(|\kk|) }{Z_h C_h |\kk|^2}+h.o.\;, \label{B.8}
\]
where $h.o.$ in the last equation means ``higher orders". Using these identities we find, up to higher order corrections:
\[
\g^{2h-\bh}\b^{\n,(2)}_h  
& =  -2\, \g^{2h} \frac{\m^2_h}{Z_h}\, \frac{1}{A_h}\,\int  \frac{d^3 \kk}{(2\pi)^3} \frac{\ff^2_0(|\kk|) +2\ff_0(|\kk|) \ff_{1}(|\kk|)}{ |\kk|^2} \non \\
& =  -2\, \g^{2h} \frac{\m^2_h}{Z_h}\, \frac{1}{A_h}\,\frac{1}{2\pi^2}\int  d\r \big(\ff^2_0(\r) + 2\ff_0(\r)\ff_1(\r))\;.
\]
Putting things together:
\[ \label{nufin}
\b^{\n}_h &=  \bigg( 6\l_h - 2\m^2_h\frac{\g^{\bh}}{Z_h}  \bigg)\frac{1}{A_h}\,   \,\frac{1}{2\pi^2} \int d\r\, \ff_0(\r)\,  \non \\
&  -2\,  \m^2_h\frac{\g^\bh}{Z_h}\, \frac{1}{A_h}\,\frac{1}{2\pi^2}\int  d\r \big(\ff^2_0(\r) + 2\ff_0(\r)\ff_1(\r) - \ff_0(\r) \big)\;.
\]
Using the definitions \eqref{3.83}--\eqref{3.83e}, this can be rewritten as 
\[\b^{\n}_h =  \bigg( 6\l_h - 2\m^2_h\frac{\g^{\bh}}{Z_h}  \bigg)\frac{1}{A_h}\, \b^{(1)}_0  -2\,  \m^2_h\frac{\g^\bh}{Z_h}\, \frac{1}{A_h}\b_0^{(1,\chi)}
\;,
\]
which implies \eqref{3.82}.

\feyn{
\begin{fmffile}{BoseFeyn/mu2d}
 \unitlength = 0.8cm
\def\myl#1{2.5cm}
\def\myll#1{2.7cm}
\begin{align*}
	 \g^{\frac h 2}\b_h^\m & = \;
	\parbox{\myll}{\centering{
			\begin{fmfgraph*}(2.8,2) 
			\fmfleft{i1}
			\fmfright{o1,o2}
			\fmftop{t}
			\fmfbottom{b}
			\fmf{phantom, tension=1.8}{t,v3}    
			\fmf{phantom, tension=1.8}{b,v4}    
            \fmf{dashes, tension=1.5}{i1,v1}			
			\fmf{plain,left=0.3}{v1,v3}
			\fmf{plain,left=0.3}{v3,v2}			
			\fmf{plain,right=0.3}{v1,v4}
			\fmf{plain,right=0.3}{v4,v2}
			\fmf{plain, tension=1.5}{o1,v2,o2}
		\fmfdot{v1,v2}
		\end{fmfgraph*} 
			}} 
			+\;
	\parbox{\myll}{\centering{  
			\begin{fmfgraph*}(2.8,2) 
			\fmfleft{i1}
			\fmfright{o1,R,o2}
                   \fmftop{T}
                   \fmfbottom{B}
			\fmf{phantom, tension=1.8}{B,vB}  
                    \fmf{phantom, tension=1.8}{T,vT}  
			\fmf{dashes,tension=1.5}{i1,v1}
			\fmf{plain, left=0.3, tension=1.2}{v1,vT}
                   \fmf{plain, left=0.2, tension=0.4}{vT,v3}
                    \fmf{phantom, tension=1}{R,vR}  
			\fmf{dashes, left=0.3, tension=1}{v3,vR}
                   \fmf{dashes, left=0.3, tension=0.4}{vR,v2}			
                  \fmf{plain, right=0.3, tension =1.4}{v1,vB}
	            \fmf{plain, right=0.2, tension=0.4}{vB,v2}           
		     \fmf{plain, tension=0.8}{o1,v2}
			\fmf{plain,tension=0.8}{o2,v3}
                   \fmfdot{v1,v2,v3}
			\end{fmfgraph*}  
			}}  
+\;
	\parbox{\myll}{\centering{ 
			\begin{fmfgraph*}(2.8,2) 
			\fmfleft{i1}
			\fmfright{o1,R,o2}
                   \fmftop{T}
                   \fmfbottom{B}
			\fmf{phantom, tension=1.8}{B,vB}  
                    \fmf{phantom, tension=1.8}{T,vT}  
			\fmf{dashes,tension=1.5}{i1,v1}
			\fmf{plain, left=0.3, tension=1.2}{v1,vT}
                   \fmf{dashes, left=0.2, tension=0.4}{vT,v3}
                    \fmf{phantom, tension=1}{R,vR}  
			\fmf{plain, left=0.3, tension=1}{v3,vR}
                   \fmf{plain, left=0.3, tension=0.4}{vR,v2}		
                  \fmf{plain, right=0.3, tension =1.4}{v1,vB}
	            \fmf{dashes, right=0.2, tension=0.4}{vB,v2}           
	             \fmf{plain, tension=0.8}{o1,v2}
			\fmf{plain,tension=0.8}{o2,v3}
                   \fmfdot{v1,v2,v3}
			\end{fmfgraph*}  
			}}  +\;
	\parbox{\myll}{\centering{ 
			\begin{fmfgraph*}(2.8,2) 
			\fmfleft{i1}
			\fmfright{o1,R,o2}
                   \fmftop{T}
                   \fmfbottom{B}
			\fmf{phantom, tension=1.8}{B,vB}  
                    \fmf{phantom, tension=1.8}{T,vT}  
			\fmf{dashes,tension=1.5}{i1,v1}
			\fmf{plain, left=0.3, tension=1.2}{v1,vT}
                   \fmf{dashes, left=0.2, tension=0.4}{vT,v3}
                    \fmf{phantom, tension=1}{R,vR}  
			\fmf{plain, left=0.3, tension=1}{v3,vR}
                   \fmf{dashes, left=0.3, tension=0.4}{vR,v2}			
                  \fmf{plain, right=0.3, tension =1.4}{v1,vB}
	            \fmf{plain, right=0.2, tension=0.4}{vB,v2}           
			\fmf{plain, tension=0.8}{o1,v2}
			\fmf{plain,tension=0.8}{o2,v3}
                   \fmfdot{v1,v2,v3}
			\end{fmfgraph*}    
			}} 
\end{align*}
\end{fmffile}	
}{ Leading order beta function for $\m_h$. }{mu2d}

\subsection{Lowest order beta function for $\m_h$}  \label{s.mu}

The lowest order contributions to $\b_h^\m$
are of the order $\m_h\l_h$ and can be represented graphically as in Fig.\ref{mu2d}. Once again, the propagators are 
associated with scale labels such that the minimum scale is equal to $h$.
We denote by $\b_h^{\m,(2) }$ (resp. $\b_h^{\m,(3)}$)  the contribution to $\b_h^{\m}$ coming from the first diagram (resp. second $+$ third $+$ fourth diagrams) in  Fig.\ref{mu2d}. We find:
\[ \label{C.m1}
\g^{\frac{h}{2}} \b_h^{\m,(2) } & = - 12\, \frac{1}{A_h C_h}\, \g^{\frac{h}{2}}  \m_h \g^{h -\bh}\l_h \,\int \frac{d^3 \kk}{(2\pi)^3} \frac{\ff^2_h(|\kk|) + 2\ff_h(|\kk|) \ff_{h+1}(|\kk|)}{|\kk|^4}    \non \\
& = - 12\, \frac{1}{A_h C_h \g^\bh}\, \g^{\frac{h}{2}}  \m_h \l_h\,\b_0^{(2)}\;,
\]
and, using the analogues of \eqref{B.7}-\eqref{B.8},
\[
\g^{\frac{h}{2}} \b_h^{\m, (3)} & = 4 \,\frac{C_h}{A_h} \g^{\frac{3}{2}h} \m_h^3 \,\int\frac{ d^3\kk}{(2\pi)^3}\,
\frac{1}{Z_h C^2_h |\kk|^4}\lft[ \ff_h^3(|\kk|) +3 \ff_h^2(|\kk|) \ff_{h+1}(|\kk|) +3 \ff_h(|\kk|) \ff^2_{h+1}(|\kk|) \rgt] \non \\
& = 4 \,\frac{1}{A_h C_h}\, \g^{\frac{h}{2}} \frac{\m_h^3}{Z_h}\,\frac{1}{2\pi^2} \int \frac{d\r}{\r^2}\, \lft[\ff^3_0(\r)  +3\ff_0^2(\r)\ff_1(\r) + 3\ff_0(\r) \ff_1^2(\r)\rgt]  
\]
In the last expression, we can rewrite
\[\frac{1}{2\pi^2} \int \frac{d\r}{\r^2}\, \lft[\ff^3_0(\r)  +3\ff_0^2(\r)\ff_1(\r) + 3\ff_0(\r) \ff_1^2(\r)\rgt]  =\b_0^{(2)}-\b_0^{(2,\chi)}\]
where $\b_0^{(2,\chi)}$ is defined as in \eqref{3.83d}, so that, putting things together,
\[
\b_h^{\m} & = \frac{1}{A_h C_h\g^\bh }\, \bigg[\m_h \Big(-12 \l_h +   4 \m_h^2  \,\frac{\g^\bh}{Z_h}\Big) \,\b_0^{(2)} -4 \, \m^3_h\, \frac{\g^\bh}{Z_h}\,\b^{(2,\chi)}_h \bigg]\;.
%
\]
which implies \eqref{3.81}


\feyn{
\begin{fmffile}{BoseFeyn/lambda2d}
 \unitlength = 0.8cm
\def\myl#1{2.7cm}
\def\myll#1{2.7cm}
\begin{align*}
	 \b_h^\l & = \;
	\parbox{\myll}{\centering{ 
			\begin{fmfgraph*}(2.8,2) 
			\fmfleft{i1,i2}
			\fmfright{o1,o2}
			\fmftop{t}
			\fmfbottom{b}
			\fmf{phantom, tension=1.8}{t,v3}    
			\fmf{phantom, tension=1.8}{b,v4}    
            \fmf{plain, tension=1.5}{i1,v1,i2}			
			\fmf{plain,left=0.3}{v1,v3}
			\fmf{plain,left=0.3}{v3,v2}			
			\fmf{plain,right=0.3}{v1,v4}
			\fmf{plain,right=0.3}{v4,v2}
			\fmf{plain, tension=1.5}{o1,v2,o2}
		 \fmfdot{v1,v2}
		\end{fmfgraph*} 
			}} 
			+\;
	\parbox{\myll}{\centering{ 
			\begin{fmfgraph*}(2.8,2) 
			\fmfleft{i1,i2}
			\fmfright{o1,R,o2}
                   \fmftop{T}
                   \fmfbottom{B}
			\fmf{phantom, tension=1.8}{B,vB}  
                    \fmf{phantom, tension=1.8}{T,vT}  
			\fmf{plain,tension=1.5}{i1,v1,i2}
			\fmf{plain, left=0.3, tension=1.2}{v1,vT}
                   \fmf{plain, left=0.2, tension=0.4}{vT,v3}
                    \fmf{phantom, tension=1}{R,vR}  
			\fmf{dashes, left=0.3, tension=1}{v3,vR}
                   \fmf{dashes, left=0.3, tension=0.4}{vR,v2}			
                  \fmf{plain, right=0.3, tension =1.4}{v1,vB}
	            \fmf{plain, right=0.2, tension=0.4}{vB,v2}           
		     \fmf{plain, tension=0.8}{o1,v2}
			\fmf{plain,tension=0.8}{o2,v3}
                    \fmfdot{v1,v2,v3}
			\end{fmfgraph*}  
			}}  
+\;
	\parbox{\myll}{\centering{ 
			\begin{fmfgraph*}(2.8,2) 
			\fmfleft{i1,i2}
			\fmfright{o1,R,o2}
                   \fmftop{T}
                   \fmfbottom{B}
			\fmf{phantom, tension=1.8}{B,vB}  
                    \fmf{phantom, tension=1.8}{T,vT}  
			\fmf{plain,tension=1.5}{i1,v1,i2}
			\fmf{plain, left=0.3, tension=1.2}{v1,vT}
                   \fmf{dashes, left=0.2, tension=0.4}{vT,v3}
                    \fmf{phantom, tension=1}{R,vR}  
			\fmf{plain, left=0.3, tension=1}{v3,vR}
                   \fmf{plain, left=0.3, tension=0.4}{vR,v2}		
                  \fmf{plain, right=0.3, tension =1.4}{v1,vB}
	            \fmf{dashes, right=0.2, tension=0.4}{vB,v2}           
	             \fmf{plain, tension=0.8}{o1,v2}
			\fmf{plain,tension=0.8}{o2,v3}
                    \fmfdot{v1,v2,v3}
			\end{fmfgraph*}    
			}}  +\;
	\parbox{\myll}{\centering{  
			\begin{fmfgraph*}(2.8,2) 
			\fmfleft{i1,i2}
			\fmfright{o1,R,o2}
                   \fmftop{T}
                   \fmfbottom{B}
			\fmf{phantom, tension=1.8}{B,vB}  
                    \fmf{phantom, tension=1.8}{T,vT}  
			\fmf{plain,tension=1.5}{i1,v1,i2}
			\fmf{plain, left=0.3, tension=1.2}{v1,vT}
                   \fmf{dashes, left=0.2, tension=0.4}{vT,v3}
                    \fmf{phantom, tension=1}{R,vR}  
			\fmf{plain, left=0.3, tension=1}{v3,vR}
                   \fmf{dashes, left=0.3, tension=0.4}{vR,v2}			
                  \fmf{plain, right=0.3, tension =1.4}{v1,vB}
	            \fmf{plain, right=0.2, tension=0.4}{vB,v2}           
			\fmf{plain, tension=0.8}{o1,v2}
			\fmf{plain,tension=0.8}{o2,v3}
                    \fmfdot{v1,v2,v3}
			\end{fmfgraph*}    
			}}  \\[12pt]
&  + \;
	\parbox{\myl}{\centering{ 
			\begin{fmfgraph*}(2.8,2) 
			\fmfleft{i1,i2}
			\fmfright{o1,o2}
			\fmf{plain}{i1,v1}
			\fmf{plain}{i2,v2}
                   \fmf{plain}{v3,o1}
			\fmf{plain}{v4,o2}
                 \fmf{plain, tension=0.4}{v1,vb}
                   \fmf{plain, tension=0.5}{vb,v3}
                    \fmf{dashes, tension=0.4}{v3,vr}
			\fmf{dashes, tension=0.5}{vr,v4}
	             \fmf{plain, tension=0.4}{v2,vt}
                    \fmf{plain, tension=0.5}{vt,v4}
			\fmf{dashes, tension=0.4}{v1,vl}
		      \fmf{dashes, tension=0.5}{vl,v2}
			 \fmfdot{v1,v2,v3,v4}
			\end{fmfgraph*}  
			}} 
+\; \parbox{\myl}{\centering{ 
			\begin{fmfgraph*}(2.8,2) 
			\fmfleft{i1,i2}
			\fmfright{o1,o2}
			\fmf{plain}{i1,v1}
			\fmf{plain}{i2,v2}
                   \fmf{plain}{v3,o1}
			\fmf{plain}{v4,o2}
                   \fmf{dashes, tension=0.4}{v1,vb}
                   \fmf{plain, tension=0.5}{vb,v3}
			\fmf{dashes, tension=0.4}{v3,vr} 
                   \fmf{dashes, tension=0.5}{vr,v4}
	             \fmf{dashes, tension=0.4}{v2,vt}
                    \fmf{plain, tension=0.5}{vt,v4}
			\fmf{plain, tension=0.4}{v1,vl} 
                    \fmf{plain, tension=0.5}{vl,v2}
		     \fmfdot{v1,v2,v3,v4}
			\end{fmfgraph*}  
			}} +
\; \parbox{\myl}{\centering{ 
			\begin{fmfgraph*}(2.8,2) 
			\fmfleft{i1,i2}
			\fmfright{o1,o2}
			\fmf{plain}{i1,v1}
			\fmf{plain}{i2,v2}
                   \fmf{plain}{v3,o1}
			\fmf{plain}{v4,o2}
                   \fmf{dashes, tension=0.4}{v1,vb}
                   \fmf{plain, tension=0.5}{vb,v3}
	            \fmf{dashes, tension=0.4}{v3,vr}
			\fmf{dashes, tension=0.5}{vr,v4}
	             \fmf{plain, tension=0.4}{v2,vt}
                    \fmf{plain, tension=0.5}{vt,v4}
			\fmf{plain, tension=0.6}{v1,vl}
                   \fmf{dashes, tension=0.3}{vl,v2}
			 \fmfdot{v1,v2,v3,v4}
			\end{fmfgraph*}     
			}} + 
\; \parbox{\myl}{\centering{ 
			\begin{fmfgraph*}(2.8,2) 
			\fmfleft{i1,i2}
			\fmfright{o1,o2}
			\fmf{plain}{i1,v1}
			\fmf{plain}{i2,v2}
                   \fmf{plain}{v3,o1}
			\fmf{plain}{v4,o2}
                   \fmf{dashes, tension=0.4}{v1,vb}
                   \fmf{plain, tension=0.8}{vb,v3}
			\fmf{dashes, tension=0.4}{v3,vr}
			\fmf{plain, tension=0.8}{vr,v4}
    			  \fmf{dashes, tension=0.4}{vt,v4}
	             \fmf{plain, tension=0.8}{v2,vt}
 			 \fmf{dashes, tension=0.4}{vl,v2}
             	\fmf{plain, tension=0.8}{v1,vl}
                 	 \fmfdot{v1,v2,v3,v4}
			\end{fmfgraph*}   
			}} 
\end{align*}
\end{fmffile}	
}{Leading order beta function for $\l_h$. }{lambda2d}

\subsection{Lowest order beta function for $\l_h$}  \label{s.lambda}

The lowest order contributions to $\b_h^\l$
are of the order $\l_h^2$ and can be represented graphically as in Fig.\ref{lambda2d}. We denote with $\b^{\l,(2)}_h$, $\b^{\l, (3)}_h$ and $\b^{\l, (4)}_h$ the contributions to $\b_h^\l$ coming from the diagrams in Fig.\ref{lambda2d} with two, three and four end-points, respectively. Proceeding as in the previous sections, we find:
\[
\g^{h-\bh}\b^{\l,(2)}_h & =  -36 \frac{1}{A_h C_h}\, \g^{h -2\bh} \l^2_h \, \b_0^{(2)} \;,\\
\g^{h-\bh}\b^{\l, (3)}_h & = 24\, \frac{1}{A_h C_h}\, \g^{h-\bh} \l_h\,\frac{\m_h^2}{Z_h} \big(\b_0^{(2)} - \b_0^{(2,\chi)} \big)\;,
\]
and
\[
\g^{h-\bh}\b^{\l, (4)}_h = -4 \frac{1}{A_h C_h }\,\g^{h}\, \frac{\m_h^4}{Z_h^2}  \frac{1}{2\pi^2} \int \frac{d\r}{\r^2}\,T_4(\r)\;,
\]
where
\[
T_{4}(\r) &= \ff^4_0(\r) +4\ff_0^3(\r)\ff_{1}(\r) +6\ff_0^2(\r)\ff^2_{1}(\r) +4\ff_0(\r)\ff^3_{1}(\r)\;.
\]
Using the definitions \eqref{3.83b}--\eqref{3.83e}, it can be checked that 
\[
 \frac{1}{2\pi^2} \int \frac{d\r}{\r^2}\,T_4(\r)=\b_0^{(2)}-\b_0^{(2,\chi)}-\b_0^{(3,\chi)}\;,
\]
so that, putting things together,
\[
\b_h^\l  = &\, \frac{1}{A_h C_h \g^\bh} \Big( -36 \,\l^2_h 
+24\,\l_h \m_h^2\frac{ \g^\bh}{Z_h} -4 \,\m_h^4  \frac{\g^{2\bh}}{Z_h^2} \Big) \b_0^{(2)} \non \\
& +\frac{1}{A_h C_h \g^\bh}\, \lft[\Big( -24\,\l_h \m_h^2 \frac{\g^\bh}{Z_h}  +4 \,\m_h^4 \frac{\g^{2\bh}}{Z_h^2} \Big)  \b_0^{(2,\chi)} +4 \, \m_h^4\frac{\g^{2\bh}}{Z_h^2}\b_0^{(3,\chi)}  \rgt]\;,
\]
which implies \eqref{3.80}.

\feyn{
\begin{fmffile}{BoseFeyn/Eh2d}
\unitlength = 0.8 cm  
\def\myl#1{2.5cm}
\[
   \beta^{E}_h  =\, \dpr_{p_0} \Biggr[ \; 
 \parbox{\myl}{\centering{ 
	\begin{fmfgraph*}(2.8,2)
			\fmfleft{i1}
			\fmfright{o1}
			\fmftop{t}
			\fmfbottom{b}
			\fmf{phantom, tension=1.8}{t,v3}    
			\fmf{phantom, tension=1.8}{b,v4}    
                   \fmf{plain, tension=1.5}{i1,v1}			
			\fmf{plain,left=0.3}{v1,v3}
			\fmf{plain,left=0.3}{v3,v2}			
			\fmf{dashes,right=0.3}{v1,v4}
			\fmf{plain,right=0.3}{v4,v2}
		       \fmf{dashes, tension=1.5}{o1,v2}
			 \fmfdot{v1,v2}
		\end{fmfgraph*}
		}}  \Biggl]_{\pp=\V0}   \non 
\]
\end{fmffile}
}{Leading order beta function for $E_h$. The graph is first computed at external momentum $\pp=(p_0,\v0)$, then derived w.r.t. $p_0$; after the action of the derivative, the external momentum is set to $\V0$.}{Eh2d}

\subsection{Lowest order beta function for $E_h$} \label{s.E}

The lowest order contribution to $\b^E_h$ (see \eqref{3.78} and \eqref{3.60e}) is of order $\l_h$ and is represented graphically in  Fig.\ref{Eh2d}. The value of the graph is (up to higher order corrections):
\[
\b^{E}_h = \b^{E}_{h,2}- \b^{E}_{h,1}
\]
with
\[
\b^{E}_{h,i} & = 4\,\frac{1}{A_h}\, \m_h^2\, \frac{E_h}{C_h Z_h}  \int \frac{d^3 \kk}{(2\pi)^3}  \frac{ k_0 T_i(\kk)}{|\kk|^2} \,\dpr_{p_0} \lft[ \frac{T_i(\kk+\pp)}{|\kk+\pp|^2} \rgt]_{\pp=\bf 0}  \non \\
& =  2\,\frac{1}{A_h}\, \m_h^2 \, \frac{E_h}{C_h Z_h}  \int \frac{d^3 \kk}{(2\pi)^3}  \, k_0 \, \dpr_{k_0}\lft(\frac{T_i(\kk)}{|\kk|^2}\rgt)^2
\]
where $T_2(\kk):= \ff_0(|\kk|) + \ff_1(|\kk|)$, $T_1(\kk):=\ff_1(|\kk|)$. By integrating by parts:
\[
\b^{E}_{h,i} & =   -2\,\frac{1}{A_h}\, \m_h^2\, \frac{E_h}{C_h Z_h}  \int \frac{d^3 \kk}{(2\pi)^3}  \, \lft(\frac{T_i(\kk)}{|\kk|^2}\rgt)^2 \;,
\]
so that 
\[
\b^{E}_{h} =   -2\,\frac{1}{A_h}\,\frac{E_h}{C_h Z_h} \m_h^2 \b^{(2)}_0\;,
\]
which gives \eqref{3.85}.

\subsection{Lowest order beta function for $B_h$} \label{s.B}

The lowest order contribution to $\b^B_h$ (see \eqref{3.77b} and \eqref{3.59b}) is of the order $\l_h$ and is represented graphically in Fig.\ref{Bh2}. The sum of the values of the
two graphs is (up to higher order corrections):
\[
\b^B_h =  \b^B_{h,2} - \b^B_{h,1} 
\]
where, defining  $\a_h=\frac{E_h^2}{Z_h C_h}$, and letting $T_1,T_2$ be the same as in the previous section, 
\[
 \b^B_{h,i} =- 2\, \frac{ \m_h^2 }{Z_hA_h }\int \frac{d^3\kk}{(2\pi)^3}  \dpr^2_{p_0} \bigg[  \frac{|\kk|^2+\a_hk_0p_0}{|\kk|^2|\kk+\pp|^2}T_{i}(\kk + \pp) T_{i} (\kk)
 \bigg]_{\pp=\V0} \;.\]
%

\feyn{
\begin{fmffile}{BoseFeyn/Bh2}
\unitlength = 1 cm  
\def\myl#1{2.5cm}
\[
   \beta^{B}_h  =\, \dpr^2_{p_0} \Biggr[ \quad 
 \parbox{\myl}{\centering{
	\begin{fmfgraph*}(2.5,2)
			\fmfleft{i1}
			\fmfright{o1}
                   \fmftop{t}
                   \fmfbottom{b}
                   \fmf{phantom, tension=1.5}{t,vt}
			\fmf{phantom, tension=1.5}{b,vb} 
                   \fmf{plain, tension=1.5}{i1,v1}			
			\fmf{plain,left=0.3}{v1,vt}	
                   	\fmf{plain,left=0.3, tension=0.8}{vt,v2}		
			\fmf{dashes,right=0.3}{v1,vb}
		      \fmf{dashes,right=0.3}{vb,v2}	
		       \fmf{plain, tension=1.5}{o1,v2}
			 \fmfdot{v1,v2}
		\end{fmfgraph*}
		}} \quad + \quad 
 \parbox{\myl}{\centering{ 
	\begin{fmfgraph*}(2.5,2)
			\fmfleft{i1}
			\fmfright{o1}
                   \fmftop{t}
                   \fmfbottom{b}
                   \fmf{phantom, tension=1.5}{t,vt}
			\fmf{phantom, tension=1.5}{b,vb} 
                   \fmf{plain, tension=1.5}{i1,v1}			
			\fmf{plain,left=0.3}{v1,vt}	
                   	\fmf{dashes,left=0.3, tension=0.8}{vt,v2}		
			\fmf{dashes,right=0.3}{v1,vb}
		      \fmf{plain,right=0.3}{vb,v2}	
		       \fmf{plain, tension=1.5}{o1,v2}
			 \fmfdot{v1,v2}
		\end{fmfgraph*}
		}} 
\quad \Biggl]_{\pp=\V0} 
\non
\]
\end{fmffile}
}{Leading order beta function for $B_h$. }{Bh2}

By computing explicitly the derivative, and then integrating by parts, we find:
\[
 \b^B_{i,h} & =  - 2\, \frac{\m_h^2 }{Z_hA_h } \int \frac{d^3\kk}{(2\pi)^3} \bigg[  2\a_h k_0 \frac{T_i(\kk)}{|\kk|^2} \dpr_{k_0}\lft(\frac{T_i(\kk )}{|\kk|^2}\rgt) 
 +T_i(\kk)\dpr^2_{k_0}\lft(\frac{T_i(\kk )}{|\kk|^2}\rgt)  \bigg]\\
& =  - 2\, \frac{ \m_h^2 }{Z_hA_h } \int \frac{d^3\kk}{(2\pi)^3} \bigg[  -\a_h \lft(\frac{T_i(\kk )}{|\kk|^2}\rgt)^2+\dpr_{k_0}T_i(\kk)\Big( \frac{2k_0T_i(\kk)}{|\kk|^4}-\frac{\dpr_{k_0}T_i(\kk)}{|\kk|^2}\Big)\Big]\;.\]
Let us now denote by $\b^{B,(1)}_{i,h}$ the contribution associated with the first term in square brackets, and by $\b^{B,(2)}_{i,h}$  the rest. We have:
\[\b_h^{B,(1)}=\b^{B,(1)}_{2,h}-\b^{B,(1)}_{1,h}=
2   \frac{\m_h^2}{A_h Z_h} \frac{E_h^2}{Z_h C_h}  \b^{(2)}_0\;,
\]
which gives the first term in the r.h.s. of \eqref{3.87}. Similarly, by passing to polar coordinates $k_0=\r\cos\th$, $|\vk|=\r\sin\th$, with $\th\in[0,\p]$, and 
by explicitly computing the integral over $\th$, we find:
\[\b_h^{B,(2)}=\b^{B,(2)}_{2,h}-\b^{B,(2)}_{1,h}=  \frac23 \frac{\m_h^2}{A_h Z_h} (\g-1) \frac{1}{2\pi^2}\int \frac{d\r}{\r}\, \Big[ \r \big(\chi'(\r)\big)^2 + 2 \chi'(\r)\, \big(1-\chi(\r)\big) \Big]\;,
\]
which gives the second term in the r.h.s.  of \eqref{3.87}.

\subsection{Lowest order beta function for $A_h$} \label{s.A}

The lowest order contribution to $\b^A_h$  is represented graphically in a way similar to Fig.\ref{Bh2}, with $\dpr_{p_0}^2$ replaced by $\dpr_{p_1}^2$. Therefore,
$\b^A_h =  \b^A_{h,2} - \b^A_{h,1}$, with 
\[
 \b^A_{h,i} =- 2\, \frac{\m_h^2 }{Z_hC_h}\, \int \frac{d^3\kk}{(2\pi)^3}T_{i} (\kk)  \dpr^2_{k_1} \Big(  \frac{T_{i}(\kk + \pp)}{|\kk+\pp|^2} \Big)\Big|_{\pp=\V0} \;.\]
On the other hand, the integral in the r.h.s. is invariant under the exchange $k_0\otto k_1$ and, therefore, using the result of the previous section, 
\[
 \b^A_{h} =\frac23 \frac{\m_h^2}{C_h Z_h} (\g-1) \frac{1}{2\pi^2}\int \frac{d\r}{\r}\, \Big[ \r \big(\chi'(\r)\big)^2 + 2 \chi'(\r)\, \big(1-\chi(\r)\big) \Big]\;,\]
 which proves \eqref{3.87a}.
 
\section{Ward identities}\label{appDd}

\subsection{Derivation of the global Ward Identities}\label{appDd.1}
In this section we derive the Ward Identities associated with a  phase transformation $\psi^\pm_\xx\to e^{\pm i\a(\xx)}\psi^\pm_\xx$, with $\a(\xx)\equiv \a$. Next, in the following 
section, we will show how to modify the computation in order to deal with a local phase transformation in which $\a(\xx)$ is a non-trivial function of $\xx$. We introduce a sequence of 
reference models, labelled by an integer $h$, defined in terms of a functional integral similar to \eqref{3.10bis}, but with an extra infrared cutoff at a pre-fixed scale $h$ (that, for definiteness, we shall assume to be $\le \bh$):
\[
e^{-|\L|\mathcal F_{h-1}-\WW^{(h-1)}(\phi)} = \int P_{\geq h}^B(d\ps) e^{-\bar V(\ps +\f)}\;.\label{C.1}
\]
Here $P_{\geq h}^B(d\ps)$ is the complex gaussian measure with propagator given by the analogue of \eqref{3.13}, modulo an extra infrared cutoff\footnote{In this section, for simplicity,
we formally write all the involved expressions in the limit $\b,L\to\infty$, as we already did in the bulk of the paper. It is implicit in the discussion that all the involved quantities have a 
finite temperature/volume counterparts, which can be easily written down, but are slightly more cumbersome than their formal $\b,L\to\infty$ limits (this is the only reason why we prefer not to write them explicitly).}:
\[g^B_{\geq h}(\xx-\yy)=\int_{\mathbb R^3} \frac{d\kk}{(2\p)^3}\hat \chi_{\geq h}(\kk)
\frac{e^{-i\kk(\xx-\yy)}}{k_0^2+(\e'(\vk))^2}\begin{pmatrix}ik_0 +F(\vk)& -\l\r_0\hat v(\vk)\\
-\l\r_0\hat v(\vk)& -ik_0+F(\vk)\end{pmatrix}\]
and $\hat \chi_{\ge h}(\kk)=1-\chi_{h-1}(\kk)$, where $\chi_{h-1}(\kk)$ was defined in \eqref{3.49c}. Moreover, $\bar V(\psi)$ is given by \eqref{3.V2} (we dropped the label $\L$, for 
simplicity). The interest of the definition \eqref{C.1} is that the {\it local parts} of the kernels of $\WW^{(h-1)}$, with $h\le \bh$, are essentially\footnote{We will see in Section 
\ref{appC.2} below that the local parts of the kernels of $\WW^{(h-1)}$ coincide with those of $\VV^{(h-1)}+\sum_{k=h}^{\bh}\LL_Q\VV^{(k)}$ up to a minor correction due to 
the ``last integration step", which is not visible at the level of the one-loop beta function, see \eqref{C.54} and the preceding discussion.} the 
same as the local part of the kernels of 
$\VV^{(h-1)}+\sum_{k=h}^{\bh}\LL_Q\VV^{(k)}$, which we computed in Section \ref{s3.C.2}
via a renormalized multiscale construction. Therefore, identities among the kernels of $\WW^{(h-1)}$ induce identities between the running coupling constants at scales $h\le \bh$, 
which are the relations we are interested in. See Section \ref{appC.2} below for a detailed discussion about the connection between $\WW^{(h-1)}$ and $\VV^{(h-1)}$.

In order to obtain the desired identities among the kernels of $\WW^{(h-1)}(\phi)$ it is convenient to preliminarily manipulate the r.h.s. of \eqref{C.1}. 
We recall that the Bogoliubov's reference gaussian measure $P^B(d\ps)$ 
was obtained by first combining the negative exponential of the first two lines in \eqref{3.V} with the free gaussian measure $P^0$, see the discussion after \eqref{3.V}, and then by 
performing the c-number substitution $\x\to\sqrt{\r_0}$ spelled out after \eqref{3.9a}. A similar connection is valid, of course, between the cut-offed measures
$P_{\geq h}^B(d\ps)$ and $P_{\geq h}^0(d\ps)$, where $P_{\geq h}^0(d\ps)$ is the free gaussian measure with propagator  
\[\int \frac{d\kk}{(2\p)^3}\hat \chi_{\geq h}(\kk)\frac{e^{-i\kk(\xx-\yy)}}{-ik_0+|\vk|^2}\;.\]
Therefore, we can write:
\[P_{\geq h}^B(d\ps)=e^{-|\L|\tilde f_h}P_{\geq h}^0(d\ps)e^{-Q^B_h(\ps)}\;,\label{C.4}\]
where, in the limit $\L\nearrow \mathbb R^3$ (and using the fact that the term in the second line of \eqref{3.V} vanishes in this limit),
\[Q^B_{\ge h}(\ps)= \frac{\l}{2}\r_0 \int\limits_{{\rm supp}\hat \chi_{\ge h}}\hskip-0.2truecm \frac{d\kk}{(2\p)^3} \,\hat v(\vk)\,\hat \chi_{\ge h}^{-1}(\kk) \,(\hat \ps^+_\kk+\hat \psi^-_{-\kk})
(\hat \ps^+_{-\kk}+\hat \ps^-_{\kk})\;.\label{C.Q}\]
Plugging \eqref{C.4} into \eqref{C.1}, we find:
\be e^{-|\L|\mathcal F_{h}-\WW_{h}(\phi)} = e^{-|\L|\tilde f_h}\int P_{\geq h}^0(d\ps) e^{-Q^B_{\ge h}(\psi)-\bar V(\ps +\f)}\;.\label{C.6}\ee
We now first perform the change of variables $\psi^\pm_\xx\to e^{\pm i\a}\psi^{\pm}_\xx$, and then we derive w.r.t. $\a$, so that:
\[ \label{A.5}
0 = \frac{\dpr}{\dpr \a} \int P_{\geq h}^0(d\ps)\, e^{- Q^B_{\ge h}(e^{i\a}\ps) -\bar V(e^{i \a }\ps+\f)} \Big|_{\a=0}\;.
\]
In order to compute the derivative explicitly, it is convenient to rewrite $\bar V=\bar W (\ps)+\bar Q(\ps)$ with
\[ 
\bar W(\ps) & =   \frac{\l}{2}\int d\xx \,d\yy
\big[|\psi_\xx+\sqrt{\r_0}|^2-\r_0\big] w(\xx-\yy) \big[|\psi_\yy+\sqrt{\r_0}|^2-\r_0\big]\\
&+ \bar \n \int  d\xx \big[|\psi_\xx+\sqrt{\r_0}|^2-\r_0\big]\;,\\
\bar Q(\ps)&=-\frac{\l}{2}\r_0\int d\xx\,d\yy\,(\psi^+_\xx+\psi^-_\xx)w(\xx-\yy)(\psi^+_\yy+\psi^-_\yy)-\bar\n\sqrt{\r_0}\int d\xx (\psi^+_\xx+\psi^-_\xx)\;,\]
and then note that 
\[  \label{d1}
\frac{\dpr}{\dpr \a} \bar W(e^{i \a}\ps + \f) \Big|_{\a=0}= -i  \int d\xx \lft[ \frac{\d \bar W(\ps+\f)}{\d \f^+_\xx}\big( \f^+_\xx + \sqrt{\r_0}\big) - \frac{\d \bar W(\ps+\f)}{\d \f^-_\xx}\big( \f^-_\xx + \sqrt{\r_0}\big) \rgt]\;,
\]
from which one gets (after an explicit computation of the contributions coming from $\bar Q$):
\[&
\frac{\dpr}{\dpr \a} \bar V(e^{i \a}\ps+ \f) \Big|_{\a=0}  
= \,- i  \int d\xx \lft[ \frac{\d \bar V(\ps+\phi)}{\d \f^+_\xx} \cdot \big( \f^+_\xx + \sqrt{\r_0}\big) - \frac{\d \bar V(\ps+\phi)}{\d \f^-_\xx} \cdot \big( \f^-_\xx + \sqrt{\r_0}\big) \rgt] \nonumber \\
&-i\l\r_0  \int d\xx\, d\yy \big(\psi^+_\xx+\f^+_\xx - \ps^-_\xx   - \f^-_\xx \big)w(\xx-\yy)\big(\ps^+_\yy + \f^+_\yy+\ps^- _\yy+  \f^-_\yy \big)\label{C.12}\\
&-i\bar\n\sqrt{\r_0}
\int d\xx  \big(\psi^+_\xx+\f^+_\xx - \ps^-_\xx   - \f^-_\xx \big)\;. \non\]
Note that the term in the last line is equal to $-i\bar\n\sqrt{\r_0}
\int d\xx  \big(\f^+_\xx   - \f^-_\xx \big)$, simply because $\psi$ has zero average, by construction. If we now use \eqref{C.12} into \eqref{A.5}, after an explicit computation of the 
contribution coming from $\bar Q^B_{\ge h}$ we find:
\[  \label{GWIx}
&  \int P_{\geq h}^B(d\ps) e^{-\bar V(\ps +\f)}\Big\{ \int d\xx\left[\frac{\d \bar V(\ps+\phi)}{\d \f_{\xx}^{+}} \left(\phi_{\xx}^{+}+\sqrt{\r_0}\right) -\frac{\d \bar V(\ps+\phi)}{\d \f_{\xx}^{-}} \left(\f_{\xx}^{-}+\sqrt{\rho_{0}}\right)\right]  + \\
&  +\l\r_0  \int d\xx\, d\yy \Big[\big(\psi^+_\xx+\f^+_\xx - \ps^-_\xx   - \f^-_\xx \big)w(\xx-\yy)\big(\ps^+_\yy + \f^+_\yy+\ps^- _\yy+  \f^-_\yy \big)-\nonumber\\
&-\big(\psi^+_\xx- \ps^-_\xx \big)w_{\ge h}(\xx-\yy)\big(\ps^+_\yy +\ps^- _\yy \big)\Big]
+\bar\n\sqrt{\r_0}
\int d\xx  \big(\f^+_\xx -  \f^-_\xx \big)\Big\}=0\;,\nonumber\]
where $w_{\ge h}(\xx)$ is the Fourier transform of $\hat v(\vk)[\hat \chi_{\ge h}(\kk)]^{-1}$. 
The equation \eqref{GWIx} can be rewritten in terms of the fields $\ps^{l},\ps^t$ and $\f^{l},\f^t$ defined via \eqref{newvar}, in terms of which we find:
\[  \label{GWI_1}
 0&= \int P_{\geq h}^B(d\ps) e^{-\bar V(\ps +\f)}\Big\{ \int d\xx\left[\frac{\d \bar V(\ps+\phi)}{\d \f_{\xx}^{l}}\phi_{\xx}^{t} -\frac{\d \bar V(\ps+\phi)}{\d \f_{\xx}^{t}} \left(\f_{\xx}^{l}+\sqrt{2\rho_{0}}\right)\right] +\\
&+  2\l \r_0  \int d\xx \,d\yy \Big[\big(\f^t_\xx + \ps^t_\xx )w(\xx-\yy)\big(\ps^l_\yy + \f^l_\yy \big) -\ps^t_\xx w_{\ge h}(\xx-\yy)\ps^l_\yy\Big]+\bar\n\sqrt{2\r_0}\int d\xx\, \phi^t_\xx\Big\}\;.
\non\]
Defining $\langle[\ \cdot\ ]\rangle_h^\phi=e^{|\L|\FF_{h-1}+\WW^{(h-1)}(\phi)}\int P_{\geq h}^B(d\ps) e^{-\bar V(\ps +\f)}[\ \cdot\ ]$, we can rewrite the last equation as
\[  \label{GWI_2}
 0&= \int d\xx\left[\frac{\d \WW^{(h-1)}(\phi)}{\d \f_{\xx}^{l}}\phi_{\xx}^{t} -\frac{\d \WW^{(h-1)}(\phi)}{\d \f_{\xx}^{t}} \left(\f_{\xx}^{l}+\sqrt{2\rho_{0}}\right)\right] 
 +\bar\n\sqrt{2\r_0}\int d\xx\, 
\langle\phi^t_\xx\rangle_h^\phi+\\
&+  2\l \r_0  \int d\xx\,d\yy \,w(\xx-\yy) \langle \big[\big(\f^t_\xx + \ps^t_\xx )\big(\ps^l_\yy + \f^l_\yy \big)-\ps^t_\xx \ps^l_\yy\big]\rangle_h^\phi
-\int d\xx\langle\D_h(\xx)\rangle_h^\phi
\;,
\non\]
where in the second line $\D_h(\xx)$ is the correction due to the presence of the cutoff:
\[\D_h(\xx):=2\l\r_0\int d\yy\, \ps^t_\xx \big(w_{\ge h}(\xx-\yy)-w(\xx-\yy)\big)\ps^l_\yy\;. \label{C.16}\]
Eq.\eqref{GWI_2} can be also rewritten in a convenient form by introducing the auxiliary functional $\tilde \WW^{(h-1)}(\phi,\Phi^1,\Phi^2,J^\D)$:
\[e^{-\tilde \WW^{(h-1)}(\phi,\Phi^1,\Phi^2,J^\D)}=e^{|\L|\FF_{h-1}}\int P^B_{\ge h}(d\psi)e^{-\bar V(\ps+\f)-\int d\xx [\Phi^1_\xx F^1_\xx(\f)+\Phi^2_\xx F^2_\xx(\ps,\f)+
J^\D_\xx \D_h(\xx)]} \label{C.17}\]
with 
\[F^1_\xx(\f)=\bar\n\sqrt{2\r_0}\phi^t_\xx\;,\qquad F^2_\xx(\ps,\phi)=2\l \r_0  \int d\yy \,w(\xx-\yy)  \big[\big(\f^t_\xx + \ps^t_\xx )\big(\ps^l_\yy + \f^l_\yy \big)-\ps^t_\xx \ps^l_\yy
\big]\;,\]
in terms of which \eqref{GWI_2} takes the form:
 \[\label{GWIbis}0&= \int d\xx\Big[\frac{\d \tilde\WW^{(h-1)}(\phi,\V0)}{\d \f_{\xx}^{l}}\phi_{\xx}^{t} -\frac{\d \tilde\WW^{(h-1)}(\phi,\V0)}{\d \f_{\xx}^{t}} \left(\f_{\xx}^{l}+\sqrt{2\rho_{0}}\right)
 \\
 &+\frac{\d \tilde\WW^{(h-1)}(\phi,\V0)}{\d \Phi^1_\xx}+\frac{\d \tilde\WW^{(h-1)}(\phi,\V0)}{\d \Phi^2_\xx}-\frac{\d \tilde\WW^{(h-1)}(\phi,\V0)}{\d J^\D_\xx}\Big]
 \non\]
and $(\phi,\V0)$ is a shorthand for $(\phi,0,0,0)$.
At this point, we can obtain infinitely many identities among the kernels of $\WW^{(h-1)}$, known as global WIs, by further deriving \eqref{GWI_2} or \eqref{GWIbis}
 w.r.t. $\phi^{l},\phi^t$, and then taking $\phi\equiv 0$. The ``formal" global WIs (discussed e.g. in \cite{CaDiC1,CaDiC2} in the framework of dimensional regularization) are those obtained by 
neglecting the effect of the cutoff, i.e., by neglecting $\langle\D_h(\xx)\rangle_h^\phi$ in the second line of \eqref{GWI_2}. 

A few global WIs that we are interested in are those obtained by: (1) deriving w.r.t. $\phi^t_\xx,\phi^l_\yy$, then integrating w.r.t. $\yy$; 
(2) deriving w.r.t. $\phi^t_\xx,\phi^t_\yy,\phi^t_\zz$ and then integrating w.r.t. $\yy,\zz$; (3) deriving w.r.t. $\phi^t_\xx$ (of course in all these cases we put $\phi\equiv 0$ after the 
derivation). Their explicit expression is (neglecting for simplicity the issue of the ``last integration scale" mentioned in footnote 15, a couple of pages above):
\[& Z_{h-1}-2 \g^{2h-\bh} \n_h -2\sqrt{2\r_0}\g^{h/2}\m_h =\int  d\xx\, d\yy \frac{\d^3 \tilde\WW^{(h)}(\V0)}{\d J^\D_\V0\d\psi^t_\xx\d \psi^l_\yy}\;, \label{C.20}\\
& \g^{h/2}\m_h-4\sqrt{2\r_0}\g^{h-\bh}\l_h=\frac1{3!}\int  d\xx\, d\yy \, d\zz\, \frac{\d^4 \tilde\WW^{(h)}(\V0)}{\d J^\D_\V0\d\psi^t_\xx\d\psi^t_\yy\d\psi^t_\zz}\;, \label{C.21}\\
& W^{(h)}_{1,0}+\sqrt{2\r_0}(\bar\n-\hat W^{(h)}_{0,2}(\V0))=\int  d\xx\,\frac{\d^2 \tilde\WW^{(h)}(\V0)}{\d J^\D_\V0\d\psi^t_\xx}\;.\]
The first two identities are clearly useful, because they relate the running coupling constants among each other. The last identity can be read as a renormalization condition,
which fixes the chemical potential to the ``right value", and is known as the Hugenholtz-Pines identity. 
It is easy to check that the correction terms in these identities are ``trivial", in the sense that they vanish in the limit of sharp cutoff function, in which case we can 
drop all these terms at once. 

\subsection{Local Ward Identities}\label{appC.2}

The local Ward identities are derived in a way completely analogous to the global ones, with the only difference that the phase factor $\a(\xx)$ appearing in the 
change of variables $\ps^\pm_\xx \arr e^{\pm i \a_\xx}\ps^\pm_\xx $ is a non-trivial function of $\xx$, and the derivatives w.r.t. $\a$ performed in the previous section should be replaced 
by functional derivatives w.r.t. $\a(\xx)$. When we derive \eqref{C.1} w.r.t. $\a(\xx)$ we produce a number of terms that are essentially the same as those discussed 
in the previous 
section, except for the fact that there is an integration over $\xx$ missing; e.g., the analogue of \eqref{d1} is
\[  \label{d1.local}
\frac{\d}{\d \a(\xx)} \bar W(e^{i \a}\ps + \f) \Big|_{\a(\xx)\equiv0}= -i \Big[ \frac{\d \bar W(\ps+\f)}{\d \f^+_\xx}\big( \f^+_\xx + \sqrt{\r_0}\big) - \frac{\d \bar W(\ps+\f)}{\d \f^-_\xx}\big( \f^-_\xx + \sqrt{\r_0}\big)\Big]\;,
\]
etc. In addition to these terms (i.e., the ``local" analogues of those of the previous section), there is an extra contribution coming from the measure $P^0_{\geq h}(d\ps)$, which is not invariant under a local gauge transformation
(while it was invariant under a global one). The gaussian weight entering the definition of $P^0_{\geq h}(d\ps)$ has the form
\[
\exp\Big\{-\int \frac{d\kk}{(2\p)^3}\hat \chi_{\ge h}^{-1}(\kk)\hat \psi_{\kk}^{+}(-ik_0+|\vk|^2)\hat \psi^-_\kk\Big\}=: \exp\Big\{-\int d\xx\, \psi^+_\xx (D_{\ge h}\psi^-_{\cdot})_\xx
\Big\}
\]
where the pseudo-differential operator $D_{\ge h}$ is defined here for the first time. Taking the functional derivative of the expression in braces, we find, after some algebra:
\[ &
-\frac{\d}{\d \a(\xx)}\int d\xx' \, e^{+i\a({\xx'})}\psi_{\xx}^{+}\Big(D_{\geq h}(e^{-i\a(\cdot)}\psi_{\cdot}^{-})\Big)_{\!\xx'}\,\Big|_{\a(\xx)\equiv0} =\label{C.18}\\
&\qquad \qquad =
-i  \Big[\dpr_{0}(\psi^+_\xx\psi^-_\xx)+\Delta(\psi^+_\xx\psi^-_\xx)-2\vec\dpr(\psi^+_\xx\vec \dpr\psi^-_\xx)\Big]-\d T_h(\xx)\;,\non\]
where $\dpr_0=\dpr_{x_0}$, $\vec\dpr=(\dpr_{x_1},\dpr_{x_2})$, and
\[ \label{C.local2}
 \d T_h(\xx)= i \int \frac{d\kk\, d\pp}{(2\pi)^6} \,e^{i\pp \xx}\hat\psi_{\kk+\pp}^{+} C_h(\kk,\pp) \hat \psi^-_\kk\;,\]
 with 
\[
  C_h(\kk,\pp) := \big(\hat \chi_{\geq h}^{-1}(\kk)-1\big) (-i k_{0} +
|\vk|^2) - \big(\hat \chi_{\geq h}^{-1}(\kk+\pp)-1\big) (-i(k_{0}+p_{0}) + |\vk+\vp|^2)
\]
The equations above can be rewritten in terms of the fields $\ps^{l},\ps^t$, in terms of which we find 
\[ & -\frac{\d}{\d \a(\xx)}\int d\xx' \, e^{+i\a({\xx'})}\psi_{\xx}^{+}\Big(D_{\geq h}(e^{-i\a(\cdot)}\psi_{\cdot}^{-})\Big)_{\!\xx'}\,\Big|_{\a(\xx)\equiv0} 
=-i(\dpr_0 j_{0,\xx}+\vec\dpr\, \vec j_\xx)-\d T(\xx)\;,\]
with 
\[ &
j_{0, \xx}=
\,\frac{1}{2}\lft[(\psi_{\xx}^{l})^2+(\psi_{\xx}^{t})^2\rgt]\;,\\
&\vec j_\xx=i\ps^l_\xx\vec\dpr \psi^t_\xx-i\ps^t_\xx\vec\dpr \psi^l_\xx\]
and
\[  \label{C.31}
 \d T_h(\xx)=  \int \frac{d\kk\, d\pp}{(2\pi)^6} \,e^{i\pp \xx}(\hat \chi_{\ge h}^{-1}(\kk)-1)\Big[&k_0(\hat \psi^l_{-\kk-\pp}\hat \psi^l_\kk+
 \hat \psi^t_{-\kk-\pp}\hat \psi^t_\kk)+\\
 &+|\vk|^2
(\hat \psi^l_{-\kk-\pp}\hat \psi^t_\kk-
 \hat \psi^t_{-\kk-\pp}\hat \psi^l_\kk)\Big]\;.\non\]
After having performed the functional derivative w.r.t. $\a(\xx)$ and having re-expressed everything in terms of the $\psi^l,\psi^t$ fields, 
we finally arrive at the analogue of \eqref{GWI_2}, which reads
\[& \label{C.33}
-i \langle\dpr_0 j_{0,\xx}+\vec\dpr\,\vec j_\xx\rangle^\phi_h =\frac{\d \WW^{(h-1)}(\phi)}{\d \f_{\xx}^{l}}\phi_{\xx}^{t} -\frac{\d \WW^{(h-1)}(\phi)}{\d \f_{\xx}^{t}} \left(\f_{\xx}^{l}+\sqrt{2\rho_{0}}\right)
 +\bar\n\sqrt{2\r_0}
\langle\phi^t_\xx\rangle_h^\phi\\
&+  2\l \r_0 \int d\yy\, w(\xx-\yy)\langle \big[\big(\f^t_\xx + \ps^t_\xx )\big(\ps^l_\yy + \f^l_\yy \big)-\ps^t_\xx \ps^l_\yy\big]\rangle_h^\phi -\langle\D_h(\xx)\rangle_h^\phi+
\langle \d T_h(\xx)\rangle_h^\phi\;.\non\]
If desired, this equation can be put in a form similar to \eqref{GWIbis}. It is enough to introduce the auxiliary potential
\[e^{-\bar \WW^{(h-1)}(\phi,{\bf J},\vec \Phi,J^\D,J^{\d T})}=e^{|\L|\FF_{h-1}}\int P^B_{\ge h}(d\psi)e^{-\bar V(\ps+\f)-({\bf J},{\bf j})-(\vec \Phi, \vec F)-
(J^\D,\D_h)-(J^{\d T},\d T)]}\label{C.barw}\]
where: ${\bf J}=(J_0,J_1,J_2)=(J_0,\vec J)$, $\vec \Phi=(\Phi^1,\Phi^2)$, and 
$({\bf J},{\bf j})=\int d\xx (J_{0,\xx}j_{0,\xx}+\vec J_\xx\,\vec j_{\xx})$, $(\vec \Phi, \vec F)=\sum_{i=1}^2\int d\xx \Phi^i_\xx F^i_\xx$, etc. 
Using the auxiliary potential \eqref{C.barw}, Eq.(\ref{C.33}) takes the form:
 \[\label{LWIbis}
& -i\,\dpr_0\, \frac{\d \bar \WW^{(h-1)}(\phi, \V0)}{\d J_{0,\xx}}-i\,\vec \dpr\, \frac{\d \bar \WW^{(h-1)}(\phi, \V0)}{\d \vec J_{\xx}} \non \\
&=\frac{\d \bar \WW^{(h-1)}(\phi,\V0)}{\d \f_{\xx}^{l}}\phi_{\xx}^{t} -\frac{\d \bar \WW^{(h-1)}(\phi,\V0)}{\d \f_{\xx}^{t}} \left(\f_{\xx}^{l}+\sqrt{2\rho_{0}}\right)
 \\
 &+\frac{\d \bar \WW^{(h-1)}(\phi,\V0)}{\d \Phi^1_\xx}+\frac{\d \bar \WW^{(h-1)}(\phi,\V0)}{\d \Phi^2_\xx}-\frac{\d \bar\WW^{(h-1)}(\phi,\V0)}{\d J^\D_\xx}
 +\frac{\d \bar \WW^{(h-1)}(\phi,\V0)}{\d J^{\d T}_\xx}\;.
 \non\]
At this point, we can obtain infinitely many identities among the kernels of $\WW^{(h-1)}$, known as local WIs, by further deriving this identity w.r.t. $\phi^{l,t}$, and then taking $\phi\equiv 0$. The ``formal" local WIs (discussed e.g. in \cite{CaDiC1,CaDiC2} in the framework of dimensional regularization) are those obtained by 
neglecting the effect of the cutoffs, i.e., by dropping the terms $\langle\D_h(\xx)\rangle_h^\phi$ and $\langle\d T_h(\xx)\rangle_h^\phi$.

Two local WIs we are interested in are those obtained by
deriving w.r.t. $\phi^l$, or w.r.t. $\phi^t$, and then taking the Fourier transform at $\pp=(p_0,\v0)$.
Their explicit expression in momentum space is (neglecting for simplicity the issue of the ``last integration scale"):
\[
p_0 \hat W^{(h)}_{1,0;J_0}(p_0,\v0)& = -\sqrt{2\r_0}\hat W^{(h)}_{1,1}(p_0,\v0) + W^{(h)}_{0,1} +\hat W^{(h)}_{1,0;J^\D}(p_0,\v0)-\hat W^{(h)}_{1,0;J^{\d T}}(p_0,\v0)\;,  \label{C.35}\\
p_0 \hat W^{(h)}_{0,1;J_0}(p_0,\v0)& = 2\sqrt{2\r_0}\hat W^{(h)}_{0,2}(p_0,\v0) - W^{(h)}_{1,0} +\hat W^{(h)}_{0,1;J^\D}(p_0,\v0)-\hat W^{(h)}_{0,1;J^{\d T}}(p_0,\v0) \;,\label{C.36}
\]
where $W^{(h)}_{1,0;J_0}(\xx,\yy)$ is the kernel of $J_{0,\xx}\psi^l_\yy$ in $\bar \WW^{(h)}(\phi,J_0,\V0)$, and $  \hat W^{(h)}_{1,0;J_0}(\pp)=\int d\xx$ $e^{i\pp\xx}$ 
$W^{(h)}_{1,0;J_0}(\xx,\V0)$
(and similarly for $\hat W^{(h)}_{1,0;J^\D}(\pp)$, etc). Eqs.\eqref{C.35}-\eqref{C.36} are the analogues of \cite[(3.11)-(3.12)]{CaDiC2}.
If we divide \eqref{C.35} by $p_0$ and then take the limit $p_0\to 0$ we find:
\[ Z_{h-1}^{J_0}=\sqrt{2\r_0}(E_{h-1}-1)+\dpr_{p_0}\hat W^{(h)}_{1,0;J^\D}(\V0)-\dpr_{p_0}\hat W^{(h)}_{1,0;J^{\d T}}(\V0)\;,\label{C.37}\]
where $Z_{h-1}^{J_0}:=\hat W^{(h)}_{1,0;J_0}(\V0)$, and the name is justified by the fact that the Feynman diagram expansion for $Z_h^{J_0}$ has the same structure 
as that for $Z_h$. In particular, inspection of perturbation theory shows that 
\[Z_h^{J_0}=\frac{Z_h - Z_\bh}{\l \hat v(\v0)\sqrt{2\r_0}}\big(1+O(\g^{h-\bh})\big)\;, \label{ZhJ0}\]
where the error terms in parentheses come from the irrelevant terms on scale $\bh$. In a similar way, if we divide \eqref{C.36} by $p_0^2$ and then take the limit $p_0\to 0$ we find,
defining $E_{h-1}^{J_0}:= -\dpr_{p_0}\hat W^{(h)}_{0,1;J_0}(\V0)$:
\[-2E_{h-1}^{J_0}=2\sqrt{2\r_0}B_{h-1}+\dpr^2_{p_0}\hat W^{(h)}_{0,1;J^\D}(\V0)-\dpr^2_{p_0}\hat W^{(h)}_{0,1;J^{\d T}}(\V0)\;,\label{C.39}\]
where 
\[E_h^{J_0}=\frac{E_h - 1}{\l \hat v(\v0)\sqrt{2\r_0}}\big(1+O(\g^{h-\bh})\big)\;. \label{C.40}\]
Contrary to the correction terms of the global Ward Identities, the corrections in \eqref{C.37},
\eqref{C.39} are not ``trivial", i.e., they do not vanish in the sharp cutoff limit. On the contrary,
they give a finite (cutoff-dependent) contribution to the beta function, which shows up as an ``anomaly" already at the level of the one-loop beta function, as discussed in Section \ref{sec3.C.4}.

\subsection{Comparison between \protect{$\WW^{(h-1)}$ and $\VV^{(h-1)}$}}\label{appDd.3}

In order to establish the exact relation between $\WW^{(h-1)}$ and $\VV^{(h-1)}$, we compute the r.h.s. of \eqref{C.1} via a multiscale integration 
analogous to the one used in the bulk of the paper for the computation of $\VV^{(h-1)}$. The integration of the ultraviolet fields on scales $>\bh$, as well as the integration of
the field $\tilde\psi^{(\bh)}$ (see \eqref{3.34}),  
is identical to the one discussed in Section 
\ref{s.BEC_UV} and at the beginning of Section \ref{sec3.c}, after which we can rewrite \eqref{C.1} as
\be e^{-|\L|\FF_{h-1}-\WW^{(h-1)}(\phi)}=e^{-|\L|\sum_{k\ge \bh}F_k}\int P_{[h,\bh]}(d\psi)e^{-\VV^{(\bh)}(\psi+\phi)}\;,\label{C.199}\ee
where $P_{[h,\bh]}(d\psi)$ is the same as the gaussian measure $P_{\le \bh}(d\psi)$ in \eqref{3.34}, modulo the presence of an infrared cutoff on scale $h$ in the 
corresponding propagator (i.e., while the cutoff function appearing in the propagator of $P_{\le \bh}(d\psi)$ is $\chi_{\bh}(\kk)$, the one in the propagator of 
$P_{[h,\bh]}(d\psi)$ is $\chi_{\bh}(\kk)-\chi_{h-1}(\kk)$). Moreover, using the same convention of Section \ref{s3.C.2}, $\VV^{(\bar h)}$ coincides with the 
function $\bar V^{(\bar h)}$ introduced after \eqref{3.34}. At this point, we start dressing the gaussian measure in \eqref{C.199}, in the same way as in Section \ref{s3.C.2}. 
Let us describe the first integration step explicitly. 
Using the notations introduced in \eqref{e3.62} and following equations, we rewrite the r.h.s. of \eqref{C.199} as
\[e^{-|\L|\sum_{k\ge\bh}F_k+\LL_Q\VV^{(\bh)}(\phi)}
\int P_{[h,\bh]}(d\psi)e^{-\LL_Q\VV^{(\bh)}(\psi)-\widehat \VV^{(\bh)}(\psi+\phi)-\big(\psi+\phi,M_Q^{(\bh)}\phi\big)}\;,\label{C.200}\]
where $\widehat \VV^{(\bh)}$ is a shorthand for $\LL_C\VV^{(\bh)}+\RR\VV^{(\bh)}$. We now combine $\LL_Q\VV^{(\bh)}(\psi)$ with the gaussian measure, as in \eqref{m1}, 
and then use the addition principle to rewrite the dressed measure as a product of a measure supported on scale $\bh$ and a measure supported on smaller scales, as in \eqref{3.68}:
\[\eqref{C.200}&=e^{-|\L|(\sum_{k\ge\bh}F_k+t_{\bh})+\LL_Q\VV^{(\bh)}(\phi)}\int P_{[h,\bh-1]}(d\psi)\cdot\\
&\cdot\int \tilde P_{\bh}(d\psi^{(\bh)})
e^{-\widehat \VV^{(\bh)}(\psi^{(\bh)}+\psi+\phi)-\big(\psi^{(\bh)}+\psi+\phi,M_Q^{(\bh)}\phi\big)}\;.\nonumber\]
At this point, we integrate the field on scale $\bh$ and define:
\[ &\label{C.22}
|\L| \tl F_{\bh} +\tilde S_\bh(\phi)+\VV^{(\bh-1)}(\ps')+\BB^{(\bh-1)}(\ps',\phi) =\\
&\qquad = -\log\int \tilde P_{h}(d\ps^{(h)}) e^{-\widehat \VV^{(\bh)}(\psi^{(\bh)}+\psi')-\big(\psi^{(\bh)}+\psi',M_Q^{(\bh)}\phi\big)}\;.
\]
where, denoting by $\hat \phi_\kk$ the two-component column vector with components $\hat \phi_\kk^l$ and $\hat \phi_\kk^t$,
\[\tilde S_\bh(\phi)=-\frac12\int\frac{d\kk}{(2\p)^3}\hat\phi^T_{-\kk}\big[\hat M_Q^{(\bh)}(\kk)\big]^T\hat{\tilde g}^{(\bh)}(\kk)\hat M_Q^{(\bh)}(\kk)\hat \phi_\kk\equiv-\frac12\big(\phi,
M_Q^{(\bh),T}{\tilde g}^{(\bh)}M_Q^{(\bh)}\phi\big)\;.\]
We now set $F_{\bh-1}=t_\bh+\tilde F_\bh$ and $S^{(\bh-1)}=\LL_Q\VV^{(\bh)}-\tilde S_\bh$, and then we iterate the same procedure. After the integration of the 
fields on scales $\ge k+1$ we rewrite \eqref{C.1} as
\[e^{-|\L|\sum_{k'\ge k}F_{k'}+ S^{(k)}(\phi)}\int P_{[h,k]}(d\psi)e^{-\VV^{(k)}(\ps+\phi)-\BB^{(k)}(\ps+\phi,\phi)}\;,\]
where one can inductively prove that, for $k<\bh$, 
\[& S^{(k)}(\phi)=\sum_{k'=k+1}^{\bh}\Big[\LL_Q\VV^{(k')}(\phi)+\frac12\big(\phi,Q^{(k'),T}\tilde g^{(k')}Q^{(k')}\phi\big)-\frac12\big(\phi,G^{(k'+1),T}M_Q^{(k')}G^{(k'+1)}\phi\big)\Big]\;,\\
& \BB^{(k)}(\ps,\phi)=\big(\psi,Q^{(k+1)}\phi\big)+\sum_{n\ge 1}\int d\xx_1\cdots d\xx_n \Big[\prod_{i=1}^n \big(
G^{(k+1)}*\phi\big)_{\xx_i}\Big]\frac{\dpr^n}{\dpr \psi_{\xx_1}\cdots \dpr \psi_{\xx_n}}\VV^{(k)}(\psi)\;,\]
and the vectorial nature of $\psi$ (i.e., the fact that $\psi$ has two components, labelled $l$ and $t$) is implicitly understood. Moreover, the functions 
$Q^{(k)}$ and $G^{(k)}$, $k\le \bh+1$, are defined by the iterative relations 
\[ Q^{(k)}=Q^{(k+1)}+M_Q^{(k)}+M_Q^{(k)}G^{(k+1)}\;,\qquad G^{(k)}=G^{(k+1)}+\tilde g^{(k)}Q^{(k)}\;,\]
with $Q^{(\bh+1)}=G^{(\bh+1)}=0$. 

The iteration goes on in the same fashion until we reach scale $h$, where a small difference from the previous scheme should be taken into account: in fact,
by proceeding as described above, one finds that the dressed propagator on the last scale, rather than being equal to $\tilde g^{(h)}$, it is equal to 
\be \tilde{\mathfrak g}^{(h)}(\xx)=
 \int  \frac{d\kk}{(2\pi)^3}\, \tl f_{h}(\kk)\,\frac{\, e^{-i\kk \cdot \xx}}{\DD_{h-1}(\kk)}\left(\begin{array}{cc}
\tl{A}_{h-1}'(\kk)\, |\vk|^{2} + \tl B_{h-1}'(\kk)\, k_0^2   & \tl E_{h-1}'(\kk)\, k_{0}\\
-\tilde E_{h-1}'(\kk)\, k_{0} &   \tl Z_{h-1}'(\kk)
\end{array}\right)
\ee
where $\tl f(\kk)=\chi_h(\kk)-\chi_{h-1}(\kk)$, as usual, and
\[
\tl A_{h-1}'(\kk) =  \tl A_h(\kk) + a_h \tilde f_h(\kk)\;, &&  \tl B_{h-1}(\kk) =  \tl B_h (\kk)+ b_h \tl f_h(\kk) \;, \non \\
 \tl E_{h-1}(\kk) =  \tl E_h (\kk) + e_h \tilde f_h(\kk) \;, && \tl Z_{h-1}(\kk) =  \tl Z_h(\kk) + z_h \tl f_h(\kk)\;.
\]
After the integration of the last scale we finally find
\[\eqref{C.1}=e^{-|\L|(\sum_{k'\ge h}F_{k'}+\tilde F_{h-1}) +\tilde S^{(h-1)}(\phi)-\tilde \VV^{(h-1)}(\phi)-\tilde \BB^{(h-1)}(\phi,\phi)}\equiv 
e^{-|\L|\FF_{h-1}-\WW^{(h-1)}(\phi)}
\;,\label{C.53}\]
where the tildes on the functions at exponent recall the fact that these functions are defined in the same fashion as their analogues without tilde, 
with the only difference that the single-scale propagator on the last scale, $\tilde g^{(h)}$, wherever it enters the definition of these objects, 
should be replaced by $\tilde{\mathfrak g}^{(h)}$. 

Eq.\eqref{C.53} provides us the desired relation between $\WW^{(h-1)}$ and $\VV^{(h-1)}$. It shows that the {\it local parts} of the kernels 
of $\WW^{(h-1)}$ (in the sense of the values of their Fourier transforms at zero external momenta, as well as the zero momenta values of their derivatives w.r.t. $\kk$),
are related in a very simple fashion with the corresponding local parts of the kernels of $\VV^{(h-1)}$, simply because the local part of the kernels of $\tilde \BB^{(h-1)}(\phi,\phi)$
is equal to $2\sum_{k=h}^\bh\LL_Q\VV^{(k)}(\phi)$, due to the compact support properties of $\tilde g^{(k)}$ and of $G^{(k)}$, and similarly the 
local part of $\tilde S^{(h-1)}$ is equal to $\sum_{k=h}^\bh\LL_Q\VV^{(k)}$. Therefore, 
\be \LL \WW^{(h-1)}(\phi)= \LL\tilde \VV^{(h-1)}(\phi)+\sum_{k=h}^{\bh}\LL_Q\VV^{(k)}\;,\label{C.54}\ee
which induces a simple, explicit, connection between the actual renormalization and running coupling constants of the model, introduced in Section \ref{s3.C.2}, and 
$\LL \WW^{(h-1)}(\phi)$.

\section{Verification of Ward Identities at lowest order}

In this section, we verify the validity of some of the global and local Ward Identities among the running coupling 
and the renormalization constants, at lowest non trivial order in perturbation theory. In particular, we shall compute the effect of the infrared cutoff and discuss its role in the different identities worked out below. 

\feyn{
\begin{fmffile}{BoseFeyn/muCorr}
 \unitlength = 0.8cm
\def\myl#1{2.5cm}
\def\myll#1{2.7cm}
\begin{align*}  
	\hat{W}^{(h-1)}_{1,1;J^\D} & = \;
		\parbox{\myll}{\centering{
			\begin{fmfgraph*}(2.8,2) 
			\fmfleft{i1}
			\fmfright{o1,R,o2}
                   \fmftop{T}
                   \fmfbottom{B}
			\fmf{phantom, tension=1.8}{B,vB}  
                    \fmf{phantom, tension=1.8}{T,vT}  
			\fmf{zigzag,tension=1.5}{i1,v1}
			\fmf{plain, left=0.3, tension=1.2}{v1,vT}
                   \fmf{plain, left=0.2, tension=0.4}{vT,v3}
                    \fmf{phantom, tension=1}{R,vR}  
			\fmf{plain, left=0.3, tension=1}{v3,vR}
                   \fmf{plain, left=0.3, tension=0.4}{vR,v2}			
                  \fmf{dashes, right=0.3, tension =1.4}{v1,vB}
	            \fmf{dashes, right=0.2, tension=0.4}{vB,v2}           
		     \fmf{plain, tension=0.8}{o1,v2}
			\fmf{dashes,tension=0.8}{o2,v3}
                   \fmfdot{v2,v3}
			\Tri{v1}
			\end{fmfgraph*}  
			}}   +\,
\parbox{\myll}{\centering{ 
			\begin{fmfgraph*}(2.8,2) 
			\fmfleft{i1}
			\fmfright{o1,R,o2}
                   \fmftop{T}
                   \fmfbottom{B}
			\fmf{phantom, tension=1.8}{B,vB}  
                    \fmf{phantom, tension=1.8}{T,vT}  
			\fmf{zigzag,tension=1.5}{i1,v1}
			\fmf{plain, left=0.3, tension=1.2}{v1,vT}
                   \fmf{plain, left=0.2, tension=0.4}{vT,v3}
                    \fmf{phantom, tension=1}{R,vR}  
			\fmf{plain, left=0.3, tension=1}{v3,vR}
                   \fmf{dashes, left=0.3, tension=0.4}{vR,v2}			
                  \fmf{dashes, right=0.3, tension =1.4}{v1,vB}
	            \fmf{plain, right=0.2, tension=0.4}{vB,v2}           
		     \fmf{plain, tension=0.8}{o1,v2}
			\fmf{dashes,tension=0.8}{o2,v3}
                   \fmfdot{v2,v3}
			\Tri{v1}
			\end{fmfgraph*}  
			}}  
+\;
	\parbox{\myll}{\centering{ 
			\begin{fmfgraph*}(2.8,2) 
			\fmfleft{i1}
			\fmfright{o1,R,o2}
                   \fmftop{T}
                   \fmfbottom{B}
			\fmf{phantom, tension=1.8}{B,vB}  
                    \fmf{phantom, tension=1.8}{T,vT}  
			\fmf{zigzag,tension=1.5}{i1,v1}
			\fmf{dashes, left=0.3, tension=1.2}{v1,vT}
                   \fmf{plain, left=0.2, tension=0.4}{vT,v3}
                    \fmf{phantom, tension=1}{R,vR}  
			\fmf{plain, left=0.3, tension=1}{v3,vR}
                   \fmf{plain, left=0.3, tension=0.4}{vR,v2}			
                  \fmf{plain, right=0.3, tension =1.4}{v1,vB}
	            \fmf{dashes, right=0.2, tension=0.4}{vB,v2}           
		     \fmf{plain, tension=0.8}{o1,v2}
			\fmf{dashes,tension=0.8}{o2,v3}
                   \fmfdot{v2,v3}
			\Tri{v1}
			\end{fmfgraph*}  
			}}   +\,
\parbox{\myll}{\centering{  
			\begin{fmfgraph*}(2.8,2) 
			\fmfleft{i1}
			\fmfright{o1,R,o2}
                   \fmftop{T}
                   \fmfbottom{B}
			\fmf{phantom, tension=1.8}{B,vB}  
                    \fmf{phantom, tension=1.8}{T,vT}  
			\fmf{zigzag,tension=1.5}{i1,v1}
			\fmf{dashes, left=0.3, tension=1.2}{v1,vT}
                   \fmf{plain, left=0.2, tension=0.4}{vT,v3}
                    \fmf{phantom, tension=1}{R,vR}  
			\fmf{plain, left=0.3, tension=1}{v3,vR}
                   \fmf{dashes, left=0.3, tension=0.4}{vR,v2}			
                  \fmf{plain, right=0.3, tension =1.4}{v1,vB}
	            \fmf{plain, right=0.2, tension=0.4}{vB,v2}           
		     \fmf{plain, tension=0.8}{o1,v2}
			\fmf{dashes,tension=0.8}{o2,v3}
                   \fmfdot{v2,v3}
			\Tri{v1}
			\end{fmfgraph*}  
			}}  
\end{align*}
\end{fmffile}	
}{Correction terms to the formal GWI relating $\m_h$ and $Z_h$, see \eqref{C.20}. The triangular vertex with the external zigzag line represents $J^\D\D_h(\xx)$, see \eqref{C.16} and \eqref{C.17}.
}{muCorr}

\subsection{Global Ward Identities}\label{appEe.1}
Let us verify  the validity of \eqref{C.20} at lowest non trivial order in $\l$ and/or in $\g^{h-\bh}$, in the region 
$h^*\le h\le \bh$. If $W^{(h)}_{1,1;J^\D}(\xx,\yy;\bz)$ is the kernel of $J^\D_{\bz}\phi^t_\xx\phi^l_\yy$ in $\tl \WW^{(h)}(\phi,J^\D, \V0)$, and \[\hat W^{(h)}_{1,1;J^\D}(\kk,\pp)=\int d\xx\, 
d\yy\, e^{-i\kk\xx+i(\kk+\pp)\yy}W^{(h)}_{1,1;J^D}(\xx, \yy;\V0),\label{ref.assurda}\] we can rewrite \eqref{C.20} as
\[  \label{D.1}
Z_{h-1}-2 \g^{2h-\bh} \n_h -2\sqrt{2\r_0}\g^{h/2}\m_h =\hat{W}^{(h)}_{1,1;J^\D}(\bz, \bz)\;.
\]
Using the beta function equations \eqref{3.75}--\eqref{3.78}, the l.h.s. can be rewritten as 
\[Z_{\bar h-1}-2 \g^{\bh} \n_\bh -2\sqrt{2\r_0}\g^{\bh/2}\m_\bh+\sum_{k=h+1}^\bh\Big(
\b^Z_k-2\g^{2k-\bh}\b^\n_k-2\sqrt{2\r_0}\g^{k/2}\b^\m_k\Big)\;.\]
Now, using the explicit expressions of $Z_{\bar h-1}, \n_\bh, \m_\bh$, one can check that the combination 
$Z_{\bar h-1}-2 \g^{\bh} \n_\bh -2\sqrt{2\r_0}\g^{\bh/2}\m_\bh$ is zero at lowest order (i.e., 
at the order $\l$, as well as at the order $\bar \n$) and, therefore,
it is (at most) of the order $O(\l^2|\bh|)$. Moreover, using the explicit expression of the beta functions for $Z_h$, $\n_h$ 
and $\m_h$, see \eqref{3.81}, \eqref{3.82}, \eqref{3.84}, as well as the replacements 
\[
Z_{h} \to 2\sqrt{2\r_0} \g^{h/2}\m_{h}\;,\qquad    \l_h \to \frac{1}{4\sqrt{2\r_0}} \g^{\bh-h/2} \m_h\,, \label{eD.3}
\]
induced by the Ward Identities \eqref{C.20}-\eqref{C.21} (which, once inserted in the expressions  \eqref{3.81}, \eqref{3.82}, \eqref{3.84} of the one-loop beta function, induce errors of higher order in $\l$ and/or in 
$\g^{h-\bh}$, as one can prove inductively in $h$), we can rewrite, for $h^*\le h\le \bh$, 
\[&\sum_{k=h+1}^\bh\Big(
\b^Z_k-2\g^{2k-\bh}\b^\n_k-2\sqrt{2\r_0}\g^{k/2}\b^\m_k\Big)=
\sum_{k=h+1}^\bh\Big(-\frac{2\m_k^2}{A_k C_k}\b_0^{(2)}\non\\
&\qquad -\frac{2\g^k}{A_k C_k}\frac1{8\r_0}(\b_0^{(1)}-2\b_0^{(1,\chi)})
+\frac{2\m_k^2}{A_k C_k}(\b_0^{(2)}+2\b_0^{(2,\chi)})\Big)\]
modulo higher order correction terms in $\l$ and/or in $\g^{h-\bh}$ (in deriving the expression of the first term in the 
second line we also used the fact that $C_kZ_k=1$, up to higher order corrections).
Now, the sum over $k$ of 
$-\frac{2\g^k}{A_k C_k}\frac1{8\r_0}(\b_0^{(1)}-2\b_0^{(1,\chi)})$ is of the order $\l^2$, which is subdominant 
with respect to the other two terms, whose sum over $k$ gives (using the fact that $\m_k=\g^{(\bh-k)/2}
\m_\bh$, up to higher order corrections)
\[\frac{\g}{\g-1}\frac{4\m_{h+1}^2}{A_{h+1} C_{h+1}}\b_0^{(2,\chi)}\;.\]
We now want to show that this expression is equal to 
$\hat{W}^{(h)}_{1,1;J^\D}(\bz, \bz) $, modulo higher order corrections. 
In fact, the lowest order contribution to $\hat{W}^{(h)}_{1,1;J^\D}(\bz, \bz) $ is given by the sum of the diagrams in Fig.\ref{muCorr}, with at least one propagator at scale $h$. After the same considerations made for the calculation of the 
beta function for $\m_{h}$ (see Section \ref{s.mu}) we find that, at lowest order, 
\[ \label{D.2}
&\hat{W}^{(h)}_{1,1;J^\D}(\bz, \bz) = 8 \l \r_0 \hat v(\v0) \frac{\m_{h+1}^2}{Z_{h+1}}\frac1{A_{h+1} C_{h+1}} \times \\
&\qquad  \times
 \int \frac{d^3\kk}{(2\pi)^3} \Big(\frac1{\sum_{k\ge 0}\ff_k(|\kk|)}
 -1 \Big) \ff_0(|\kk|) \frac{\ff_0^2(|\kk|) +3 \ff_0(|\kk|) \ff_{1}(|\kk|) +3 \ff^2_{1}(|\kk|) }{|\kk|^4}\;.
\non \]
Note that, for $j\ge 0$, 
\[\Big(\frac1{\sum_{k\ge 0}\ff_k(|\kk|)}
 -1 \Big) \ff_j(|\kk|) = u_0(|\kk|)\d_{0,j}\]
 where 
 \[u_0(|\kk|):=
\begin{cases}
\;1 - \ff_0(|\kk|)\;, &{\rm if}\  \g^{-1} \leq |\kk| \leq 0 \\
\;0\;, & \text{otherwise}\;.
\end{cases}
\]
Note that,  if $\g^{-1} \leq|\kk|< 1$, then $\ff_1(|\kk|)=0$. Therefore, 
\[
\hat{W}^{(h)}_{1,1;J^\D}(\bz, \bz) &= 8 \l \r_0 \hat v(\v0) \frac{\m_{h+1}^2}{Z_{h+1}} \frac{1}{A_{h+1} C_{h+1}} \frac{1}{2\pi^2} 
\int_{\g^{-1}}^1 \frac{d\r}{\r^2}\,(1 - \ff_0(\r))\ff^2_0(\r)  \non \\
& =  4\,\m_{h+1}^2 \frac{1}{A_{h+1} C_{h+1}}\, \frac{\g }{2\pi^2} \int_{1}^\g \frac{d\r}{\r^2}\,\chi(\r) (1 - \chi(\r))^2   \non \\
& = 4\,\m_{h+1}^2 \frac{1}{A_{h+1} C_{h+1}} \frac{\g}{\g -1}\, \b_0^{(2,\chi)}\;,  
\]
where we used that $Z_h=2 \l \r_0 \hat v(\v0)$, up to higher order corrections. 

A similar discussion can be repeated for proving the validity of \eqref{C.21}, but we will not belabor the 
details here.

\feyn{
\begin{fmffile}{BoseFeyn/EhZh2}
\unitlength = 0.8 cm  
\def\myl#1{2.6cm}
\def\myll#1{2.6cm}
\begin{align*}
   \beta^{E}_h &=\, \dpr_{p_0} \Biggr[ \; 
 \parbox{\myl}{\centering{ 
	\begin{fmfgraph*}(2.8,2)
			\fmfleft{i1}
			\fmfright{o1}
			\fmftop{t}
			\fmfbottom{b}
			\fmf{phantom, tension=1.8}{t,v3}    
			\fmf{phantom, tension=1.8}{b,v4}    
                   \fmf{plain, tension=1.5}{i1,v1}			
			\fmf{plain,left=0.3}{v1,v3}
			\fmf{plain,left=0.3}{v3,v2}			
			\fmf{dashes,right=0.3}{v1,v4}
			\fmf{plain,right=0.3}{v4,v2}
		       \fmf{dashes, tension=1.5}{o1,v2}
			 \fmfdot{v1,v2}
		\end{fmfgraph*} 
		}}  \Biggl]_{\pp=\V0}  &  \hskip 0.5cm
 \b_h^{Z_{J_0}}  &= \;
	\parbox{\myl}{\centering{  
			\begin{fmfgraph*}(2.8,2) 
			\fmfleft{i1}
			\fmfright{o1}
			\fmftop{t}
			\fmfbottom{b}
			\fmf{phantom, tension=1.8}{t,v3}    
			\fmf{phantom, tension=1.8}{b,v4}    
            \fmf{wiggly, tension=1.5}{i1,v1}			
			\fmf{plain,left=0.3}{v1,v3}
			\fmf{plain,left=0.3}{v3,v2}			
			\fmf{plain,right=0.3}{v1,v4}
			\fmf{plain,right=0.3}{v4,v2}
			\fmf{dashes, tension=1.5}{o1,v2}
		\fmfdot{v1,v2}
		\end{fmfgraph*}   
			}}    \\
 \dpr_{p_0}\hat W^{(h)}_{1,0;J^\D}(\bz)  &= \; \dpr_{p_0} \Bigg[
	\parbox{\myll}{\centering{ 
			\begin{fmfgraph*}(2.8,2) 
			\fmfleft{i1}
			\fmfright{o1}
			\fmftop{t}
			\fmfbottom{b}
			\fmf{phantom, tension=1.8}{t,v3}    
			\fmf{phantom, tension=1.8}{b,v4}    
            \fmf{zigzag, tension=1.5}{i1,v1}			
			\fmf{plain,left=0.3}{v1,v3}
			\fmf{plain,left=0.3}{v3,v2}			
			\fmf{dashes,right=0.3}{v1,v4}
			\fmf{plain,right=0.3}{v4,v2}
			\fmf{dashes, tension=1.5}{o1,v2}
		\fmfdot{v2}  \Tri{v1}
		\end{fmfgraph*} 
			}} \Bigg]_{\pp=\V0} & \hskip -0.2cm
  \dpr_{p_0}\hat W^{(h)}_{1,0;J^{\d T}}(\bz)  &= \; \dpr_{p_0} \Bigg[
	\parbox{\myll}{\centering{ 
			\begin{fmfgraph*}(2.8,2) 
			\fmfleft{i1}
			\fmfright{o1}
			\fmftop{t}
			\fmfbottom{b}
			\fmf{phantom, tension=1.8}{t,v3}    
			\fmf{phantom, tension=1.8}{b,v4}    
            \fmf{wiggly, tension=1.5}{i1,v1}			
			\fmf{plain,left=0.3}{v1,v3}
			\fmf{plain,left=0.3}{v3,v2}			
			\fmf{plain,right=0.3}{v1,v4}
			\fmf{plain,right=0.3}{v4,v2}
			\fmf{dashes, tension=1.5}{o1,v2}
		\fmfdot{v2} \Square{v1}
		\end{fmfgraph*} 
			}}   \Bigg]_{\pp=\V0} 
\end{align*}
\end{fmffile}
}{Diagrams involved in the local WI relating $E_h$ and $Z_h$. On the first line the lowest order beta function for $E_h$ and $Z^{J_0}_h$. On the second line, the correction terms coming from the cutoffs. The triangular vertex with the external zigzag line represents $J^\D_\xx\D_h(\xx)$, see \eqref{C.16}-\eqref{C.17}. The diamond vertex with wiggly external line represents $J_{\xx}^{\d T} \d T_h(\xx)$, see \eqref{C.31}-\eqref{C.barw}.}{EhZh2}

\subsection{Local WI relating $E_h$ and $Z_h$}\label{appE.2}

In this section we prove the validity of the local WI \eqref{C.37}, that is
\[ Z_{h-2}^{J_0}=\sqrt{2\r_0}(E_{h-2}-1)+\dpr_{p_0}\hat W^{(h-1)}_{1,0;J^\D}(\V0)-\dpr_{p_0}\hat W^{(h-1)}_{1,0;J^{\d T}}(\V0)\;,\label{C.37bis}\]
at lowest non trivial order. Now, using \eqref{3.85}, we find that at lowest order
\[
E_{h-2}- 1 = -2 \frac{1}{A_h C_h Z_h} \m_h^2  \frac{\g}{\g-1}\b_0^{(2)}\;,
\]
while, using \eqref{ZhJ0} and \eqref{3.84},
\[
Z_{h-2}^{J_0} = \sqrt{2\r_0}\,  \frac{Z_{h-2} -Z_\bh}{ Z_\bh} = - 2 \sqrt{2\r_0} \frac{1}{A_h C_h Z_\bh}  \m_h^2 \frac{\g}{\g-1} \b_0^{(2)}\;,
\]
as can also be checked by an explicit calculation of the second diagram on the first line of Fig.\ref{EhZh2}. 
Therefore, at lowest order $Z_{h-2}^{J_0}=\sqrt{2\r_0}(E_{h-2}-1)$. In fact, it can be checked that 
the two correction terms in the r.h.s. cancel among each other: at lowest order their values are 
\[
\dpr_{p_0}\hat W^{(h-1)}_{1,0;J^\D}(\bz)  &= - 2 \g^{\frac{h}{2}} \m_h \frac{2\l \r_0  \hat v (\v0)E_h}{A_hC_hZ_h}\, \dpr_{p_0}  \int \frac{d^3\kk}{(2\pi)^3}\,  \big(\hat \chi^{-1}_{\geq h}(\kk) -1\big) \, \frac{\ff_{h}(|\kk+\pp|)}{|\kk+\pp|^2} \frac{k_0 \ff_{h}(|\kk|)}{|\kk|^2}\Big|_{\pp=\V0} \non  \\
\dpr_{p_0}\hat W^{(h-1)}_{1,0;J^{\d T}}(\bz)& = -2  \g^{\frac{h}{2}}\m_h \frac{1}{A_h C_h}  \dpr_{p_0}\int \frac{d^3\kk}{(2\pi)^3}\, k_0 \big(\hat \chi_{\geq h}^{-1}(\kk) -1 \big)  \frac{\ff_{h}(|\kk+\pp|)}{|\kk+\pp|^2} \frac{\ff_{h}(|\kk|)}{|\kk|^2} 
\Big|_{\pp=\V0}\;,
\]
which are the same, since $E_h=1$ up to higher order corrections.

\feyn{
\begin{fmffile}{BoseFeyn/BhEh}
\unitlength = 0.8 cm  
\def\myl#1{2.5cm}
\[
& \b^B_h =\,\dpr^2_{p_0} \Biggr[ \; 
\parbox{\myl}{\centering{ 
	\begin{fmfgraph*}(2.8,2.2)
			\fmfleft{i1}
			\fmfright{o1}
			\fmftop{t}
			\fmfbottom{b}
			\fmf{phantom, tension=1.8}{t,v3}    
			\fmf{phantom, tension=1.8}{b,v4}    
                   \fmf{plain, tension=1.5}{i1,v1}			
			\fmf{plain,left=0.3}{v1,v3}
			\fmf{plain,left=0.3}{v3,v2}			
			\fmf{dashes,right=0.3}{v1,v4}
			\fmf{dashes,right=0.3}{v4,v2}
		       \fmf{plain, tension=1.5}{o1,v2}
			\fmfdot{v2}
			 \fmfdot{v1,v2}	
		\end{fmfgraph*} 
		}}
\; + \;
 \parbox{\myl}{\centering{ 
	\begin{fmfgraph*}(2.8,2.2)
			\fmfleft{i1}
			\fmfright{o1}
			\fmftop{t}
			\fmfbottom{b}
			\fmf{phantom, tension=1.8}{t,v3}    
			\fmf{phantom, tension=1.8}{b,v4}    
                   \fmf{plain, tension=1.5}{i1,v1}			
			\fmf{plain,left=0.3}{v1,v3}
			\fmf{dashes,left=0.3}{v3,v2}			
			\fmf{dashes,right=0.3}{v1,v4}
			\fmf{plain,right=0.3}{v4,v2}
		       \fmf{plain, tension=1.5}{o1,v2}
			\fmfdot{v2}
			 \fmfdot{v1,v2}	
		\end{fmfgraph*} 
		}}\; \Biggl]_{\pp=\V0}
\hskip0.5cm
\b_h^{E_{J_0}}=\dpr_{p_0} \Biggr[ \; 
 \parbox{\myl}{\centering{ 
	\begin{fmfgraph*}(2.8,2.2)
			\fmfleft{i1}
			\fmfright{o1}
			\fmftop{t}
			\fmfbottom{b}
			\fmf{phantom, tension=1.8}{t,v3}    
			\fmf{phantom, tension=1.8}{b,v4}    
                   \fmf{wiggly, tension=1.5}{i1,v1}			
			\fmf{plain,left=0.3}{v1,v3}
			\fmf{dashes,left=0.3}{v3,v2}			
			\fmf{plain,right=0.3}{v1,v4}
			\fmf{plain,right=0.3}{v4,v2}
		       \fmf{plain, tension=1.5}{o1,v2}
			\fmfdot{v1,v2}
		\end{fmfgraph*} 
		}}\; \Biggl]_{\pp=\V0} \non \\
%
	& \hskip 2.1cm \dpr^2_{p_0}\hat W_{0,1;J^\D}(\V0)  = \, \dpr^2_{p_0}\,\Bigg[\,
		\parbox{\myl}{\centering{ 
	\begin{fmfgraph*}(2.8,2.2)
			\fmfleft{i1}
			\fmfright{o1}
                   \fmftop{t}
                   \fmfbottom{b}
                   \fmf{phantom, tension=1.5}{t,vt}
			\fmf{phantom, tension=1.5}{b,vb} 
                   \fmf{zigzag, tension=1.5}{i1,v1}			
			\fmf{plain,left=0.3}{v1,vt}	
                   	\fmf{plain,left=0.3, tension=0.8}{vt,v2}		
			\fmf{dashes,right=0.3}{v1,vb}
		      \fmf{dashes,right=0.3}{vb,v2}	
		       \fmf{plain, tension=1.5}{o1,v2}
			\Tri{v1}
                    \fmfdot{v2}
		\end{fmfgraph*} 
		}}    +\,
\parbox{\myl}{\centering{  
	\begin{fmfgraph*}(2.8,2.2)
			\fmfleft{i1}
			\fmfright{o1}
                   \fmftop{t}
                   \fmfbottom{b}
                   \fmf{phantom, tension=1.5}{t,vt}
			\fmf{phantom, tension=1.5}{b,vb} 
                   \fmf{zigzag, tension=1.5}{i1,v1}			
			\fmf{plain,left=0.3}{v1,vt}	
                   	\fmf{dashes,left=0.3, tension=0.8}{vt,v2}		
			\fmf{dashes,right=0.3}{v1,vb}
		      \fmf{plain,right=0.3}{vb,v2}	
		       \fmf{plain, tension=1.5}{o1,v2}
			\Tri{v1}
                    \fmfdot{v2}
		\end{fmfgraph*} 
		}}\, \Bigg]_{\pp=\V0}   \non \\
 & \hskip 2cm \dpr^2_{p_0}\hat W_{0,1;J^{\d T}}(\V0)  =\, \dpr_{p_0}^2\, \Biggr[ \,
 \parbox{\myl}{\centering{ 
	\begin{fmfgraph*}(2.8,2.2)
			\fmfleft{i1}
			\fmfright{o1}
			\fmftop{t}
			\fmfbottom{b}
			\fmf{phantom, tension=1.5}{t,v3}    
			\fmf{phantom, tension=1.5}{b,v4}    
                   \fmf{wiggly, tension=1.5}{i1,v1}			
			\fmf{plain,left=0.3}{v1,v3}
			\fmf{plain,left=0.3}{v3,v2}			
			\fmf{plain,right=0.3}{v1,v4}
			\fmf{dashes,right=0.3}{v4,v2}
		       \fmf{plain, tension=1.5}{o1,v2}
			\fmfdot{v2}
                    \Square{v1}
		\end{fmfgraph*} 
}}
+\,
\parbox{\myl}{\centering{ 
	\begin{fmfgraph*}(2.8,2.2)
			\fmfleft{i1}
			\fmfright{o1}
			\fmftop{t}
			\fmfbottom{b}
			\fmf{phantom, tension=1.5}{t,v3}    
			\fmf{phantom, tension=1.5}{b,v4}    
                   \fmf{wiggly, tension=1.5}{i1,v1}			
			\fmf{plain,left=0.3}{v1,v3}
			\fmf{dashes,left=0.3}{v3,v2}			
			\fmf{plain,right=0.3}{v1,v4}
			\fmf{plain,right=0.3}{v4,v2}
		       \fmf{plain, tension=1.5}{o1,v2}
			\fmfdot{v2}
                    \Square{v1}
		\end{fmfgraph*} 
		}}\,  \Biggl]_{\pp=\V0}\non
\]
\end{fmffile}  \vskip -0.5cm
}{Diagrams involved in the local WI relating $B_h$ and $E_h$.  On the first line the lowest order beta function for $B_h$ and $E^{J_0}_h$. On the second and third lines, the correction terms coming from the cutoffs.}{BhEh}

\subsection{Local WI relating $B_h$ and $E_h$}\label{appE.3}

In this section we prove the validity of the local WI 
\eqref{C.39}, that is
\[-2E_{h-2}^{J_0}=2\sqrt{2\r_0}B_{h-2}+\dpr^2_{p_0}\hat W^{(h-1)}_{0,1;J^\D}(\V0)-\dpr^2_{p_0}\hat W^{(h-1)}_{0,1;J^{\d T}}(\V0)\;,\label{C.39bis}\]
Let us discuss the various terms involved in this equation separately. Using \eqref{3.87} and the fact that $B_\bh=0$, we find
\[B_{h-2} = 2 \frac{\m^2_h}{A_h Z_h} \frac{\g}{\g-1} \Big(\frac{E^2_h}{C_h Z_h} \b_0^{(2)}+\frac13 \b_0^{(\chi')}\Big)\;. \]
Using \eqref{C.40} and \eqref{3.85} we find
\[E_{h-2}^{J_0}=-\frac{\g}{\g-1}\frac{2}{\l\hat v(\v0)\sqrt{2\r_0}}\,\frac{1}{A_h C_h }\, \m_h^2\, \frac{E_h}{Z_h}  \b^{(2)}_0\;,\]
so that, recalling that at lowest order $E_h=1$ and $Z_h=2\l\r_0\hat v(\v0)$,
\[-2E_{h-2}^{J_0}-2\sqrt{2\r_0}B_{h-2}=-\frac43\sqrt{2\r_0}\frac{\m^2_h}{A_h Z_h} \frac{\g}{\g-1}\b_0^{(\chi')}\;.\label{eD.17}\]
We now want to verify that the sum of the two correction terms in the r.h.s. of \eqref{C.39bis} gives exactly the same  contribution.

The leading order contribution to the correction term $\dpr^2_{p_0}\hat W_{0,1;J^\D}^{(h-1)}(\V0)$ is represented on the second line of Fig.\ref{BhEh}. Using the same notations as in Section \ref{s.B}, we find
\[
\dpr^2_{p_0}\hat W_{0,1;J^\D}^{(h-1)}(\V0) & = - 4\l\r_0\hat v(\v0) \,\g^{-h/2}\m_h\,\frac{1}{A_hZ_h} \int \frac{d^3 \kk}{(2\pi)^3}  \Big(\frac1{\sum_{k\ge 0}\ff_k(|\kk|)} -1 \Big)\ff_0(|\kk|) \cdot\non \\ 
& \hskip 1cm \cdot \dpr^2_{p_0} \bigg[\,
\Big(1+\a_hp_0 \frac{k_0}{|\kk|^2}\Big)\frac{\ff_0(|\kk+\pp|)}{|\kk+\pp|^2} \bigg]_{\pp={\bf 0}}\;,
\]
that is, using also the fact that at lowest order $\a_h=1$ and $Z_h=2\l\r_0\hat v(\v0)$,
\[
\dpr^2_{p_0}\hat W_{0,1;J^\D}^{(h-1)}(\V0) & = - 2 \,\frac{\g^{-h/2}\m_h}{A_h} \int \frac{d^3 \kk}{(2\pi)^3}  u_0(|\kk|) \bigg[ \frac{2k_0}{|\kk|^2}\dpr_{k_0}\Big(\frac{\ff_0(|\kk|)}{|\kk|^2}\Big)+\dpr^2_{k_0}
\Big(\frac{\ff_0(|\kk|)}{|\kk|^2}\Big)\Big]\;.
\]
We now compute the derivatives and pass to polar coordinates, thus finding
\[
& \dpr^2_{p_0}\hat W_{0,1;J^\D}^{(h-1)}(\V0) = \non\\
&= - \frac{2\g^{-h/2}\m_h}{A_h} \int \frac{d^3 \kk}{(2\pi)^3}  u_0(|\kk|) \Big[
  \frac{2 \ff_0(|\kk|) }{|\kk|^6}(k_0^2- |\vec k|^2)  +   
\frac{\ff'_0(|\kk|) }{|\kk|^5} (|\vec k|^2-2k_0^2)  + \ff''_0(|\kk|) \frac{k_0^2}{|\kk|^4} \bigg]  \non\\
&=\frac{\g^{-h/2}\m_h}{A_h}\frac1{2\p^2} \int d\r\, u_0(\r) \bigg[\frac{4}{3}\frac{\ff_0(\r)}{\r^2}  -  \frac{2}{3} \ff''_0(\r) \bigg]\;.\label{D.21}
\]
The leading order contribution to the correction term $\dpr^2_{p_0}\hat W_{0,1;J^{\d T}}^{(h-1)}(\V0)$ is represented on the third line of Fig.\ref{BhEh}, and is equal to 
\[
& \dpr^2_{p_0}\hat W_{0,1;J^{\d T}}^{(h-1)}(\V0) =  -2 \frac{\g^{-\frac{h}{2}}\m_hE_h}{A_hC_h Z_h}
 \int \frac{d^3\kk}{(2\pi)^3} k_0\Big(\frac1{\sum_{k\ge 0}\ff_k(|\kk|)} -1 \Big)\frac{\ff_0(|\kk|)}{|\kk|^2}\dpr_{p_0}^2\Big[p_0\frac{\ff_0(|\kk+\pp|)}{|\kk+\pp|^2}\Big]_{\pp=\V0}
\non \\
&=\frac{\g^{-h/2}\m_h}{A_h}\frac{E_h}{C_hZ_h}\frac1{2\p^2} \int d\r\, u_0(\r) \bigg[\frac{8}{3}\frac{\ff_0(\r)}{\r^2}  -  \frac{4}{3} \frac{\ff'_0(\r)}{\r} \bigg]\;.\label{D.22}\]
Using the fact that at lowest order $E_h/(C_hZ_h)=1$, we find that the difference between \eqref{D.21} and \eqref{D.22} is
\[\dpr^2_{p_0}\hat W^{(h-1)}_{0,1;J^\D}(\V0)-\dpr^2_{p_0}\hat W^{(h-1)}_{0,1;J^{\d T}}(\V0)=
\frac{\g^{-h/2}\m_h}{A_h}\frac1{2\p^2} \int d\r\, u_0(\r) \bigg[-\frac{4}{3}\frac{\ff_0(\r)}{\r^2}+\frac{4}{3}\frac{\ff_0'(\r)}{\r}  -  \frac{2}{3} \ff''_0(\r) \bigg]\;.
\]
Using the first of \eqref{eD.3}, we can rewrite this equation as
\[\dpr^2_{p_0}\hat W^{(h-1)}_{0,1;J^\D}(\V0)-\dpr^2_{p_0}\hat W^{(h-1)}_{0,1;J^{\d T}}(\V0)=-\frac43\sqrt{2\r_0}
\frac{\m_h^2}{A_h}\frac1{2\p^2} \int d\r\, u_0(\r) \bigg[2\frac{\ff_0(\r)}{\r^2}-2\frac{\ff_0'(\r)}{\r}  + \ff''_0(\r) \bigg]
\]
that, after an integration by parts, can be easily recognized to be the same as \eqref{eD.17}.
}  

\bibliographystyle{abbrv}  
\bibliography{biblioBOSE2}  
\end{document}